\documentclass{eethesis}

\usepackage{amsmath,amsthm,amssymb}
\usepackage{graphicx}
\usepackage{nccfoots}

\newcommand{\ie}{{\em i.e.}}
\newcommand{\eg}{{\em e.g.}}

\newcommand{\dd}{\mathrm{d}}

\newcommand{\lsim}{\lesssim}
\newcommand{\gsim}{\gtrsim}

\renewcommand{\type}{Dissertation}

\renewcommand{\degree}{DOCTOR OF PHILOSOPHY}	

\renewcommand{\major}{Physics}	

\begin{document}
\pagenumbering{roman}
\maketitlepage
{Charmonium in Hot Medium}   
{Xingbo Zhao} 
{Doctor of Philosophy}                
{December 2010}
{Physics}         

\approvaltwo
{Charmonium in  Hot Medium}
{Xingbo Zhao}
{Ralf Rapp}
{Che-Ming Ko}
{Joseph Natowitz}
{Bhaskar Dutta}
{Valery Pokrovsky}
{Edward Fry}
{December 2010}




\absone
{Charmonium in Hot Medium}
{December 2010}
{Xingbo Zhao}
{B.S., University of Science and Technology of China}  
{Dr. Ralf Rapp}

{We investigate charmonium production in the hot medium created by
heavy-ion collisions by setting up a framework in which in-medium
charmonium properties are constrained by thermal lattice QCD (lQCD)
and subsequently implemented into kinetic approaches. A Boltzmann
transport equation is employed to describe the time evolution of the
charmonium phase space distribution with the loss and gain term
accounting for charmonium dissociation and regeneration (from charm
quarks), respectively. The momentum dependence of the charmonium
dissociation rate is worked out. The dominant process for in-medium
charmonium regeneration is found to be a 3-to-2 process. Its
corresponding regeneration rates from different input charm-quark
momentum spectra are evaluated. Experimental data on $J/\psi$
production at CERN-SPS and BNL-RHIC are compared with our numerical
results in terms of both rapidity-dependent inclusive yields and
transverse momentum ($p_t$) spectra. Within current uncertainties from
(interpreting) lQCD data and from input charm-quark spectra the
centrality dependence of $J/\psi$ production at SPS and RHIC (for both
mid- and forward rapidity) is reasonably well reproduced. The $J/\psi$
$p_t$ data are shown to have a discriminating power for in-medium
charmonium properties as inferred from different interpretations of lQCD
results.}


\dedicate{my parents}

\acknow{First, I would like to thank my advisor, Prof. Ralf Rapp, for
his patience, enthusiasm and dedication. His guidance helped me during
the research and writing of this dissertation. I look forward to our
continued collaboration.

I wish to express my gratitude to Prof. Che-Ming Ko for his
suggestions and help throughout the past five years.

Prof. Valery Pokrovsky, Prof. Bhaskar Dutta, and Prof. Joseph Natowitz
deserve my special thanks as my dissertation committee members.

My sincere thanks also go to Prof. Pengfei Zhuang and Mr. Yunpeng Liu
in Tsinghua University. I learned a lot from the discussions with
them.

I am grateful to Dr. Grandchamp Loic for providing us with his
codes. 

I acknowledge fruitful discussions with Dr. Hendrik van Hees, Dr. Riek
Felix and Dr. Min He.}
\pagestyle{headings}
\setlength{\headheight}{36pt}
\tableofcontents
\listoffigures
\clearpage

\pagenumbering{arabic}
\setlength{\headheight}{12pt}
\pagestyle{myheadings}
\graphicspath{{./1fg/}}
\chapter[Introduction]{Introduction\Footnotemark{}}
\begin{figure}[htp]
  \vspace{2ex}
  \centering
   \includegraphics[width=0.39\textwidth,clip=]{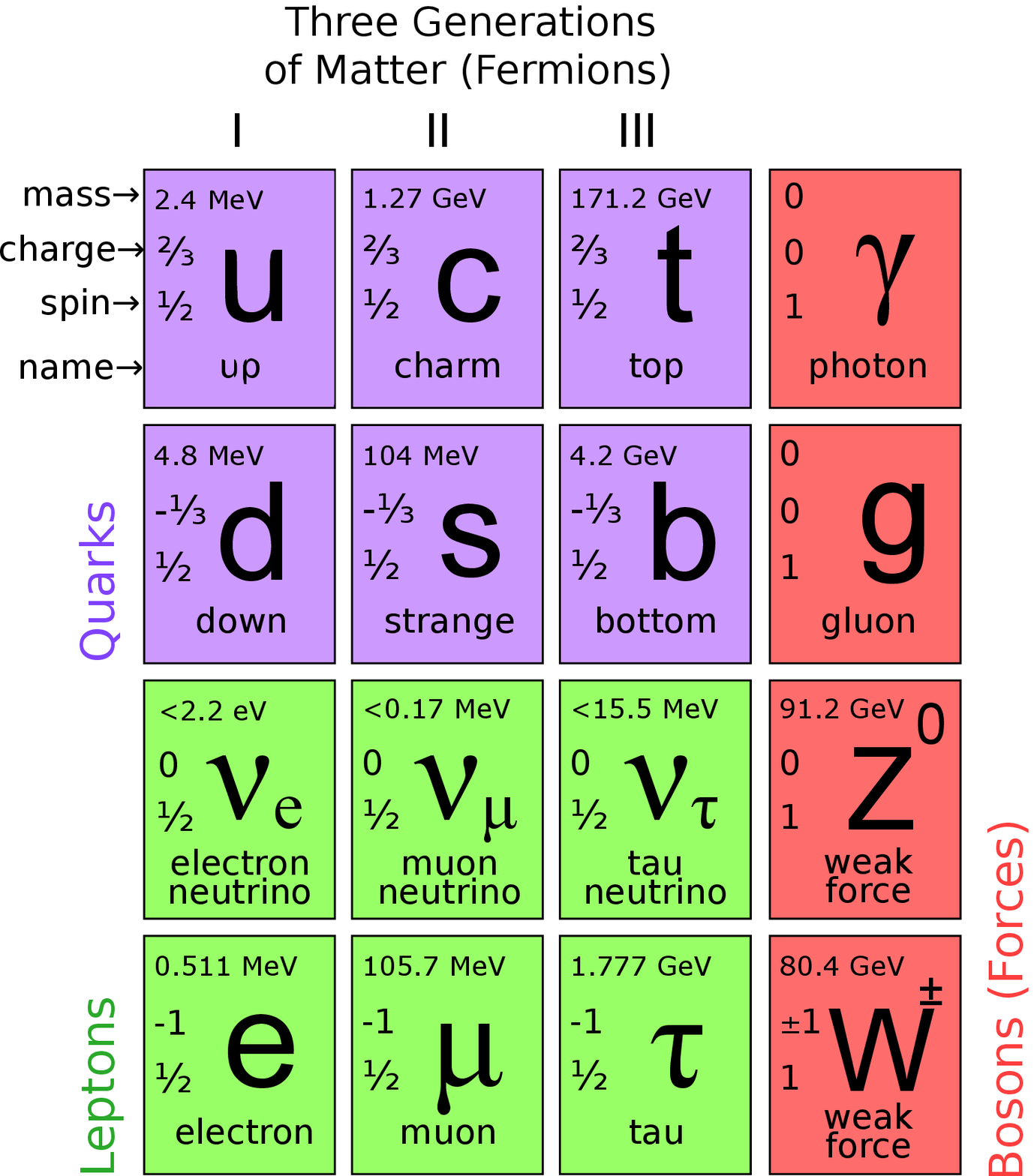}
  \caption[Elementary particles in the standard model]{Elementary particles in the standard model.}
\label{fg:std_md}
\end{figure}

\Footnotetext{}{This dissertation follows the style of Physical Review C.}
\label{ch:intro}
The standard model has been established as the underlying theory of modern particle physics. It describes the three fundamental interactions (electromagnetic, weak and strong) in a unified framework. The standard model includes 12 spin-1/2 particles (fermions). They are six quarks: up ($u$), down ($d$); charm ($c$), strange ($s$); top ($t$), bottom ($b$) and six leptons: electron ($e$), electron neutrino ($\nu_e$), muon ($\mu$), muon neutrino ($\nu_\mu$), tau ($\tau$), tau neutrino ($\nu_\tau$). It also includes spin-1 particles (bosons) mediating the interaction between fermions. They are the photon mediating the electromagnetic interaction,  $W^+$, $W^-$ and $Z$ bosons mediating the weak interactions and 8 gluons mediating the strong interaction. The elementary particles in the standard model are summarized in Fig.~\ref{fg:std_md}.

The interaction among elementary particles is mathematically described by gauge field theories with the (local) gauge symmetry group of $SU(3)_C\otimes SU(2)_L\otimes U(1)_Y$, where $SU(3)_C$  and $SU(2)_L\otimes U(1)_Y$ are the gauge groups governing the strong and the electro-weak interaction, respectively. The gauge field theory with the $SU(3)_C$ gauge group is quantum chromodynamics (QCD)~\cite{Gross:1973id,Politzer:1973fx}. Although QCD has been established as the underlying theory for the strong interaction, many aspects of it are still not well understood, especially in regimes where the strong coupling becomes large. The ultimate goal of this work is to improve our understanding about the strong interaction in these regimes. Therefore let us begin with a review of properties of QCD.


\section{Quantum Chromodynamics}
\label{sec:qcd}

The fundamental degrees of freedom in QCD are quarks and gluons. Their interaction is described by the following Lagrangian,
\begin{equation}
  \label{eq:86}
  \mathcal{L}_{QCD} = \sum\limits_{f}^{N_f}
  \bar{\psi}_f(i\gamma^{\mu}D_{\mu}-m_f)\psi_f -
  \frac{1}{4}F_a^{\mu\nu}F_{\mu\nu}^a\ .
\end{equation}
Here the gluon field strength tensor $F_{\mu\nu}^a$ reads
\begin{equation}
  \label{eq:88}
  F_{\mu\nu}^a = \partial_{\mu}A_{\nu}^a - \partial_{\nu}A_{\mu}^a +
  i g f_{abc}  A_{\mu}^b A_{\nu}^c\ , 
\end{equation}
in terms of the gluon gauge fields $A_a^{\mu}$ (a$=1\cdots8$). 
The colored quark fields $\psi_f$ ($f = u,d,s,c,b,t$) are coupled to
the gluons through the gauge covariant derivative
\begin{equation}
  \label{eq:89}
  D_{\mu} = \partial_{\mu} - i g \frac{\lambda_a}{2}A_{\mu}^a\ , 
\end{equation}
where $\lambda_a$ are the Gell-Mann matrices, which are generators of the
$SU(3)_c$ group satisfying 
\begin{equation}
  \label{eq:90}
  \left[\lambda_a , \lambda_b\right] = f_{abc} \lambda_c\ ,
\end{equation}
$f_{abc}$ being the structure constants of
$SU(3)_c$.

The important properties of QCD include:

1. Asymptotic freedom. One of the most notable differences between the
strong and electromagnetic interaction comes from the fact that not
only quarks but also gluons themselves carry color charge, reflected
in the last term of Eq.~(\ref{eq:88}). This gives rise to gluon
self-interactions, which in turn lead to the QCD coupling
constant,$\alpha_s=g^2/4\pi$ decreasing logarithmically with the
momentum transfer in the process,
\begin{align}
  \alpha_s(Q) =\frac{g^2}{4\pi}
  =\frac{1}{\beta_0\ln(Q^2/\Lambda_{QCD}^2)}\ ,
\label{eq:alpha_s_evo}
\end{align}
\begin{figure}[htp]
  \centering
  \vspace{3ex}
  \includegraphics[width=0.49\textwidth,clip=]{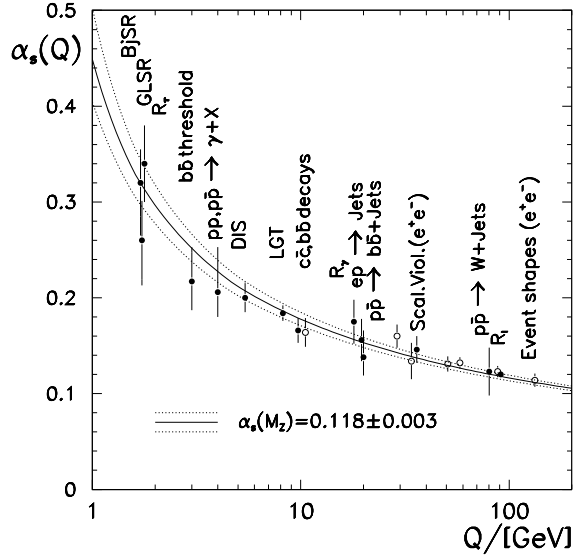}
  \caption[Running coupling constant $\alpha_s$ of QCD]{Running coupling constant $\alpha_s$ of QCD. Dependence of the QCD coupling constant, $\alpha_s=g^2/4\pi$, on the momentum transfer, $Q$, of the interaction. Figure taken from Ref.~\cite{Schmelling:1996wm}.}
\label{fg:alph_s_evo}
\end{figure}
where $\Lambda_{QCD}\simeq$200MeV is introduced as a
``non-perturbative'' scale where $\alpha_s(Q)$ formally diverges, see
Fig.~\ref{fg:alph_s_evo}. Note that the logarithmic behavior in
Eq.~(\ref{eq:alpha_s_evo}) is based on perturbative calculations, which
break down at momentum transfer well above $\Lambda_{QCD}$. In
practice, the scale for the onset of non-perturbative effects is
typically given by the hadronic mass scale of $\sim$1\,GeV. For
interactions with large momentum transfer $Q\gg$1\,GeV (or
equivalently at small distance $r\ll 1$fm according to the uncertainty
principle), theoretical calculations can be organized in a converging
series of terms characterized by increasing powers of $\alpha_s$. In
this ``perturbative'' regime, QCD is well tested, attaining excellent
agreement with experiment. In the opposite limit, due to the growing
strong coupling constant toward small $Q$, the perturbative expansion
breaks down and the QCD enters the ``strong'' regime where
``non-perturbative'' phenomena occur.

2. Confinement. A striking non-perturbative phenomenon is the confinement of color charges, which refers to the fact that isolated colored particles, quarks and gluons, have never been observed. They are always confined in ``bags'' of color-neutral baryons or mesons. For example, in deep inelastic scattering (DIS) experiments, when a colored quark is knocked out of a proton, $q\bar{q}$ pairs form along the trajectory of the outgoing quark to neutralize color. The knocked-out quark is thus accompanied by large concentrations of hadrons along its direction of propagation, so-called ``jets''.

3. Spontaneous Breaking of Chiral Symmetry (SBCS). Usually the $u$ and
$d$ (and sometimes $s$) quarks are referred to as ``light quarks'';
$c$, $b$, $t$ quark are referred to as heavy quarks. In the massless
limit of light quarks the QCD Lagrangian with two light flavors can be
written as 
\begin{equation}
  \label{eq:lag_chi}
  \mathcal{L}_{QCD} = \bar{\psi}_L i \gamma^{\mu} D_{\mu} \psi_L +
  \bar{\psi}_R i \gamma^{\mu} D_{\mu} \psi_R -
  \frac{1}{4}F_a^{\mu\nu}F_{\mu\nu}^a\ ,   
\end{equation}
where
\begin{equation}
  \psi= \left(\begin{array}{c} 
      u\\
      d 
    \end{array} \right).
\end{equation}
The $\psi_L$ and $\psi_R$ are the left-handed and right-handed components of the quark field. The Lagrangian (\ref{eq:lag_chi}) is unchanged under a rotation in flavor space ($u\leftrightarrow d$) ``independently'' with respect to $\psi_L$ or $\psi_R$. This symmetry of the Lagrangian is called the chiral symmetry. This symmetry is, however, spontaneously broken by the complex structure of QCD vacuum. The latter is filled with various condensates of quark-antiquark and gluon fields. In particular, the scalar quark condensate of up and down quarks can be quantified by a vacuum expectation value, 
\begin{equation}
  \label{eq:96}
  \langle \bar{\psi}\psi \rangle = \langle 0 | \bar{\psi}_R\psi_L +
  \bar{\psi}_L\psi_R | 0 \rangle \simeq (-250\mathrm {MeV})^3\ ,
\end{equation}
translating into a total pair density of about 4/fm$^3$ (for two flavors). As a result the quarks inside the hadrons propagating through the QCD vacuum acquire an effective mass, $m^*_q\simeq$350\,MeV, which is much larger than their bare mass, $m^0_{u,d}\simeq$ 5-10\, MeV. The QCD condensates are thus the main source of the visible mass in the Universe.


Currently the theoretical efforts of studying QCD in the non-perturbative regime are mainly pursued in two directions:


1. Effective theory/model. Various effective theories are developed
with degrees of freedom appropriately adapted to specific
problems. They are rather successful in describing the physics
within their applicable energy range. For example in the low-energy
regime of QCD, chiral perturbation theory was developed with
hadrons, rather than quarks and gluons, as the effective degrees of freedom.
The interaction term is dictated by the chiral symmetry, which is
approximately respected by the original QCD Lagrangian.  After the
effective coupling constants are determined by fitting to experimental
data, chiral perturbation theory gains predictive power for low
energy hadronic reactions/decays. Another example is Non-Relativistic
QCD, in which the heavy quarks are described by a Schr\"odinger field
theory while the gluons and light quarks are modelled by the usual
relativistic Lagrangian of QCD.

2. Lattice QCD. The basic idea of lattice QCD (lQCD) is to study the QCD Lagrangian in discretized euclidean (imaginary-time) spacetime, with the lattice points, called sites, separated by the lattice spacing, $a$. This effectively introduces an ultraviolet cutoff, $\Lambda=1/a$  on any momentum component. Fermion fields, $\psi$, reside on the lattice sites, while the gauge fields, $A$, are associated with the links joining neighboring sites. The lQCD partition function can then be calculated on lattice as
\begin{equation}
  \label{eq:lqcd_part}
  \mathcal{Z} =
 \int
  [dA][d\bar{\psi}][d\psi]e^{-\int\mathcal{L}d^4x} \ .  
\end{equation}
Here functional integration denotes summing over all possible field configurations with every possible value of the gauge fields, the antifermion fields and the fermion fields on each link and sites, respectively. In practice, Monte-Carlo simulations are employed to make this summation process possible. 
In terms of the partition function the thermal average values of observables are given by 
\begin{equation}
  \label{eq:lqcd_operator}
  \langle \mathcal{O}(A,\bar{\psi},\psi)\rangle =
  \frac{1}{\mathcal{Z}} \int
  [dA][d\bar{\psi}][d\psi]e^{-\int\mathcal{L}d^4x}
  \mathcal{O}(A,\bar{\psi},\psi)\ .  
\end{equation}

Due to the ``fermion sign problem'' lQCD is currently applicable only in the regime of low baryon density.

As the computer technology and power have advanced in recent years lattice QCD has yielded many important results and thus provided profound insights into the non-perturbative regime of QCD, see Ref.~\cite{Bazavov:2009bb} for a recent review. One of the central goals in this work is to establish the link between the lQCD results and heavy-ion phenomenology.


\section{QCD Phase Diagram}
\label{sec:qcd_phase}

\begin{figure}[tp]
  \centering
  \includegraphics[width=0.48\textwidth,clip=]{e_vs_t.eps}
  \includegraphics[width=0.51\textwidth,clip=]{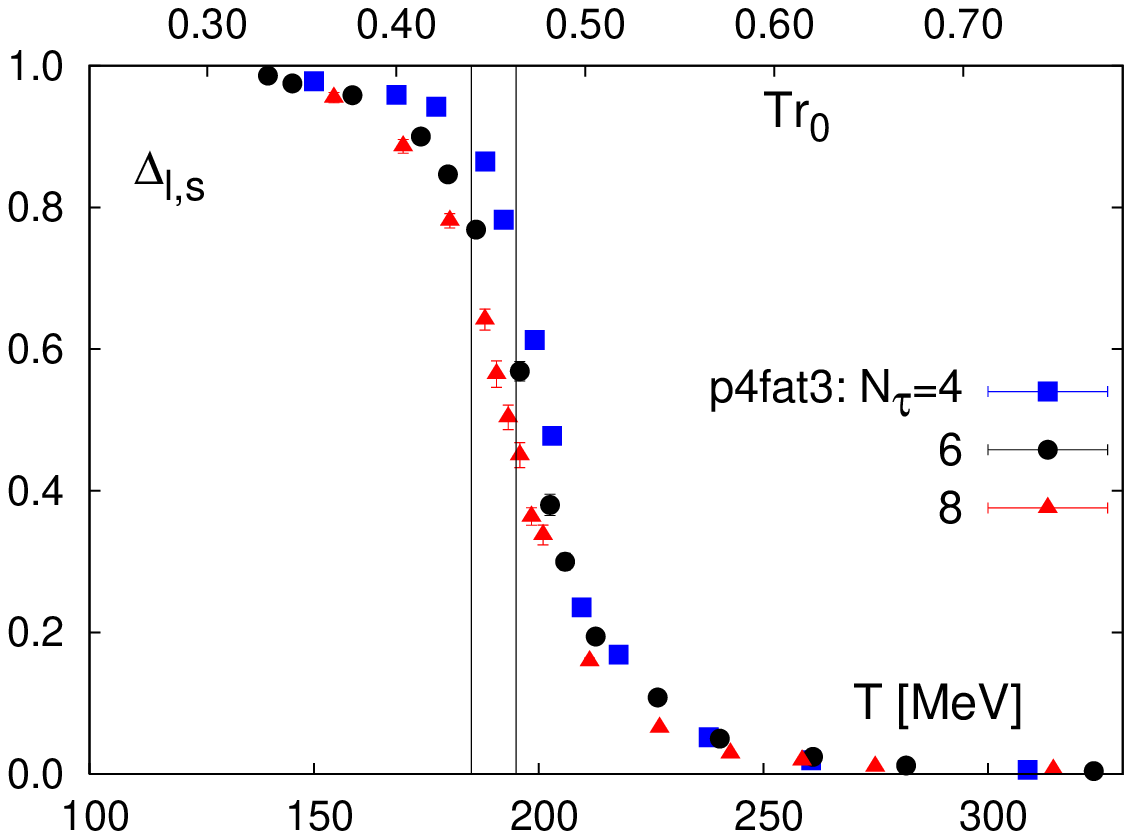}
  \caption[Energy density and chiral condensate from lQCD]{Energy density and chiral condensate from lattice QCD. Left panel: Lattice calculation of the energy density and pressure of QCD matter as a function of temperature, taken from Ref.~\cite{Petreczky:2009at}. Right panel: Strength of the chiral condensate as a function of temperature, taken from Ref.~\cite{Cheng:2007jq}.}
\label{fg:lat_e_vs_t} 
\end{figure}

A question of fundamental importance is what happens if the hadronic matter is compressed so that the distance between hadrons is smaller than the radius of hadrons. Intuitively one expects that the boundary of hadrons disappears and the quarks and gluons can move freely inside the entire nuclear matter and become the relevant degrees of freedom (deconfinement).  It turns out that in the low baryon density region this picture is supported by recent lQCD calculations: The energy density shows a rapid rise in the temperature region around 170-190\,MeV,  as shown in the left panel of Fig.~\ref{fg:lat_e_vs_t}. At high temperatures the energy density $\epsilon$ is within 15\% of the values expected for an ideal gas of quarks and gluons, known as the Stefan-Boltzmann limit, which implies the relevant degrees of freedom have indeed transitioned into quarks and gluons, forming the quark-gluon plasma (QGP).
Lattice QCD calculations also show that the rapid increase of $\epsilon$ is accompanied by a sudden decrease of quark condensate, see the right panel of Fig.~\ref{fg:lat_e_vs_t}, implying its evaporation at high temperature similar to the evaporation of Cooper pairs above the critical temperature.


  
\begin{figure}[tp]
  \centering
  \includegraphics[width=0.59\textwidth,clip=]{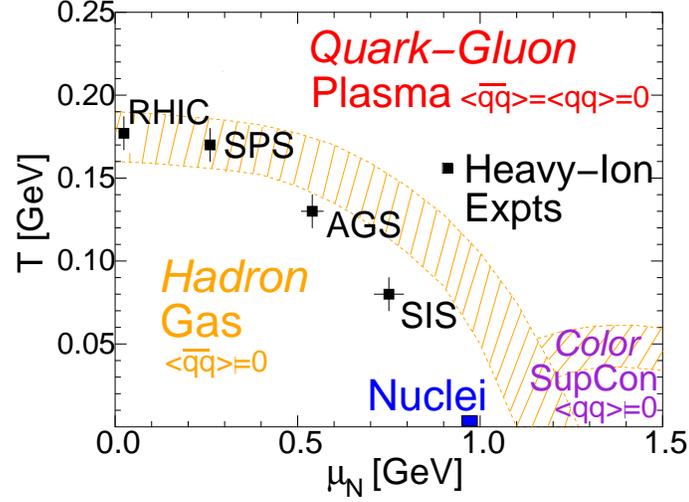}
  \caption[Schematic view of QCD phase diagram in terms of the baryon chemical potential $\mu_B$ and temperature $T$]{Schematic view of QCD phase diagram in terms of the baryon chemical potential $\mu_B$ and temperature $T$. Figure taken from Ref.~\cite{Rapp:2008qc}.}
\label{fg:qcd_pdiag}
\end{figure}

In the regime with low temperature and high baryon density (characterized by large baryon chemical potential, $\mu_B$), theoretical studies~\cite{Rapp:1997zu,Alford:1997zt} reveal the existence of another deconfined phase, see Fig.~\ref{fg:qcd_pdiag}, where the high density quarks form Cooper pairs which condense ($\langle qq\rangle\neq$0) and result in superconducting of color charge. 


Since the formation of deconfined matter (also called quark matter)
requires either extremely high temperature or density the natural
occurrence is expected in the early universe or inside compact
stars. According to the Big Bang theory the universe is believed to
have passed through the QGP phase at the age of a few
microseconds. Inside compact stars such as neutron star the
temperature is far lower than 180MeV, but the high density makes it
possible that the inside matter is in the color-superconducting
phase.  Since the properties of quark matter are 
characterized by the strong interactions, new insights about the
strong interaction is expected to be obtained by studying these new
states of matter.


\section{Heavy-Ion Collisions}
\label{sec:hic}

Since quark matter in nature is far from reach scientists are
motivated to create it in laboratories, through ultra-relativistic
heavy-ion collisions (URHICs). 

The idea is that by accelerating heavy nuclei to very high speed and
then colliding them a large amount of energy is deposited into a small
spatial region and converted to thermal energy resulting in extremely
high temperature. The currently running experimental facilities
include the Super-Proton-Synchrotron (SPS) at CERN, Relativistic
Heavy-Ion Collider (RHIC) at BNL and newly constructed Large Hadron
Collider (LHC) at CERN. The new Facility for Antiproton and Ion
Research (FAIR) at GSI (Germany) will be completed within a few
years. These experimental facilities come with one of two typical
setups: the first category is the fixed-target experiments, in which
only one beam of ions (the projectiles) is accelerated, and its
colliding partner is placed in a stationary target into the path of
the beam. The SPS and FAIR are fixed target experiments; the second
category is the collider, in which both beams (projectile and targets)
are accelerated and directed to collide with each other. The RHIC and
LHC fall into this category. Usually the fixed target accelerators
have higher luminosity (leading to a larger number of collision events
per unit time) so that more rare reactions can be studied, whereas the
advantage of colliders is that higher collision energy can be reached (in fixed
target accelerators a large amount of the energy of the projectile is
``wasted'' on the kinetic energy of the center of mass of two
colliding nuclei). 

So far at SPS collisions have been conducted between various ion beams, such
as proton (p), deuteron (d), O, S, Pb, and different targets such as S, Si, Cu,
W, Pb, U, at different energies from 20 AGeV to 158 AGeV (for a proton
beam it can reach up to 450 GeV). 
RHIC has produced collisions between p+p, d+Au, Cu+Cu and
Au+Au at different energies ranging from $\sqrt{s}$=22.4AGeV to
200AGeV. In near future Pb+Pb collisions
will be performed at LHC with up to $\sqrt{s}$=5.5\,ATeV. FAIR will
carry out heavy-ion collisions with $\sqrt{s}$ close to
10AGeV. Different heavy-ion experiments, with different beam energies,
probe different regions in the QCD phase diagram,
Fig.~\ref{fg:qcd_pdiag}: the matter created in the central region of
collisions with higher beam energies is more symmetric between baryons
and antibaryons, while the lower energy experiments (such as FAIR)
enable to study the properties of dense baryonic matter. In this work we
mainly focus on charmonium production in Pb(158\,AGeV)+Pb
collisions at SPS and $\sqrt{s}$=200\,AGeV Au+Au collisions at
RHIC.

\begin{figure}[htp]
  \centering
  \vspace{3ex}
  \includegraphics[width=0.99\textwidth,clip=]{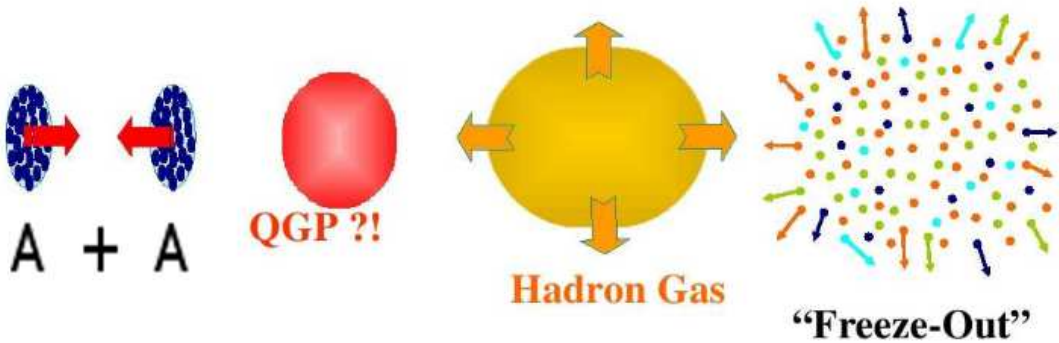}
  \caption[Schematic picture of the various stages of a heavy-ion collision]{Schematic picture of the various stages of a heavy-ion collision. Picture taken from Ref.~\cite{Rapp:2008qc}.}
\label{fg:hic_evo}
\end{figure}

The time evolution of a typical heavy-ion collision is sketched in
Fig.~\ref{fg:hic_evo}. Two Lorentz-contracted nuclei approach each
other at close to the speed of light until primordial nucleon-nucleon
collisions occur. After subsequent reinteractions for
$\tau_0$=0.5-1fm/$c$ a Quark-Gluon Plasma (QGP) is supposedly
created. Driven by the pressure gradient the QGP expands and cools
(for a duration of $\tau_{QGP}\sim$3-5fm/$c$). The hadronization then
follows with further expansion in the hadronic phase until the
``chemical freeze-out'' point when inelastic interactions cease with
particle abundances fixed; after further expansion/cooling until
``kinetic freeze-out'' elastic interactions stop with particle
transverse momentum spectra fixed. The total fireball lifetime is
approximately 10-15fm/$c$ depending on the beam energy.

\begin{figure}[htp]
  \centering
  \includegraphics[width=0.49\textwidth]{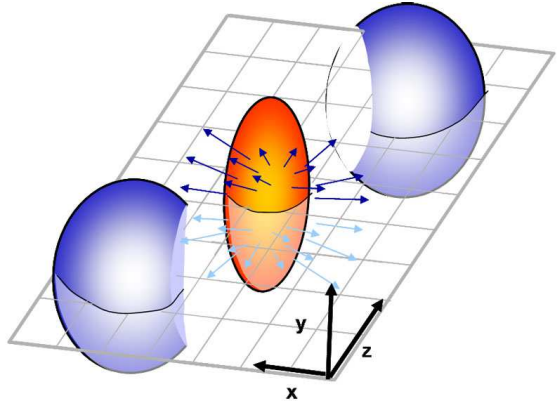}
  \caption[Schematic representation of a noncentral heavy-ion
  collision, characterized by an almond-shaped initial overlap zone,
  and a subsequent pressure-driven build-up of elliptic
  flow]{Schematic representation of a noncentral heavy-ion collision,
    characterized by an almond-shaped initial overlap zone, and a
    subsequent pressure-driven build-up of elliptic flow. Picture
    taken from Ref.~\cite{{Rapp:2008qc}}.}
\label{fg:ell_flow}
\end{figure}
\begin{figure}[tp]
  \centering
  \vspace{3ex}
  \includegraphics[width=0.49\textwidth]{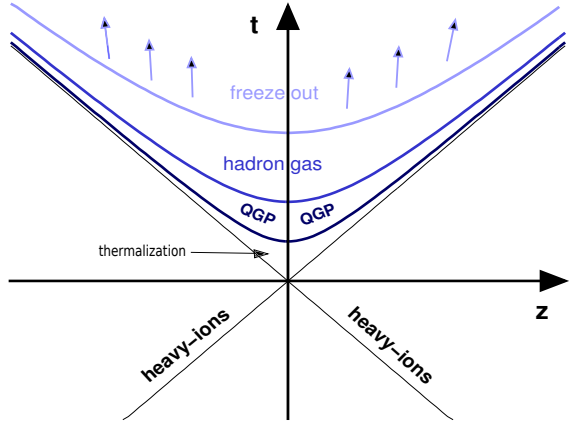}
  \caption[Spacetime diagram of ultra-relativistic nuclear collisions]{The spacetime diagram of ultra-relativistic nuclear collisions. Figure taken from~\cite{Hirano:2008hy}.}
\label{fg:spacetime_evo}
\end{figure}

It is convenient to introduce the standard coordinate system for heavy-ion collisions: The $z$ axis is parallel to the beam line. Since most nucleus-nucleus (A-A) collisions are not head-on collisions, there exists a two-dimensional vector connecting centers of the colliding nuclei in the plane transverse to $z$ axis, which is called the impact vector, $\vec b$, its length is the impact parameter, $b$. The $x$-axis is chosen to be parallel to the impact vector, $\vec b$, see Fig.~\ref{fg:ell_flow}. The $x$- and $z$-axes span the ``reaction plane'' of a given collision. The $x$- and $y$-axes span the ``transverse plane''. The component of the 3-momentum of produced particles parallel to $z$-axis is denoted by $p_z$, and the transverse component is $\vec p_t$. For relativistic particles it is convenient to use the (longitudinal) rapidity instead of the (longitudinal) velocity. The former is defined as
\begin{equation}
  \label{eq:rapidity}
  y=\tanh^{-1}\left(\frac{p_z}{E}\right)=\tanh^{-1} v_z\ .
\end{equation}
Here $E=\sqrt{m^2+\vec p^2}$ is the energy of a particle. Due to the time-dilation effect particles with larger $v_z$ in center of mass frame are ``younger'' than the ones with smaller $v_z$. This means that the above mentioned evolution of the matter is ``measured'' by the longitudinal proper time $\tau$=$\sqrt{t^2-z^2}$ rather than by the lab time $t$, as illustrated in Fig.~\ref{fg:spacetime_evo}.

Due to the short lifetime of the medium, special probes are needed to access the properties of the medium. The only probes turn out to be the produced particles themselves. According to their energies the probes are divided into two categories: soft probes and hard probes.

The soft probes are associated with the particles with relatively low energy (e.g., $\lesssim$2GeV), which constitute the bulk medium ($>$95\% at RHIC) created in heavy-ion collisions.  The soft probes reflect the collective properties of the medium, such as thermal and transport properties. One of the key measurements at RHIC is the elliptic flow, $v_2$, for the bulk particles ($\pi$,$K$,$p$). The elliptic flow, $v_2$, characterizes the azimuthal asymmetry of these particles in the transverse plane in terms of the second harmonic coefficient of an azimuthal Fourier decomposition of the momentum spectra,
\begin{equation}
\left.\frac{d N}{d^2p_t d y}\right|_{y=0}=\frac{d N}{\pi d p_t^2 d y} 
\left[ 1 + 2 v_2(p_t) \cos(2\phi) + \dots \right].
\label{eq:v2}
\end{equation}
Here $\phi$ is the azimuthal angle in the transverse plane with $\phi$=0 for $x$-axis. At mid-rapidity the system is symmetric about $y-z$ plane, so there is no $\cos \phi$ term.

\begin{figure}[tp]
  \centering
   \includegraphics[width=0.59\textwidth,clip=]{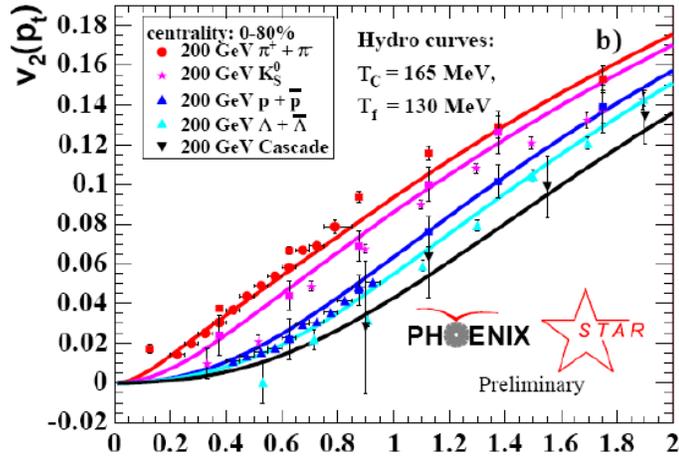}
  \caption[Elliptic flow measured by PHENIX and STAR compared to hydrodynamic calculations]{Elliptic flow measured by PHENIX and STAR~\cite{Adams:2005dq,Adcox:2004mh} compared to hydrodynamic calculations~\cite{Huovinen:2001cy}.}
\label{fg:v2_rhic}
\end{figure}

For soft particles ($p_t<$2\,GeV) the elliptic flow arises because, in
semi-central collisions, the geometry of the initial interaction
region has the shape of an ellipse, see Fig.~\ref{fg:ell_flow}. Once
the system thermalizes this initial geometrical anisotropy translates
into stronger pressure gradients in the direction of the smaller axis
of the ellipse. This induces momentum correlations among
particles which \textit{flow} preferentially along the small axis of
the ellipse, leading to a positive $v_2$, see Fig.~\ref{fg:v2_rhic}.
Since the spatial anisotropy is largest at the beginning of the
evolution, a measurement of $v_2$ provides access to the
thermalization time scale $\tau_0$ of the system. Applications of
ideal relativistic hydrodynamics have shown that the experimentally
measured $v_2(p_T)$ for various hadrons ($\pi$, $K$, $p$, $\Lambda$)
is best described when implementing a thermalization time of
$\tau_0$=0.5-1~fm/$c$.

Other valuable soft probes include particle ratios and the HBT
(Hanbury-Brown-Twiss) particle interferometry. The former is used to
estimate the temperature $T$ and chemical potential $\mu_B$ of
``chemical freeze-out'' when the inelastic interactions cease and the
particle ratios are fixed. The latter provides clues about the size,
shape and time evolution of the medium~\cite{Pratt:1984su}.

To further study the properties of the medium, in particular the more
microscopic aspects, the second category of the probes, namely the
hard probes, are required.

The hard probes are associated with the particles with relatively high energy ($>$2GeV) including either light particles with large momentum or heavy particles irrespective of their momenta. Usually a hard probe can only be generated in initial hard collisions (their energy scale is usually much larger than the typical temperature of the medium) and their initial production can be estimated from p+p collisions. By measuring the modification of hard probes after traversing the QGP medium, one can obtain information on the microscopic interaction between strongly interacting medium and the probe particle. To quantitatively describe the modification due to the medium it is convenient to define the nuclear modification factor,
\begin{equation}
  R_{AA}(b;p_T)=\frac{d N^{AA}/d p_T}{N_{\rm coll}(b) \ d N^{NN}/d p_T} \ ,  
\end{equation} 
where $N_{\rm coll}(b)$ is number of binary collisions for a given impact parameter, $b$. The relation between $N_{\rm coll}$ and $b$ will be worked out in Section~\ref{ch:pre-eq}.\ref{sec:psi_aa}.\ref{ssec:glauber}. The production of hard probe (high $p_t$, large mass) particles usually scales with $N_{\rm coll}(b)$, implying $R_{AA}$=1 if there are no medium induced modifications. For soft particles $R_{AA}$ is usually less than 1 because the production of these particles scales with number of participant nucleons, $N_{\rm part}(b)$. ($N_{\rm part}(b)$ is smaller than $N_{\rm coll}(b)$ in A-A collisions).

One of the key measurements of hard probes at RHIC is the observation of ``jet-quenching''~\cite{Wang:1991xy} for high $p_t$ particles.
\begin{figure}[htp]
  \centering
  \includegraphics[width=0.49\textwidth,clip=]{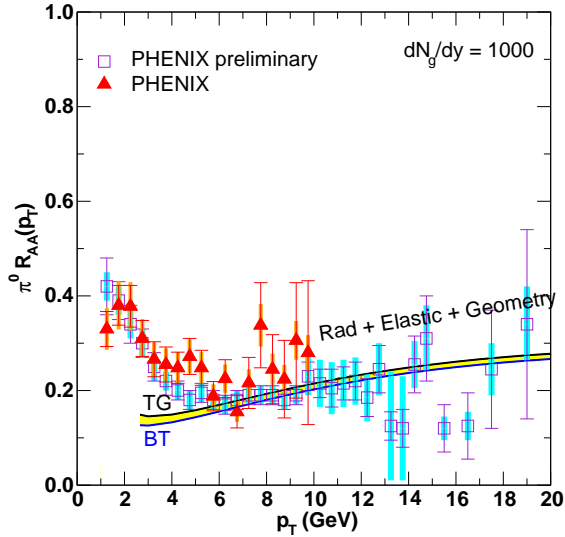}
  \caption[$R_{AA}(p_t)$ ratio measured by PHENIX for neutral pions, compared to jet-quenching calculations]{$R_{AA}(p_t)$ ratio measured by PHENIX for neutral pions~\cite{Adcox:2004mh}, compared to jet-quenching calculations~\cite{Wicks:2005gt}.}
  \label{fg:jet_quench}
\end{figure}
A jet is a narrow cone of hadrons produced by the hadronization of a high momentum parton. If these partons traverse the QGP they are expected to undergo collisional and medium-induced radiational energy loss. The energy loss will be reflected in the suppression of high $p_t$ hadron multiplicities. The data on $\pi^0$ production in central Au+Au collisions at RHIC have shown a large suppression by a factor of 4-5 up to $p_t\simeq$ 20\,GeV, see Fig.~\ref{fg:jet_quench}. The attenuation of the parton energy allows one to extract the initial parton density in QGP in terms of the transport coefficient $\hat{q}$=$Q^2/\lambda$, which characterizes the (squared) momentum transfer per mean free path of the fast parton.

Another important hard probe is the heavy quark, which is expected to only partially thermalize in the medium considering its large mass and the limited fireball lifetime. One advantage of heavy quarks over high $p_t$ light particles is that they are always distinguishable from bulk particles even if they are slowed down. Therefore the final heavy-quark (HQ) spectra may be taken as ``witnesses'' carrying  a memory of the interaction history throughout the evolving fireball, by operating in between the limits of thermalization and free streaming.


 Typical hard probe particles have only one ``hard'' scale characterized by their high energy, therefore they are not particularly sensitive to physics at the energy scale of medium temperature, $T$. However there exists one special hard probe particle which has an additional (softer) energy scale (on the order of $T$) making it very sensitive to physics at the medium temperature. This probe is the charmonium. 

\section{Charmonium}
\label{sec:charmo}

\begin{figure}[htp]
  \centering
  \vspace{3ex}
  \includegraphics[width=0.99\textwidth,clip=]{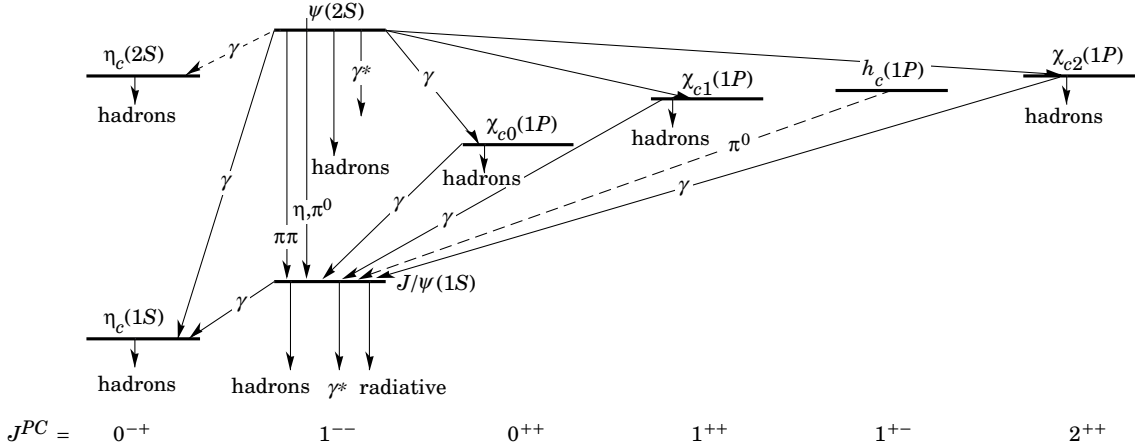}
  \caption[Charmonium spectrum]{Charmonium spectrum. Figure taken from Ref.~\cite{Nakamura:2010zzi}.}
\label{fg:charmo_spe}
\end{figure}

A quarkonium is a bound state of a quark and its own antiquark. A quarkonium made of a pair of heavy quarks ($c$,$b$) is called heavy quarkonium. Heavy quarkonium includes charmonium ($c\bar c$) and bottomonium ($b\bar b$). The heaviest toponium does not exist because the top quark decays through the electroweak interaction ($\tau_t$=$1/\Gamma_t\simeq$0.1fm/$c$) before a bound state can form. In this dissertation we focus on charmonium which can be rather abundantly produced at SPS and RHIC energies. The charmonium spectrum in vacuum is summarized in Fig.~\ref{fg:charmo_spe}.
Since only vector mesons can couple to virtual photons and have
dilepton as the decay product, which allow for quite accurate
measurements, in this work we mainly focus on the productions of
vector charmonium, such as, $J/\psi$ and $\psi'$. However we keep in
mind that 32\% (8\%) of observed $J/\psi$ are from feeddown of
$\chi_c$($\psi'$)~\cite{Antoniazzi:1992af,Abt:2002vq}, which happens
at around 1000fm/$c$, much later than typical thermal medium lifetime
($\sim$10fm/$c$). Throughout this dissertation we will use $\Psi$ to denote
$J/\psi$, $\chi_c$ and $\psi'$. The vacuum masses for $J/\psi$,
$\chi_c$ and $\psi'$ are $m_{J/\psi}$=3.1\,GeV,
$m_{\chi_{c0}}$=3.4\,GeV, $m_{\chi_{c1}}$=3.5\,GeV,
$m_{\chi_{c2}}$=3.55\,GeV and $m_{\psi'}$=3.7\,GeV~\cite{Nakamura:2010zzi}.

Unlike light quarkonium states the heavy quarks move inside the heavy
quarkonium with a speed significantly smaller than the speed of light,
with, \eg, $\langle v^2/c^2\rangle\sim$0.25 for $J/\psi$~\cite{Karsch:1987zw}. As a result the heavy quark bound system can be described
with non-relativistic Schroedinger approach with static heavy quark
potentials, in a similar way of describing, \eg, the hydrogen atom. An
early ansatz yet successful in describing the charmonium spectrum is
the Cornell-potential~\cite{Eichten:1978tg,Eichten:1979ms},
\begin{equation}
  \label{eq:cornell}
  V(r;T=0)=-\frac{4}{3}\frac{\alpha_s}{r}+\sigma r\ ,
\end{equation}
with $\alpha_s\simeq$0.35 and $\sigma\simeq$1\,GeV/fm~\cite{Jacobs:1986gv}. The first term corresponds to a Coulombic part which originates from one-gluon exchange and is dominant at small distance ($r$), the second term linear in $r$ reflects the confining interaction. 

As mentioned in the previous section, aside from its large mass, the charmonium has another characteristic energy scale, which is its binding energy $\epsilon_B$. The charmonium binding energy is usually counted as the difference between the charmonium mass and the open-charm threshold,
\begin{equation}
  \label{eq:psi_binding_vac}
\epsilon^0_B=2m_D-m_\Psi\ ,  
\end{equation}
with $m_D\simeq$1.87\,GeV. In vacuum the $D\bar D$ pair is usually
considered as the open charm threshold for charmonium states. The
charmonium states typically have binding energies of the order of
several hundred MeV, e.g., $\epsilon^{J/\psi}_B$=640MeV, which is on
the the same order of typical medium temperatures at URHICs.

If a charmonium is put inside the deconfined QGP medium the color force between $c$ and $\bar c$ is subject to screening by the surrounding colored partons, see Fig.~\ref{fg:debye_sc}, in a way similar to screening of the electric field in dielectric materials: 
\begin{figure}[tp]
  \centering
  \includegraphics[width=0.89\textwidth,clip=]{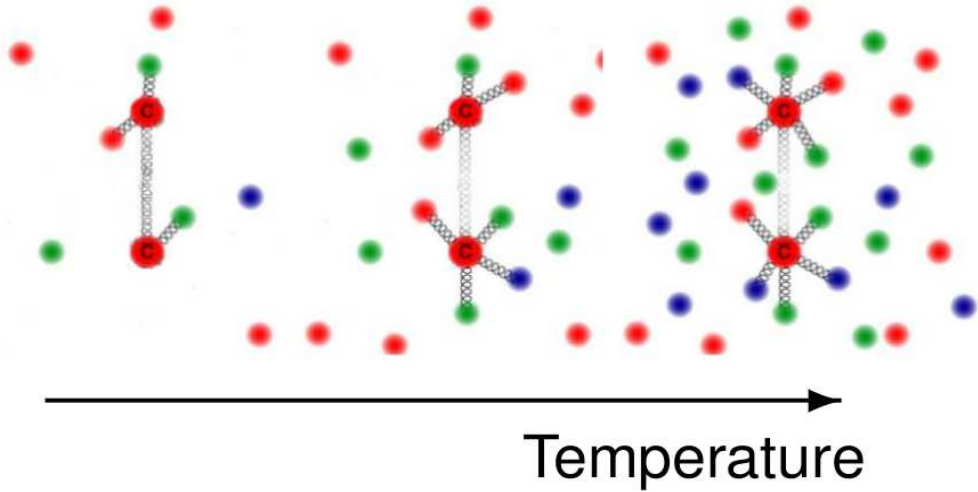}
  \caption[Schematic representation of color-Debye screening in a deconfined medium]{Schematic representation of color-Debye screening in a deconfined medium.}
\label{fg:debye_sc}
\end{figure}
The $c$($\bar{c}$) quark attracts partons in the medium with opposite
color charge and forms the ``Debye cloud'' which screens the color
electric field of the $c$($\bar{c}$) quark. The screening effect in
the Coulombic part of the Cornell potential can be evaluated with
thermal perturbative QCD (pQCD). In the confining parts it is usually
described by a phenomenological ansatz in early calculations, leading
to the following form of the screened Cornell potential at finite
temperature ~\cite{Karsch:1987pv},
\begin{equation}
V_{\bar QQ}(r;T)=\frac{\sigma}{\mu_D(T)} 
\left( 1-{\rm e}^{-\mu_D(T)r}\right) -\frac{4\alpha_s}{3r} \ 
{\rm e}^{-\mu_D(T)r}  \ . 
\label{eq:Cornell-T}
\end{equation}
A direct consequence of the color Debye-screening is the lowering of charmonium binding energies, see Fig.~\ref{fg:eb_mu}.
\begin{figure}[tp]
  \centering
  \includegraphics[width=0.49\textwidth,clip=]{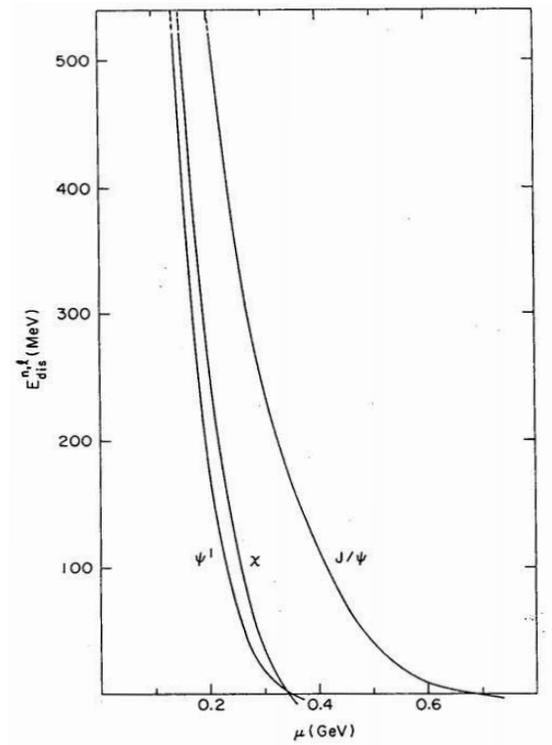}
  \caption[Charmonium binding energies as a function of Debye mass]{Charmonium binding energies as a function of Debye mass. They are estimated from Eq.~(\ref{eq:Cornell-T}). Figure taken from~\cite{Karsch:1987pv}.}
\label{fg:eb_mu}
\end{figure}
According to thermal pQCD calculations the Debye mass is related to the temperature of the medium,$T$, via
\begin{equation}
  \label{eq:debye_m}
  \mu_D(T)\sim gT.
\end{equation}
Here $g$ is the strong coupling constant. Inserting $g\sim$2 (corresponding to $\alpha_s\sim$0.3), we see that above a temperature of $T\sim$ 350MeV the $J/\psi$ is not bound any more and is expected to dissolve into separate $c$ and $\bar{c}$ quarks.  
Based on this mechanism $J/\psi$ suppression was first suggested in 1986 as a signature of QGP~\cite{Matsui:1986dk}.
Similar phenomena are expected for the excited charmonium states, such as $\psi'$ and $\chi_c$.  Their smaller binding energies imply lower dissociation temperatures. Therefore the entire charmonium spectra could provide a ``thermometer'' for the matter created in heavy-ion collisions.

\begin{figure}[tp]
  \centering
  \includegraphics[angle=-90,width=0.49\textwidth,clip=]{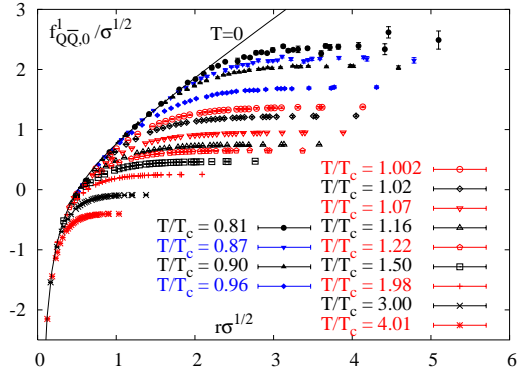}
  \caption[Free energy of a static $c\bar{c}$ pair, as computed in lattice QCD]{Free energy of a static $c\bar{c}$ pair, as computed in lattice QCD~\cite{Kaczmarek:2005ui}.}
\label{fg:f_qq}
\end{figure}

Recently\footnote{The discussions in this and next paragraph mostly follow Ref.~\cite{Zhao:2010nk}.} quantitative lQCD computations of the free energy,
$F_{Q\bar{Q}}(r;T)$, of a static heavy quark pair at finite
temperatures, became
available~\cite{Kaczmarek:2005ui,Petreczky:2004pz}, see
Fig.~\ref{fg:f_qq}.
This has been used as the main input for recent potential
models. However, it remains controversial to date whether the free
energy or the internal energy,
\begin{equation}
U_{Q\bar Q}(r;T)=F_{Q\bar Q}(r;T)
-T \frac{\partial F_{Q\bar Q}(r;T)}{\partial T} \ , 
\label{eq:e_int}
\end{equation}
or any combination thereof, should be identified with a static $Q\bar{Q}$ potential at finite temperature, $T$. Usually the internal energy leads to stronger binding than the free-energy. We refer to the former ($V_{Q\bar{Q}}=U_{Q\bar Q}$) and latter ($V_{Q\bar{Q}}=F_{Q\bar Q}$) as strong- and weak-binding scenario, respectively. 

\begin{figure}[tp]
  \centering
  \includegraphics[width=0.49\textwidth]{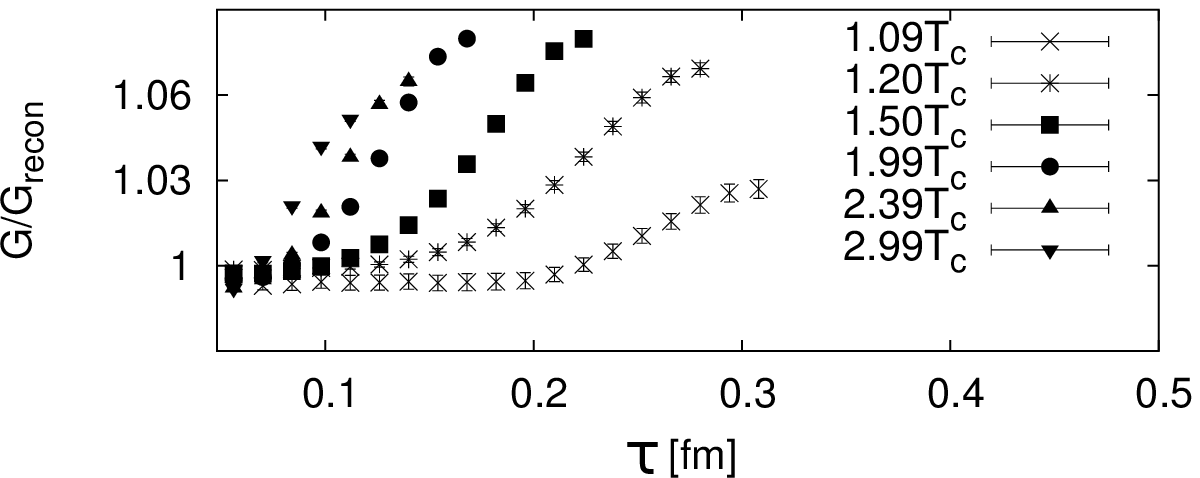}
  \includegraphics[width=0.49\textwidth,height=0.137\textheight]{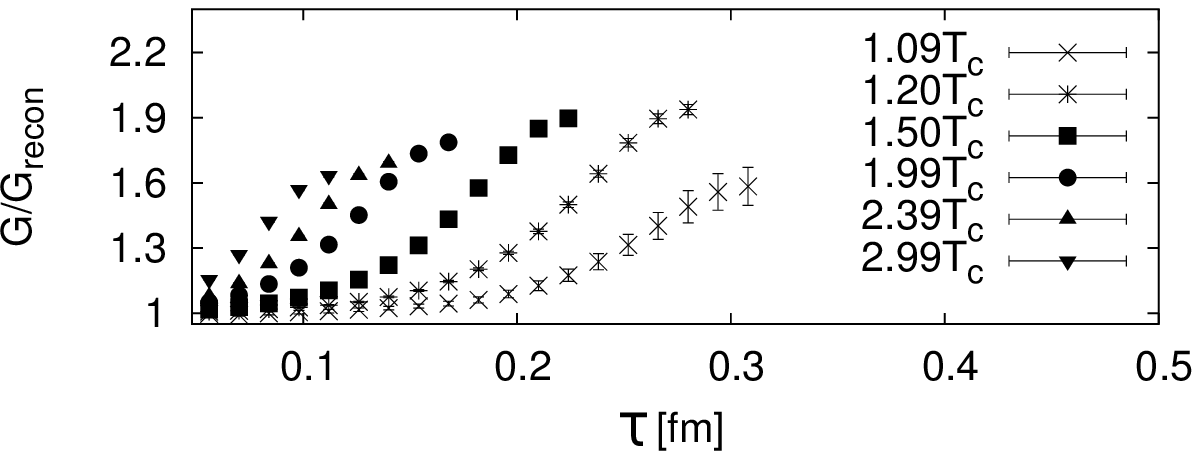}
  \caption[Charmonium correlator ratio in the vector and scalar
  channel as computed in lattice QCD]{Charmonium correlator ratio in
    the vector (left panel) and scalar (right panel) channel as
    computed in lattice QCD~\cite{Jakovac:2006sf}.}
\label{fg:gratio}
\end{figure}

Further progress in thermal lQCD came with the computation of two-point correlation functions of a quarkonium current, $j_\alpha$, with hadronic quantum number $\alpha$,
\begin{equation}
G_{\alpha}(\tau,\vec r)=
\langle\langle j_\alpha(\tau,\vec r)j^\dagger_\alpha (0,\vec 0)\rangle\rangle
 \ ,
\label{eq:Gtau}
\end{equation}
as a function of imaginary (euclidean) time, $\tau$ (also called 
temporal correlator). The imaginary part of the Fourier transform of the 
correlation function, $G_\alpha (\tau,\vec r)$, is commonly referred
to as the spectral function,
\begin{equation}
\sigma_\alpha (\omega,p)=-\frac{1}{\pi}\mathrm{Im}G_{\alpha}(\omega,p) \ ,
\label{eq:spectral}
\end{equation}
which is related to the temporal correlator via
\begin{equation}
G_{\alpha}(\tau,T)
=\int^\infty_0 d\omega \sigma_\alpha (\omega,T)K(\omega,\tau,T)
\label{eq:Gtau-sig}
\end{equation}
with the finite-$T$ kernel
\begin{equation}
K(\omega,\tau,T)=\frac{\cosh[(\omega(\tau-1/2T)]}{\sinh[\omega/2T]} \ .
\label{kernel}
\end{equation}
Lattice QCD results for two-point correlation functions are usually
normalized to a ``reconstructed" correlator evaluated with the kernel at 
temperature $T$,
\begin{equation}
G_\alpha^{\rm rec}(\tau,T)=\displaystyle\int^{\infty}_0 d\omega \ 
\sigma_\alpha(\omega,T^*) \ K(\omega,\tau,T) \ ,
\label{corr_rec}
\end{equation}
but with a spectral function at low temperature, $T^*$, where no
significant medium effects are expected. The correlator ratio,
\begin{equation}
R_\alpha(\tau,T)=G_\alpha(\tau,T)/G_\alpha^{\rm rec}(\tau,T) \ ,
\label{eq:corr_ratio}
\end{equation}
is then an indicator of medium effects in $G_\alpha(\tau,T)$ through
deviations from one. Current lQCD calculations find that the
correlator ratio, $R_\alpha(\tau,T)$, in the pseudoscalar ($\eta_c$)
and vector ($J/\psi$) channel are close to 1 (within ca.~10\%) at
temperatures up to
2-3\,$T_c$~\cite{Jakovac:2006sf,Datta:2003ww,Aarts:2007pk}, see the
left panel of Fig.~\ref{fg:gratio}. In the $P$-wave channels (scalar
and axialvector) the correlators ratios are substantially enhanced
over 1 at large $\tau$, see the right panel of
Fig.~\ref{fg:gratio}. This feature is believed to be due to
``zero-mode" contributions (at $\omega$=0) which are related to the
scattering of a charm (or anti-charm) quark, $c\to c$ (or $\bar c\to
\bar c$), rather than to $c\bar c$ bound-state
properties~\cite{Umeda:2007hy}. This interpretation is supported by
studies of the $\tau$-derivative of $P$-wave correlator ratios, which
exhibits a much smaller variation (in the limit that the zero-mode
part is a $\delta$-function, $\sigma_{\rm
  zm}(\omega)\propto\delta(\omega)$, its contribution to the temporal
correlator is a constant)~\cite{Umeda:2007gk,Petreczky:2008px}.

In principle, the in-medium properties of charmonia, such as pole
mass, in-medium width and dissociation temperature, are fully encoded
in their spectral function. However, the finite number of data points
for the two-point correlator computed in lQCD severely hampers the
inversion of the transform in Eq.~(\ref{eq:Gtau-sig}), rendering the
determination of the spectral function difficult, although various
attempts are made, such as the Maximum Entropy Method
(MEM)~\cite{Asakawa:2000tr}.

Although associated with large uncertainties, these lQCD data suggest
that the $J/\psi$ bound states might still survive above the critical
temperature. As a consequence the $c$ and $\bar{c}$ pairs produced in
initial hard collisions may coalesce and regenerate $J/\psi$ in
QGP~\cite{Andronic:2006ky,Grandchamp:2001pf}, rendering the original
picture with $J/\psi$ suppression as the signal of QGP more
complicated. Quantitative calculations disentangling primordial
$J/\psi$ production and regeneration are thus necessary for utilizing
$J/\psi$ to assess the basic properties of the hot and dense medium
created in URHICs. These constitute the main part of this dissertation.

\section{Outline of Dissertation}
\label{sec:outline}


In Chapter~\ref{ch:trans} we first give an overview of the physical processes related to charmonium production in each stage of heavy-ion collisions. Then we introduce our main framework: the Boltzmann transport equation approach. The Boltzmann transport equation describes the time evolution of the charmonium phase space distribution function due to dissociation, regeneration and drifting. If one is only interested in the inclusive yield of charmonium the Boltzmann transport equation can be reduced to a simpler rate-equation. Next we discuss the important relation between the dissociation rate and the regeneration rate, which is the principle of detailed balance. We discuss its microscopic origin and show its connection with the equilibrium limit of the charmonium phase distribution. We also examine the experimental fact that the $c$ and $\bar{c}$ are always produced in pair and discuss its consequence on the charmonium regeneration. Finally we consider the correlation between the $c$ and $\bar{c}$ in coordinate space and estimate its influence on charmonium regeneration.

In Chapter~\ref{ch:psi_hot} we discuss the microscopic interaction
between the charmonium states and the hot medium. In the QGP we
calculate the dissociation rates from a ``quasifree'' mechanism with
realistic charmonium in-medium binding energies estimated from lQCD
potentials. We work out the momentum dependence of the dissociation
rates. Based on the detailed balance we calculate and compare the
charmonium regeneration rates with different input charm-quark
momentum spectra ranging from the hard pQCD spectra to the thermal
spectra. With the in-medium width of charmonia estimated from
``quasifree'' dissociation rates we construct a model spectral
function and calculate the correlator ratios. By comparing these ratios
with lQCD data we extract the $\Psi$ dissociation temperatures, which
determine the start time of the regeneration process in the kinetic
approach. 

In Chapter~\ref{ch:fireball} we present our description of the space-time evolution of a heavy-ion collision, which is modeled by an isentropically and cylindrically expanding fireball with input parameters determined from the observed hadro-chemistry and the collective flow velocities. The resulting thermal evolution scenario is in line with basic features of hydrodynamic calculations. In particular we extend the fireball description to lower energies, which will be explored by future FAIR experiments. In this regime it turns out that, despite the low initial temperature, the deconfined phase still lasts for about 4fm/$c$, with a notable reheating process in the mixed phase driven by the latent heat. Finally, based on the flow field of the fireball model, we estimate the transverse momentum spectra of locally thermalized charmonia.

In Chapter~\ref{ch:pre-eq} we estimate the charmonium phase space distribution at the moment of thermalization which serves as the initial condition of the Boltzmann transport equation. We discuss the primordial production of charmonium in initial nucleon-nucleon collisions and its subsequent interaction with the pre-equilibrium medium. Emphases are placed on the deviation of primordial charmonium production in nucleus-nucleus (A+A) from scaled p+p due to the so-called cold nuclear matter (CNM) effects. 

In Chapter~\ref{ch:nume_results} we present the numerical results of our model of charmonium production in heavy-ion collisions. We introduce the procedure of applying the kinetic equations to calculate charmonium production in the strong and weak binding scenarios. We first compare our results for both inclusive $J/\psi$ yields and their transverse momentum spectra with data at SPS and RHIC. Within current theoretical uncertainties we find that both scenarios can reproduce SPS and RHIC data for the centrality  dependence of  inclusive $J/\psi$ production reasonably well. However the partition of primordial and regeneration yields is quite different in the two scenarios: the former dominates for strong binding, while for weak binding regeneration largely prevails at RHIC energies (except for peripheral collisions). This difference entails the strong binding scenario to be slightly favored by the $p_t$ data. We also study the effects specifically relevant for high $p_t$ $J/\psi$ production, including formation time effects and $B$-meson feeddown. These effects lead to a moderate enhancement of $J/\psi$ $R_{AA}$ at high $p_t$. Then we explicitly evaluate $J/\psi$ regeneration from different input charm-quark spectra. Next we compare the results of $\psi'$ to $\psi$ ratio to experimental data at SPS and provide the prediction of $\chi_c$ and $\psi'$ production at RHIC. Last we present our prediction for charmonium production at FAIR energies.

We conclude in Chapter~\ref{ch:end} and give a few directions along which we plan to expand the work undertaken in this dissertation. The input charm-quark phase space distribution based on Langevin simulations, hydrodynamic simulations of the medium evolution as well as a microscopic model for primordial $c\bar{c}$ and charmonium production need to be implemented to improve charmonium toward a more quantitative probe of the hot and dense QCD matter.

\graphicspath{{./2fg/}}
\chapter{Charmonium Transport Theory}
\label{ch:trans}
In this chapter we discuss the main framework in which charmonium
production in heavy-ion collisions is studied. In
Section~\ref{ch:trans}.\ref{sec:overview} we start with an overview of charmonium
production in different stages of heavy-ion collisions and relevant
production/dissociation mechanisms in each stage. In
Section~\ref{ch:trans}.\ref{sec:Boltz} we introduce the Boltzmann transport equation to
describe the evolution of charmonium phase space distribution
functions. The Boltzmann transport equation plays a central role in
our framework. Finally we investigate constraints imposed by the
pair production of charm-anticharm quarks and study their influence on charmonium
regeneration in Section~\ref{ch:trans}.\ref{sec:cc_corr}.


\section{Overview of Charmonium Production in Relativistic Heavy-Ion Collisions}
\label{sec:overview}

The conventional picture of charmonium ($\Psi$=$J/\psi$, $\chi_c$, $\psi'$) production in heavy-ion collisions
involves three stages: 1) Hard production stage: The two
ultra-relativistically moving nuclei collide with each other and charm
quark pairs are created from the hard collisions between
partons. This is a fast process due to the rather large momentum transfer
($\sim 2m_c$=2.5GeV) involved. The typical time scale is estimated as
$\tau \sim \frac{1}{2m_c} \sim 0.1 \mbox{ fm/c}$.  2) Pre-equilibrium
stage: If the charm and anti-charm quarks produced in initial hard
collisions are sufficiently close to each other in phase space they
can further develop into a charmonium state through non-perturbative
interactions. The charmonium formation processes are rather slow
compared to the charm quark production processes. The typical
estimates for the charmonium formation time are $\sim$1 fm/$c$~\cite{Karsch:1987zw}. In the mean
time the partons created in initial hard collisions rescatter and
are in the process of thermally equilibrating. When the
``pre-charmonium'' $c\bar c$ pairs travel in this medium some of them
may collide with passing-by nucleons and be subsequently dissociated.
The process is usually called ``nuclear absorption'', which will be
discussed in more detail in Chapter~\ref{ch:pre-eq}. The overall
duration of the pre-equilibrium stage is estimated to be about $0.3-1.0$fm/$c$
depending on the center-of-mass energy of the collision. 3) Equilibrium stage:
When the medium reaches thermal equilibrium, it is supposed to be
either in the deconfined Quark-Gluon Plasma (QGP) phase, if the energy
deposited in heavy-ion collision is large enough, or otherwise in the
hadronic gas (HG) phase. The medium continues to expand and
cool until it becomes so dilute that the particles in it stop
interacting with each other. This moment is usually called
``freeze-out'', which signifies the end of the entire evolution of the
medium. After ``freeze-out'' the produced particles stream freely to
the detectors with their yields and transverse momentum spectra fixed
at the moment of ``freeze-out''. The lifetime of the thermal medium
depends largely on initial energy densities of the medium. Typically
it lasts for about 10(12)fm/$c$ for central Pb+Pb (Au+Au)
collisions at SPS (RHIC). The description of the thermal medium will be
detailed in Chapter~\ref{ch:fireball}. In the equilibrium stage there exist two
competing effects on the charmonium abundance: On the one hand
charmonia undergo inelastic collisions with particles in the thermal
bath and are subsequently destroyed, on the other hand the charm
quarks or $D$ mesons in the medium may recombine and regenerate
charmonia. Since these two competing processes are inverse processes of
each other, their relative strength is governed by the principle of detailed
balance which ensures that the charmonium abundances approach their
thermal equilibrium values. The rates of approaching equilibrium
strongly depend on the temperature and particularly the effective
degrees of freedom, \ie, the phase, of the medium. In QGP, both the significantly reduced $\Psi$ binding energies (due to
color-Debye screening) and the thermal parton density (due to the large color degeneracy) lead to
dissociation/regeneration rates much larger than those
in HG phase.


\section{Boltzmann Transport Equation}
\label{sec:Boltz}


\subsection{General Setup}
\label{ssec:gsetup}

The charmonia in the thermal medium form a system off thermodynamic
equilibrium. The Boltzmann transport equation is an ideal tool for
studying such a system. 
In this work we employ the Boltzmann transport equation to
describe the time evolution of the charmonia phase space distribution
function $f_\Psi(\vec p,\vec x,t)$ in the thermal medium~\cite{Polleri:2003gj,Polleri:2003kn,Yan:2006ve,Zhang:2002ug,Linnyk:2008uf},
\begin{equation}
\partial f_\Psi/\partial t+\vec v_\Psi \cdot \vec \nabla f_\Psi
=-\alpha_\Psi f_\Psi+\beta_\Psi\ .
\label{eq:boltz}
\end{equation}
Here, the distribution function $f_\Psi(\vec x,\vec p,t)$ is the
number of charmonia ($\Psi=J/\psi,\chi_c,\psi'$) per unit phase space volume at a given time $t$,
\begin{equation}
\label{eq:dist_psi}
  f_\Psi(x,p,t)=\frac{(2\pi)^3d N_\Psi}{d^3 x\, d^3 p}\ .
\end{equation}
The first term on the left-hand side of Eq.~(\ref{eq:boltz}) describes
the rate of change of the charmonium distribution function at a
specific position in the phase-space. The second (drift) term reflects
the charmonium diffusion from one position to another with velocity
$\vec v_\Psi=\vec p/\sqrt{p^2+m^2_\Psi}$. Since the mass of $\Psi$ is
much larger than typical medium temperature ($m_\Psi\gg T)$ we neglect
the elastic scattering between $\Psi$ and the medium and assume the
$\Psi$ to move on a straight line.  Therefore the left-hand side of
Eq.~(\ref{eq:boltz}) represents the change of charmonium density other
than charmonium diffusion.  Other than diffusion the change in $\Psi$
distribution function can only be due to the dissociation and
regeneration of charmonia, which are accounted for by the first
(``loss'') and second (``gain'') term on the right-hand side,
respectively. 
The loss term consists of the product of the charmonium phase space
density and its dissociation rate, $\alpha_\Psi(\vec p,\vec x)$. The
gain term, $\beta_\Psi(\vec p,\vec x)$, reflects the charmonium
regeneration rate from charm quark coalescence. 

One of the premises of Boltzmann transport equation is that the medium
should be ``dilute'' enough so that a subsequent collision occurs well
after the end of the first one (no quantum interference between
successive collisions). The typical duration for one collision can be
estimated by the reciprocal of the virtuality, $\sqrt{Q^2-m^2_\Psi}$, of the $\Psi$ after
scattering with thermal particles, which can be estimated as follows:
Denote the four momentum of the $\Psi$ before a collision as $(p^2/(2m_\Psi)+m_\Psi, \vec p)$, and the four momentum of a thermal parton as $(k, \vec k)$ with $k$=$|\vec k|$. Then the four momentum of $\Psi$ after scattering becomes $(q_0,\vec q)$=$(p^2/(2m_\Psi)+m_\Psi+k, \vec p+\vec k)$. The virtuality is given by
\begin{align}
  \label{eq:virtuality}
  Q^2-m^2_\Psi&=q^2_0-\vec q^2-m^2_\Psi=\left (\frac{p^2}{2m_\Psi}+m_\Psi+k\right)^2-(\vec p+\vec k)^2-m^2_\Psi\nonumber \\
  &=2m_\Psi k-2\vec{p}\cdot\vec k+\frac{p^2k}{m_\Psi}+\frac{p^4}{4m^2_\Psi}\ .
\end{align}
Assuming thermal momenta for $\Psi$ and the parton, so that
$p^2/(2m)\sim T$, $p\sim \sqrt{m_\Psi T}$, $k\sim T$, the virtuality
$\sqrt{Q^2-m^2_\Psi}$ to leading order in $m_\Psi(\gg T)$ is on the
order of $\sqrt{m_\Psi T}$. Therefore with a typical charmonium mass
$\sim$3\,GeV and a typical temperature $T\sim$\,0.3GeV, the estimated
duration for a dissociation process is around 0.2fm/$c$. On the other
hand the typical $\Psi$ dissociation rate at RHIC energy is on the
order of 100MeV, as will be evaluated in
Section~\ref{ch:psi_hot}.\ref{sec:charm-diss-qgp}, corresponding to an
average time of 2fm/$c$ between successive collisions. Therefore the
average interval between $\Psi$ inelastic collisions is indeed much
longer than typical collision time thus satisfying the premise of
Boltzmann transport equation.

We proceed to work out the microscopic expression for the dissociation
and gain rates, $\alpha_\Psi$ and $\beta_\Psi$. From now on we will
omit the vector sign ``$\,\vec{~}~$'' on top of $x$ and $p$ if no ambiguity
arises. For simplicity let us first consider a 2-body
dissociation/regeneration process, \eg, $\Psi+g\leftrightarrows c+\bar
c$, in which a $\Psi$ scatters with a gluon and is subsequently
dissociated into a $c\bar{c}$ pair. The dissociation rate is expressed
as follows:
\begin{align}
  \alpha_\Psi(p,x)&=\frac{1}{2E_\Psi}\int d \Phi_1(p_g)\ d\Phi_2(p_c,p_{\bar c}) \nonumber \\&\qquad \times (2\pi)^4
\delta^{(4)}(p_c+p_{\bar c}-p-p_g)\,d_g\,f_g(p_g)\,\lvert\overline{\mathcal M_{\Psi g\to c\bar{c}}(s,t)}\rvert^2 \ ,
\label{eq:diss_rate}
\end{align}
where $E_\Psi=\sqrt{p^2+m^2_\Psi}$ is the charmonium energy,
\begin{equation}
d_gf_g(p_g)=\frac{(2\pi)^3dN_g}{d^3x\,d^3p_g}\ 
\label{eq:gluon_phase_dist}
\end{equation}
is the gluon phase space distribution function with $d_g$=16 being the
color-spin degeneracy factor for gluons, and $
\lvert\overline{\mathcal M_{c\bar c\to \Psi g}(s,t)}\rvert^2$ is the
transition matrix element between the initial state, $\Psi+g$, and the
final state, $c+\bar c$, as a function of $s=(p+p_g)^2$ and
$t=(p-p_c)^2$. The ``bar'' on top of ${\lvert\mathcal M\rvert^2}$
stands for summing over color and spin degeneracies of final states
and averaging over color and spin degeneracy of initial states, \eg,
$ {\lvert\overline{\mathcal M_{\Psi g\to c\bar
      c}}\rvert^2}$=$\Sigma\lvert\mathcal M_{\Psi g\to c\bar
  c}\rvert^2/(d_gd_\Psi)$, with $d_g$ and $d_\Psi$ being color and
spin degeneracies of the gluon and the charmonium. The
$\Sigma\lvert\mathcal M_{\Psi g\to c\bar c}\rvert^2$ denotes the
transition matrix element with both initial and final state degeneracy
summed over. In Eq.~(\ref{eq:diss_rate}) $d \Phi_1(p_g)$ and $d
\Phi_2(p_c,p_{\bar c})$ are Lorentz invariant 1-body and 2-body phase
space integration measures,
\begin{align}
  \label{eq:1-2-body}
  d \Phi_1(p_g)&=\frac{d^3p_g}{(2\pi)^32E_g}\ ,\\
  d \Phi_2(p_c, p_{\bar c})&=\frac{d^3p_c}{(2\pi)^32E_c}\frac{d^3p_{\bar c}}{(2\pi)^32E_{\bar c}}\ ,
\end{align}
where $E_g$, $E_c$ and $E_{\bar c}$ are the gluon, $c$ and $\bar c$
energies, respectively. 

It is convenient to express the dissociation rate $\alpha_\Psi$
in terms of the charmonium dissociation cross section $\sigma_\Psi$, which
can be expressed in terms of the transition probability, $
{\lvert\overline{\mathcal M_{c\bar c\to \Psi g}}\rvert^2}$, as follows,
\begin{equation}
  \label{eq:diss_xsec}
  \sigma_\Psi(s)=\frac{1}{2E_\Psi}\int d \Phi_2(p_c,p_{\bar c}) (2\pi)^4
  \delta^{(4)}(p_c+p_{\bar c}-p-p_g) \frac{1}{2E_g\lvert v_\Psi-v_g\rvert} \lvert\overline{\mathcal M_{\Psi g\to c\bar c}(s,t)}\rvert^2\ ,
\end{equation}
where $v_\Psi=p/\sqrt{p^2+m^2_\Psi}$ and $v_g=p_g/\sqrt{p^2_g+m^2_g}$ are the velocities of the charmonium and gluon, respectively. By comparing Eq.~(\ref{eq:diss_xsec}) with Eq.~(\ref{eq:diss_rate}) one can identify the relation between $\alpha_\Psi$ and $\sigma_\Psi$,
\begin{equation}
  \label{eq:rate_xsec}
  \alpha_\Psi(p,x)=\int \frac{d^3p_g}{(2\pi)^3}\,d_g\,f_g( p_g)\,v_{rel}\,\sigma_\Psi(s)\ ,
\end{equation}
where $v_{rel}$=$\lvert v_\Psi-v_g\rvert$ is the relative velocity between the gluon and the charmonium.

Next we proceed to the microscopic expression of the gain rate, $\beta_\Psi$, given by
\begin{align}
  \label{eq:gain_rate}
    \beta_\Psi(p,x)=\frac{1}{2E_\Psi}\int d \Phi_1(p_g)d \Phi_2(p_c,p_{\bar c}) (2\pi)^4
\delta^{(4)}(p_c+p_{\bar c}-p-p_g)\nonumber\\ 
\times \lvert\overline{\mathcal M_{c\bar c\to \Psi g}(s,t)}\rvert^2\,
d_c\,f_c(p_c)\,d_{\bar{c}}\,f_{\bar c}(p_{\bar c})\,(1+ f_g(p_g))\ ,
\end{align}
where $d_cf_c(p_c)$ and $d_{\bar{c}}f_{\bar c}(p_{\bar c})$ are charm and
anti-charm phase space distribution functions,
\begin{equation}
d_cf_c(p_c)=\frac{(2\pi)^3dN_c}{d^3x\,d^3p_c}\ ,
\label{eq:c_phase_dist} 
\end{equation}
with $d_c$=$d_{\bar{c}}$=6 being the color-spin degeneracy factors for
$c$ and $\bar{c}$, respectively. The factor of $1+ f_g(p_g)$ is the
Bose-enhancement factor for the final state gluon. The ``bar'' on top
of $\lvert\overline{\mathcal M_{c\bar c\to \Psi g}}\rvert$ denotes
summing over final states and averaging over the initial states, 
$\lvert\overline{\mathcal M_{c\bar c\to \Psi
    g}}\rvert$=$\Sigma\lvert\mathcal M_{c\bar c\to \Psi
  g}\rvert/(d_cd_{\bar c})$.  We note that the underlying dynamics of
the transition between $g+\Psi$ and $c+\bar c$, QCD, is time-reversal
symmetric, therefore the transition probability satisfies the
principle of detailed balance as
\begin{equation}
  \label{eq:d_balance}
  \Sigma\lvert\mathcal M_{c\bar c\to \Psi g}\rvert^2=\Sigma\lvert\mathcal M_{\Psi g\to c\bar c}\rvert^2\ .
\end{equation}

It is straightforward to extend the above expressions for the
dissociation and gain rates to 2-to-3 processes, $i + \Psi \to
i+c+\bar c$ ($i$=$g$,$q$,$\bar q$) and its inverse 3-to-2 processes,
$i+c+\bar c\to i + \Psi$ . What needs to be modified are: 1) Replacing the
2-body integration measures in
Eqs.~(\ref{eq:diss_rate}), (\ref{eq:diss_xsec}) and (\ref{eq:gain_rate})
by 3-body integration measures,
\begin{equation}
\label{eq:3-body}
  d\Phi_3(p_c, p_{\bar c}, \bar{p}_i)=\frac{d^3p_c}{(2\pi)^32E_c}\frac{d^3p_{\bar c}}{(2\pi)^32E_{\bar c}}\frac{d^3\bar{p}_i}{(2\pi)^32E_i}\ ,
\end{equation}
accounting for the phase space of the extra light parton in the final
state of dissociation (or the initial state of regeneration), 2)
Inserting the four-momentum of the extra light parton into the
4-momentum conserving delta function. 3) Supplying the phase space
distribution function, $d_if_i(\bar{p}_i)$, for the extra parton if it
appears in the initial state or the Bose-enhancement/Pauli-blocking
factors, $1\pm f_i(\bar{p}_i)$, if it appears in the final state. The
``$\pm$'' takes $+$($-$) for $i=g\ (i=q,\bar q)$. The explicit
expressions of $\sigma_\Psi$, $\alpha_\Psi$ and $\beta_\Psi$ for
two-to-three processes are
\begin{align}
  \label{eq:diss_xsec3}
  \sigma_\Psi(s)&=\frac{1}{2E_\Psi}\sum_{i}\int d\Phi_3(p_c, p_{\bar c},\bar{p}_i) (2\pi)^4
  \delta^{(4)}(p_c+p_{\bar c}+\bar{p}_i-p-p_i) \nonumber\\
  &\qquad\times\frac{1}{2E_g\lvert v_\Psi-v_i\rvert} \lvert\overline{\mathcal M_{\Psi i\to c\bar{c}i}(s,t)}\rvert^2(1\pm f_i(\bar{p}_i))\ ,\\[4mm]
   \alpha_\Psi(p,x)&=\frac{1}{2E_\Psi}\sum_{i}\int d\Phi_1(p_i) \Phi_3(p_c,p_{\bar c},\bar{p}_i) (2\pi)^4
\delta^{(4)}(p_c+p_{\bar c}+\bar{p}_i-p-p_i)\label{eq:diss_rate3} \nonumber\\
&\qquad\times\lvert\overline{\mathcal M_{\Psi i\to c\bar{c}i}(s,t)}\rvert^2\, d_i\,f_i(p_i)\,(1\pm f_i(\bar{p}_i))\ ,\\[4mm]
\beta_\Psi(p,x)&=\frac{1}{2E_\Psi}\sum_{i}\int d \Phi_1(p_i)d \Phi_3(p_c,p_{\bar c},\bar{p}_i) (2\pi)^4
\delta^{(4)}(p_c+p_{\bar c}+\bar{p}_i-p-p_i)\label{eq:gain_rate3}\nonumber\\
&\qquad\times \lvert\overline{\mathcal M_{c\bar{c}i\to \Psi i}(s,t)}\rvert^2\,
d_c\,f_c(p_c)\,d_{\bar{c}}\,f_{\bar c}(p_{\bar c})\,d_i\,f_i(\bar{p}_i)\,(1\pm f_i(p_i))\ .
\end{align}
Here $\bar{p}_i$ denotes the momentum of the light parton for the three-body state
(together with $c$ and $\bar c$). 

We will illustrate in Section~\ref{ch:psi_hot}.\ref{sec:open_charm_med} that in the
case where the charmonium binding energies are small compared to the medium
temperature, a 3-to-2 regeneration process, $i+c+\bar c \to i + \Psi$ 
$(i=g, q, \bar q)$, can be readily factorized into a 2-to-2
(perturbative) quasi-elastic scattering process, $i + c(\bar{c}) \to
i+c'(\bar{c}')$ and 2-to-1 (non-perturbative) coalescence process,
$c$+$\bar{c}\to \Psi$ owing to the different time scales for these two
processes. The four-momentum is conserved for the entire 3-to-2
regeneration process.

So far we have established the connection between the charmonium transport equation and the microscopic dynamics of $\Psi$ dissociation/regeneration. The required inputs from microscopic calculations are 1) The transition matrix element, $\lvert\mathcal M_{ic\bar{c}\to \Psi i}\rvert^2$; 2) Charmonium binding energies, $\epsilon_\Psi$, which determine the initial state phase space for the dissociation processes and the final state phase space for the regeneration processes; 3) The in-medium charm quark spectra, $f_c(p_c)$, $f_{\bar c}(p_{\bar c})$, determining the charmonium regeneration rates. We will discuss these input quantities based on microscopic calculations in Chapter~\ref{ch:psi_hot}. With all these quantities supplied by microscopic calculations we are ready to solve the Boltzmann transport equation (\ref{eq:boltz}).

The Boltzmann transport equation, Eq.~(\ref{eq:boltz}), is a first-order, linear partial differential equation. It can be solved using the method of change of variables. Introducing the new variable
\begin{equation}
  \label{eq:vb_ch}
  u=x-v_\Psi t\ ;
\end{equation}
we have
\begin{align}
  \label{eq:vb_ch_fab}
  f_\Psi(x,t)&=f_\Psi(u+v_\Psi t,t)\ ,\\
  \alpha_\Psi(x,t)&=\alpha_\Psi(u+v_\Psi t,t)\ ,\\
  \beta_\Psi(x,t)&=\beta_\Psi(u+v_\Psi t,t)\ .
\end{align}
Since in Eq.~(\ref{eq:boltz}) there is no derivative with respect to
$p$, we consider $p$ as a parameter and suppress its dependence in
$f_\Psi$, $\alpha_\Psi$, $\beta_\Psi$ for the moment. Substituting Eq.~(\ref{eq:vb_ch}) into Eq.~(\ref{eq:boltz}), we obtain
\begin{equation}
  \label{eq:bv_ch_boltz}
  \partial f_\Psi(u+v_\Psi t,t)/\partial t
=-\alpha_\Psi(u+v_\Psi t,t) f_\Psi(u+v_\Psi t,t)+\beta_\Psi(u+v_\Psi t,t)\ .
\end{equation}
Now Eq.~(\ref{eq:bv_ch_boltz}) is a first-order ordinary differential equation with $u$ as a parameter. It has the following solution,
\begin{align}
  \label{eq:solu_rate}
  f_\Psi(u+v_\Psi t,t)&=f_\Psi(u+v_\Psi t_0,t_0)e^{-\int^t_{t_0} d t' \alpha_\Psi(u+v_\Psi t',t')}\nonumber \\
&\quad+\int^t_{t_0}d t'\beta_\Psi(u+v_\Psi t',t')e^{-\int^t_{t'} d t'' \alpha_\Psi(u+v_\Psi t'',t'')}\ .
\end{align}
Finally we substitute Eq.~(\ref{eq:vb_ch}) into Eq.~(\ref{eq:solu_rate}) and obtain the solution for the Boltzmann transport equation.
\begin{align}
  \label{eq:solu_boltz}
    f_\Psi(p,x,t)&=f_\Psi(p,x-v_\Psi(t-t_0),t_0)e^{-\int^t_{t_0} d t' \alpha_\Psi(p, x-v_\Psi(t-t'),t')}\nonumber\\
&\quad+\int^t_{t_0}d t'\beta_\Psi(p,x-v_\Psi(t-t'),t')e^{-\int^t_{t'} d t'' \alpha_\Psi(p,x-v_\Psi(t-t''),t'')}\ ,
\end{align}
where $t_0$ is the start time of the evolution. The initial phase space
distribution of charmonia, $f_\Psi(p,x, t_0)$, is
determined by initial hard production and cold nuclear matter effects,
which are the main topic of Chapter~\ref{ch:pre-eq}. The solution of the Boltzmann
transport equation, Eq.~(\ref{eq:solu_boltz}), consists of two terms:
The first term decays with time  and describes the dissociation
process of the initially produced charmonia. This component of the
solution is usually referred to as the ``primordial'' component. The second term
increases with time and describes the regeneration process of charmonia
from coalescence of charm quarks in the medium. This component is
referred to as the ``regeneration'' component.

\subsection{Rate-Equation}
\label{ssec:rate-eq}
If one is only interested in the inclusive yield of charmonia, $N_\Psi$, it is convenient to employ a rate-equation instead of the more differential Boltzmann equation. In this section we derive the widely employed rate-equation from the Boltzmann equation.

We start by integrating over the entire charmonium phase space on both sides of the Boltzmann equation~(\ref{eq:boltz}). The left-hand side becomes
\begin{align}
  \label{eq:rate-eq_left}
  &\int d^3x\, \frac{d^3p}{(2\pi)^3} \,(\partial f_\Psi(p,x,t)/\partial t
+v_\Psi \cdot \nabla f_\Psi(p,x,t))\nonumber\\
&=\frac{\partial}{\partial t} \int d^3x\, \frac{d^3 p}{(2\pi)^3} f_\Psi(p,x,t)
+\int \frac{d^3p}{(2\pi)^3} \int d^3x \nabla \cdot (f_\Psi(p,x,t)v_\Psi\nonumber)\\
&=d N_\Psi(t)/d t + \int \frac{d^3p}{(2\pi)^3} \oint_S d\vec S\cdot (f_\Psi(p,x,t)\vec v_\Psi)\nonumber\\
&=d N_\Psi(t)/dt\ .
\end{align}
In the second equality we used the definition of charmonium phase
space distribution function, Eq.~(\ref{eq:dist_psi}) and Gauss's law,
where $S$ is a large surface enclosing the integration volume of
coordinate space. In the third equality we used the fact that the
charmonium phase space distribution $f_\Psi(p,x,t)$ drops
sufficiently fast as $|\vec x|\to\infty$. Next we proceed to the
right-hand side, which consists of the loss and the gain terms. For
the loss term, if spatial homogeneity is assumed, namely,
$\alpha_\Psi(p,x,t)\to \alpha_\Psi(p,t)$, we have
\begin{align}
  \label{eq:rate-eq_loss}
\int d^3 x\, \frac{d^3p}{(2\pi)^3}\, \alpha_\Psi(p,x,t) f_\Psi(p,x,t)
&=\int \frac{d^3p}{(2\pi)^3}\, \alpha_\Psi(p,t) \int d^3 x\, f_\Psi(p,x,t)\nonumber\\
&\approx \alpha_\Psi(\langle p\rangle,t)\int \frac{d^3p}{(2\pi)^3}\ d^3x\, f_\Psi(p,x,t)\nonumber\\
&\approx \Gamma_\Psi(\langle p\rangle,t)N_\Psi(t)\ ,
\end{align}
where $\langle p\rangle$ is the average momentum of $\Psi$, and the $\alpha_\Psi(\langle p\rangle,t)$ is conventionally denoted as $\Gamma_\Psi(\langle p\rangle,t)$. Similar manipulations can be applied to the gain term. With the assumption of spatial homogeneity the integration over coordinate space reduces to the multiplication with the volume of the medium (fireball), $V_{FB}$. The integration over charmonium momentum can also be performed. The resulting integrated gain term is conventionally denoted by $G_\Psi(t)$,
\begin{align}
  \label{eq:rate-eq_gain}
G_\Psi(t)& \equiv \int d^3x \frac{d^3p}{(2\pi)^3}\ \beta_\Psi(p,x,t)\nonumber\\
& =V_{FB}\int \frac{d^3p}{(2\pi)^3}\ \beta_\Psi(p,t)\ .
\end{align}
Putting Eqs.~(\ref{eq:rate-eq_left}), (\ref{eq:rate-eq_left}) and (\ref{eq:rate-eq_gain}) together we
obtain the rate equation~\cite{Thews:2005vj} describing the time-evolution of the
inclusive charmonium yield,
\begin{equation}
  \label{eq:rate-eq}
  \frac{d N_\Psi(t)}{dt}=-\Gamma_\Psi(t)N_\Psi(t)+G_\Psi(t)\ .
\end{equation}

\subsection{Equilibrium Limit}
\label{ssec:eq_limit}
Let us consider the equilibrium limit of the solution of the Boltzmann
transport equation, Eq.~(\ref{eq:boltz}), which is defined
by
\begin{align}
  \label{eq:eq_limit}
  f^{\rm{eq}}_\Psi(p)& =\frac{\beta_\Psi(p)}{\alpha_\Psi(p)}\ ,
\end{align}
where $f^{\rm{eq}}_\Psi(p)$, $\beta_\Psi(p)$ and $\alpha_\Psi(p)$ are also
assumed to be homogeneous in space. It is easy to verify that the equilibrium distribution $f^{\rm{eq}}_\Psi(p)$
solves the Boltzmann equation. In this limit
the (integrated) gain term in the rate-equation, $G_\Psi$, can be
written in terms of $f^{\rm{eq}}_\Psi(p)$ as
\begin{align}
  \label{eq:gain-eq}
  G_\Psi(t)& =V\int \frac{d^3p}{(2\pi)^3}\, \beta_\Psi(p,t)\nonumber \\
&=V\int \frac{d^3p}{(2\pi)^3}\, \alpha_\Psi(p,t)f^{\rm{eq}}_\Psi(p)\nonumber \\
&\approx \alpha_\Psi(\langle p\rangle,t)V\int \frac{d^3p}{(2\pi)^3} f^{\rm{eq}}_\Psi(p)\nonumber \\
&=\Gamma_\Psi(t)N^{\rm{eq}}_\Psi(t)\ .
\end{align}
Substituting Eq.~(\ref{eq:gain-eq}) into Eq.~(\ref{eq:rate-eq}), we obtain another common form of the rate-equation~\cite{Grandchamp:2003uw},
\begin{equation}
  \label{eq:rate-eq-eq}
  \frac{\dd N_\Psi(t)}{\dd t}=-\Gamma_\Psi(t)[N_\Psi(t)-N^{\rm{eq}}_\Psi(t)]\ .
\end{equation}
This form is particularly convenient in the special case where the
momentum spectra for both the charm-quark and the light partons are
thermal. In this case the inclusive yield of charmonium in the
equilibrium limit, $f^{\rm{eq}}_\Psi(p)$, is independent of the
transition probability, $\lvert\mathcal M\rvert^2$. It is instructive
to further discuss this special case.

For simplicity we first consider a 2-to-2 process,
$\Psi+g\leftrightarrows c+\bar c$. We denote thermal (Boltzmann)
charm-quark spectra as
\begin{align}
f_c(p_c)=\gamma_c\,
e^{-\sqrt{m^2_c+p^2_c}/T}=\gamma_c\, e^{-E_c/T}\ , 
\label{eq:charm_spectra_th}
\end{align}
where $\gamma_c$ is a charm quark fugacity reflecting its total yield
deviating from chemical equilibrium (since $m_c\gg T$ we cannot expect
charm quarks to reach chemical equilibrium). The gluons follow the
thermal Bose distribution,
\begin{align}
f_g(p_g)=\frac{1}{\exp{\left(\frac{\sqrt{m^2_g+p^2_g}}{T}\right)}-1}=\frac{1}{\exp{\left(\frac{E_g}{T}\right)}-1}\ , 
\label{eq:gluon_th}
\end{align}
where $E_g$ is the energy of the gluon. Plugging these spectra into
Eq.(\ref{eq:gain_rate}), we obtain
\begin{align}
    \label{eq:gain-th}
    \beta_\Psi(p)& =\frac{1}{2E_\Psi}\int d \Phi_1(p_g)d \Phi_2(p_c,p_{\bar c}) (2\pi)^4
\delta^4(p_c+p_{\bar c}-p-p_g) \nonumber\\ &\qquad\times\lvert\overline{\mathcal M_{c\bar c\to \Psi g})}\rvert^2\,
d_c\,f_c(p_c)\,d_{\bar{c}}\,f_{\bar c}(p_{\bar c})\,(1+ f_g(p_g))\nonumber\\
& =\frac{\gamma^2_cd^2_c}{2E_\Psi}\int d \Phi_1(p_g)d \Phi_2(p_c,p_{\bar c}) (2\pi)^4
\delta^4(p_c+p_{\bar c}-p-p_g)\nonumber\\ &\qquad\times\lvert\overline{\mathcal M_{c\bar c\to \Psi g}}\rvert^2\,
e^{-(E_c+E_{\bar c})/T}\,(1+ f_g(p_g))\nonumber\\
& =\frac{\gamma^2_cd^2_c}{2E_\Psi}\int d \Phi_1(p_g)d \Phi_2(p_c,p_{\bar c}) (2\pi)^4
\delta^4(p_c+p_{\bar c}-p-p_g)\nonumber\\ &\qquad\times \lvert\overline{\mathcal M_{c\bar c\to \Psi g}}\rvert^2\,
 e^{-(E_\Psi+E_g)/T}\,(1+ f_g(p_g))\nonumber\\
& =\frac{\gamma^2_cd_\Psi}{2E_\Psi}\int d \Phi_1(p_g)d \Phi_2(p_c,p_{\bar c}) (2\pi)^4
\delta^4(p_c+p_{\bar c}-p-p_g)\nonumber\\ &\qquad\times \lvert\overline{\mathcal M_{\Psi g\to c\bar c}}\rvert^2\,
 e^{-\sqrt{m^2_\Psi+p^2}/T}\,d_g\,f_g(p_g)\nonumber\\
& =\gamma^2_c d_\Psi e^{-\sqrt{m^2_\Psi+p^2}/T}\, \alpha_\Psi(p)\ ,
\end{align}
where in the third and fourth equality we used four-momentum
conservation imposed by the 4-D $\delta$-function, and an identity for
the Bose-distribution for the gluon, $e^{-E_g/T} (1+
f_g(p_g))=f_g(p_g)$. In the fifth equality the detailed balance
between $\Sigma\lvert\mathcal M_{\Psi g\to c\bar c}\rvert^2$ and
$\Sigma\lvert\mathcal M_{c\bar c\to \Psi g}\rvert^2$,
Eq.~(\ref{eq:d_balance}), and the definition of dissociation rate
$\alpha_\Psi$, Eq.~(\ref{eq:diss_rate}), are used, The color-spin
degeneracy factors for $\Psi$, gluon, $c$ and $\bar c$ are $d_\Psi$=3,
$d_g$=16, $d_c$=$d_{\bar c}$=6. Although this derivation is made for
2-to-2 processes, it is straightforward to verify that this relation
also holds for 2-to-3 processes as long as the detailed balance and
four-momentum conservation hold. Comparing Eq.~(\ref{eq:gain-th}) with Eq.~(\ref{eq:eq_limit}), we obtain
\begin{equation}
    \label{eq:f_psi_eq}
    f^{\rm{eq}}_\Psi(p)=\gamma^2_c d_\Psi e^{-\sqrt{m^2_\Psi+p^2}/T}\ .
\end{equation}
This relation shows that as long as both charm quarks and 
light partons are in thermal equilibrium the regenerated
charmonia have the thermal momentum spectra as their equilibrium
limit which does not depend on the microscopic details of the
dissociation/regeneration mechanisms. Different microscopic transition
mechanisms will only affect the transition rates, namely the speed with which
charmonium spectra approach the equilibrium limit.

Since the thermal production and annihilation rates of $c\bar{c}$ are
believed to be small in heavy-ion collisions at SPS and RHIC energies,
$c\bar{c}$ pairs are assumed to be exclusively produced in primordial
N+N collisions and conserved thereafter~\cite{Andronic:2006ky}. The charmonium
equilibrium limit at a given time can be conveniently evaluated by
solving the following charm conservation relation for $\gamma_c$,
\begin{align}
  \label{eq:Ncc_conserve}
  \frac{N_{c\bar c}}{V}& =\frac{1}{2}\int \frac{d^3p_c}{(2\pi)^3} (f_c(p_c)+f_{\bar c}(p_{\bar{c}}))+\int \frac{d^3p}{(2\pi)^3} f^{\rm{eq}}_\Psi(p)\nonumber\\
& =\gamma_c n^{\rm{th}}_c+\gamma^2_c n^{\rm{th}}_\Psi\ ,
\end{align}
where $N_{c\bar c}$ is the total number of $c\bar c$ pairs from
primordial production and, $n^{\rm{th}}_c$ and $n^{\rm{th}}_\Psi$ are thermal spatial
densities for $c$ and $\Psi$,
\begin{equation}
  \label{eq:therml_density_c}
  n^{\rm{th}}_c=d_c\int \frac{d^3p_c}{(2\pi)^3}e^{-\frac{\sqrt{m^2_c+p^2_c}}{T}}\ ,
\end{equation}
\begin{equation}
  \label{eq:therml_density_psi}
  n^{\rm{th}}_\Psi=d_\Psi\int \frac{d^3p}{(2\pi)^3}e^{-\frac{\sqrt{m^2_\Psi+p^2}}{T}}\ .
\end{equation}
This relation shows that in thermal equilibrium the partition of all charm quarks between the open-charm and charmonium states is solely determined by their respective masses and degeneracy factors~\cite{BraunMunzinger:2000px}.

We should keep in mind that the charmonium equilibrium limits
have this nice feature only if the charm quark spectra are
thermal, which, however, may not be the case in
heavy-ion collisions, since charm quarks are heavy and take a rather
long time to thermalize, compared to the typical lifetime of the
medium (delayed by a factor of $m_c/T$ compared to light partons). We will discuss more realistic charm quark spectra and their
implications for charmonium production in the medium in Chapter~\ref{ch:psi_hot}.


\section{$c$-$\bar{c}$ Correlations for Charmonium Production}
\label{sec:cc_corr}


\subsection{Charmonium Production in the Canonical Ensemble}
\label{ssec:canonical}

Before we finish this chapter and discuss the input to the kinetic
equations from microscopic calculations we need to address another issue
associated with $\Psi$ regeneration. The problem originates from a
simple experimental fact: The charm and anti-charm quarks are always
produced in pairs. Its impact on charmonium production can be easily
seen in a situation where few charm quark pairs are produced per
event. For example, at SPS energy there is one charm quark pair
produced every ten events in central Pb+Pb collisions, namely on average $N_c$=$N_{\bar c}$=0.1
per event. Accordingly one expects the average number of regenerated
$\Psi$ to be 1/10 of what we would get if there is one charm quark
pair created in each event. However Eq.~(\ref{eq:gain_rate}) naively
suggests that the average number of regenerated $\Psi$ is proportional
to $N_cN_{\bar c}$=0.01, one hundredth of what we would get for the
$N_c$=$N_{\bar c}$=1 case, an underestimate by a factor of 10!  This
example illustrates the importance of the correlation between charm and
anti-charm production in charmonium regeneration.

To systematically solve this issue we employ the statistical description of charm quark pair production in a grand-canonical ensemble, where all thermodynamic properties of the charm pair system can be derived from its grand-partition function $Z$~\cite{Rafelski:1980gk}. 
Let us start with the 1-body partition function (for 1 charm or anti-charm quark),
\begin{equation}
  \label{eq:1-part}
  Z_1=\gamma_cn^{\rm{th}}_cV_{FB}=\gamma_c\, V_{FB}\,d_c\int \frac{d^3p_c}{(2\pi)^3}e^{-\sqrt{m^2_c+p^2_c}/T}\ ,
\end{equation}
where $V_{FB}$ and $T$ are the volume and temperature of the system, the charm quark fugacity $\gamma_c$ is to account for the chemical off-equilibrium of charm quarks. $K_2(x)$ is the modified Bessel function, $n^{\rm{th}}_c$ is the thermal density of charm quarks, defined in Eq.~(\ref{eq:therml_density_c}).
The partition function with $k$ charm quarks is
\begin{equation}
  \label{eq:k-part}
  Z_k=\frac{Z^k_1}{k!}\ ,
\end{equation}
where $k!$ in the denominator results from the indistinguishability of $k$ charm quarks.
If we now impose the constraint that for each charm quark there exists its partner - anti-charm quark, we obtain the partition function for $k$ charm pairs,
\begin{equation}
  \label{eq:k-pair-part}
  Z^{pair}_k=Z_k\cdot Z_k=\frac{Z^k_1}{k!}\frac{Z^k_1}{k!}\ .
\end{equation}
The $k$ (anti-)charm quarks are indistinguishable from each other, however the charm quarks are distinct from anti-charm quarks.

Next we sum over the contributions from arbitrary $k$ pairs of charm quarks and obtain the grand-partition function $Z$, satisfying the constraint $N_c-N_{\bar c}$=0 for each individual event.
\begin{align}
  \label{eq:partition}
  Z=\sum^\infty_{k=0}Z^{pair}_k=\sum^\infty_{k=0}\frac{Z^k_1Z^k_1}{k!k!}=I_0(2Z_1)\ ,
\end{align}
where $I_0$ is the modified Bessel function.  Sometimes such an ensemble is called ``canonical ensemble''. We keep in mind however that this canonicality is only referred to the strict constraint of the net charm number ($N_c-N_{\bar c}=$0) rather than the ``total'' charm number, $N_c+N_{\bar c}$, which can still fluctuate event by event.

With grand-partition function obtained we are ready to evaluate average values over events (denoted by ``$\langle \cdots \rangle$" in this section) of 
any thermodynamic quantity as 
\begin{equation}
  \label{eq:o_aver}
  \langle \mathcal{O}\rangle\equiv \sum^\infty_{k=0}\mathcal{O}(k)P(k)\ ,
\end{equation}
where 
\begin{equation}
  \label{eq:k_dist}
  P(k)=\frac{Z^{pair}_k}{Z}\ ,
\end{equation}
is the probability of an event with an integer ($k$) number of pairs of charm quarks.
For example, the average number of (open) charm pairs is
\begin{align}
  \label{eq:ncc_aver}
  \langle N_{op}\rangle=\frac{1}{Z}\sum^\infty_{k=0}kZ^{pair}_k=Z_1\frac{I_1(2Z_1)}{I_0(2Z_1)}\ .
\end{align}
Matching $\langle N_{op}\rangle$ to the experimentally measured $\langle
N_{c\bar c}\rangle$ allows the determination of $Z_1$,
Eq.~(\ref{eq:1-part}) and therefore $P(k)$ (the number of the
charmonium states in the system is numerically negligible compared to
$\langle N_{op}\rangle$).

\begin{figure}
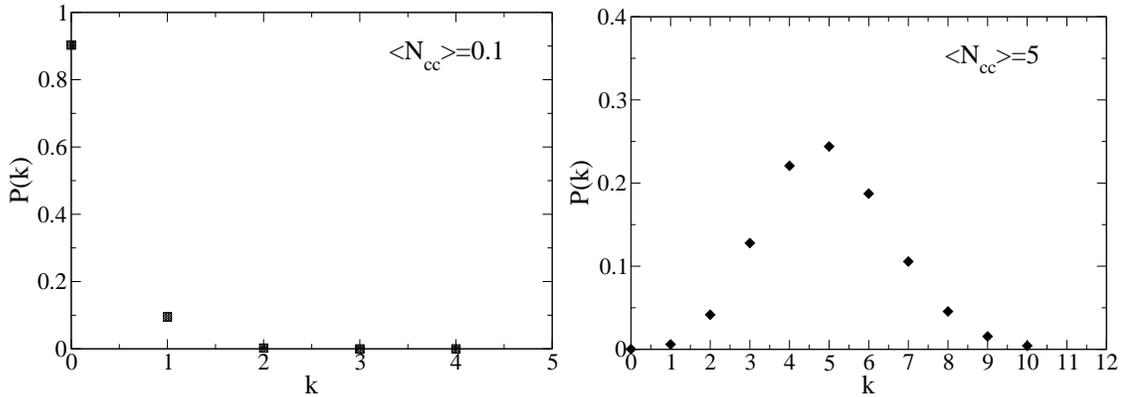

  \centering
  \includegraphics[width=0.48\textwidth]{pk_dist_sps_1019.eps}
  \includegraphics[width=0.48\textwidth]{pk_dist_rhic.eps}
  \caption[The probability distribution of integer number ($k$) pairs of charm quarks in one event]{The probability distribution of integer number ($k$) pairs of charm quarks in one event. Left panel:  $ \langle N_{c\bar c}\rangle$=0.1 corresponding to central Pb+Pb collisions (within $\Delta y$=1.8 around $y$=0) at SPS ($\sqrt s$=17.3AGeV). Right panel: $ \langle N_{c\bar c}\rangle$=5.0, corresponding to central Au+Au collisions (within $\Delta y$=1.8 around $y$=0) at RHIC ($\sqrt s$=200AGeV).}
\label{fig:ncc_distrib}
\end{figure}

In central Pb+Pb collisions at SPS ($\sqrt s$=17.3AGeV) and in central Au+Au collisions at RHIC ($\sqrt s$=200AGeV) there are on average 0.1 and 5.0 charm-quark pairs produced in the rapidity window of $\Delta y$=1.8 around $y$=0 produced, respectively. The corresponding probability distributions $P(k)$ are shown in Fig.~\ref{fig:ncc_distrib}.

Eq.~(\ref{eq:solu_boltz}) states that the number of regenerated $\Psi$ is proportional
to the gain rate, $\beta_\Psi$, which, in turn, is proportional to the
product of the charm and anti-charm quark phase space distribution,
$f_cf_{\bar{c}}$. In events with exactly $k$ pairs of charm quarks
produced, the regeneration component can be written as
$N^{\rm{reg}}_\Psi(k)=k^2N^{\rm{reg}}_1$, with $N^{\rm{reg}}_1$ being the number of $\Psi$
regenerated from 1 $c\bar{c}$ pair. Therefore the average number of
charmonia regenerated in the canonical ensemble over $k$ is
\begin{align}
  \label{eq:psi_reg_can}
  \langle N^{\rm{reg}}_\Psi\rangle=\sum^\infty_{k=0}N^{\rm{reg}}_\Psi(k)P(k)=\frac{N^{\rm{reg}}_1}{I_0(2Z_1)}{\sum^\infty_{k=0}k^2Z^{pair}_k}=N^{\rm{reg}}_1Z^2_1\ .
\end{align}
Specifically, in the equilibrium limit we have
\begin{align}
  \label{eq:psi_eq_1}
  N^{\rm{reg}}_1=N^{\rm{eq}}_1=\gamma^2_{c(1)}V_{FB}\int \frac{d^3p}{(2\pi)^3}\ d_\Psi e^{-\sqrt{m^2_\Psi+p^2}/T}=\gamma^2_{c(1)}V_{FB}n_\Psi\ ,
\end{align}
where $\gamma_{c(1)}$ satisfies 
\begin{align}
  \label{eq:gama_c1}
  1=\gamma_{c(1)}V_{FB}\int \frac{d^3p}{(2\pi)^3}\ d_c e^{-\sqrt{m^2_c+p^2}/T}=\frac{\gamma_{c(1)}Z_1}{\gamma_c}\ .
\end{align}
Substituting Eq.~(\ref{eq:gama_c1}) into Eqs.~(\ref{eq:psi_eq_1}) and (\ref{eq:psi_reg_can}) we obtain 
\begin{align}
  \label{eq:psi_reg_eq}
\langle N^{\rm{eq}}_\Psi\rangle=N^{\rm{eq}}_1Z^2_1=\gamma^2_cV_{FB}n_\Psi\ .
\end{align}
With $\langle N_{op}\rangle$ and $\langle N^{\rm{eq}}_\Psi\rangle$ known we are ready to write down the charm conservation equation
in the ``canonical'' ensemble~\cite{Gorenstein:2000ck},
\begin{align}
  \label{eq:psi_reg_can2}
  \langle N_{c\bar{c}}\rangle&=Z_1\frac{I_1(2Z_1)}{I_0(2Z_1)}+N^{\rm{eq}}_1Z^2_1\nonumber \\
&=\gamma_cn_cV_{FB}\frac{I_1(\gamma_cn_cV_{FB})}{I_0(\gamma_cn_cV_{FB})}+\gamma^2_cV_{FB}n_\Psi\ .
\end{align}
This equation allows us to solve for
$\gamma_c$ and obtain $\langle N^{\rm{eq}}_{\Psi}\rangle$. Again the second term on the r.h.s is numerically negligible due to $m_\Psi\gg m_c$.

It is instructive to examine two limits where a large
(small) number of charm-quark pairs are produced. We first note the
following property of Bessel functions,
\begin{equation}
  \label{eq:Bessel}
  \frac{I_1(x)}{I_0(x)} \rightarrow \left\{\begin{array}{l} 
      1\ , \quad x \gg 1 \\
      \frac{x}{2}\ , \quad x \ll 1 \ .
    \end{array} \right.
\end{equation}
Therefore, in these two limits, Eq.~(\ref{eq:ncc_aver}) reduces to 
\begin{equation}
  \label{eq:ncc_Bessel}
   \langle N_{c\bar c}\rangle \rightarrow \left\{\begin{array}{l} 
      Z_1\ , \quad  \langle N_{c\bar c}\rangle \gg 1 \\
      Z^2_1\ , \quad  \langle N_{c\bar c}\rangle \ll 1 \ .
    \end{array} \right.
\end{equation}
Plugging Eq.~(\ref{eq:ncc_Bessel}) into Eq.~(\ref{eq:psi_reg_can}), one obtains
\begin{equation}
  \label{eq:npsi_Bessel}
  \langle N^{\rm{reg}}_\Psi\rangle \rightarrow \left\{\begin{array}{l} 
      \langle N_{c\bar c}\rangle^2N^{\rm{reg}}_1\ , \quad  \langle N_{c\bar c}\rangle \gg 1 \\
      \langle N_{c\bar c}\rangle\, N^{\rm{reg}}_1\ , \quad  \langle N_{c\bar c}\rangle \ll 1 \ .
    \end{array} \right.
\end{equation}
In the limit of $\langle N_{c\bar c}\rangle\gg$1, the yield of the
regeneration component is proportional to $\langle N_{c\bar
  c}\rangle^2$, the same result one would get if the canonical
constraint $N_c\equiv N_{\bar c}$ was neglected; therefore the this
limit is often called the ``grand-canonical'' limit.  In the opposite
(``canonical'') limit, where $\langle N_{c\bar c}\rangle\ll$1, the
regeneration component is proportional to $\langle N_{c\bar
  c}\rangle$: the canonical constraint $N_c\equiv N_{\bar c}$
effectively enhances the charmonium regeneration by a factor of
$1/\langle N_{c\bar c}\rangle$ over the grand-canonical limit. As a
side remark, particle production in the ``canonical ensemble'' can
also be conveniently described with a kinetic master equation approach
as developed in Ref.~\cite{Ko:2000vp}, which has been applied to study
strange particle production in low energy heavy-ion
collisions~\cite{Pal:2001nk} where the net ``strangeness'',
($N_s-N_{\bar s}$), is constrained to zero in each event.

\subsection{Charm-Quark Correlation Volume}
\label{ssec:corr_vol}
It turns out that the correlation between the charm and anti-charm
quarks goes beyond ``production-in-pair'' in each event. The charm
and anti-charm quarks are produced at the same spatial point in hard
N+N collisions and then recede from each other. Before the medium
evolution stops (freeze-out) they can only diffuse into a limited
portion of the fireball volume. The volume they explore is referred to as
charm-quark correlation volume~\cite{Grandchamp:2003uw}, $V_{co}$, which in general is smaller than the
full fireball volume, $V_{FB}$. A schematic estimate of $V_{co}(t)$
will be presented in Section~\ref{ch:nume_results}.\ref{sec:trans_rate_eq}. The restriction of $c\bar{c}$
pairs to within the correlation volume effectively increases the $c\bar
c$ coalescence probability: One charm quark can more easily find its
partner nearby inside the correlation volume $V_{co}$ with the effective spatial
density, $n^{eff}_{\bar c}\equiv d_{\bar{c}}\int \frac{d^3p}{(2\pi)^3} f^{eff}_{\bar c}(p)$, of its partner given by
\begin{equation}
\label{eq:fc_vco}
 n^{eff}_{\bar c}=\frac{k}{kV_{co}}=\frac{1}{V_{co}}\ ,
\end{equation}
for an event in which $k$ charm quark pairs are produced. The
effective density, $n^{eff}_{\bar c}$, is larger than $k/V_{FB}$ (the
$\bar{c}$ density without correlation volume effect) by a factor of
$V_{FB}/(kV_{co})$, if we assume $k$ anti-charm quarks residing in $k$
correlation volume ``bubbles'' (no merging of correlation volumes).

It is instructive to work out the parametric dependence of the
regeneration component, $N^{\rm{reg}}(k)$, on the charm quark pair number $k$ in
two limits:

(a) In the limit of $\langle N_{c\bar
  c}\rangle\ll V_{FB}/V_{co}$ the overlap of correlation
``bubbles'' can be neglected; we have $N^{\rm{reg}} (k)= [V_{FB}/(kV_{co})]\cdot k^2N^{\rm{reg}}_1$\ .  

(b) In the opposite limit, $\langle N_{c\bar c}\rangle \gg
V_{FB}/V_{co}$, the $k$ correlation volume ``bubbles'' maximally overlap with
each other with their total volume filling the entire fireball volume
$V_{FB}$, leading to an effective anti-charm quark density 
$n^{eff}_{\bar c}=k/V_{FB}$. In this limit there is no correlation volume effect so we have $N^{\rm{reg}} (k)=k^2N^{\rm{reg}}_1$.
In these two limits the average number of regenerated $\Psi$ is
\begin{equation}
   \label{eq:ncc_vco}
  \langle N^{\rm{reg}}_\Psi\rangle =\sum^\infty_{k=0}N^{\rm{reg}}(k)P(k)\rightarrow \left\{\begin{array}{l} 
      N^{\rm{reg}}_1Z_1\frac{I_0(2Z_1)}{I_1(2Z_1)}\frac{V_{FB}}{V_{co}}\ , \quad\quad  \langle N_{c\bar c}\rangle \ll V_{FB}/V_{co} \\
      N^{\rm{reg}}_1Z^2_1 , \quad\quad\quad\quad\qquad\,  \langle N_{c\bar c}\rangle \gg V_{FB}/V_{co} \ .
    \end{array} \right.
\end{equation}

Comparing Eq.~(\ref{eq:ncc_vco}) with Eq.~(\ref{eq:ncc_aver}) one sees
that in the case of $1\ll\langle N_{c\bar c}\rangle \ll V_{FB}/V_{co}$
the correlation volume effect renders $\langle N^{\rm{reg}}_\Psi\rangle$
to depend linearly on $\langle N_{c\bar c}\rangle$ even if the
$c\bar{c}$ system is in the ``grand-canonical'' limit. In other words,
the correlation volume effect ``delays'' the transition between the
``canonical'' and ``grand-canonical'' ensembles from $\langle N_{c\bar
  c}\rangle\sim 1$ to $\langle N_{c\bar c}\rangle\sim
V_{FB}/V_{co}$. In the general case where $\langle N_{c\bar c}\rangle$ is
comparable to $V_{FB}/V_{co}$ the merge of correlation volumes needs
to be taken care of on a term-by-term basis in the series of
$N^{\rm{reg}}(k)$. Whenever two correlation volume bubbles merge their new
total volume (smaller than the sum of their individual volumes) should
be used for estimating the effective anti-charm density
$n^{eff}_c$. In Section~\ref{ch:nume_results}.\ref{sec:full_boltz}.\ref{ssec:raa_centra_cquark}, we will study the sensitivity of
$J/\psi$ regeneration on different correlation volume sizes based on a
rather schematic prescription of merging correlation volumes: If at any
given time, $t$, the total correlation volume of individual ``bubbles''
$kV_{co}(t)$ is larger than the fireball volume $V_{FB}(t)$,
$kV_{co}(t)$ is set to $V_{FB}(t)$ for determination of the effective
anti-charm density $n^{eff}_c$ in the subsequent evolution.

We conclude this section by pointing out that a more rigorous and
systematic way to account for canonical-ensemble and
correlation volume effects is to use the joint phase space
distribution function, $\langle f_cf_{\bar{c}}\rangle$, as the input
for $\Psi$ regeneration rate, $\beta_\Psi$. The $\langle
f_cf_{\bar{c}}\rangle$ can be obtained, \eg, from high statistics
Langevin simulations of the evolution of charm quark pairs in the medium.

\graphicspath{{./3fg/}}

\chapter{Charmonium in the hot medium}
\label{ch:psi_hot}

In this chapter we discuss the microscopic mechanisms for the
reactions of charmonia, $\Psi$ ($\Psi$=$J/\psi$, $\chi_c$, $\psi'$), in
the hot medium created in heavy-ion collisions. In
Section~\ref{ch:psi_hot}.\ref{sec:qgp}, we review the relevant properties of $\Psi$ in
the QGP relevant for charmonium production. In
Section~\ref{ch:psi_hot}.\ref{sec:charm-diss-qgp}, we discuss the charmonium
dissociation mechanisms for $\Psi$ with reduced in-medium binding energy
$\epsilon_B<T$, the quasifree-dissociation process of charmonium
dissociation, and compare to the more traditional
mechanism, gluo-dissociation process. In
Section~\ref{ch:psi_hot}.\ref{sec:open_charm_med}, we explicitly calculate charmonium
regeneration rates in QGP with several input charm-quark spectra. In
Section~\ref{ch:psi_hot}.\ref{sec:charm-diss-hm}, we briefly discuss charmonium
dissociation in the hadronic matter (at $T$ below the critical temperature $T_c$).  In
Section~\ref{ch:psi_hot}.\ref{sec:charmo_spf}, we construct charmonium spectral
functions based on the quasifree dissociation rates and compare to
lattice calculations, through which we extract charmonium dissociation
temperatures, $T^\Psi_{diss}$.


\section{Charmonium in QGP}
\label{sec:qgp}

\begin{figure}[!t]
\centering
\includegraphics[width=0.59\textwidth]{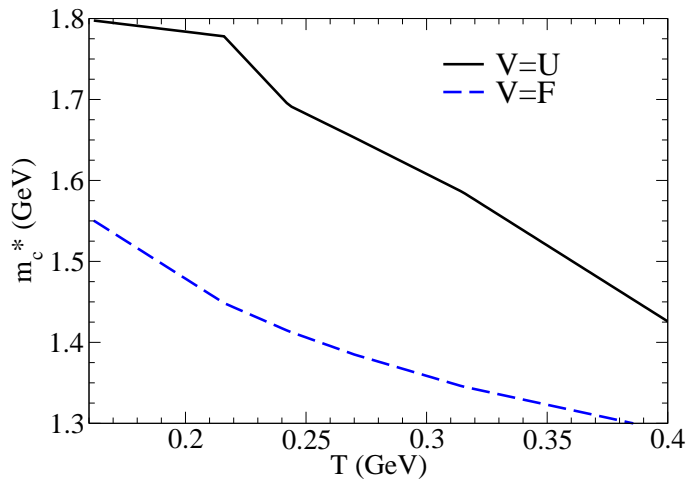}
\caption[Temperature dependence of in-medium charm quark mass in the
strong and weak binding scenarios]{Temperature dependence of in-medium
  charm quark mass in the strong (solid line) and weak binding (dashed
  line) scenarios. Figure taken from Ref.~\cite{Zhao:2010nk}. }
\label{fg:mc}
\end{figure}

On the microscopic level the medium affects charmonium states in
three ways: 

1) Debye screening: The QGP is a deconfined medium
therefore the freely moving colored partons screen the binding force
of a charmonium state and decrease its binding energy,
$\epsilon_B(T)$.  The in-medium $\epsilon_B(T)$ is defined as 
\begin{equation}
  \label{eq:psi_binding}
\epsilon_B(T)=2m^*_c(T)-m_\Psi(T)\ ,  
\end{equation}
where $m^*_c$ and $m_\Psi$ are in-medium charm-quark and charmonium
masses. The in-medium part of $m^*_c$ is identified with the
asymptotic value of the heavy-quark (HQ) potential, as displayed in
Fig.~\ref{fg:mc}.
\begin{equation}
m^*_c(T)\equiv m^0_c+V_{Q\bar Q}(r\to\infty;T)/2 \ . 
\end{equation}
In this work we approximate the HQ potential, $V_{Q\bar Q}(r;T)$, based on
the lQCD heavy-quark free energy, $F_{Q\bar{Q}}(r;T)$, (recall Section~\ref{ch:intro}.\ref{sec:charmo}) within either the ``strong''
($V_{Q\bar{Q}}=U_{Q\bar{Q}}(r;T)$) or ``weak'' binding
($V_{Q\bar{Q}}=F_{Q\bar{Q}}(r;T)$) scenario. 
The in-medium masses decrease with temperature appreciably, while the 
magnitude of $m^*_c(T)$ is significantly smaller in the
weak-binding compared to the strong-binding scenario. 

\begin{figure}[!t]
\centering
\includegraphics[width=0.59\textwidth]{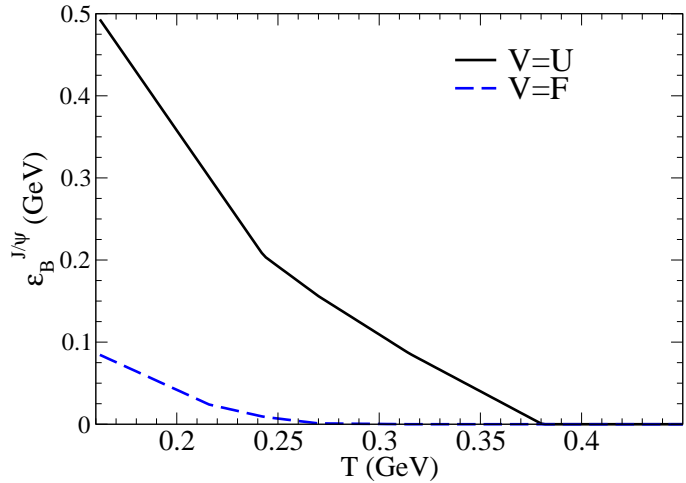}
\caption[Temperature dependence of $J/\psi$ binding energy in the
strong and weak binding scenarios]{Temperature dependence of $J/\psi$
  binding energy in the strong (solid line) and weak binding (dashed
  line) scenarios. }
\label{fg:epsB}
\end{figure}

For a quantitative estimate of $\epsilon_B(T) $ we take recourse to the
potential model of Ref.~\cite{Riek:2010fk} where quarkonium spectral
functions and correlators (recall Eq.~(\ref{eq:spectral}) and Eq.~(\ref{eq:Gtau})) have been calculated in a thermodynamic
$T$-matrix approach, consistent with vacuum spectroscopy and including
relativistic corrections for a proper description of scattering
states. The calculations in there have been carried out for both free
and internal energies as potential, and for two different lQCD
inputs~\cite{Petreczky:2004pz,Kaczmarek:2007pb}. Since the internal
energy leads to stronger binding than the free energy we refer to the
former and latter as strong- and weak-binding scenario,
respectively. In both scenarios (and for both potentials), an
approximate constancy (within $\pm$15\%) of the correlator ratios, Eq.~(\ref{eq:corr_ratio}), for
pseudoscalar charmonium has been found (see lower panels of Fig.~12
and 14 in Ref.~\cite{Riek:2010fk}).  We believe that these results
provide a reasonable bracket for potential-model
results. 
Similar to $m^*_c(T)$, the resulting binding energies (plotted in
Fig.~\ref{fg:epsB}) also decrease with $T$, again being significantly
smaller in the weak-binding scenario. These features are, in fact, the
main reason that both scenarios can be compatible with the small
variations found in the lQCD correlator ratios: for weak/strong
binding, a small/large constituent mass combines with a small/large
binding energy, respectively, leading to an approximate compensation
in the bound-state mass, $m_\Psi(T)$, recall
Eq.~(\ref{eq:psi_binding}).

  
2) Collisional dissociation/regeneration: Charmonia are subject to dissociation through collisions with particles in the medium even if they have finite binding energies. We assume both chemical and thermal equilibrium for light partons, so that their $p_t$ spectra follow the thermal Bose/Fermi distribution,
 \begin{eqnarray}
 & f^i_B(k,T) & = \frac{1}{\exp\left(\frac{\sqrt{k^2+m^2_i}}{T}\right) -1}\
 , \qquad i=g,\label{eq:parton_th_dist_b} \\
 & f^i_F(k,T) & = \frac{1}{\exp\left(\frac{\sqrt{k^2+m^2_i}}{T}\right) +1}\
 , \qquad i=u,\bar{u},d,\bar{d},s,\bar{s}\label{eq:parton_th_dist_f}\ .
\end{eqnarray}
Their thermal masses $m_i$ are guided by perturbative QCD (pQCD) calculations,
\begin{eqnarray}
  m_{u,d}^2 & = &\frac{g^2T^2}{3}\ , \label{eq:m_ud} \\
  m_s^2 & = & m_0^2 + \frac{g^2T^2}{3}\ , \label{eq:m_s} \\
  m_g^2 & = & \left(1+\frac{N_f}{6}\right)\frac{g^2T^2}{2} \ , \label{eq:m_g}
\end{eqnarray}
where $N_f$=2.5 is the number of flavors, and the coupling constant
$g$ is adjusted so that the resulting energy density of the QGP
medium, $e(T)$, reproduces the lattice-QCD results, see
Chapter~\ref{ch:fireball} for more
details. 
The microscopic mechanisms of the scattering between $\Psi$ and
particles in the heatbath will be detailed in the next section.

3) Bose-enhancement/Pauli blocking: In charmonium dissociation or
regeneration processes light partons may be produced in the final
state, \eg, $i+\Psi \rightarrow i + c + \bar{c}(i={g,q,\bar{q}})$.
In the QGP the final state phase space of the light particle is
altered due to the Bose-enhancement $(1+f^i_B(k,T))$ and Pauli blocking
$(1-f^i_F(k,T))$ factors for gluons and quarks, respectively, recall Eqs.~(\ref{eq:diss_xsec3}), (\ref{eq:diss_rate3}) and (\ref{eq:gain_rate3}).

\section{Charmonium Dissociation in the QGP}
\label{sec:charm-diss-qgp}

\begin{figure}[tp]
  \centering
  \includegraphics[width=0.3\textwidth,clip=]{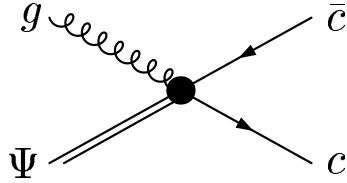}
  \caption[Diagram of the gluo-dissociation process]{Diagrams of the gluo-dissociation process. A charmonium absorbs a gluon and is subsequently dissociated.}
\label{fg:gdiss}
\end{figure}

\begin{figure}[!t]
  \centering
  \includegraphics[width=0.59\textwidth,clip=]{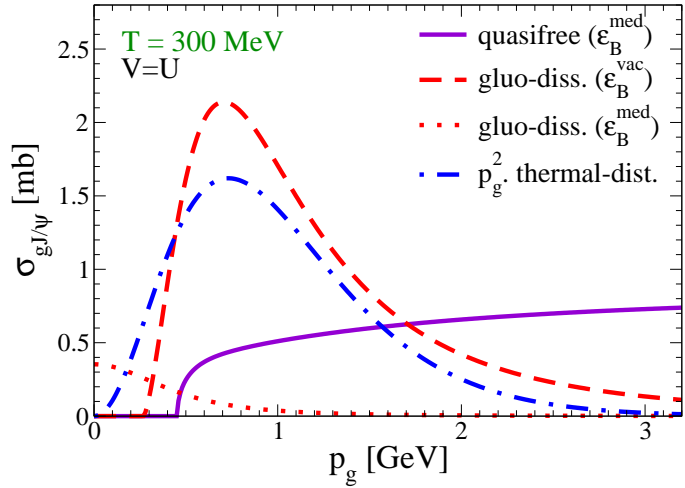}
  \caption[Comparison of the parton-induced dissociation cross section]{Comparison of the parton-induced dissociation cross section. Solid line: quasifree dissociation mechanism; dashed line: gluo-dissociation with vacuum $J/\psi$ binding energy; dotted line: gluo-dissociation cross section with in-medium $J/\psi$ binding energy of 110MeV; dot-dashed line: gluon thermal distribution $p^2_g\,f_g(p_g)$ at $T$=300MeV.}
\label{fg:rate_xsec}
\end{figure}

Let us start with the traditional gluo-dissociation process
($\Psi+g\to c+\bar c$) as illustrated in Fig.~\ref{fg:gdiss}, proposed
by Bhanot and Peskin in the 1970s~\cite{Peskin:1979va}. They evaluated the
interaction between a charmonium state and a gluon within an Operator
Product Expansion (OPE) formalism. The leading order in $\alpha_s$
turns out to be a dipole interaction ($\vec r \cdot \vec E$) between
the $\Psi$ and the gluon. For $J/\psi$ dissociation cross section it results in the following expression,
\begin{equation}
  \label{eq:gluo-diss}
  \sigma_{gJ/\psi}(E_g) = \frac{2\pi}{3} \left(\frac{32}{3}\right)^2
  \left(\frac{m_c}{\epsilon_B}\right)^{1/2} \frac{1}{m_c^2}
  \frac{(E_g/\epsilon_B-1)^{3/2}}{(E_g/\epsilon_B)^5}\ , 
\end{equation}
where the $m_c$ is the mass of the charm quark, $E_g$ is the energy of
incoming gluon and $\epsilon_B$ is the binding energy of $J/\psi$, see
Eq.~(\ref{eq:psi_binding}). We note that the cross section
$\sigma_{gJ/\psi}(E_g)$ exhibits a pronounced peak at
$E_g=\frac{10}{7}\epsilon_B$, see Fig.~\ref{fg:rate_xsec}. Note that
since we adopt a massive gluon with $m_g|_{T=300\rm{MeV}}\simeq$580MeV (see
Eq.~(\ref{eq:m_g})), the peak appears at gluon momentum
$p_g\simeq$700MeV for vacuum $J/\psi$ binding energy of 640MeV.
The limitations of the gluo-dissociation process include: (1) 
As mentioned in Ref.~\cite{Peskin:1979va} 
the OPE procedure is valid only if the energy scale of the incoming gluon
is much less the charmonium binding $E_g\ll\epsilon_B$. (2) Since
Eq.~(\ref{eq:gluo-diss}) is only to the leading order in $\alpha_s$, it
does not include the interaction between $J/\psi$ and light quarks.

Indeed, with in-medium binding energy
$\epsilon_B<T$ the gluo-dissociation mechanism turns out to be not numerically efficient in
destroying $J/\psi$. To see this we calculate 
the dissociation rate, $\Gamma_\Psi$, which is a
convolution of the density of incoming particles (gluons in this
context), $f_g(p_g)$, with the dissociation cross section,
$\sigma_{g\Psi}(s)$ and the relative velocity $v_{rel}$ between the
incoming particle and the charmonium (cf. Eqs.~(\ref{eq:diss_rate}) and (\ref{eq:rate_xsec})),
\begin{equation}
  \label{eq:rate_xsec4}
  \Gamma_\Psi(p,x)=\int \frac{d^3p_g}{(2\pi)^3}\,d_gf_g(p_g)\,v_{rel}\,\sigma_{g\Psi}(s)\ .
\end{equation}
Here the gluon momentum distribution $f_g(p_g)$ follows from Eq.~(\ref{eq:parton_th_dist_b}).

Figure~\ref{fg:rate_xsec} shows that the inefficiency of the
gluo-dissociation for $J/\psi$ with in-medium binding energy
originates from the peak structure of the gluo-dissociation cross
section. Note that since we use a massive gluon with
$m_g|_{T=300\rm{MeV}}\simeq$580\,MeV, the gluon energy at $p_g$=0 is
already significantly larger than
$\frac{10}{7}\epsilon_B|_{T=300\rm{MeV}}$$\sim$160\,MeV, therefore
only the ``tail'' of the gluo-dissociation cross section appears in
Fig.~\ref{fg:rate_xsec}, leading to a small overlap between the
thermal gluon distribution $p^2_g\,f_g(p_g)$ and the dissociation
cross section, $\sigma_{gJ/\psi}(p_g)$. The gluo-dissociation cross
sections for $\chi_c$ and $\psi'$ show similar peak structures,
resulting in small in-medium dissociation rates.

On the other hand we know that in the limit of incoming gluon energy large compared to the charmonium binding energies, $\epsilon_B>E_g$, the interaction between the $c$ and $\bar c$ inside the charmonium, which occurs on a time scale of $1/\epsilon_B>1/E_g$, cannot interfere with the interaction between the incoming gluon and the $c(\bar c)$. In this limit the $g-\Psi$ scattering cross section should approach the value of $2\sigma_{gc}$, the sum of the probability of the scattering between the gluon and one of the constituent charm quarks. Therefore the peak structure of gluo-dissociation cross section will be superseded by continuum-like ``quasifree'' dissociation cross section.

\begin{figure}[tp]
  \centering
  \includegraphics[width=0.3\textwidth,clip=]{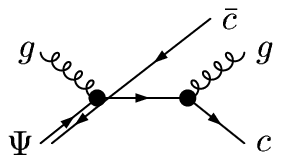}
  \hspace{12mm}
  \includegraphics[width=0.3\textwidth,clip=]{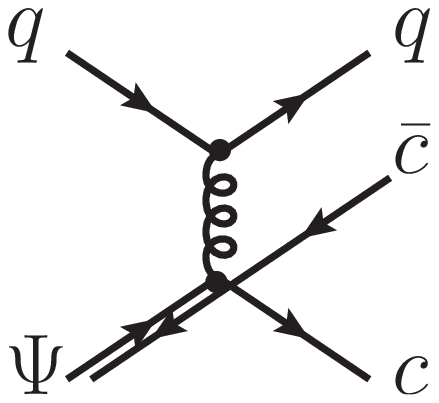}
  \caption[Diagrams of the quasifree process]{Diagrams of the quasifree process. A charmonium can be dissociated by either a quark or a gluon.}
\label{fg:qfree}
\end{figure}

Following this idea an alternative (quasifree) dissociation mechanism, $i+\Psi \rightarrow i + c
+ \bar{c}(i={g,q,\bar{q}} )$, 
was proposed to describe the inelastic scattering between the
charmonium and partons~\cite{Grandchamp:2001pf}. In these processes the incoming parton (gluon or
quark) collides with the $c$ or $\bar{c}$ quark in the
$\Psi$. Since in the QGP charmonia are loosely bound states one such 
collision would be enough to dissociate the $\Psi$, as illustrated in
Fig.~\ref{fg:qfree}.
The transition matrix element for the quasifree dissociation can therefore be factorized as
\begin{align}
  \label{eq:m_qfree}
&\lvert\overline{\mathcal M_{\Psi i\to c\bar{c}i}(p_i,p_\Psi;\bar{p}_i,p_c,p_{\bar{c}})}\rvert^2\delta^{(4)}(p_i+p_{\Psi}-p_c-p_{\bar{c}}-\bar{p}_i)\nonumber\\[1mm]
&\qquad\qquad\rightarrow 2\ \left\lvert\overline{\mathcal M_{ci\to ci}\left(p_i,\frac{m_{c'}}{m_\Psi}p_\Psi;\bar{p}_i,p_c\right)}\right\rvert^2\delta^{(4)}\left(\frac{m_{c'}}{m_\Psi}p_\Psi+p_i-p_c-\bar{p}_i\right)\nonumber\\
&\qquad\qquad\qquad\times(2\pi)^3\ (2E_{\bar{c}})\ \delta^{(3)}\left(\frac{m_c}{m_\Psi}p_\Psi-p_{\bar{c}}\right)\ ,
\end{align}
where $\lvert\overline{\mathcal M_{\Psi i\to c\bar{c}i}}\rvert^2$ is the quasifree dissociation transition matrix element, and $\lvert\overline{\mathcal M_{ci\to ci}}\rvert^2$ the transition matrix element for the elastic scattering between $i$ and $c$, calculated in Ref.~\cite{Combridge:1978kx}. The momenta $p_i$, $p_\Psi$, $\bar{p}_i$, $p_c$ and $p_{\bar{c}}$ correspond to the initial state parton $i$, initial $\Psi$, final state parton $i$, final state $c$ and final state $\bar{c}$.
The overall factor of 2 is accounting for the fact that the elastic scattering with either $c$ or the $\bar{c}$ can dissociate the $\Psi$. 


To account for the leading kinematic correction from the residual
binding energy, the incoming parton needs to be energetic enough to
break up the bound state, which sets a lower limit for the incoming
parton momentum.  Overall 4-momentum conservation for the process
$i+\Psi \to i+c+\bar c$ is maintained by assigning the binding energy
to a decrease in mass of the initial-state charm-quark, $c'$, \ie,
$m_{c'} = m_c -\epsilon_B$. In addition, we have introduced a Debye
mass, $m_D=gT$, into the denominator of $t$-channel gluon-exchange
propagator, $1/t\to 1/(t-m^2_D)$, to regulate the divergence for
forward scattering (the strong coupling in $m_D$ is taken consistently
with the coupling constant $\alpha_s$ used for the quasifree process).

Inserting the quasifree transition probability Eq.~(\ref{eq:m_qfree})
into Eq.~(\ref{eq:diss_rate3}) we can evaluate the corresponding
dissociation cross section and rate. The comparison of the
dissociation cross sections between the quasifree dissociation
processes and the gluo-dissociation process is shown in
Fig.~\ref{fg:rate_xsec}. In contrast to the gluo-dissociation cross
section the quasifree dissociation cross section saturates for partons
with large incoming momentum ($k\gg\epsilon_B$).
Since the quasifree process may not be the only dissociation mechanism for $\Psi$, for practical applications we effectively parameterize other dissociation mechanism into the quasifree processes by using the strong coupling constant $\alpha_s$ (figuring into $|\mathcal M_{ci\to ci}|^2$) as an adjustable parameter. We adjust it to the $J/\psi$ suppression data measured at SPS/RHIC, see Chapter~\ref{ch:nume_results} for the detailed procedure. The resulting value of $\alpha_s$ turns out to be $\simeq$0.3, quite compatible with the short-distance (color-Coulomb) term in the effective potential used to extract the binding energies by the $T$-matrix potential model. Note that the strong coupling constant in the quasifree process is decoupled from that figuring into parton thermal masses in Eqs.~(\ref{eq:m_ud}), (\ref{eq:m_s}) and (\ref{eq:m_g}).

\begin{figure}[tp]
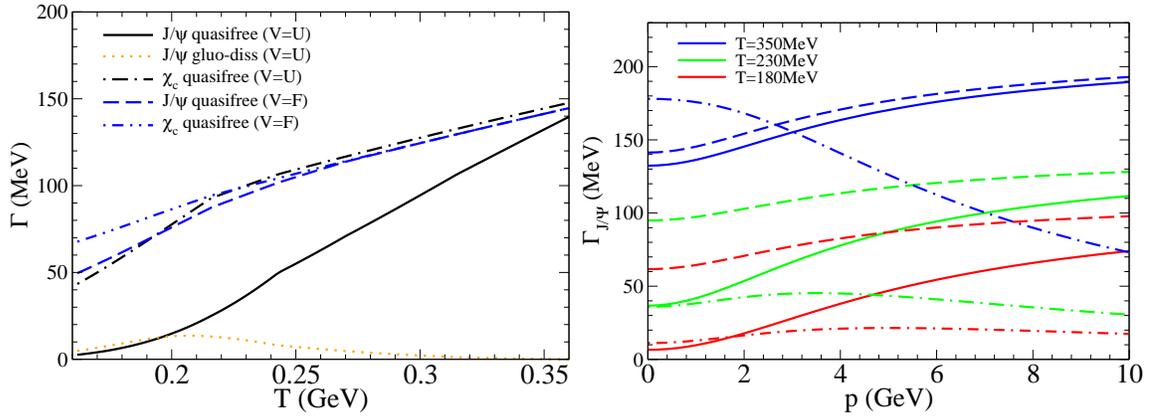

  \centering
  \includegraphics[width=0.49\textwidth]{rate_vs_T_0923.eps}
  \includegraphics[width=0.49\textwidth]{rate_vs_p_0923.eps}
  \caption[Temperature and momentum dependence of $\Psi$ dissociation
  rates]{Temperature and momentum dependence of $\Psi$ dissociation
  rates. Left panel: Temperature dependence of dissociation rates (at
    $p$=0) for $J/\psi$ and $\chi_c$. Solid line: $J/\psi$ quasifree
    dissociation rate in the strong binding scenario; dashed line:
    $J/\psi$ quasifree dissociation rate in the weak-binding scenario;
    dot-dashed line: $\chi_c$ quasifree dissociation rate in the
    strong-binding scenario; double-dot-dashed line: $\chi_c$
    quasifree dissociation rate in the weak-binding scenario; dotted
    line: $J/\psi$ gluo-dissociation rate in the strong binding
    scenario with $\alpha_s$=0.3. Right panel: Momentum dependence of
    $J/\psi$ quasifree dissociation rates for strong binding scenario
    (solid line); weak binding scenario (dashed line) and
    gluo-dissociation rate with vacuum $J/\psi$ binding energy
    (dot-dashed line).}
  \label{fg:rate_temp}
\end{figure}

The temperature dependence of the quasifree dissociation rates 
for the $J/\psi$ and $\chi_c$ in the QGP is plotted for $\vec p$=0
 in the left panel of Fig.~\ref{fg:rate_temp}, and for the
$J/\psi$ as a function of $p$ at selected temperatures in the right panel.
 
 
In the weak-binding scenario, there is rather little difference
between the dissociation rates of $J/\psi$ and $\chi_c$, especially
above $T=200$\,MeV. Only in the strong-binding scenario the larger
$J/\psi$ binding energy makes a large difference, suppressing its
destruction by, \eg, a factor of $\sim$5 at $T\simeq$200\,MeV relative
to the $\chi_c$ and $\psi'$ (not shown); this difference becomes
larger (smaller) at smaller (larger) $T$. For comparison we also
calculated the rate due to the gluo-dissociation mechanism employing
the expression derived in Ref.~\cite{Peskin:1979va} with the same
$\alpha_s\simeq$0.3 as in the quasifree rate and with $\epsilon_B$
obtained from the strong-binding scenario (note that the Coulombic
binding is much smaller), which turns out to be inefficient for
dissociating $J/\psi$'s (and even more so for the excited states) and
is thus neglected in the following. We also note that to achieve a
comparable dissociation rate, say, $\Gamma_{J/\psi}\sim 100$\,MeV at
$T\sim 300$\,MeV in a hadronic medium, one needs a hadron density of about
5/fm$^3$ for typical thermal averaged hadronic dissociation cross section of around
\,1mb~\cite{Lin:1999ad}.

The 3-momentum dependence of
the rates shows a monotonous increase with increasing $p$, which
becomes more pronounced with increasing binding energy (for larger
$\epsilon_B$ a finite 3-momentum facilitates the break-up since, on
average, a larger center-of-mass energy is available in the collision
of the bound state with thermal partons). This increase is a simple
kinematic consequence of a monotonously increasing (or even constant)
cross section with finite threshold and an increasing parton flux
encountered by a moving $J/\psi$. 


\section{Charm-Quark Spectra and Charmonium Regeneration in the QGP }
\label{sec:open_charm_med}
In the previous section we discussed in-medium dissociation of
charmonium states on the microscopic level. The inverse process,
the regeneration of charmonium states from charm quarks, should also
occur and obey the principle of detailed balance with the
dissociation processes. The expressions for the regeneration rates, Eq.~(\ref{eq:gain_rate}) and Eq.~(\ref{eq:gain_rate3}), are
similar to those of the dissociation rates, Eq.~(\ref{eq:diss_rate}) and Eq.~(\ref{eq:diss_rate3}). The only difference is
that the charm-quark phase space distributions, $f_{c(\bar{c})}(
x,p_c,t)$, appear in the initial phase space. Below we first give a
brief review of the typical time evolution of $f_{c(\bar{c})}(x,p_c,
t)$:

Since the $c$ and $\bar c$ quarks are produced in hard collisions
their initial transverse momentum spectra are rather ``hard'', with
$\langle p^2_t\rangle\simeq$3\,GeV$^2$. (As an estimate for a typical
temperature of $T$=250\,MeV in heavy-ion collisions, the thermal
momentum spectra have $\langle p^2_t\rangle\simeq$1\,GeV$^2$, much
smaller than the initial spectra from hard collision.) Later they
collide with particles in the (thermalized) medium and gradually
equilibrate their momentum with the heatbath. If the medium does not
cool down they will eventually be thermalized and their momentum
spectra follow the thermal spectra, $f_{c(\bar{c})}(p_c)\propto
e^{-E_c/T}$ with $E_c=\sqrt{m^2_c+p^2_c}$. However, compared to light
quarks the charm quark thermalization is a slow process, ``delayed''
relative to light quarks by a factor of $\sim m_c/T\simeq$ 5. Since the
thermalization time for the bulk medium is of order $\sim$
1fm/$c$, the thermal relaxation time for charm quarks is expected be
on the order of $\sim$ 5fm/$c$~\cite{Rapp:2009my}, which is comparable
to the lifetime of QGP at RHIC. Therefore, during the lifetime of the
hot medium the charm-quark spectra are expected to be partially
thermalized. Quantitative calculations of charm quark thermalization
can be performed with a Boltzmann transport equation for charm quarks.

The Boltzmann equation for charm quarks is an intego-differential
equation, which is numerically difficult to handle. Therefore
Fokker-Planck equations and Langevin simulation techniques are often
employed as approximations to the Boltzmann transport equation to
describe the time evolution of charm-quark phase space distribution
function, see Ref.~\cite{Rapp:2009my} for a recent review.

In this work we compare charmonium regeneration rates resulting from two limiting
cases of charm-quark spectra: \\
1) fully thermalized charm-quark
spectra. In the local rest frame, they have the following form,
\begin{equation}
 \frac{dN^{\rm{th}}_c}{d^3p}\propto e^{-\sqrt{m^2_c+p^2}/T}\ .
\label{eq:f_c_th}
\end{equation}
\\
2) spectra from initial hard production (pQCD spectra). We
employ a parameterization given in Ref.~\cite{vanHees:2004gq},
\begin{equation}
 \frac{dN^{\rm{pqcd}}_c}{d^3p}\propto  \frac{(p+A)^2}{(1+p/B)^\alpha}\ ,
\label{eq:f_c_pqcd}
\end{equation}
with $A$=0.5\,GeV, $B$=6.8\, GeV, $\alpha$=21. This parameterization is based on the charm-quark $p_t$ spectra generated in 200\,GeV proton-proton (p+p) collisions by PYTHIA~\cite{Sjostrand:2000wi}.\\
3) pQCD spectra (\ref{eq:f_c_pqcd}) in transverse direction, and a
thermal spectrum in longitudinal ($p_z$) direction,
\begin{equation}
 \frac{dN^{\rm{pqcd+th}}_c}{d^2p_tdp_z}\propto  \frac{(\sqrt{p^2_t+p^2_z}+A)^2}{N(p_z)(1+\sqrt{p^2_t+p^2_z}/B)^\alpha}\times e^{-\sqrt{m^2_c+p^2_z}/T}\ ,
\label{eq:f_c_pqcd21}
\end{equation}
where 
\begin{equation}
 N(p_z)=\int d^2 p_t  \frac{(\sqrt{p^2_t+p^2_z}+A)^2}{(1+\sqrt{p^2_t+p^2_z}/B)^\alpha}\ .
\label{eq:f_c_pqcd21_npz}
\end{equation}
The thermal distribution in longitudinal direction is to mimic the
longitudinal smearing of the center of the mass momentum due to
various momentum fractions, $x$, carried by the two colliding primordial partons in
$c\bar{c}$ production process $g(q)+g(\bar{q})\to c+\bar{c}$. We
denote these spectra as ``pQCD+thermal'' spectra.

\begin{figure}[tp]
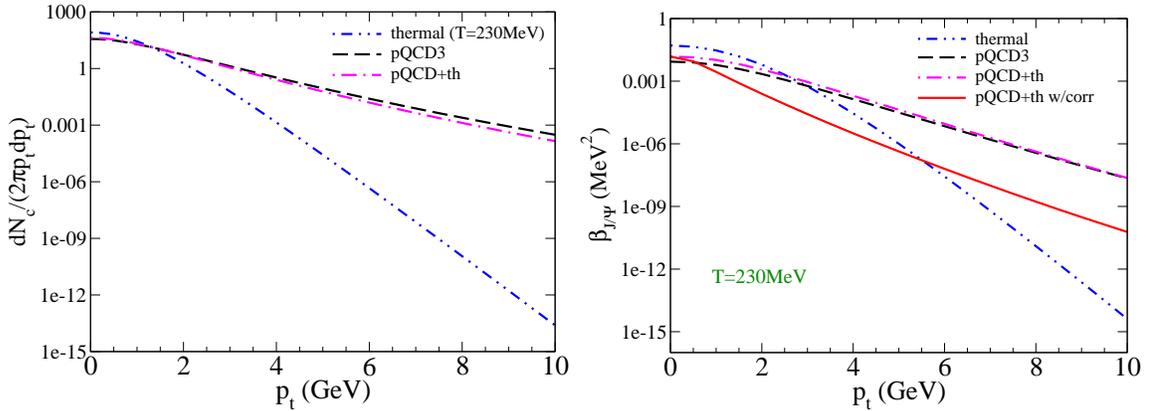

  \centering
  \includegraphics[width=0.49\textwidth]{charm_spectra_0923.eps}
  \includegraphics[width=0.49\textwidth]{reg_rate_0923.eps}
  \caption[Charm-quark $p_t$ spectra and $J/\psi$ regeneration rates]{Charm-quark $p_t$ spectra and $J/\psi$ regeneration rates. Left panel: Comparison of different input charm-quark $p_t$ spectra. Right panel: their resulting charmonium regeneration rates evaluated in the strong binding scenario. Dashed line: 3-dimensional isotropic pQCD charm-quark spectra; dot-dashed line:transversely pQCD + longitudinally thermal spectra; double-dot-dashed line: thermal charm-quark spectra with $m_c$=1.73\,GeV; solid line: transversely pQCD with angular correlation + longitudinally thermal spectra.}
\label{fg:rates_reg}
\end{figure}

The three types of charm-quark distribution are compared in the left
panel of Fig.~\ref{fg:rates_reg} in terms of the $p_z$-integrated
$p_t$ spectra. They are all normalized to the same total charm pair
number. The thermal spectra are the softest ($\langle
p^2_t\rangle$=1.1\,GeV$^2$), with most of the yield concentrated at low
$p_t$. The 3-dimensionally isotropic pQCD spectra ($\langle
p^2_t\rangle$=3.5\,GeV$^2$) and the transversely pQCD + longitudinally
thermal spectra ($\langle p^2_t\rangle$=3.0\,GeV$^2$) are almost of
comparable hardness in the transverse plane, with the 2+1-dimensional
pQCD spectra being slightly softer due to their thermalization in $z$
direction.

As discussed in Section~\ref{ch:psi_hot}.\ref{sec:charm-diss-qgp} the quasifree process
is the dominant dissociation mechanism for in-medium $\Psi$. According
to the principle of detailed balance the inverse quasifree process, $
i + c + \bar{c}\rightarrow i+\Psi\ (i={g,q,\bar{q}})$, is the
corresponding regeneration processes. Its transition matrix element
can be obtained by applying detailed balance,
Eq.~(\ref{eq:d_balance}), to the quasifree dissociation matrix element,
Eq.~(\ref{eq:m_qfree}),
\begin{align}
  \label{eq:m_qfree_reg}
&\lvert\overline{\mathcal M_{c\bar{c}i\to \Psi i}(p_i,p_c,p_{\bar{c}};\bar{p}_i,p_\Psi)}\rvert^2\delta^{(4)}(p_c+p_{\bar{c}}+p_i-\bar{p}_i-p_{\Psi})\nonumber\\[4mm]
&\qquad\qquad\rightarrow 2\ \left(\frac{d_\Psi}{d_cd_{\bar{c}}}\right)\ \left\lvert\overline{\mathcal M_{ci\to ci}\left(p_i,p_c\,;\, \bar{p}_i,\frac{m_{c'}}{m_\Psi}\,p_\Psi\right)}\right\rvert^2\ \delta^{(4)}\left(p_i+p_c-\bar{p}_i-\frac{m_{c'}}{m_\Psi}\,p_\Psi\right)\nonumber\\
&\qquad\qquad\qquad\times(2\pi)^3\ (2E_{\bar{c}})\ \delta^{(3)}\left(p_{\bar{c}}-\frac{m_c}{m_\Psi}\,p_\Psi\right)\ .
\end{align}
Here, $\lvert\overline{\mathcal M_{ci\to ci}}\rvert^2$ is the
transition matrix for the elastic scattering between $i$ and $c$. The
momenta of initial state parton $i$, initial state $c$, initial state
$\bar{c}$, final state $i$ and final state $\Psi$ are $p_i$, $p_c$,
$p_{\bar{c}}$, $\bar{p}_i$, and $p_\Psi$. $m_{c'} = m_c -\epsilon_B$
is the mass of the charm quark in the final state of the elastic
scattering process. Note that the factor of
$\frac{d_\Psi}{d_cd_{\bar{c}}}$ comes from summing over final state
color-spin degeneracy and averaging over initial state degeneracy. In
the quasifree approximation the 3$\to$2 coalescence process is reduced
to a 2$\to$2 (quasi-elastic) scattering process.

We proceed to calculate the charmonium regeneration rate by plugging the
charm-quark spectra and
the transition matrix element, Eq.~(\ref{eq:m_qfree_reg}), into
Eq.~(\ref{eq:gain_rate3}). The resulting $J/\psi$ regeneration rates,
$\beta_\Psi(p_t)$, from different charm-quark spectra are compared in
the right panel of Fig.~\ref{fg:rates_reg}. The $\beta_\Psi(p_t)$ can
be interpreted as the regenerated number of $\Psi$ with a given $p_t$
per unit volume per unit time. Here we have integrated over the $p_z$
dependence of $\beta_\Psi(p)$.


As expected, the $p_t$ dependence of the regeneration rate from thermal
(pQCD) charm spectra follows a thermal (pQCD) trend. 
The thermal charm spectra, with the largest charm-quark phase overlap in the
low-$p_t$ region, lead to the largest inclusive regeneration rate,
$\beta_{tot}=\int d^3p\ \beta (p)$. The 3-dimensional isotropic pQCD
spectra and the transverse pQCD + longitudinal thermal spectra lead to
significantly smaller inclusive regeneration rates, amounting to 28\%
and 47\% of that from thermal charm-quark spectra, respectively. This
is similar to what is found in Ref.~\cite{Greco:2003vf}, where the
dynamics of $c\bar{c}$ coalescence was encoded in a Gaussian Wigner
function and the resulting number of $J/\psi$ from pQCD charm-quark spectra is
smaller than that from thermal spectra by a factor of 3. However in
Ref.~\cite{Yan:2006ve} where the (inverse) gluo-dissociation mechanism
was employed for $J/\Psi$ regeneration it is found that the inclusive
yield of regenerated $J/\psi$ from pQCD charm spectra is quite
comparable with that from thermal spectra (within $\sim$30\%). 
Further investigations are needed to clarify the discrepancy.

For pQCD+thermal charm spectra we also consider a possible angular
correlation between $c$ and $\bar{c}$ in momentum space: it is
expected from pQCD that in initial hard collisions back-to-back charm
pair production is favored. We study its consequence on $J/\psi$
regeneration by including a schematic ansatz of
$dN_{c\bar{c}}/d\theta$$\sim(1-\cos{\theta})$ in
Eq.~(\ref{eq:diss_rate3}) (with $\theta$ being the relative angle
between the $p_t$ of $c$ and $\bar c$ in the transverse plane). It
turns out that the angular correlation of $c$ and $\bar{c}$
significantly reduces the regeneration rate for high $p_t$ $J/\psi$ as
seen in Fig.~\ref{fg:rates_reg}. The reason is that, compared to low
$p_t$ $J/\psi$s, high $p_t$ ones are more likely to be regenerated
from a $c\bar{c}$ pair with small angle between them, the probability
of which is suppressed by the angular correlation factor
$(1-\cos{\theta})$. The inclusive regeneration rate from the angular
correlated pQCD+thermal spectra is 11\% of that from thermal charm
spectra. We subsequently plug $\beta_\Psi(p_t)$ into the Boltzmann
transport equation and the numerical results for the inclusive yield
and $p_t$ spectra of the regenerated $J/\psi$ will be compared in
Chapter~\ref{ch:nume_results}.


\section{Charmonium Dissociation in Hadronic Matter}
\label{sec:charm-diss-hm}

As the fireball expands, the medium keeps cooling. When the medium temperature decreases to the critical temperature, $T_c$, the deconfined QGP undergoes a phase transition to the confined hadronic phase. The charmonium dissociation rate in the hadronic phase is expected to be small compared to QGP due to the smaller light particle density and the larger charmonium binding energy. However, for a quantitative calculation of charmonium yields in heavy-ion collisions, dissociation/regeneration in hadronic phase needs to be taken into account.

\begin{figure}[tp]
  \centering
      \includegraphics[width=0.59\textwidth,clip=]{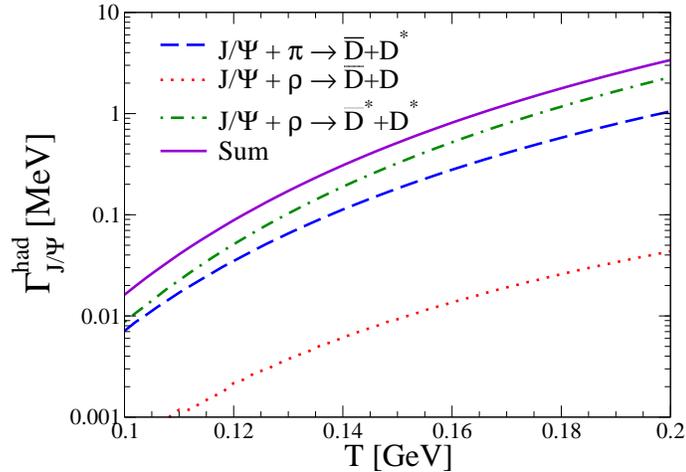}
      \caption[$J/\psi$ dissociation rates by pions and rhos as a function 
      of the temperature $T$ of the hadron gas]{$J/\psi$ dissociation
      rates by pions and rhos as a function 
      of the temperature $T$ of the hadron gas. Figure taken from Ref.~\cite{GrandchampDesraux:2003ca}. The full line is the sum
      of the three contributions, $J/\psi + \pi$ (dashed line),
      $J/\psi +\rho \rightarrow D^{\star} + \bar{D^{\star}}$
      (dot-dashed line) and $J/\psi + \rho \rightarrow D + \bar{D}$
      (dotted line).} 
\label{fg:had_rate_psi}
\end{figure}

The microscopic approaches for the hadronic charmonium dissociation
can be divided into two categories, based on either quark or hadronic
degrees of freedom. The approaches based on hadronic degrees of
freedom are often less cumbersome yet very effective in assessing many
dissociation processes, see Ref.~\cite{Rapp:2008tf} for a recent
review. In this work we adopt dissociation rates based on an effective
meson Lagrangian with a local $SU$(4) flavor
symmetry~\cite{Lin:1999ad,Haglin:2000ar,GrandchampDesraux:2003ca}. The
considered interactions are $J/\psi$ with pions and rho mesons which
are the most abundant particles in the hadronic medium,
\begin{eqnarray}
  & \pi  + J/\psi & \rightarrow D + \bar{D}^{\star}, \bar{D} + D^{\star}
  \label{eq:psi_pi} \\
  & \rho + J/\psi & \rightarrow D + \bar{D}
  \label{eq:psi_rho_D} \\
  & \rho + J/\psi & \rightarrow D^{\star} + \bar{D}^{\star}\ .
  \label{eq:psi_rho_Dstar}
\end{eqnarray}
We neglect the interactions between $J/\psi$ and kaons 
(In Ref.~\cite{Azevedo:2003qh}, the total cross section of $J/\psi$
dissociation by kaons is found to be much smaller than that by
pions). The resulting dissociation rates are shown in
Fig.~\ref{fg:had_rate_psi}, with the main contribution given by the
$\rho + J/\psi \rightarrow D^{\star} + \bar{D}^{\star}$ process. The
vacuum masses are assumed for all the hadrons including $J/\psi$.

For dissociation rates of $\chi_c$ and $\psi'$ we assume a geometrical scaling~\cite{Grandchamp:2002wp} of the dissociation rates for $J/\psi$ by the respective charmonium radii, namely, 
\begin{align}
  \label{eq:rate_chi_psi}
  \Gamma_{\chi}^{HAD} = \left(\frac{r_{\chi}}{r_{J/\psi}}\right)^2
  \Gamma_{J/\psi}^{HAD} \simeq 2.36\ \Gamma_{J/\psi}^{HAD}\ , \\
  \Gamma_{\psi'}^{HAD} = \left(\frac{r_{\psi'}}{r_{J/\psi}}\right)^2
  \Gamma_{J/\psi}^{HAD} \simeq 3.73\ \Gamma_{J/\psi}^{HAD}\ .
\end{align}
The radii of the excited charmonia states are estimated by non-relativistic potential models from Ref.~\cite{Karsch:1987pv}. 

As expected, the overall charmonium dissociation rates in the hadronic phase are significantly smaller than in the QGP phase, resulting in a rather mild charmonium suppression in the hadronic medium compared to the QGP.  As in the QGP, $\Psi$ can also be regenerated in the hadronic matter~\cite{Ko:1998fs}; their regeneration rates can be obtained from their dissociation rates using the principle of detailed balance.  


\section{Charmonium Spectral Function}
\label{sec:charmo_spf}

So far we have obtained the charmonium binding energy $\epsilon_B$ and
in-medium charm quark mass $m^*_c$ (from the potential model) and its
dissociation rate (from quasifree approximation in QGP and $SU$(4)
effective theory in HG). The charmonium binding energy $\epsilon_B$
and in-medium charm quark mass $m^*_c$ allow us to infer the
charmonium pole mass via $m_\Psi$=2$m^*_c$-$\epsilon_B$. These
pieces of information figure into the charmonium spectral function,
Eq.~(\ref{eq:spectral}), which is the imaginary part of the Fourier
transform of the charmonium current-current correlation function,
Eq.~(\ref{eq:Gtau}). Due to current limitations mentioned in
Section~\ref{ch:intro}.\ref{sec:charmo}, we can not reliably extract the charmonium
spectral function from the charmonium current-current correlation
function calculated by lattice QCD. However, we can still utilize the
rather precise correlator ratios, Eq.~(\ref{eq:corr_ratio}), as a constraint on the consistency
among the charmonium pole mass, binding energy and its dissociation
rate.

To be specific we adopt the following strategy:
We ``reconstruct" in-medium charmonium spectral functions using a
relativistic Breit-Wigner + continuum ansatz, where the $\Psi$ width
and mass figure into the Breit-Wigner part while the continuum is
determined by the open-charm threshold (2$m_c^*$). For a more
realistic evaluation, we include a polestrength factor, $Z_\Psi(T)$, for
the Breit-Wigner strength and a non-perturbative rescattering
enhancement in the continuum~\cite{Cabrera:2006wh,Mocsy:2007yj}.  The
vanishing of the polestrength factor furthermore serves to estimate
the dissociation temperature of the ground state in each channel.

We first construct a model spectral function in vacuum, consisting of a
zero-width bound-state and a perturbative (leading order) continuum part,
\begin{multline}
\sigma_\Psi (\omega)=A_\Psi \ \delta(\omega-m_{\Psi})
+\frac{B_\Psi N_c}{8\pi^2}\Theta(\omega-\sqrt{s_0})\omega^2
\sqrt{1-\frac{s_0}{\omega^2}}(a+b\frac{s_0}{\omega^2}) \ .
\label{spf}
\end{multline}
Here, $N_c$=3 is the number of colors and the coefficients 
$(a,b)=(1,-1), (2,1)$ characterize the scalar and vector channel, 
respectively~\cite{Mocsy:2005qw}. The open-charm threshold in vacuum, 
$\sqrt{s_0}$, is assumed to be given by twice the free $D$-meson mass,
$\sqrt{s_0}\equiv 2m_D=3.74$\,GeV. The coefficient $A_\Psi$ is related to the 
overlap of the wave-function, $R_{J/\psi}(0)$, or its derivative, 
$R'_{\chi_{c}}(0)$, at the origin~\cite{Mocsy:2005qw,Bodwin:1994jh},
\begin{equation}
  A_{J/\psi}=\frac{3N_c}{2\pi}|R_{J/\psi}(0)|^2 \ , \ \ 
  A_{\chi_{c}}=\frac{36N_c}{2\pi M^2_{\chi_{c}}}|R'_{\chi_c}(0)|^2 \ .
\label{norm_res}
\end{equation}
These quantities can be estimated from 
the electromagnetic decays widths via~\cite{Bodwin:1994jh}
\begin{equation}
\Gamma_{ee}=\frac{4e_Q^2\alpha^2N_c}{3m^2_{J/\psi}}|R_{J/\psi}(0)|^2 
\ , \ 
\Gamma_{\gamma\gamma}=\frac{144e_Q^4\alpha^2N_c}{m^4_{\chi_{c}}}
|R'_{\chi_c}(0)|^2
\label{gamma_em}
\end{equation}
where $\alpha$=1/137 is the electromagnetic coupling constant and
$e_Q=2/3$ the charge of the charm quark (we use 
$\Gamma_{ee}$=5.55\,keV for the $J/\psi$ and 
$\Gamma_{\gamma\gamma}$=2.40\,keV for the $\chi_{c0}$). The resulting 
relations between $A_\Psi$ and $\Gamma_{\Psi\to ee,\gamma\gamma}$
are
\begin{equation}
A_{J/\psi}=\frac{81m^2_{J/\psi}}{32\pi\alpha^2} \ \Gamma_{ee} \ , \ \ 
A_{\chi_{c}} =\frac{81m^2_{\chi_{c0}}}{128\pi\alpha^2}  
\ \Gamma_{\gamma\gamma} \ .
\label{norm_r}
\end{equation}
The free $J/\psi$ and $\chi_{c0}$ masses are taken at their empirical 
vacuum values.
The coefficient $B_\Psi$ in the continuum part of Eq.~(\ref{spf}) equals to
one in the non-interacting limit. To account for rescattering, which
is particularly important close to threshold, we scale it up to
match the continuum as calculated from the vacuum $T$-matrix in 
Ref.~\cite{Riek:2010fk}, amounting to $B_{J/\Psi}\simeq2$ and 
$B_{\chi_{c}}\simeq4$ in the vector and scalar channel, respectively.
For simplicity we neglect $\psi'$, $\chi_{c}'$ and higher excited 
states which play little role in the correlator ratios, Eq.~(\ref{eq:corr_ratio}).

\begin{figure}[t]
\centering
\includegraphics[width=0.59\textwidth]{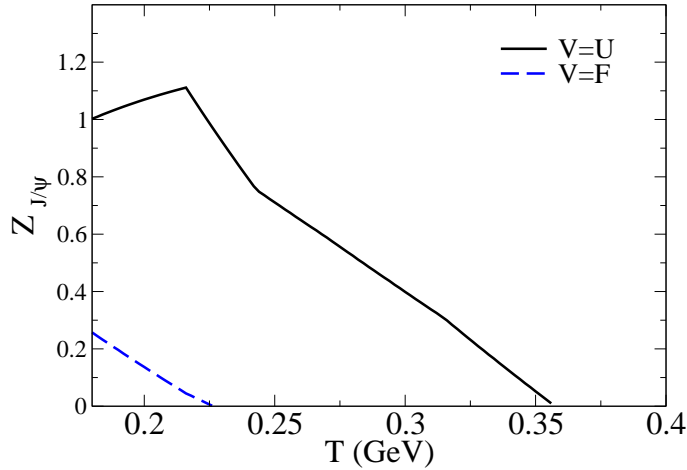}
\caption[Temperature dependence of the strength of the resonance part
of the $S$-wave spectral function, $Z_\Psi(T)$, for the strong and
weak binding scenarios]{Temperature dependence of the strength of the
  resonance part of the $S$-wave spectral function, $Z_\Psi(T)$, for
  the strong (solid line) and weak (dashed line) binding scenarios.}
\label{fig_zpsi}
\end{figure}
At finite temperature we replace the $\delta$-function bound-state
part by a relativistic Breit-Wigner (RBW) distribution while the
continuum part is assumed to be of the same form as in the vacuum,
\begin{align}
\label{spfT} 
\sigma_\Psi(\omega)&=A_\Psi \ Z_\Psi(T) \frac{2\omega}{\pi}
\frac{\omega\Gamma_\Psi(T)}{(\omega^2-m_{\Psi}^2(T))^2+\omega^2
\Gamma_\Psi(T)^2} \qquad
\nonumber\\[2mm]
&\quad +\frac{B_\Psi N_c}{8\pi^2}\Theta(\omega-\sqrt{s(T)})\omega^2 
\sqrt{1-\frac{s(T)}{\omega^2}}(a+b\frac{s(T)}{\omega^2}) \ .
\end{align}
The in-medium continuum edge, $s(T)$, is now taken as the charm quark
threshold at finite temperature, $\sqrt{s(T)}\equiv2m^*_c(T)$,
consistent with the potential model, see Fig.~\ref{fg:mc}. The RBW
term includes: (i) the in-medium charmonium mass, $m_\Psi(T)$,
extracted from Eq.~(\ref{eq:psi_binding}) based on Figs.~\ref{fg:mc}
and \ref{fg:epsB}; (ii) the width $\Gamma_\Psi$ identified with the
inelastic dissociation width discussed in the previous section; (iii)
the aforementioned polestrength factor, $Z_\Psi(T)$, representing the
modification of the strength of the bound-state part at finite
temperatures relative to its vacuum value ($A_\Psi$), with
$Z_\Psi|_{T=0}$=1. The $Z_\Psi(T)$ is adjusted to minimize the deviation
of the correlator ratios from one.

\begin{figure}[tp]
\includegraphics[width=0.49\textwidth]{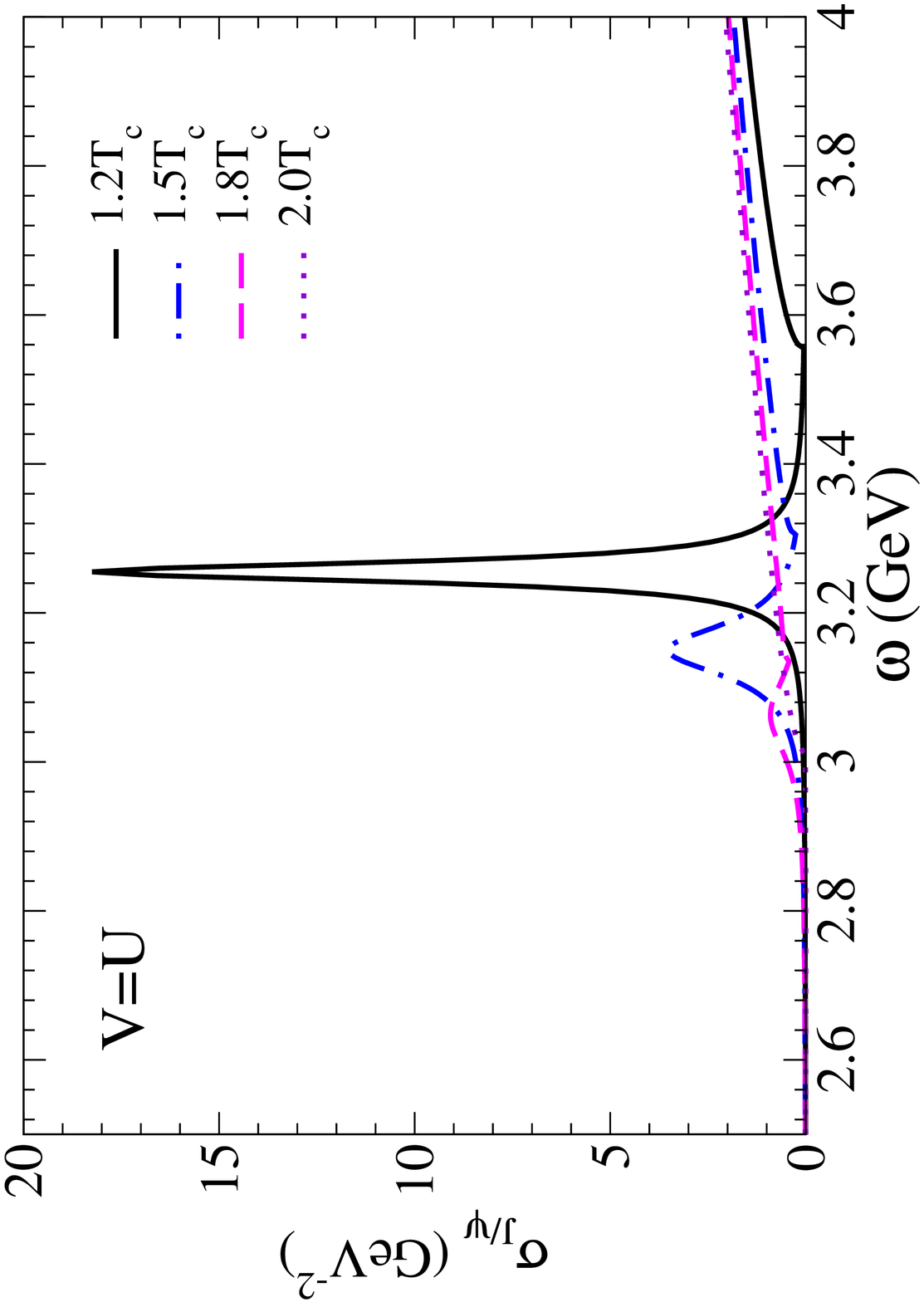}
\includegraphics[width=0.49\textwidth]{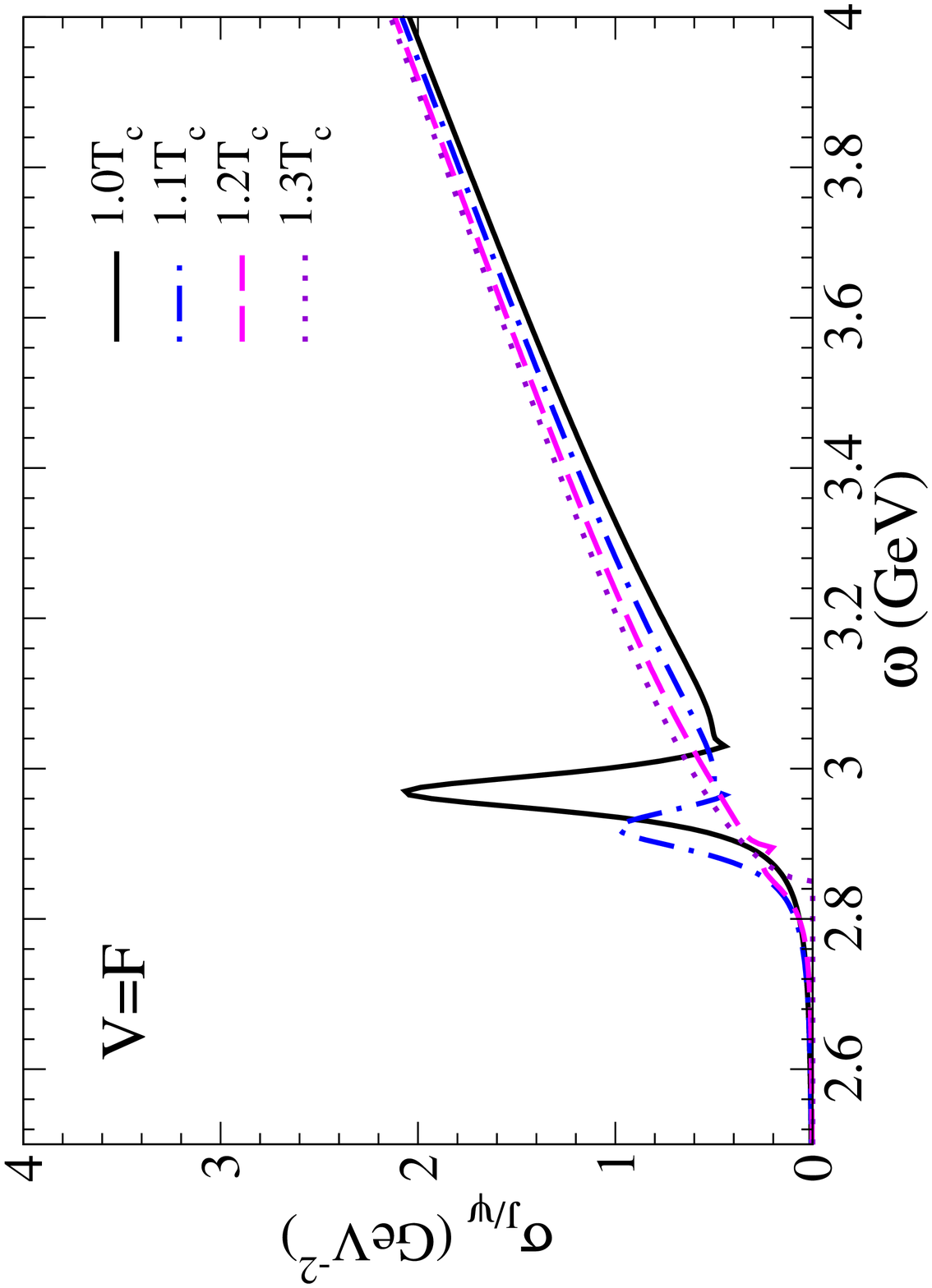}
\caption[Spectral functions in the vector channel for the strong and
weak binding scenarios]{Spectral functions in the vector channel for
  the strong (left panel) and weak (right panel) binding scenarios.}
\label{fig_sf-jpsi}
\end{figure}
\begin{figure}[bp]
\centering
\includegraphics[width=0.49\textwidth]
{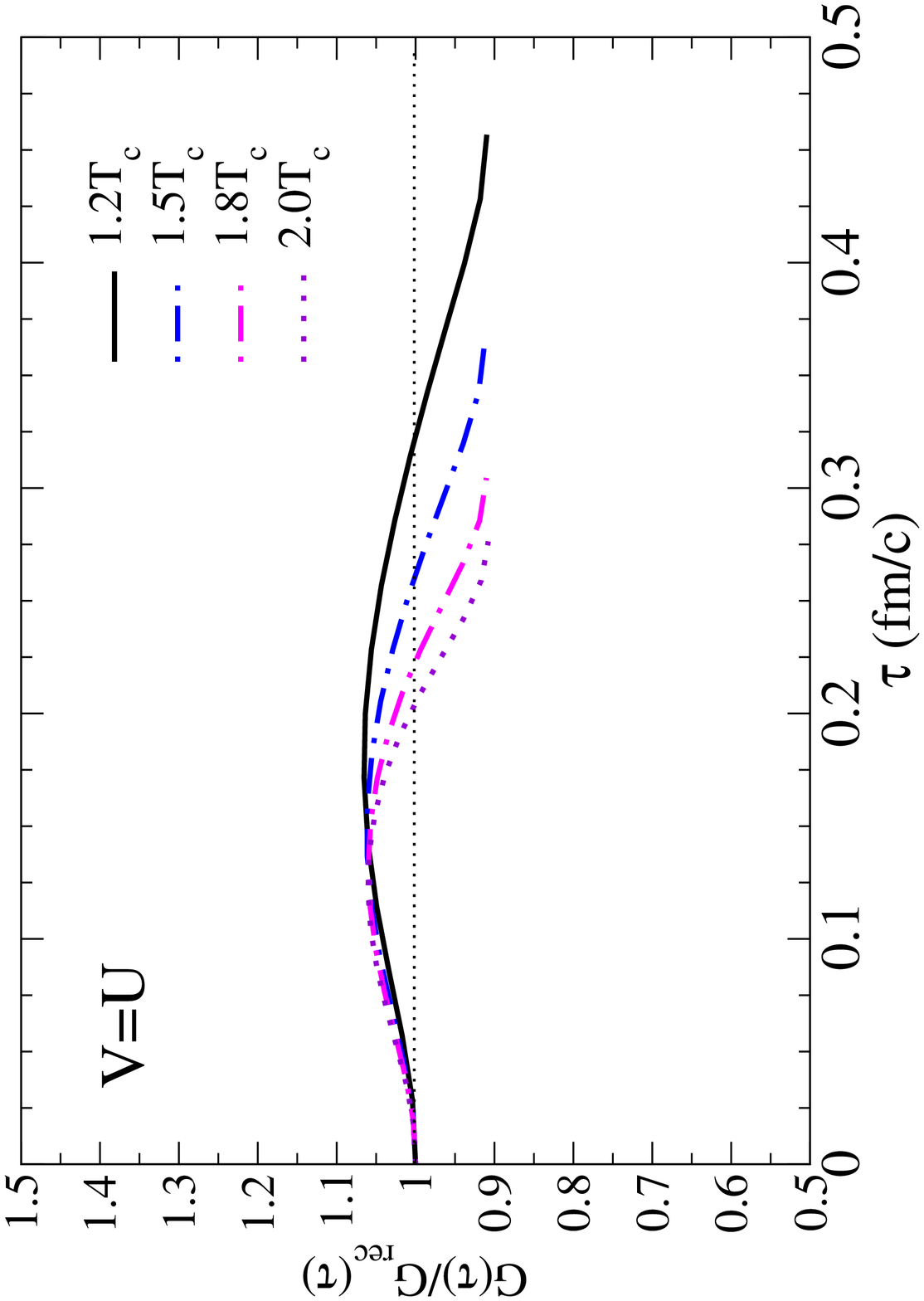}
\includegraphics[width=0.49\textwidth]
{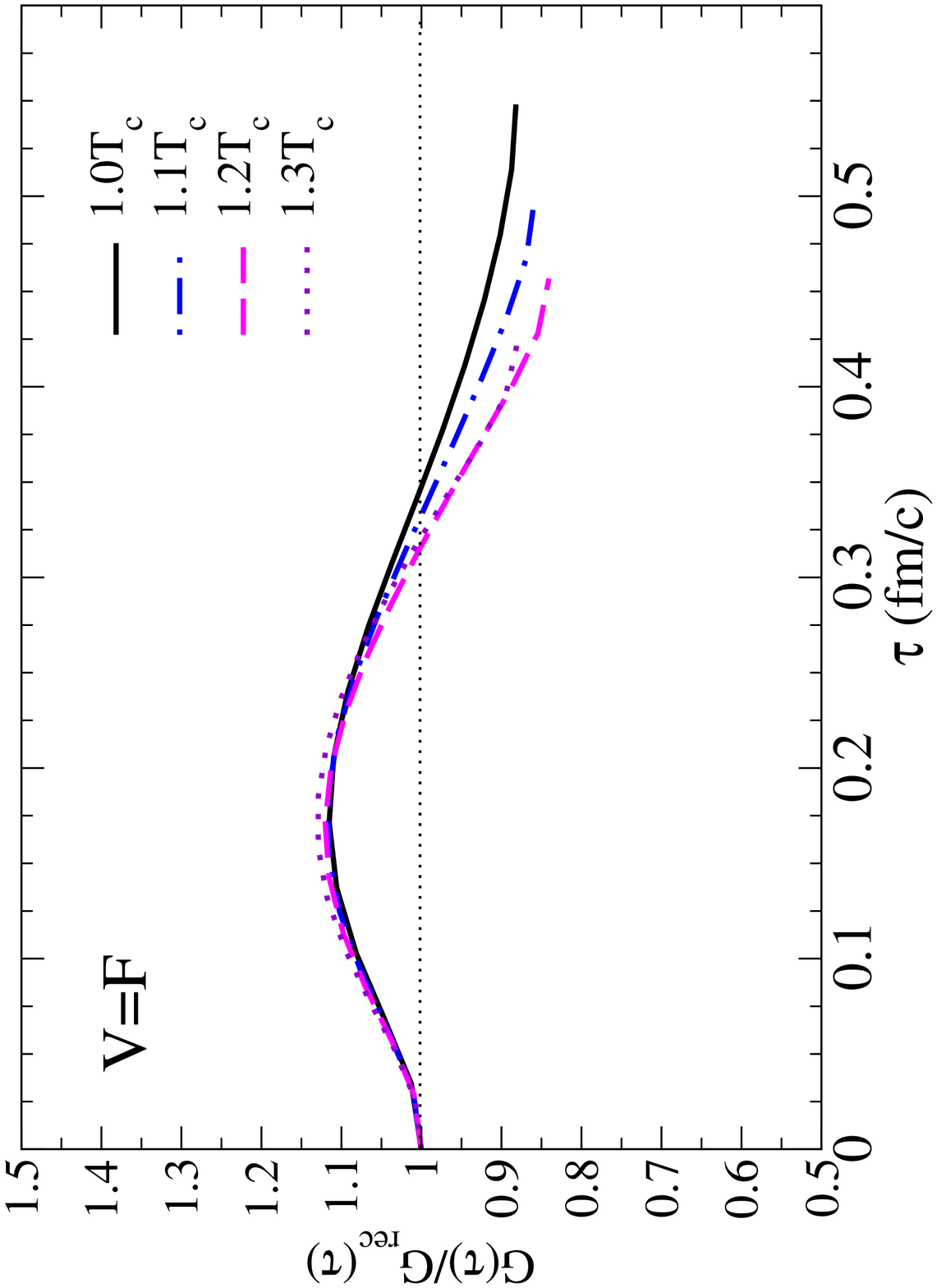}
\caption[Ratio of vector channel correlator to the reconstructed
correlator for the strong and weak binding scenarios]{Ratio of vector
  channel correlator to the reconstructed correlator for the strong
  (left panel) and weak (right panel) binding scenarios. }
\label{fig_corr-jpsi}
\end{figure}
\begin{figure}[tp]
\centering
\includegraphics[width=0.49\textwidth]{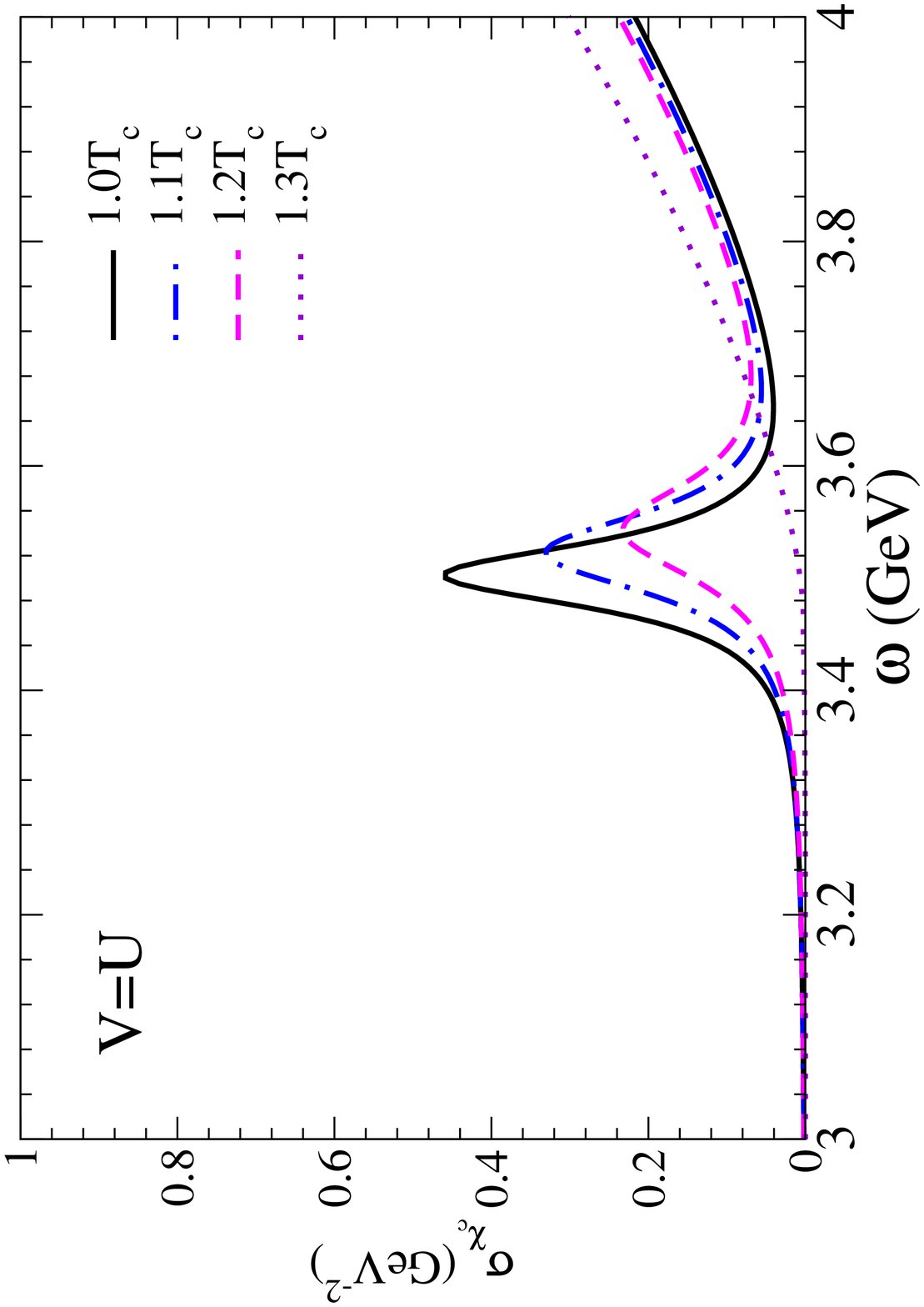}
\includegraphics[width=0.49\textwidth]{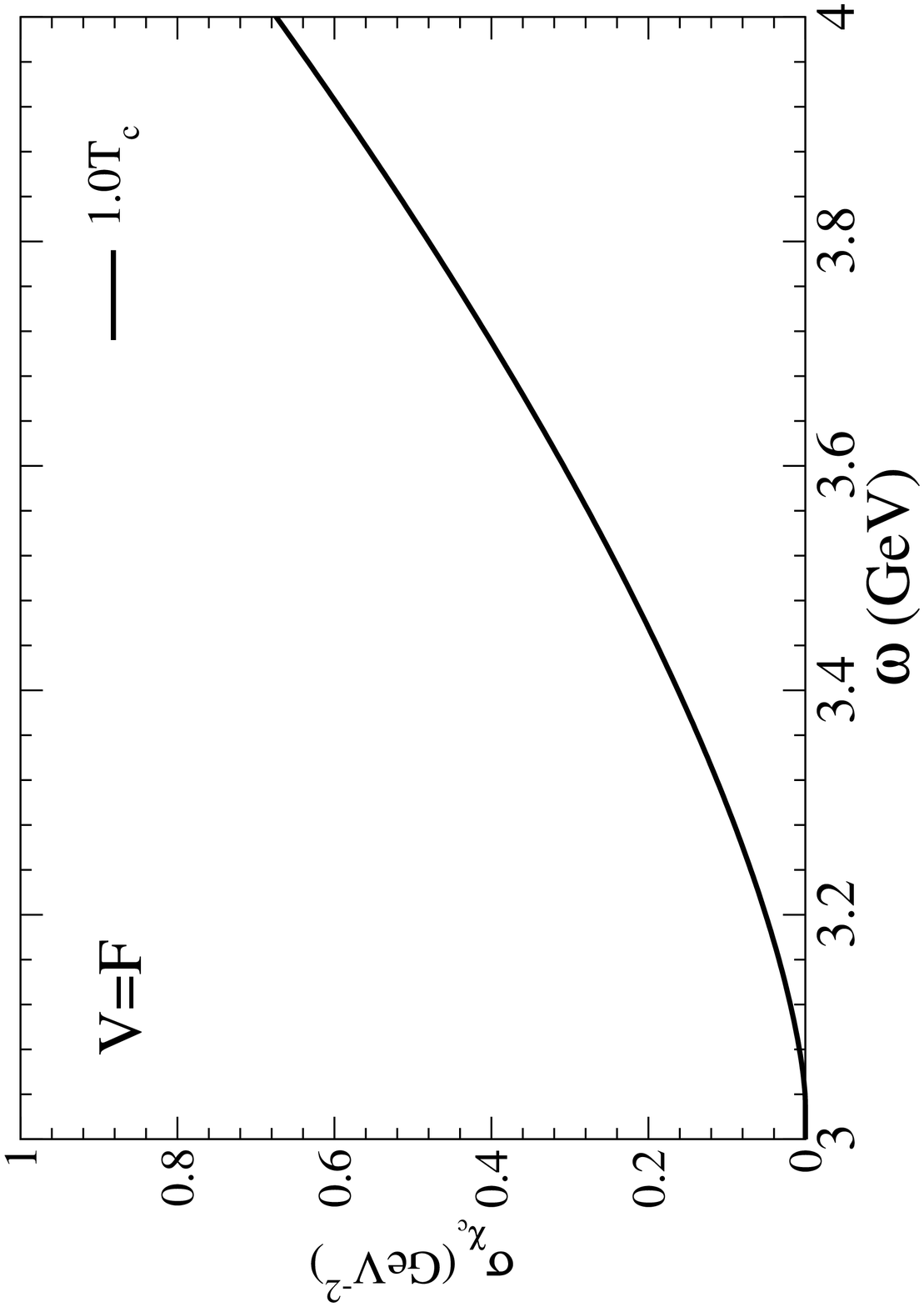}
\caption[Spectral functions in the scalar channel]{Spectral functions in the scalar channel. The
  left (right) panel is for the strong (weak) binding scenario. In
  the weak-binding scenario $\chi_{c0}$ has already melted at $T_c$.}
\label{fig_sf-chi}
\end{figure}
The resulting $Z_\Psi(T)$ for $J/\psi$ (vector channel) is plotted in 
Fig.~\ref{fig_zpsi}, from which we extract its dissociation temperature
$T_{J/\psi}^{\rm diss}$=2.0(1.25)$T_c$ in the strong (weak) binding 
scenario. Similar analysis in the scalar channel yields $\chi_{c}$ 
dissociation temperatures of  $T_{\chi_{c}}^{\rm diss}$=1.3(1.0)$T_c$ 
in the strong (weak) binding scenarios. We assume that $\chi_{c1}$ and
$\chi_{c2}$ have the same dissociation temperatures as the
$\chi_{c0}$. For $\psi'$ we simply assume its dissociation temperature
to be $T_c$ for both the strong- and weak-binding scenarios.

To comprehensively illustrate the medium effects we plot the final
spectral functions for the vector channel in the strong- and
weak-binding scenario in the QGP in Fig.~\ref{fig_sf-jpsi}, and their
corresponding correlator ratios in Fig.~\ref{fig_corr-jpsi}; the
spectral functions for the scalar channel are displayed in
Fig.~\ref{fig_sf-chi}.

We see that the correlator ratios are indeed close to one, as found in
lQCD~\cite{Aarts:2007pk,Datta:2003ww,Jakovac:2006sf}. In the hadronic
phase (not shown), we assume vacuum masses for both charmonia and
open-charm hadrons, which automatically ensures that the correlator
ratios are close to one (deviations due to small charmonium widths in
hadronic matter are negligible).

\graphicspath{{./4fg/}}

\chapter{Thermal Fireball Description of Medium Evolution}
\label{ch:fireball}

To solve the kinetic equation for the charmonium ($\Psi$=$J/\psi$, $\chi_c$, $\psi'$) evolution we need a description of the expanding medium in heavy-ion collisions. In particular, the temperature and the volume of the medium at any given time are required for estimating parton density and charm quark density, respectively, which are essential components for calculating the $\Psi$ dissociation and regeneration rates. 
The traditional approaches describing the medium evolution include transport models, see Refs.~\cite{Bass:1998ca,Cassing:1999es} for reviews, and relativistic hydrodynamics, see~\cite{Heinz:2009xj} for a review. 
In this work we however describe the medium evolution with a fireball model~\cite{Rapp:1999us,Rapp:1999zw} which is simpler yet captures the basic features of relativistic hydrodynamics. In Section~\ref{ch:fireball}.\ref{sec:fb_exp} we discuss the fireball description of the spatial expansion of the medium. In Section~\ref{ch:fireball}.\ref{sec:fb_eos} we discuss the equation of the state of the medium and extract the temperature of the fireball. In Section~\ref{ch:fireball}.\ref{sec:blastwave} we review the popular blastwave formula for estimating particle $p_t$ spectra with local thermal distributions boosted by the collective flow of the expanding source.


\section{Fireball Expansion Profile}
\label{sec:fb_exp}

The fireball model approximates the medium created in heavy-ion
collisions as a boost-invariant thermal fireball. The
``boost-invariance'' originates from the experimental fact that the
rapidity ($y$) distribution of the charged particles, $dN_{ch}/dy$, is
constant in the mid-rapidity ($y\sim 0$) region. This means that
the medium in the central region is invariant under Lorentz boosts in the
longitudinal ($z$) direction, which further implies that all
thermodynamic quantities characterizing the central region
depend only on the longitudinal proper time
$\tau$=$\sqrt{t^2-z^2}$, recall Fig.~\ref{fg:spacetime_evo}. 

The fireball volume, $V_{FB}(\tau)$, expands
cylindrically according to
\begin{equation}
\label{eq:Vfb}
V_{FB}(\tau)=(z_0+v_z\tau+\frac{1}{2}a_z \tau^2) \ \pi \ 
\left(r_0+\frac{\sqrt{1+a^2_\perp\tau^2}-1}{a_\perp}\right)^2\
\end{equation} 
with expansion parameters ${v_z,a_z,a_{\perp}}$ chosen so that the
results are consistent with the experimental data on the final
light-hadron flow and resemble the evolution by hydro-dynamical
calculations. The relativistic form of transverse acceleration
$(\sqrt{1+a^2_\perp\tau^2}-1)/a_\perp$ limits the surface speed,
$v_s(\tau)$, to below the speed of light in the large $\tau$
limit. For small $\tau$ it recovers the non-relativistic form with
constant transverse acceleration, $v_s(\tau)\sim a_{\perp}\tau$. Note
that the fireball expansion profile, Eq.~(\ref{eq:Vfb}), depends only
on the longitudinal proper time, $\tau$=$\sqrt{t^2-z^2}$, and not
separately on $t$ or $z$, as required by the boost invariance of the
medium.

The initial transverse radius $r_0$ represents the initial transverse
overlap of the two colliding nuclei at a given impact parameter $b$,
while the initial longitudinal length, $z_0$, is related to
thermalization time $\tau_0$ through $z_0\simeq\Delta y\tau_0$ where
$\Delta y$=1.8 represents the typical longitudinal rapidity coverage
of a thermal fireball. We assume that at a formation time of
$\tau_0$=1.0 (0.6)\,fm/$c$ the medium at SPS (RHIC) first thermalizes
with all the entropy, $S_{\rm tot}(b)$, being built up. The latter is
estimated from the multiplicities of observed charged particles and
assumed to be conserved during the adiabatic expansion.

\section{Equation of State of the Medium}
\label{sec:fb_eos}

To determine the temperature of the system we utilize the
equation of state (EoS) of the medium.  An important insight from the
success of the ideal hydrodynamic description of the medium evolution is
that the total entropy of the system is conserved. Together with the information of the fireball
expansion profile the entropy density of the fireball at each given
moment can be obtained. This quantity allows us to estimate the
temperature of the fireball through comparison with the thermodynamic
equation of state (the entropy density as a function of
the temperature, $T$). In QGP we model the medium
by an ideal gas of massive quarks and gluons, with the entropy density as a
function of temperature given by
\begin{equation}
  \label{eq:eos_s}
  s(T) = \sum_i \frac{d_i}{(2\pi)^3}\int d^3k [\pm(1\pm f_i(k;T))\log (1\pm f_i(k;T))- f_i(k;T)\log f_i(k;T)]\ ,
\end{equation}
where $d_i$ is the color-spin degeneracy of partons, the $f_i(k;T)$
are the thermal distributions for massive quarks and gluons, given by
Eqs.~(\ref{eq:parton_th_dist_f}) and (\ref{eq:parton_th_dist_b})
respectively, and the plus sign in ``$\pm$'' is taken for gluons ($i$=$g$)
and the minus sign is taken for quarks
($i$=$q,\bar{q}$).
\begin{figure}[tp]
  \centering
  \includegraphics[width=0.59\textwidth,clip=]{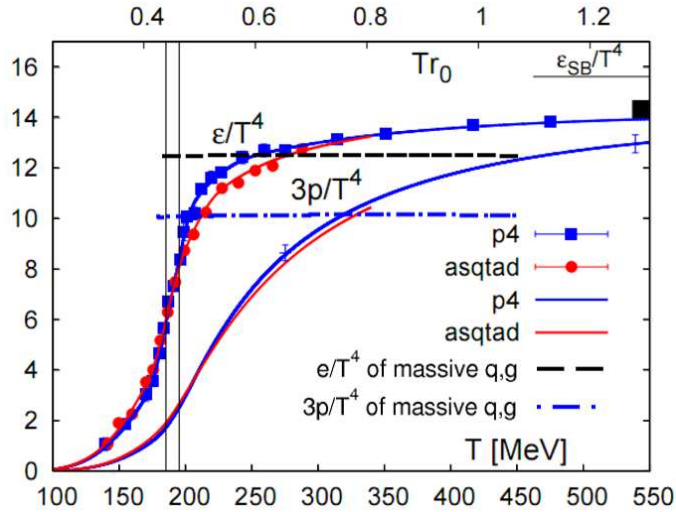}
  \caption[Comparison of energy density and pressure from an ideal massive parton gas with data from lattice QCD]{Comparison of energy density from an ideal massive parton gas with data from lattice QCD~\cite{Petreczky:2009at}. Dashed (dot-dashed) line: energy density (pressure) calculated from an ideal massive quark and gluon gas with $g_s$=2.3.}
\label{fg:e_lQCD}
\end{figure}
For the temperature dependence of thermal quasiparticle masses $m_i(T)$
we take guidance from pQCD calculation, see Eqs.~(\ref{eq:m_ud}),
(\ref{eq:m_s}) and (\ref{eq:m_g}). The strong coupling constant
$g$ is estimated by matching the resulting energy density of the
parton gas to lQCD calculations, see Fig.~\ref{fg:e_lQCD}, 
resulting in $g_s\sim$2.3. Note that this $g$ is decoupled from the
$\alpha_s$ in the quasifree dissociation cross
section. 
In the hadronic medium an EoS similar to Eq.~(\ref{eq:eos_s}) is
employed but including 76 mesonic and baryonic states up to masses of
2\,GeV. For particles which do not decay strongly, \eg, pion, kaon,
antiproton, we include their corresponding chemical potentials to
maintain their abundances when the system cools down from chemical to
thermal freeze-out~\cite{Rapp:1999us}. The critical temperature,
$T_c$=170 (180)\,MeV at SPS (RHIC), is roughly consistent with
thermal-model fits to observed particle
ratios~\cite{BraunMunzinger:2003zd} and predictions of lattice
QCD~\cite{Cheng:2006qk}. A freeze-out temperature of $T_{\rm
  fo}\simeq120$\,MeV terminates the evolution and results in a total
fireball lifetime of $\tau_{\rm fo}$=10-12\,fm/$c$ for central A+A
collisions. The resulting temperature evolution as a function of time
$\tau$ is displayed in Fig.~\ref{fg:fig_fb} for SPS and RHIC. Note
that there is little difference between mid ($|y|<0.35$) and forward
rapidity ($|y|\in[1.2,2.2]$) for Au+Au collisions at RHIC due to the
slowly varying rapidity density of charged particles over this $y$
range~\cite{Arsene:2004fa}, cf.~Ref.~\cite{Zhao:2008pp} for more
details.
\begin{figure}[tp]
\centering
\includegraphics[width=0.59\textwidth]{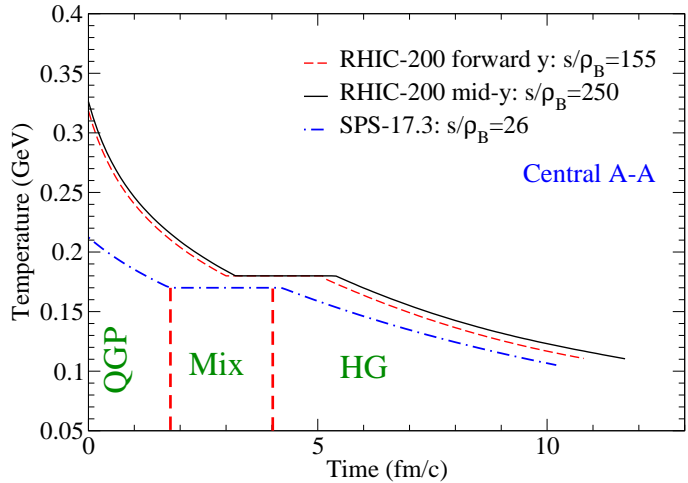}
\caption[Time profiles of temperature for central collisions of heavy
nuclei (participant number $N_{\rm part}$=380) at RHIC and SPS]{Time profiles of temperature for central collisions of heavy
  nuclei (participant number $N_{\rm part}$=380) at RHIC
  ($\sqrt{s}$=200\,AGeV; solid line: mid rapidity; dashed line:
  forward rapidity) and SPS ($\sqrt{s}$=17.3\,AGeV; dot-dashed line).}
\label{fg:fig_fb}
\end{figure} 

\begin{figure}[tp]
  \centering
  \includegraphics[width=0.59\textwidth]{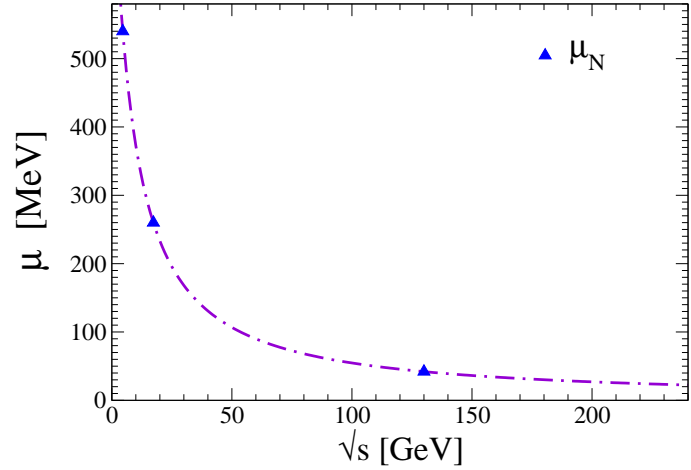}
  \caption[Excitation function of the baryon chemical potential at the
  assumed phase transition line]{Excitation function of the baryon
    chemical potential $\mu_B(T_c)$ at the assumed phase transition
    line. The points correspond to values extracted from particle
    ratios.}
\label{fg:mub_vs_s}
\end{figure}

\begin{figure}[tp]
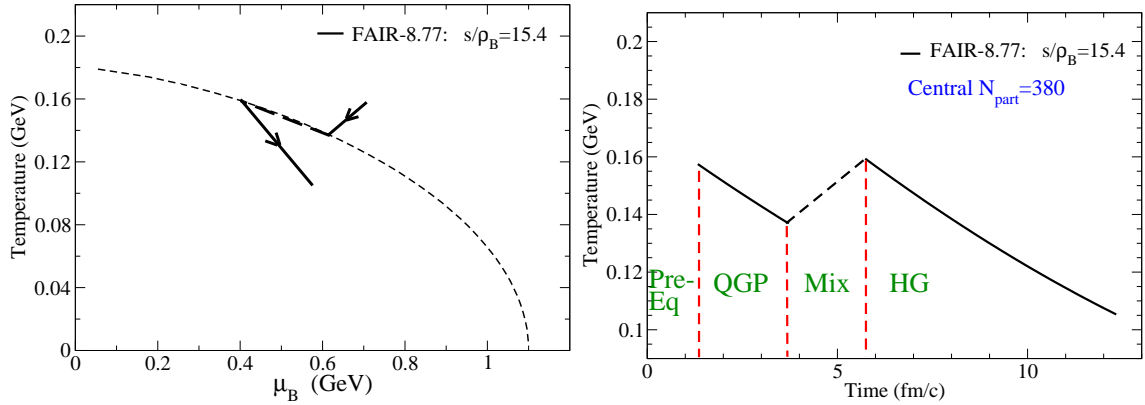

  \centering
  \includegraphics[width=0.49\textwidth]{Temp_vs_muB_1018.eps}
  \includegraphics[width=0.49\textwidth]{Temp_vs_time_1018.eps}
  \caption[Temperature evolution of heavy-ion collisions at FAIR]{Temperature evolution of heavy-ion collisions at FAIR. Left panel: Trajectory of heavy-ion collisions on the QCD phase diagram at FAIR ($\sqrt{s}$=8.8AGeV). Right panel: Time profile of temperature for central collisions of heavy nuclei (participant number $N_{part}$ =380) at FAIR ($\sqrt{s}$=8.8AGeV). }
\label{fg:evo_traj}
\end{figure}

For heavy-ion collisions with lower center-of-mass (cms) energies, \eg, at
the future FAIR facility, stronger stopping of incoming nuclei entails
the medium to be more asymmetric in terms of the net baryon number,
corresponding to a larger baryon chemical potential $\mu_B$, see Fig.~\ref{fg:mub_vs_s}, and lower
initial temperature, $T$. In this regime the thermodynamic properties
of the medium are characterized by both $T$ and $\mu_B(\gtrsim T)$.
In order to determine the temperature evolution in such systems we resort to a second conserved quantity during the fireball evolution, which is the net baryon number, $N_B$. With both baryon density, $n_B$, and entropy density, $s$, given at each time we are able to solve for $\mu_B(\tau)$ and $T(\tau)$ by using
\begin{align}
  \label{eq:eos_mub}
  s(T,\mu_B) &= \sum_i \frac{d_i}{(2\pi)^3}\int d^3k [\pm(1\pm f_i(k;T,\mu_B))\log (1\pm f_i(k;T,\mu_B))\nonumber \\
   &\qquad\qquad - f_i(k;T,\mu_B)\log f_i(k;T,\mu_B)]\ ,\\
  n_B(T,\mu_B) &= \sum_i \frac{d_i}{3\times(2\pi)^3}\int d^3k [(f_q(k;T,\mu_B)-f_{\bar{q}}(k;T,\mu_B)]\ ,
\end{align}
The thermal quark distributions at finite $\mu_B$ are
 \begin{align}
  \label{eq:parton_th_dist_2_mub}
  f_q(k;T,\mu_B)  &= \frac{1}{\exp\left(\frac{\sqrt{k^2+m^2_i}-\mu_B/3}{T}\right) + 1}\,\nonumber \\
  f_{\bar{q}}(k;T,\mu_B) &= \frac{1}{\exp\left(\frac{\sqrt{k^2+m^2_i}+\mu_B/3}{T}\right) + 1}\ .
\end{align}
As part of the EoS of the system, the phase boundary $\mu_B(T_c)$ (instead of a single $T_c$) is determined with guidance from thermal-model fits to observed particle
ratios at various center-of-mass energies~\cite{BraunMunzinger:2003zd}, see Figs.~\ref{fg:mub_vs_s} and \ref{fg:evo_traj}.

The resulting time evolution of the fireball is characterized by a trajectory in the QCD phase diagram at fixed $s/n_B$ ratio, see the left panel of Fig.~\ref{fg:evo_traj}. 
We note the zigzag structure of the trajectory, which reflects the fact that the medium is reheated by the latent heat when it undergoes the phase transition, as first pointed out in Ref.~\cite{Hung:1997du}.

The resulting temperature evolution as a function of time $\tau$ for central Pb+Pb collisions at
FAIR energies is displayed in the right panel of Fig.~\ref{fg:evo_traj}.
The thermalization time $\tau_0$ is assumed to be 1.3fm/$c$ at FAIR
energies. With an initial temperature lower than 160\,MeV the QGP phase
still lasts for 1-2fm/$c$, which is followed by the mixed phase with
duration of around 2fm/$c$. The overall fireball lifetime is again around
10fm/$c$.


\section{Blastwave Description of Charmonium $p_t$ Spectra}
\label{sec:blastwave}

In addition to the temperature evolution profile the thermal fireball
model allows to define a flow field, which enables us
to estimate the transverse momentum ($p_t$) spectra of
locally (kinetically) thermalized particles, boosted by the flow field. 
Historically this analysis was developed by Siemens and
Rasmussen~\cite{Siemens:1978pb}, and is referred to as the blastwave
model. Later the blastwave model in a boost-invariant medium was
developed by Schnedermann, Sollfrank and
Heinz~\cite{Schnedermann:1993ws}.  Here we follow
Ref.~\cite{Schnedermann:1993ws} and briefly review the derivation of
the ``blastwave formula''.

Since the medium is not static in the lab frame we employ the
Cooper-Frye formula~\cite{Cooper:1974mv} to count the particle number
in the freeze-out hyper-surface,
\begin{equation}
  \label{eq:cooper_frye}
  E\frac{dN}{d^3p}=\frac{1}{(2\pi)^3}\int d\Sigma_\mu(x) p^\mu f(x,p)=\frac{d}{(2\pi)^3}\int d\Sigma_\mu(x) p^\mu e^{-p\cdot u/T}\ .
\end{equation}
Here the $\Sigma_\mu$ is the freeze-out hyper-surface, $u^{\mu}$ is
the four velocity of a given medium cell, and $f(x,p)$=$d\,e^{-p\cdot
  u/T}$ is the thermal particle distribution in the local rest frame
with $d$ being the degeneracy factor.

Since the freeze-out is commonly assumed to occur at fixed longitudinal proper time, $\tau$=$\sqrt{t^2-z^2}$, it is convenient to parameterize the freeze-out hyper-surface by 
\begin{equation}
  \label{eq:hyper_s_fo}
  x^\mu=(\tau_{fo}\cosh \eta,r\cos \phi,r\sin \phi,\tau_{fo}\sinh \eta)\ ,
\end{equation}
where $\eta$=$\tanh^{-1}(z/t)$ is the space-time rapidity. The volume element of the hypersurface is therefore
\begin{equation}
  \label{eq:dhyper_s_fo}
  d\Sigma^\mu=\tau_{fo}\left(\cosh\eta,0,0,\sinh\eta\right)r\ dr\ d\eta\ d\phi\ .
\end{equation}
The particle four-momentum can be parameterized as
\begin{equation}
  \label{eq:four_momentum}
  p^\mu=\left(m_t\cosh y,p_t\cos \phi_p,p_t\sin \phi_p,m_t\sinh y\right)\ ,
\end{equation}
where $y$=$\tanh^{-1}(p_z/E)$ is the momentum rapidity of the particle and $m_t$=$\sqrt{m^2+p^2_t}$ its transverse mass.
Therefore we have
\begin{equation}
  \label{eq:p_dot_sigma}
  p\cdot d\Sigma=\tau_{fo} m_t\cosh (y-\eta) rdr d\eta d\phi\ .
\end{equation}
 A parameterization of the four-velocity $u^\mu$ of a medium cell can be written as
\begin{equation}
  \label{eq:four_velocity}
  u^\mu=\left(\cosh \rho \cosh y^{cell},\sinh \rho\cos \phi_p,\sinh \rho\sin \phi_p,\cosh \rho \sinh y^{cell}\right)\ .
\end{equation}
Here $y^{cell}=\tanh^{-1}(v^{cell}_z)$ is the longitudinal
rapidity of a medium cell.  According to boost-invariant
condition ($v^{cell}_z=z/t$), it is equal to the longitudinal space-time rapidity,
namely, $y^{cell}=\eta$; $\rho$=$\tanh^{-1}(v^{cell}_\perp)$ is
the transverse rapidity of the medium cell. Therefore the four product $p\cdot u$ can be expressed as
\begin{equation}
  \label{eq:p_dot_u}
  p\cdot u=m_t\cosh \rho\cosh(\eta-y)-p_t\sinh \rho \cos(\phi-\phi_p)\ .
\end{equation}
Plugging Eqs.~(\ref{eq:p_dot_sigma}) and (\ref{eq:p_dot_u}) into Eq.~(\ref{eq:cooper_frye}) we obtain the one particle momentum spectrum in the lab frame
\begin{align}
  \label{eq:blastwave}
&E\frac{dN}{d^3p}=\frac{d}{(2\pi)^3}\int^{2\pi}_0 d\phi\int^{\infty}_{-\infty}d\eta\int^R_0rdr\nonumber \\
&\qquad\times\tau_{fo} m_t\cosh (y-\eta)\exp [(-m_t\cosh \rho\cosh(\eta-y)+p_t\sinh \rho \cos(\phi-\phi_p))/T]\ .
\end{align}
After the integration over $\eta$ and $\phi$ is performed we arrive at the blastwave formula,
\begin{equation}
  \label{eq:blastwave_2}
  E\frac{dN}{d^3p}=\frac{dN}{dyd^2p_t}=\frac{d}{2\pi^2}\tau_{fo}m_t \int^R_0rdr K_1(\frac{m_t \cosh \rho}{T})I_0(\frac{p_t \sinh \rho}{T})\ .
\end{equation}
Here $K_1$ and $I_0$ are modified Bessel functions. The transverse
rapidity of a medium cell $\rho$=$\tanh^{-1}(v^{cell}_\perp)$ is given
by the fireball model by assuming a linear transverse flow profile
with
\begin{equation}
  \label{eq:flow_vel}
  v^{cell}_\perp(\vec r)=\frac{r}{R} v_{s}\ ,
\end{equation}
where $R$ is the radius of the fireball and $v_{s}(\tau)$=$a_\perp
\tau$ is the transverse flow velocity at the surface of the fireball.

In Chapter~\ref{ch:nume_results} we will utilize the blastwave formula to
estimate the transverse momentum spectra of $\Psi$ regenerated from
thermal charm quark spectra.

\graphicspath{{./5fg/}}

\chapter{Primordial Charmonium Production}
\label{ch:pre-eq}

The central purpose of studying charmonium production in heavy-ion
collisions is to utilize charmonium as a probe of the hot and dense
medium. Therefore charmonium production in the pre-equilibrium stage
is necessary as a baseline for assessing any modifications due to the
hot medium. Since charm-anticharm production is a hard process at SPS
and RHIC (the production time of charm quark pair,
$\tau^{c\bar{c}}\sim$0.07fm/$c$, is shorter than nucleus passage time,
$\tau_{pass}\sim$0.13fm/$c$ at RHIC), it can be approximated as
superposition of production in elementary nucleon-nucleon
collisions. Thus, in a first approximation, the initially
produced charmonia in A+A collisions can be estimated from their
production in p+p collisions (scaled by the number of binary
nucleon-nucleon collisions in an A+A collision). The main purpose of
this chapter is to apply further corrections specific for p+A and
A+A collisions. These corrections are usually referred to as cold
nuclear matter (CNM) effects. We first give an overview of charmonium
production in p+p collisions in Section~\ref{ch:pre-eq}.\ref{sec:psi_pp}. Then we
proceed to the CNM effects in Section~\ref{ch:pre-eq}.\ref{sec:psi_aa}.


\section{Charmonium Production in p+p Collisions}
\label{sec:psi_pp}

In p+p collisions charmonia are produced in two steps:\\ 1) The
$c\bar c$ pairs are produced through hard collisions (with large
momentum transfer) between the partons from the two colliding
protons. The leading order processes in pQCD include the gluon fusion
(dominating at high energies) and quark annihilation,
\begin{equation}
  \label{eq:ccbar}
 g+g \rightarrow c + \bar{c}, \qquad q+\bar{q}\rightarrow c+\bar{c}\ .  
\end{equation}
As hard processes the $c\bar{c}$ production cross section in p+p
collisions can be factorized into a convolution of incoming parton
distribution function, $f_i(x,Q^2)$, and the parton scattering
cross section, $\sigma^{ij\to [c\bar c]}(x_1,x_2,Q^2)$, to yield,
\begin{equation}
\sigma^{NN\to [c\bar c]}(Q^2)=\int^1_0 dx_1\int^1_0 dx_2\sum_{i,j}
 f_i(x_1,Q^2)f_j(x_2,Q^2)
\sigma^{ij\to [c\bar c]}(x_1,x_2,Q^2)\ .
 \label{eq;jpsi}
\end{equation}
Here, $x_1$ and $x_2$ are the momentum fractions of the two colliding
partons within the two colliding protons. The large virtuality, $Q^2$,
allows us to compute the partonic cross-section,
$\sigma(x_1,x_2,Q^2)$, as a perturbative expansion in powers of
$\alpha_s(Q^2)$. According to the uncertainty principle the time scale
of the perturbative partonic process is $\sim
1/Q$. 
\\ 2) $c$ and $\bar c$ quarks which are close to each other in the
phase space (``pre-resonance'' $c\bar{c}$ states) may develop into a
charmonium state through non-perturbative ``final-state''
interactions. Compared to charm quark production this step is much
slower, with a typical formation time for charmonium states on the
order of $1/\epsilon^\Psi_B\gg 1/(2m_c)$. Calculations based on a
non-relativistic Schr\"odinger equation suggest the typical formation
time of charmonium states to be
$\tau_\Psi\sim$1-2\,fm/$c$~\cite{Karsch:1987zw}. Due to the large
separation in the time scale the quantum interference between charm
quark production and charmonium production is suppressed. Therefore
the factorization between the charm quark production and charmonium
production is usually assumed,
\begin{equation}
\sigma_{A+B\rightarrow \Psi+X} 
\approx 
\sum_{n} \int d\Phi_{c\bar{c}}\
\sigma_{A+B\rightarrow c\bar{c}[n]+X}(\Phi_{c\bar{c}}, m_c) \
F_{c\bar{c}[n]\rightarrow \Psi}(\Phi_{c\bar{c}}) \ ,
\label{qq-fac}
\end{equation}
with a sum over possible $c\bar{c}[n]$ states and an integration over
available $c\bar{c}$ phase space $d\Phi_{c\bar{c}}$; $F$ represents
a non-perturbative transition probability for a pair of off-shell
$c\bar{c}$ to a charmonium state, $\Psi$. The microscopic mechanisms
for this transition are still under debate. Three widely discussed
models in the literature are the color evaporation model (CEM), the color
singlet model (CSM) and the non-relativistic QCD (NRQCD) model.

The CEM~\cite{Einhorn:1975ua,Amundson:1996qr,Gavai:1994in} assumes
that all $c\bar{c}$ pairs with invariant mass less than the threshold
of producing a pair of open-charm mesons, regardless of their color,
spin, and invariant mass, have the same probability to become a
charmonium.  That is, the $F_{c\bar{c}[n]\to \Psi}$ in
Eq.~(\ref{qq-fac}) is a constant for a given quarkonium state, which
is usually obtained by a fit to the data.

The
CSM~\cite{Berger:1980ni,Chang:1979nn,Baier:1981uk,Baier:1983va,Schuler:1994hy}
assumes that only a color-singlet charm quark pair with the right
quantum number can become a charmonium of the same quantum number and
the transition from the pair to a meson is given by the charmonium
wave function overlap at the origin.

The NRQCD model~\cite{Bodwin:1994jh,Braaten:1996pv,Beneke:1997av} allows every $c\bar{c}[n]$ state to become a bound
charmonium, while the probability is determined by corresponding
non-perturbative matrix elements $F_{c\bar{c}[n]\rightarrow \Psi}$. The
$F_{c\bar{c}[n]\rightarrow \Psi}$ encoding the non-perturbative
dynamics is expanded in terms of local matrix elements in a power series of
the heavy quark velocity, $v$.

Each of these models has its advantages and limitations in explaining
available experimental data, see Ref.~\cite{Brambilla:2004wf} for a
review. The $\Psi$ production mechanisms in p+p collisions
also have influence on $\Psi$ production in A+A collisions. For example,
recent RHIC measurements~\cite{Adare:2008sh,Abelev:2009qaa} found a
reduced suppression for high $p_t$ $J/\psi$'s compared to low $p_t$
ones in A+A collisions, which lends support to the picture that the
``pre-resonance'' $c\bar{c}$ states are in a color-singlet state, see
Section~\ref{ch:nume_results}.\ref{sec:num_results_sps_rhic}.\ref{ssec:psi_pt}.\ref{sssec:high_pt_psi}
for a more detailed discussion.

Since our purpose is to utilize charmonium as a probe to the hot
medium we rely on experimental values on $J/\psi$ and $c\bar c$
production in p+p collisions as the baseline for assessing in-medium
physics in A+A collisions.
For the p+p charmonium
production cross per unit rapidity we take the values $\dd
\sigma_{pp}^{\Psi}/dy$=37\,nb~\cite{Abt:2005qr} for $\sqrt
s$=17.3\,AGeV Pb-Pb~\cite{Abt:2005qr} (with ca.~40\% uncertainty) and
$\dd \sigma_{pp}^{\Psi}/dy$=750(500)\,nb for $\sqrt s$=200\,AGeV
Au-Au~\cite{Adare:2006kf} at mid and forward rapidity (with
ca.~10(20)\% uncertainty). The input charm quark N+N cross section at
SPS energy, $\dd \sigma_{\bar cc}/\dd y$ ($y$=0)=2.2~$\mu$b, is taken
from a recent compilation of experimental data in
Ref.~\cite{Lourenco:2006vw}. For full RHIC energy, we use $\dd
\sigma_{\bar cc}/\dd y$ ($y$=0)=123$\pm$40$\mu$b, in line with recent
PHENIX measurements~\cite{Adare:2006hc}. We reduce the input charm
quark cross section at forward rapidity by 1/3, $\dd \sigma^{\bar
  cc}_{pp}/\dd y$ ($y$=1.7)=$\frac{2}{3}\dd \sigma^{\bar cc}_{pp}/\dd
y$($y$=0), according to recent experimental data~\cite{Zhang:2008kr}.

\section{Charmonium Production in A+A Collisions}
\label{sec:psi_aa}

After having fixed charmonium production in p+p collisions the next
step is to scale it by the number of binary nucleon-nucleon (N+N)
collisions, $N_{\rm{coll}}$, to estimate charmonium primordial
production in A+A collisions, which is a standard procedure for a hard
probe. $N_{\rm{coll}}$ is usually calculated with the Glauber model,
which plays an important role in connecting p+p and A+A
collisions. Below we briefly review the main results from the
(optical) Glauber model which are needed in this work. The emphasis is
placed on concepts relevant for $\Psi$ production, for a more
comprehensive review of the Glauber model, see
Ref.~\cite{Miller:2007ri}.
\subsection{Brief Review of Glauber Model}
\label{ssec:glauber}
\begin{figure}[tp]
  \centering
  \includegraphics[width=0.89\textwidth,clip=]{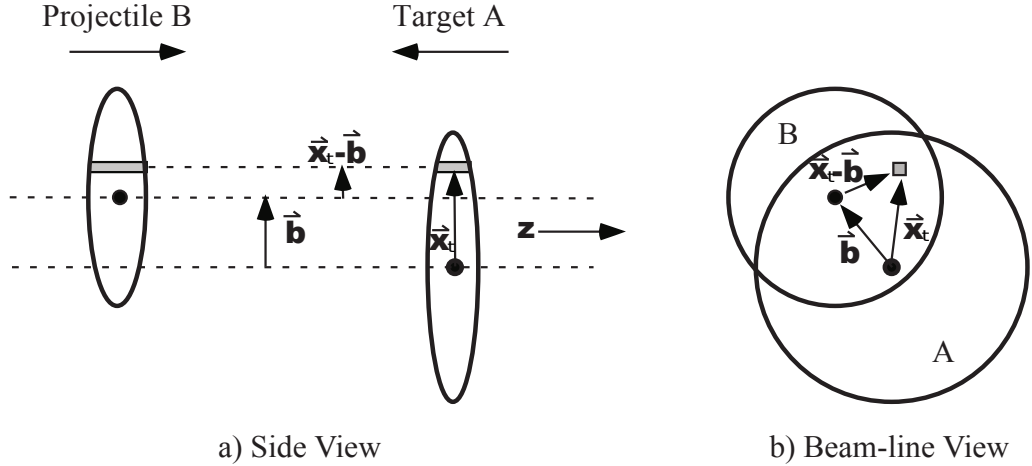}
  \caption[Schematic representation of the geometry of Glauber model]{Schematic representation of the geometry of Glauber model. Left panel: transverse view. Right panel: longitudinal view. The picture is taken from Ref.~\cite{Miller:2007ri}.}
\label{fg:glauber}
\end{figure}

In this section we give a brief review of the Glauber model and
introduce two important quantities linking A+A collisions with
elementary nucleon-nucleon (N+N) collisions: 1) the number of
binary collisions, $N_{\rm{coll}}$, 2) the number of participants
or wounded nucleons, $N_{\rm{part}}$, which are nucleons from 
the projectile or the target which suffer at least one inelastic
collision. In the Glauber model the collision between two nuclei, A
and B, consisting of $A$ and $B$ nucleons, respectively, is
considered as a superposition of (binary) collisions of the individual incoming
nucleons. The geometry of the Glauber model is
schematically sketched in Fig.~\ref{fg:glauber}.

The inputs of the Glauber model are 1) nuclear charge densities,
usually taken as a Wood-Saxon density profile,
 \begin{equation}
\rho (r) = \frac{\rho_0}{1 + \exp \left( \frac{r-R}{a} \right) } \ ,
\label{eq:wood_saxon}
\end{equation}
where $\rho_0$=0.17\,fm$^{-3}$ is the nucleon density in the center of the nucleus,
$R$ is the nuclear radius and $a$ is the thickness of the nuclear
skin. For $^{197}$Au, $R$=6.38\,fm, a=0.535\,fm. For $^{208}$Pb,
$R$=6.62\,fm, a=0.549\,fm~\cite{De Jager:1974dg}; 2) Inelastic nucleon-nucleon cross section,
$\sigma^{NN}_{inel}$, which can be obtained from
experimental measurements, \eg, $\sigma^{NN}_{inel}$=30(42)\, mb at
SPS (RHIC)~\cite{Nakamura:2010zzi}.

The nuclear thickness function, 
\begin{equation}
  T_A(\vec{x}_t) = \int\limits_{}^{} dz \; \hat{\rho}_{A}(\vec{x}_t,z)\ ,
\end{equation}
can be interpreted as the probability of finding a nucleon within
a unit transverse area at $\vec{x}_t$ of nucleus A.  The $\hat{\rho}_{A}$ is the nuclear
density in Eq.~(\ref{eq:wood_saxon}) normalized to 1, namely,
\begin{equation}
  \int \hat{\rho}_{A} d^2x_tdz= \frac{1}{A}\int \rho_{A} d^2x_tdz\ =1\ .
\label{eq:ws_norm}
\end{equation}
Therefore $T_A(\vec{x}_t)T_B(\vec{x}_t-\vec{b})$ is the joint
probability of finding a pair of nucleons from nuclei A and B,
respectively, within the common unit transverse area at
$\vec{x}_t$. Their corresponding number of collisions is given by
$T_A(\vec{x}_t)T_B(\vec{x}_t-\vec{b})\sigma^\mathrm{NN}_\mathrm{inel}\,d^2x_t$. Integrating
over the transverse plane we obtain the total number of collisions
contributed by this pair,
\begin{equation} 
  N^{pair}_{\mathrm{coll}}(b)=\sigma^{\mathrm{NN}}_{\mathrm{inel}}\,{T}_{AB}(b) = \sigma^\mathrm{NN}_\mathrm{inel}\int {T}_A
    (\vec{x}_t ){T}_B (\vec{x}_t - \vec{b})\ d^2 x_t\ ,
\label{eq:nuc_overlap}
\end{equation}
where $T_{AB}(b)$ is called the nuclear overlap function.
Since from the two nuclei A and B a total number of $AB$ such
pairs can be found and each of them contributes an equal number of
collisions, we obtain the following expression for the total number of
binary collisions
\begin{equation}
\label{eq:ncoll}
N_{\mathrm{coll}} (b)  =AB\,T_{AB}(b)\,\sigma^\mathrm{NN}_\mathrm{inel}\ . 
\end{equation}


The centrality of heavy-ion collisions is often expressed in terms of
the number wounded nucleons (participants), $N_\mathrm{part}$. In the
Glauber model, $N_\mathrm{part}$ can be estimated as follows: The
probability for a given nucleon from nucleus A to be located at
transverse position $\vec{x}_t$ is $T_A(\vec{x}_t)$, and the
probability for this nucleon to collide with a nucleon from nucleus B
(located at $(\vec{x}_t-\vec{b})$) is $T_A(\vec{x}_t)\,T_B (\vec{x}_t - {\vec
  b})\,\sigma^\mathrm{NN}_\mathrm{inel}$. The probability of not
colliding is thus $T_A(\vec{x}_t)\,(1-T_B( \vec{x}_t - {\vec b})\,
\sigma^\mathrm{NN}_\mathrm{inel})$. The probability of not
colliding with any of the $B$ nucleons from nucleus B is thus
$T_A(\vec{x}_t)\,[1-T_B(\vec{x}_t - {\vec
  b})\,\sigma^\mathrm{NN}_\mathrm{inel}]^B$. Therefore the probability
for the nucleon at $\vec{x}_t$ suffering at least one collision is $T_A(\vec{x}_t)\,(1
- [ {1 - T_B ( { \vec{x}_t - {\vec b} }
  )\,\sigma^\mathrm{NN}_\mathrm{inel} } ]^B)$.  Integrating over the
transverse plane we obtain the probability for a given nucleon in
nucleus A suffering at least one collision,
\begin{align} 
  P^A_{wo}(b)= \int T_A \left( \vec{x}_t \right)\left\{ {1 - \left[ {1 - T_B
          ( { \vec{x}_t - {\vec b} }
          )\,\sigma^\mathrm{NN}_\mathrm{inel} } \right]^B }
  \right\}d^2 x_t\ .
\end{align}
Since there are $A$($B$) nucleons in nucleus A(B) we obtain the total number of wounded nucleons
(participants) in A+B collisions at impact parameter $b$ as
\begin{align}
  N_\mathrm{part}(b) &= A\ P^A_{wo}(b) + B\ P^B_{wo}(b)\nonumber  \\&=A \int {T_A
    \left( \vec{x}_t \right)\left\{ {1 - \left[ {1 - T_B \left( {
                \vec{x}_t - {\vec b} }
            \right)\sigma^\mathrm{NN}_\mathrm{inel} } \right]^B }
    \right\}d^2 x_t
  } \nonumber  \\
  &\qquad +B\int {T_B \left( {\vec{x}_t - \vec{b} } \right)\left\{ {1
        - \left[ {1 - T_A \left( {\vec{x}_t }
            \right)\sigma^\mathrm{NN}_\mathrm{inel} } \right]^A }
    \right\}d^2 x_t}\ .
\label{eq:n_part}
\end{align}

\subsection{Cold Nuclear Matter Effects}
\label{ssec:cnm}
The notion that $\Psi$ production in A+A collisions can be viewed as superposition of independent N+N collisions is only approximately true. The deviation of primordial $\Psi$ production in A+A from $N_{\mathrm{coll}}$-scaled p+p collisions is usually attributed to the so-called cold nuclear matter (CNM) effects.

In this section we examine the following three aspects of CNM effects: 1) Nuclear shadowing, 2) Cronin effect 3) Nuclear absorption. These CNM effects can, in principle, be estimated from p+A collisions where no hot medium is expected to form.
\subsubsection{Nuclear Shadowing}
\label{sssec:shadow}
\begin{figure}[tp]
  \centering
  \includegraphics[width=0.59\textwidth,clip=]{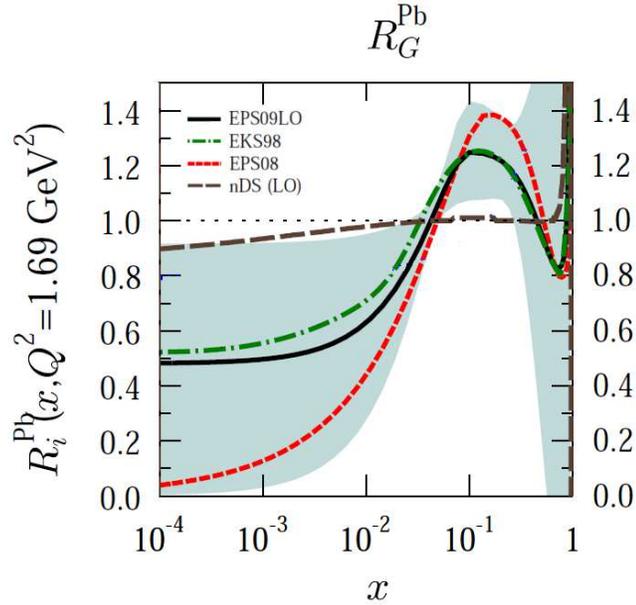}
  \caption[Modification of parton distribution function in lead nucleus]{Modification of parton distribution function in nucleus. Figure taken from Ref.~\cite{Eskola:2009uj}.}
\label{fg:npdf}
\end{figure}

It is a well-established fact that the partonic structure of high-energy nuclei is different from the incoherent superposition of the constituent nucleons, see Ref. \cite{Armesto:2006ph} for a recent review. This modification is usually parameterized by
\begin{equation}
\label{eq4}
R^A_i (x,Q^2) = \frac{f^A_i (x,Q^2)}{ A f_i (x,Q^2)}\ , \ \
i = q, \bar{q}, g \ ,
\end{equation}
defined as the ratio of the parton distribution function for a
nucleon inside the nucleus (nPDF), $f_i^A(x,Q^2)$, to the
corresponding one for a free proton (PDF), $f_i(x,Q^2)$, where $x$ is the
longitudinal momentum fraction of the parton within the nucleon. As
illustrated in Fig.~\ref{fg:npdf} different names have
been assigned to these modifications depending on the relevant range
of $x$ under consideration: 1) {\it Shadowing} for the suppression
observed at small ($x\lesssim 0.05$). 2) {\it Antishadowing} for the
enhancement at moderate values of $0.05\lesssim x\lesssim 0.3$. 3)
{\it EMC effect} for the suppression observed in the region
$0.3\lesssim x\lesssim 0.7$; and 4) {\it Fermi motion} for the
enhancement when $x\to 1$.
Let us estimate the impact on charmonium production from the
modification of nuclear parton distribution function by assuming that the
initially produced charm quark pair has the same rapidity as the
charmonium into which they evolve. Then the momentum fraction of the incoming
partons $x$ is
\begin{equation}
x_{1,2} = \frac{m_t}{\sqrt{s_{NN}}} \exp{(\pm y)}\ ,
\label{eq:intr-x1-x2-expr}
\end{equation}
with the transverse mass $m_t=\sqrt{m^2_\Psi+p_t^2}$ of $\Psi$, and $y$ being its momentum rapidity.  In
$\sqrt{s}$=17.3AGeV Pb+Pb collisions at SPS the relevant
$x\sim$0.2 is in the anti-shadowing region. In $\sqrt{s}$=200AGeV
Au+Au collisions at RHIC, for charmonia produced at mid-rapidity,
the relevant $x$ for partons from both colliding nuclei is around
0.02, close to the transition from the shadowing to the
anti-shadowing region.  For charmonia produced at forward rapidity,
$y\sim$1.7, the $x$ of the parton from the forward going (in the same
direction with the produced $\Psi$) nucleus is around 0.1
(anti-shadowing region), while the $x$ of the backward going parton is
around 0.003 (shadowing region). In most parameterizations of nPDF,
the shadowing of the backward parton is generally considered to be
stronger than the anti-shadowing of the forward going parton, see
Fig.~\ref{fg:npdf}. So the overall effect from the modification of
parton distribution function is expected to cause suppression of
$\Psi$ production relative to p+p
collisions. 
\begin{figure}[tp]
  \centering
  \includegraphics[width=0.49\textwidth,clip=]{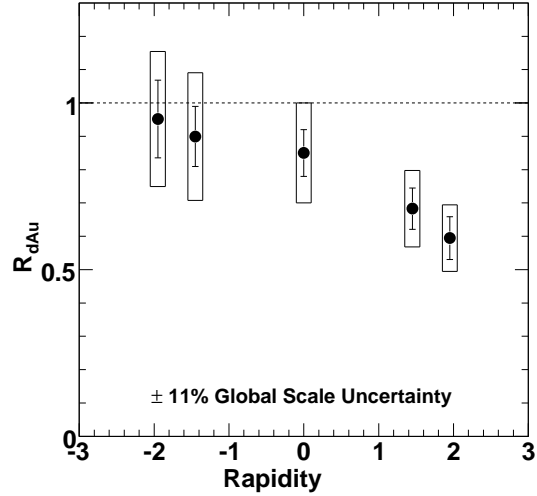}
  \caption[Rapidity dependence of $J/\psi$ $R_{dAu}$ in d+Au collisions at $\sqrt{s}$=200AGeV measured by PHENIX]{Rapidity dependence of $J/\psi$ $R_{dAu}$ in d+Au collisions at $\sqrt{s}$=200AGeV measured by PHENIX~\cite{Adare:2007gn}. }
  \label{fg:shdw_dAu}
\end{figure}
However, for high $p_t$ $\Psi$ produced at forward $y$ at RHIC the $x$
of the backward going parton could be shifted to 0.01 (for
$p_t$=10\,GeV), where the shadowing is small and could well be
compensated by the anti-shadowing of the forward going
parton. Therefore it is expected that for high $p_t$ $\Psi$ produced
at forward $y$ the suppression due to nuclear shadowing should be
suppressed.  The shadowing / anti-shadowing effects are experimentally
observed in d+Au collisions at RHIC, see
Fig.~\ref{fg:shdw_dAu}. The enhancement (suppression) of charmonia
produced at backward (forward) rapidity is due to anti-shadowing
(shadowing).
\subsubsection{Cronin Effect}
\label{sssec:cronin}
\begin{figure}[tp]
  \centering
  \includegraphics[width=0.49\textwidth,clip=]{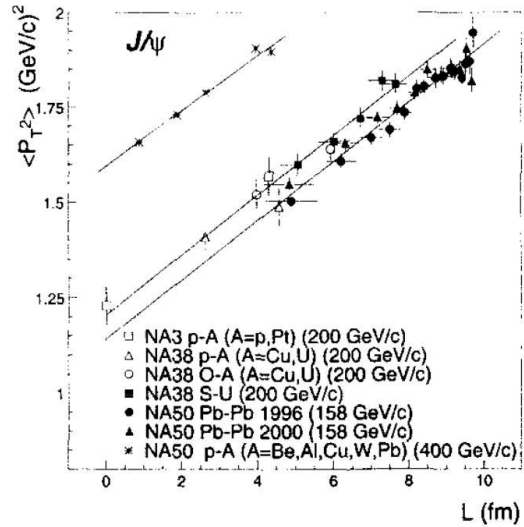}
  \caption[$\langle p^2_t\rangle$ of the $J/\psi$ as a function of the geometric length of matter, $L$, traversed by partons in the initial state]{$\langle p^2_t\rangle$ of the $J/\psi$ as a function of the geometric length of matter, $L$, traversed by partons in the initial state. The figure is taken from Ref.~\cite{Topilskaya:2003iy}.}
\label{fg:pt2_sps}
\end{figure}

The Cronin effect refers to an enhancement of hadron production at
intermediate and high $p_t$ in p+A
relative to p+p collisions (scaled by $N_{\mathrm{coll}}$). This
effect is generally attributed to multiple soft scatterings of the
projectile partons propagating through the target nucleus before the hard
scattering. From the transverse kicks in the soft scatterings the
partons acquire additional $\langle p^2_t\rangle$ and therefore the
$\langle p^2_t\rangle$ of the finally produced charmonia increases
correspondingly.  Fig.~\ref{fg:pt2_sps} illustrates the Cronin effect
at SPS energies, where the $\langle p^2_t\rangle$ of charmonium
produced in p+A collisions increases with size of the colliding
nucleus (approximated by the path length travelled by the colliding
parton). The Cronin effect for $J/\psi$ is currently difficult to quantify at
RHIC energies due to large uncertainties associated with d+Au
data.
\subsubsection{Nuclear Absorption}
\label{sssec:nucabs}
\begin{figure}[tp]
  \centering
  \includegraphics[width=0.49\textwidth,clip=]{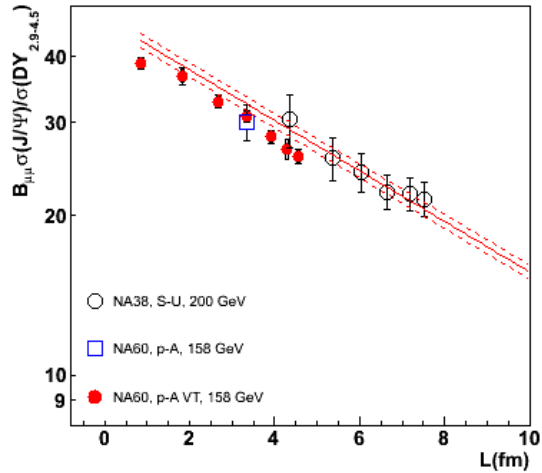}
  \caption[$J/\psi$ $R_{AA}$ (normalized to Drell-Yan pairs) vs. $L$ for p+A, S+U, Pb+Pb
  systems]{$J/\psi$ $R_{AA}$ (normalized to Drell-Yan pairs)
    vs. $L$ for various p+A and S+U systems. Note that this ``$L$'' is
    the effective average length travelled by (pre-) charmonium
    states, not by the initial state partons. This is different from
    the ``$L$'' in Fig.~\ref{fg:pt2_sps}. The figure is taken from
    Ref.~\cite{Arnaldi-priv}.}
\label{fg:nuabs_sps}
\end{figure}

In p+A or A+A collisions the ``pre-resonance'' $c\bar{c}$
states (those $c\bar c$ pairs close to each other in phase space which
would form charmonium if there were no rescattering off surrounding
particles, \ie, in p+p collisions) are subject to
dissociation through inelastic collisions with passing-by nucleons
before they are fully developed into charmonia. Nuclear absorption is
observed in p+A collisions at SPS, where the energy deposited is
too small to create a hot medium: charmonium production is
substantially suppressed relative to ($N_{\rm{coll}}$ scaled)
p+p collisions, see Fig.~\ref{fg:nuabs_sps}.
\begin{figure}[tp]
  \centering
  \includegraphics[width=0.49\textwidth,height=0.7\textheight]{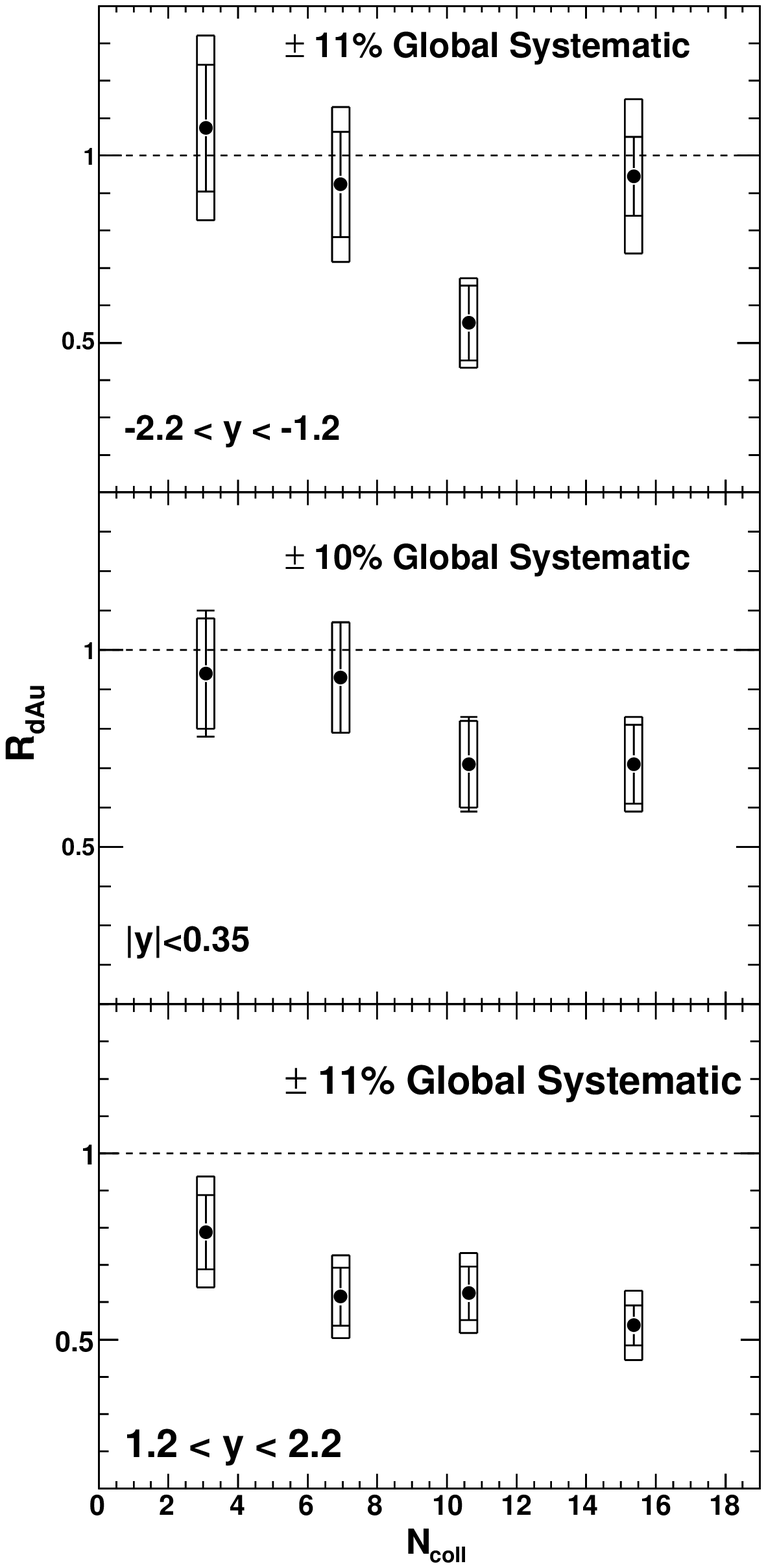}
  \caption[Centrality dependence of $J/\psi$ $R_{dAu}$ in 200AGeV d+Au collisions measured by PHENIX]{Centrality dependence of $J/\psi$ $R_{dAu}$ in 200AGeV d+Au collisions measured by PHENIX~\cite{Adare:2007gn}.}
\label{fg:nucabs_rhic}
\end{figure}
At RHIC energy nuclear absorption is observed in d+Au collisions, see
Fig.~\ref{fg:nucabs_rhic}, which is weaker than at SPS, presumably
because the time scale over which the two nuclei pass through each
other (2$R/\gamma$, $\gamma$: Lorentz factor) is ten times shorter at
RHIC. Therefore the ``pre-resonance'' $c\bar{c}$ states collide with
passing-by nucleons shortly after their pair production, when they are
still small in size (not far from their production vertex) and thus
have a smaller dissociation cross section compared to at SPS energy.

\subsection{Implementation of CNM Effects}
\label{ssec:imple}

\begin{figure}[tp]
  \centering
  \includegraphics[width=0.89\textwidth,clip=]{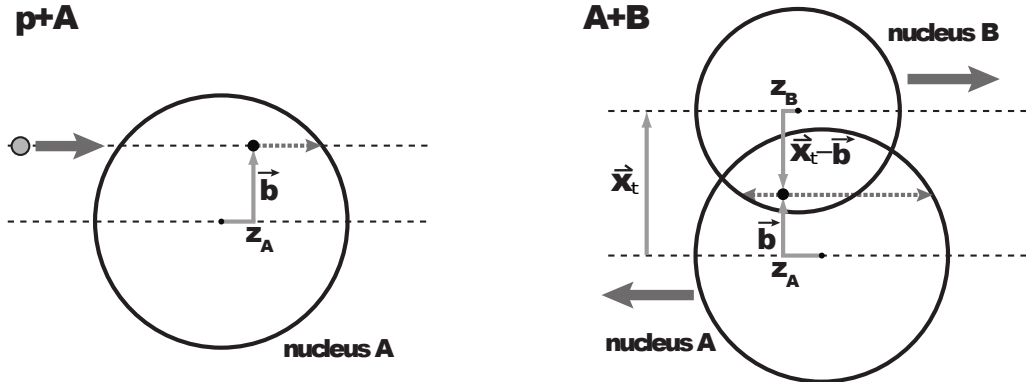}
  \caption[Schematic representation of the geometry of $J/\psi$
    nuclear absorption]{Schematic representation of the geometry of $J/\psi$
    nuclear absorption. The dot represents a (pre)-charmonium. The
    dashed arrows are their absorption trajectories. Figure taken from Ref.~\cite{Miller:2007ri}.}
\label{fg:nuc_abs}
\end{figure}

In our transport approach charmonium production in the pre-equilibrium
stage serves as the initial condition for the evolution in the hot
medium. For simplicity we assume that the initial phase space distribution
of charmonia, $f(\vec x,\vec p, \tau_0)$, determined from $\Psi$
production in the pre-equilibrium stage, can be factorized into
coordinate and momentum spaces. According to boost-invariance we only
need to consider the phase space distribution in the transverse
plane,
 \begin{align}
 f_\Psi(b,\vec{x}_t,\vec{p}_t,\tau_0)= f_\Psi(b,\vec{x}_t,\tau_0)f_\Psi(b,\vec{p}_t,\tau_0)\ .
\label{eq:cnm_x_p}
\end{align} 
Let us start with the spatial part, $f_\Psi(b,\vec{x_t},\tau_0)$. The Glauber model tells us the initially produced charmonia distribution according to the nuclear overlap function, $f_\Psi(b,\vec{x_t},\tau_0)\propto {T}_A \left(\vec{x}_t \right){T}_B (\vec{x}_t  - \vec{b})$. We augment it with the nuclear absorption: The probability for the dissociation of a pre-resonance $c\bar c$ state created at ($\vec{x}_t$,$z_A$) inside a nucleus A through passing-by nucleons in the same nucleus A is
 \begin{align}
   P^A_{\mathrm{abs}}({\vec x_t},z_{\mathrm{A}}) = 
   \sigma_{\mathrm{abs}} \, T_{\mathrm{A>}}({\vec x_t}, z_{\mathrm{A}})
   \quad \mathrm{with} \quad
   {T}_{\mathrm{A>}}({\vec x_t}, z_{\mathrm{A}}) = \int \limits_{z_\mathrm{A}}^{\infty}
   \hat{\rho}_\mathrm{A}({\vec x_t},z) \, \mathrm{d}z \ , 
   \label{eq:p_abs_A}
 \end{align}
 where $\hat{\rho}$ is normalized to 1, see
 Eq.~(\ref{eq:ws_norm}). The effective absorption cross section
 $\sigma_{\mathrm{abs}}$ parameterizes the inelastic scattering
 between $\Psi$ and nucleons. The integration limits are determined by
 the absorption trajectories, as illustrated in Fig.~\ref{fg:nuc_abs}.
Therefore the corresponding survival probability is 
\begin{align}
  P_\mathrm{surv}^\mathrm{A}({\vec x}_t,z_\mathrm{A}) =
  \left( 1 - \sigma_\mathrm{abs} \, 
  {T}_\mathrm{A>}({\vec x}_t, z_\mathrm{A}) \right)^{A-1}
  \approx 
  \exp\left(-(A-1)\, \sigma_\mathrm{abs} \, 
  {T}_\mathrm{A>}({\vec x}_t, z_\mathrm{A}) \right)\ ,
  \label{eq:p_surv_A}
\end{align}
where the factor of ($A-$1) reflects the fact that the $c\bar c$ producing nucleon does not participate in the absorption.
Similarly the probability for a pre-resonance $c\bar{c}$ state created at $(\vec x_t-\vec b,z_\mathrm{B})$ surviving from collisions with nucleons in nucleus B is
\begin{align}
  P_\mathrm{surv}^\mathrm{B}({\vec x}_t-\vec{b},z_\mathrm{B})
  \approx 
  \exp\left(-(B-1)\, \sigma_\mathrm{abs} \, 
  {T}_\mathrm{B>}({\vec x}_t-\vec{b}, z_\mathrm{B}) \right)\ .
  \label{eq:p_surv_B}
\end{align}
Therefore the number of surviving charmonia at $\vec x_t$ for impact parameter $b$ is
\begin{align}
  f_\Psi(\vec{b},\vec{x}_t,\tau_0)=&\Delta
  y\frac{\dd\sigma^\Psi_{pp}}{\dd y}AB\int \ dz_\mathrm{A}\
  dz_\mathrm{B}\hat{\rho}_A(\vec{x}_t,z_\mathrm{A})\
  \hat{\rho}_B(\vec{x}_t-\vec{b},z_\mathrm{B})\nonumber \\ &\qquad \times P_\mathrm{surv}^\mathrm{A}({\vec
    x}_t,z_\mathrm{A})P_\mathrm{surv}^\mathrm{B}({\vec
    x}_t-\vec{b},z_\mathrm{B})\nonumber\\[1mm] =&\Delta
  y\frac{\dd\sigma^\Psi_{pp}}{\dd y}AB\int \ dz_\mathrm{A}\
  dz_\mathrm{B}\hat{\rho}_A(\vec{x}_t,z_\mathrm{A})\
  \hat{\rho}_B(\vec{x}_t-\vec{b},z_\mathrm{B})\nonumber\\
  &\qquad\times\exp\left\{-(A-1)\int^\infty_{z_{\mathrm{A}}}
    dz\hat{\rho}_A(\vec{x}_t,z)\sigma_{abs}\right\}\nonumber\\
  &\qquad\times\exp\left\{-(B-1)\int^\infty_{z_{\mathrm{B}}}
    dz'\hat{\rho}_B(\vec{x}_t-\vec{b},z')\sigma_{abs}\right\}\ ,
\label{eq:glauber}
\end{align}
where $\sigma_{pp}^{\Psi}$ is the charmonium production cross
section in elementary nucleon-nucleon collisions.
The total number of surviving charmonia is thus
\begin{align}
N_{\Psi}(b)=\int f_\Psi(\vec{b},\vec{x}_t,\tau_0)\,d^2 \vec{x}_t\ .
\label{eq:npsi_nuc_supp}
\end{align}
It is convenient to define a nuclear suppression factor,
\begin{align}
\label{eq:nuc_supp}
S_{\rm nuc}(b)&=\frac{1}{T_{AB}(b)}\int d^2 \vec{x}_t \, dz_\mathrm{A}\,
  dz_\mathrm{B}\,\hat{\rho}_A(\vec{x}_t,z_\mathrm{A})\
  \hat{\rho}_B(\vec{b}-\vec{x}_t,z_\mathrm{B})\nonumber\\
  &\qquad\times\exp\left\{-(A-1)\int^\infty_{z_{\mathrm{A}}}
    dz\hat{\rho}_A(\vec{x}_t,z)\sigma_{abs}\right\}\nonumber\\
  &\qquad\times\exp\left\{-(B-1)\int^\infty_{z_{\mathrm{B}}}
    dz'\hat{\rho}_B(\vec{x}_t-\vec{b},z')\sigma_{abs}\right\}\ ,
\end{align}
which can be used to express the number of surviving charmonia as
\begin{align}
N_{\Psi}(b)=\Delta y\frac{\dd\sigma^{\Psi}_{pp}}{\dd y}\,AB{T}_{AB}(b)S_{\rm nuc}(b)\ .
\label{eq:npsi_nuc_supp_factor}
\end{align}
Eqs.(\ref{eq:glauber}), (\ref{eq:npsi_nuc_supp}), (\ref{eq:nuc_supp}) and (\ref{eq:npsi_nuc_supp_factor}) can be reduced to p+A collisions by setting $\hat{\rho}_B(\vec{x}_t,z)$=$\delta^{(2)}(\vec{x}_t)\delta(z)$ and $B$=1. For example, for p+A collisions, Eq.~(\ref{eq:nuc_supp}) reduces to
\begin{align}
\label{eq:nuc_supp_pa}
S_{\rm nuc}(b)&=\frac{1}{T_{A}(b)}\int \ dz_\mathrm{A}\
  \hat{\rho}_A(\vec{b},z_\mathrm{A})\nonumber\\
  &\qquad\times\exp\left\{-(A-1)\int^\infty_{z_A}
    dz\hat{\rho}_A(\vec{b},z)\sigma_{abs}\right\}\ .
\end{align}

\begin{figure}[tp]
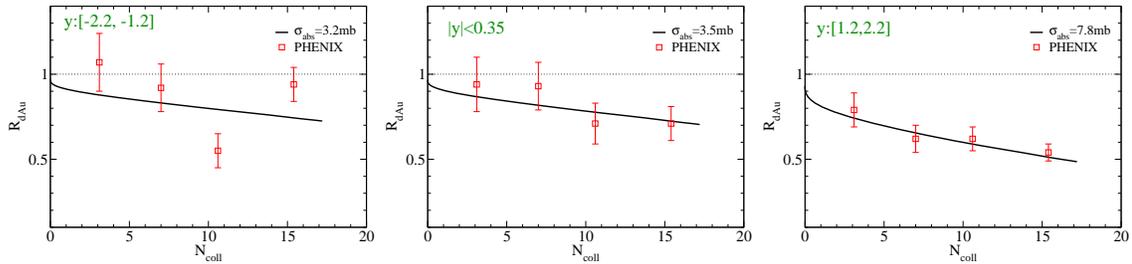

  \centering
  \includegraphics[width=0.32\textwidth,clip=]{rdau_centra_back_1004.eps}
  \includegraphics[width=0.32\textwidth,clip=]{rdau_centra_mid_0930.eps}
  \includegraphics[width=0.32\textwidth,clip=]{rdau_centra_forw_1004.eps}
  \caption[Centrality dependence of $J/\psi$ $R_{dAu}$ in 200\,AGeV d+Au collisions measured by PHENIX together with a Glauber fit with a nuclear absorption cross section $\sigma_{abs}$=3.2mb, 3.5mb, and 7.8mb for backward, mid- and forward rapidity]{Centrality dependence of $J/\psi$ $R_{dAu}$ in 200\,AGeV d+Au collisions measured by PHENIX~\cite{Adare:2007gn} together with a Glauber fit with a nuclear absorption cross section $\sigma_{abs}$=3.2mb, 3.5mb, and 7.8mb for backward, mid- and forward rapidity.}
\label{fg:glauber_nucabs_dau}
\end{figure}

We parameterize both nuclear shadowing and nuclear absorption with
the effective $\Psi$-N absorption cross section, $\sigma_{\rm
  abs}$. Applying Eq.~(\ref{eq:nuc_supp_pa}) to p+A collisions at
SPS we obtain $\sigma_{\rm abs}^{J/\psi}$=7.3$\pm$1\,mb from the
recent NA60 data at $E_{\rm lab}$=158\,GeV (corresponding to
$\sqrt{s_{NN}}$=17.3\,GeV)~\cite{Arnaldi:2010ky}. This updated
measurement at 158\,GeV gives a significantly larger value
than previously available for 400\,GeV proton projectiles,
$\sigma_{\rm abs}\simeq$4.4\,mb~\cite{Alessandro:2006jt} (the latter
has been confirmed by NA60~\cite{Arnaldi:2010ky}, \ie, at 400\,GeV), which has been used
in our previous calculations~\cite{Grandchamp:2003uw,Zhao:2007hh}.
The comparison with recent PHENIX
data~\cite{Adare:2007gn,Frawley-priv} yields $\sigma_{\rm
  abs}\simeq3.5$\,mb (5.5\,mb) for $\sqrt{s}$=200\,AGeV Au-Au
collisions at mid rapidity, $|y|<0.35$ (forward rapidity,
$|y|\in[1.2,2.2]$), see, \eg,
Fig.~\ref{fg:glauber_nucabs_dau}. For simplicity, we assume the same absorption
cross sections for the $\chi_c$ as for the $J/\psi$.  However, for
excited states $\sigma_{\rm abs}$ is expected to be significantly
larger, even if they are not fully formed when the dissociation
occurs. Taking guidance from the NA50 measurement with 400\,GeV
protons, we use $\sigma_{\rm abs}^{\psi'}\simeq13$\,mb at
$\sqrt{s}$=17.3\,AGeV and $\sigma_{\rm abs}^{\psi'}\simeq6.5(10)$\,mb
at $\sqrt{s}$=200\,AGeV for mid (forward) $y$.

The rather pronounced rapidity dependence of $\sigma_{\rm abs}$ at RHIC 
casts doubt on interpreting this quantity as an actual absorption cross 
section. It seems more reasonable to associate its increase at forward 
$y$ with nuclear shadowing~\cite{Ferreiro:2009ur} since the dissociation 
kinematics is very similar between mid and forward rapidity. While this 
does not affect the use of our ``effective" $\sigma_{\rm abs}$,
it does imply a nuclear shadowing effect on the open-charm cross
section in A+A collisions (which is an important ingredient in the
calculation of regeneration). As a ``minimal" scheme we therefore 
associate the additional absorption of the $J/\psi$ yield at forward $y$ 
(relative to mid rapidity) with a suppression of open charm production
caused by shadowing, while we assume no shadowing corrections at mid
rapidity. Thus, at both SPS and RHIC the number of primordially 
produced $c\bar c$ pairs at mid-rapidity is calculated from 
the p+p cross section as
\begin{equation}
N^{\rm mid}_{c\bar c}(b)=
\left.\Delta y\frac{\dd\sigma^{c\bar c}_{pp}}{\dd y}\right\vert_{y=0}
ABT_{AB}(b) \ ,
\label{Ncc_mid}
\end{equation}
while for forward $y$ at RHIC we use 
\begin{equation}
N^{\rm for}_{c\bar c}(b)=
\left.\Delta y\frac{\dd\sigma^{c\bar c}_{pp}}{\dd y}\right\vert_{y=1.7} 
ABT_{AB}(b)\frac{S_{\rm nuc}^{\rm for}}{S_{\rm nuc}^{\rm mid}} \ . 
\label{Ncc_forw}
\end{equation}
Here, $T_{AB}(b)$ is the usual nuclear overlap function, Eq.~(\ref{eq:nuc_overlap}) and 
$S_{\rm nuc}$, defined in Eq.~(\ref{eq:nuc_supp}), denotes the $J/\psi$ suppression factor due to CNM 
effects, parameterized by $\sigma_{\rm abs}$ in the Glauber formula, 
Eq.~(\ref{eq:glauber}). In particular, the ratio 
$S_{\rm nuc}^{\rm for}/S_{\rm nuc}^{\rm mid}$ represents the extra 
suppression associated with nuclear shadowing, operative for both 
$J/\psi$ and $c\bar c$ production. 

For the momentum dependent part $f_\Psi(b,\vec{x}_t,\tau_0)$ we take
the charmonium $p_t$ spectra in p+p collisions as the baseline and
apply a $p_t$ broadening associated with the Cronin effect. The $J/\psi$ $p_t$ spectra in
p+p collisions at SPS are parameterized with an exponential
distribution,
\begin{equation}
  \label{eq:psi_pt_sps}
  \frac{d\sigma^{J/\psi}_{pp}}{2\pi p_tdp_t}\propto \frac{1}{\langle p^2_t\rangle}e^{-p^2_t/\langle p^2_t\rangle}\ ,
\end{equation}
where $\langle p^2_t\rangle$=1.15GeV$^2$~\cite{Topilskaya:2003iy}. At full RHIC energy the
$J/\psi$ $p_t$ spectra are parameterized with a power-law
distribution,
\begin{equation}
  \label{eq:psi_pt_rhic}
  \frac{d\sigma^{J/\psi}_{pp}}{2\pi p_tdp_t}\propto \frac{C}{[1+(p_t/D)^2]^6}\ ,
\end{equation}
where $D$ is adjusted so that $\langle
p^2_t\rangle$=4.14\,(3.56)GeV$^2$ for $J/\psi$ produced at mid
(forward) rapidity~\cite{Adare:2006kf} and $C$ is a normalization factor.

The Cronin effect is readily implemented into the 3-momentum dependent
part $f_\Psi(b,\vec{p_t},\tau_0)$ via a Gaussian smearing of the
charmonium $p_t$ distribution in p+p collisions,
$f_\Psi^{pp}(p_t)$,
\begin{equation}
  f_\Psi(b,\vec{p_t},\tau_0)=\int\frac{d^2q_t}{2\pi\langle\Delta p_t^2\rangle}
\exp{\left(-\frac{q_t^2}{2 \langle \Delta p_t^2\rangle}\right )}\
  f_\Psi^{pp}(|\vec{p}_t-\vec{q}_t|) \ .  
\label{Cronin}
\end{equation}
The nuclear increase of the average $p_t^2$, $\langle \Delta
p_t^2\rangle = \langle p_t^2\rangle_{AA}-\langle p_t^2\rangle_{pp}$,
is estimated within a random-walk treatment of parton-nucleon
collisions~\cite{Hufner:2001tg} as being proportional to the mean
parton path length, $\langle l^{ab} \rangle$, in the cold medium: $ \langle
\Delta p^2_t\rangle=a_{gN}~\langle l^{ab}\rangle$. The coefficient $a_{gN}$
is estimated from p+A data at SPS~\cite{Topilskaya:2003iy} and
d-Au data at RHIC~\cite{Adare:2007gn}. We use
$a_{gN}$=0.076\,GeV$^2$/fm for $\sqrt s$=17.3\,AGeV Pb-Pb collisions
and $a_{gN}$=0.1(0.2)\,GeV$^2$/fm for $\sqrt s$=200\,AGeV Au-Au
collisions at mid (forward) rapidity. The mean parton path length $\langle l^{ab} \rangle$ is determined from the geometric
lengths $l^a$ and $l^b$ which two partons $a$ and $b$ travel before they collide, weighted
according to the survival probability of final state charmonia~\cite{Hufner:2001tg}
\begin{equation}
\label{eq:lab}
l^{ab}(\vec b)=\frac{\int \  d^2x_t\ dz_a\
dz_b\left(l^a(\vec x_t,z_a)+l^b(\vec b -\vec
x_t,z_b)\right)K(\vec b,\vec x_t,z_a,z_b)} {\int d^2x_t\ dz_a\
dz_b\ K(\vec b,\vec x_t,z_a,z_b)}\ , 
\end{equation}
where 
\begin{align}
\label{eq:la} 
&l^a(\vec x_t,z_a)=A\int^{z_a}_{-\infty}dz\hat{\rho}_A(\vec x_t,z)/\rho_0\ ,
\nonumber\\
&l^b(\vec b -\vec x_t,z_b)=B\int^{\infty}_{z_b}dz\hat{\rho}_B(\vec
x_t-\vec b ,z)/\rho_0\ ,
\end{align}
are the respective path lengths travelled by parton $a$ and $b$
weighted according to the nuclear density $\rho(\vec s,z)$. 
The kernel $K(\vec b,\vec x_t,z_A,z_B)$ is given by
\begin{align}
\label{eq:kernel} 
K(\vec b,\vec x_t,z_a,z_a)&=\rho_A(\vec x_t,z_a)\rho_B (\vec x_t-\vec
b,z_b)\nonumber\\
&\qquad\times\exp\left(-\sigma_{abs}\left[(A-1)\int^\infty_{z_a}dz\hat{\rho}_A(\vec
    x_t,z)\right.\right.\nonumber\\
&\qquad\left.\left.+(B-1)\int_{-\infty}^{z_b}dz\hat{\rho}_B(\vec
    x_t-\vec b,z)\right]\right)\ ,
\end{align}
reflecting  the survival probability of final state charmonia, recall Eq.~(\ref{eq:glauber}).

\graphicspath{{./6fg/}}

\chapter[Application to Heavy-Ion Collisions]{Application to Heavy-Ion Collisions}
\label{ch:nume_results}

In previous chapters we have introduced all the components of the
kinetic rate equation/transport approach. In this section we elaborate
its application to heavy-ion collisions and compare our numerical
results with experimental data.

In Section~\ref{ch:nume_results}.\ref{sec:trans_rate_eq} we discuss the procedure of
applying the kinetic approach to heavy-ion collisions. For the
suppression of primordial charmonium ($\Psi$=$J/\psi$, $\chi_c$,
$\psi'$) we employ the Boltzmann equation to evaluate the time
evolution of the $\Psi$ phase space distribution function. To solve
the time evolution of regenerated $\Psi$ using the Boltzmann 
equation requires calculating $\Psi$ regeneration rates from the full
time-dependent phase distribution function of charm quarks, which is
numerically rather involved and is still a work in progress. For the
most parts we adopt a simplified procedure: We estimate the inclusive
yield from regeneration with a rate equation and estimate their
$p_t$-spectra with the blastwave
formula. 

In Section~\ref{ch:nume_results}.\ref{sec:num_results_sps_rhic} we present the numerical
results of the inclusive $J/\psi$ yield and its transverse momentum
($p_t$) spectra and compare with SPS and RHIC data. We specifically
compare two scenarios with the internal (``strong binding'') or free
(``weak binding'') energy identified as the heavy quark 2-body
potential. The effects specifically relevant for high $p_t$ ($>$5GeV)
$J/\psi$ production are discussed. The discussions in
Section~\ref{ch:nume_results}.\ref{sec:trans_rate_eq} and
Section~\ref{ch:nume_results}.\ref{sec:num_results_sps_rhic} mostly follow
Ref.~\cite{Zhao:2010nk}.


In Section~\ref{ch:nume_results}.\ref{sec:full_boltz} we study the impact on charmonium
regeneration due to off-equilibrium effects in charm-quark phase space
distributions. Specifically, we compare the inclusive yield and $p_t$
spectra of regenerated $J/\psi$ from limiting charm-quark spectra
including: (1) thermal charm-quark spectra; (2) pQCD charm
spectra. Also, the impact of a charm-quark correlation volume is
discussed within the framework of the Boltzmann equation.

In Section~\ref{ch:nume_results}.\ref{sec:psip} the production of excited $\Psi$ states such
as $\chi_c$ and $\psi'$ at SPS and RHIC is briefly discussed.

Finally, in Section~\ref{ch:nume_results}.\ref{sec:psi_fair} we present predictions for
charmonium production at FAIR energies, where the medium is expected
to have a lower initial temperature and higher baryon density compared
to SPS energies.


\section{The Rate Equation Approach}
\label{sec:trans_rate_eq}

Throughout this chapter we solve kinetic equations (Boltzmann or rate
equation) separately for $\Psi$=$J/\psi$, $\chi_c$, $\psi'$.  In the
hot medium the dissociation and regeneration are the two main
processes affecting the $\Psi$ yield. It is often desirable to
disentangle these two effects and study their respective strength. For
this purpose we decompose the charmonium distribution in the medium at
any time $\tau$,
\begin{equation}
f_\Psi(p_t,x_t,\tau)=f^{\rm prim}_{\Psi}(p_t,x_t,\tau)+f^{\rm reg}_{\Psi}(p_t,x_t,\tau) \ , 
\label{eq:fpsi-tau}
\end{equation}
into a (suppressed) primordial component and a regenerated one by
exploiting the linearity of the Boltzmann or 
rate equation.  According to boost-invariance we only need to
solve for the time evolution of $f_\Psi(p_t,x_t,\tau)$ in the
transverse plane. We define $f^{\rm prim}_{\Psi}(p_t,x_t,\tau)$ as the
solution of the homogeneous Boltzmann equation,
\begin{equation}
\label{eq:boltz-eq_dir}
\partial f^{\rm prim}_\Psi/\partial \tau+v_t \cdot \nabla_t f^{\rm prim}_\Psi
=-\alpha_\Psi f^{\rm prim}_\Psi\ ,
\end{equation}
with the same initial condition as for the full Boltzmann equation,
(\ref{eq:boltz}), $f_{\Psi}^{\rm prim}(p_t,x_t,\tau_0)$ =
$f_{\Psi}(p_t,x_t,\tau_0)$, which is obtained from
Eq.~(\ref{eq:cnm_x_p}).  The explicit expression for the solution of
Eq.~(\ref{eq:boltz-eq_dir}) is
\begin{equation}
  \label{eq:solu_boltz_prim}
    f^{\rm prim}_\Psi(p_t,x_t,\tau)=f_\Psi(p_t,x_t-v_t(\tau-\tau_0),\tau_0)e^{-\int^\tau_{\tau_0} d \tau' \alpha_\Psi(p_t,x_t-v_t(\tau-\tau'),\tau')}\ .
\end{equation}
We account for the ``leakage effect'', \ie, charmonia escaping the
fireball volume no longer being subject to suppression, by setting
$\alpha_\Psi\equiv0$ whenever
$|\vec{x}_t-\vec{v}_t(\tau-\tau')|>R(\tau')$, where $R(\tau')$ is the
fireball radius at time $\tau'$.  Due to the leakage effect the
suppression of high $p_t$ charmonia is reduced compared to low $p_t$
ones, since the former are more likely to escape from the fireball.
  
The regeneration component, $f^{\rm reg}_{\Psi}$, follows as the
difference between the solution of the full and the homogeneous
Boltzmann equation, which can be expressed as
\begin{equation}
\label{eq:boltz-eq_reg}
 \partial f^{\rm reg}_\Psi/\partial \tau+v_t \cdot \nabla_t f^{\rm reg}_\Psi
=-\alpha_\Psi f^{\rm reg}_\Psi+\beta_\Psi\ ,
\end{equation} 
with vanishing initial condition, 
$f_{\Psi}^{\rm reg}(\vec x,\vec p,\tau<\tau_0^\Psi)=0$. The onset time of 
regeneration processes, $\tau_0^\Psi$, is 
defined by the dissociation temperature $T(\tau_0^\Psi)= T^{\rm diss}_\Psi$ for each state $\Psi$. 
The explicit expression for the solution of Eq.~(\ref{eq:boltz-eq_reg}) is
\begin{equation}
  \label{eq:solu_boltz_reg}
    f^{\rm reg}_\Psi(p_t,x_t,\tau)=\int^\tau_{\tau_0^\Psi}d \tau'\beta_\Psi(p_t,x_t-v_t(\tau-\tau'),\tau')e^{-\int^\tau_{\tau'} d \tau'' \alpha_\Psi(p_t,x_t-\vec v_t(\tau-\tau''),\tau'')}.
\end{equation}

Due to the complication mentioned in the introduction of this chapter
we adopt the following approximation: Instead of the Boltzmann
transport equation~(\ref{eq:boltz-eq_reg}) we solve the following rate
equation for the inclusive yield of the regeneration component,
\begin{equation}
\label{eq:rate-eq_reg}
\frac{\dd N_{\Psi}^{\rm reg}}{\dd \tau}=
-\Gamma_{\Psi} \ (N_{\Psi}^{\rm reg}-N_{\Psi}^{\rm eq}) \ .
\end{equation} 
For the dissociation rate, $\Gamma_{\Psi}$, we employ a 3-momentum average whose precise value
is obtained by matching the final yield of the loss term to the exact
result obtained from solving the momentum-dependent Boltzmann equation for the primordial component, Eq.~(\ref{eq:boltz-eq_dir}).
For the $\Psi$ equilibrium limit we adopt the statistical model mentioned in Section~\ref{ch:trans}.\ref{sec:cc_corr}.
Since the thermal production and annihilation rates of $c\bar c$ are 
believed to be small at SPS and RHIC energies, $c\bar c$ pairs are 
assumed to be exclusively produced in primordial N+N collisions
and conserved thereafter. The open and hidden charm states are then
populated in relative chemical equilibrium according to the
canonical charm-conservation equation, (\ref{eq:psi_reg_can2}),
\begin{equation}
N_{c\bar c}=Z_1\frac{I_1(2Z_1)}
{I_0(2Z_1)} + N_{\rm hid} \ ,
\label{eq:Ncc}
\end{equation}
with $N_{c\bar c}$: total number of charm-quark pairs from initial
production; $Z_1 =\gamma_c V_{\rm FB} n_{\rm op}$: 1-body open charm
partition function (recall Eq.~(\ref{eq:1-part})), with pertinent
equilibrium density $n_{\rm op}$; $N_{\rm hid}=\gamma_c^2 V_{\rm FB}
n_{\rm hid}$: total number of all charmonium states with pertinent
equilibrium density $n_{\rm hid}$; and $\gamma_c$: charm-quark
fugacity accounting for the deviation of chemical equilibrium with the
heatbath ($\gamma_c$=1 in full equilibrium). The ratio of modified
Bessel functions, $I_1(2Z_1)/I_0(2Z_1)$, on the right-hand-side of
Eq.~(\ref{eq:Ncc}) is the characteristic canonical suppression factor
which accounts for the exact conservation of net-charm number,
$N_c-N_{\bar c}$, in each event~\cite{Cleymans:1990mn,Gorenstein:2000ck}: for $Z_1\ll
1$, one has $I_1(2Z_1)/I_0(2Z_1)\to Z_1$,
which acts as an additional (small) probability to enforce a vanishing
net charm content in the system (\ie, both $c$ and $\bar c$ have to be
present simultaneously), see Section~\ref{ch:trans}.\ref{sec:cc_corr}.\ref{ssec:canonical} for a more
detailed discussion.

The 1-body open charm partition function, $Z_1$, is evaluated as follows.  For the
QGP phase in the weak-binding scenario only charm quarks are counted
as open-charm states. In the strong-binding scenario, the $T$-matrix
calculations of Ref.~\cite{Riek:2010fk} suggest that $c\bar q$ and
$\bar cq$ (charm-light) bound states ($D$-mesons) survive in QGP up to
$\sim$1.3\,$T_c$; therefore, we count both charm quarks and the
lowest-lying $S$-wave $D$-mesons ($D$, $D^*$, $D_s$ and $D^*_s$), as
open charm states for $T<1.3T_c$. The charm-quark masses in the QGP
correspond to the temperature-dependent ones displayed in
Fig.~\ref{fg:mc}, while for the meson resonances above $T_c$ we
estimate from Ref.~\cite{Riek:2010fk} $m_{D}=m_{D^*}\simeq2.0$\,GeV
and $m_{D_s}=m_{D^*_s}\simeq2.1$\,GeV (hyperfine splitting has been
neglected).  For the hadronic phase all charmed hadrons listed by the
particle data group~\cite{Yao:2006px} are counted as open-charm
states, with their vacuum masses.  The number of hidden charm states, $N_{\rm
  hid}$, is evaluated in line with the existing charmonium states and
their masses at given temperature $T$, but its contribution to
$N_{c\bar c}$ is numerically negligible.

\begin{figure}[!t] 
\centering
\includegraphics[width=0.49\textwidth]{gammac_b0_sm_1016.eps}
\includegraphics[width=0.49\textwidth]{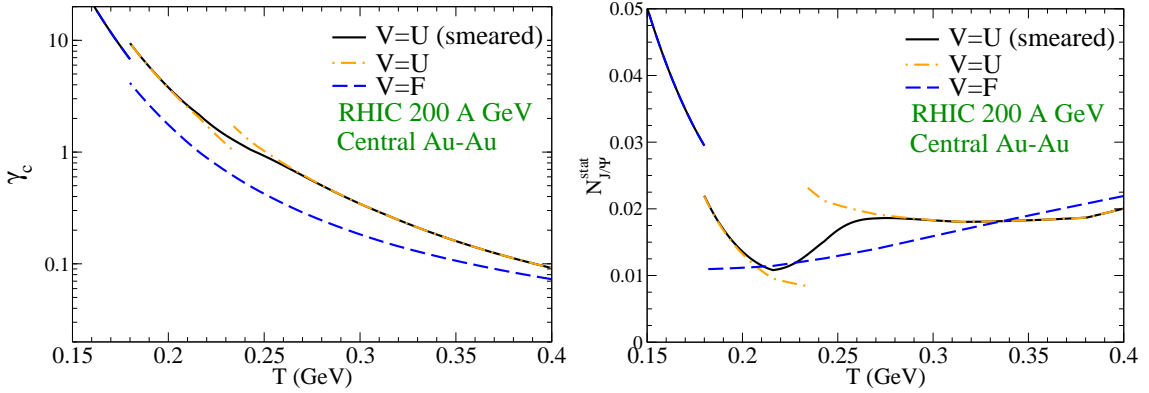}
\caption[Temperature dependence of charm-quark fugacity and in-medium
$J/\psi$ equilibrium limit using the statistical model in the QGP
within the strong-binding scenario, the weak-binding scenario and
in the hadronic phase for temperatures below $T_c$=180\,MeV]{Temperature
  dependence of the charm-quark fugacity (left panel) and the
  in-medium $J/\psi$ equilibrium limit (right panel) using the
  statistical model in the QGP within the strong-binding scenario
  (dot-dashed lines: with and without $D$-meson resonances below and
  above 1.3\,$T_c$$\simeq$234\,MeV, respectively; solid line: smooth
  interpolation of the previous two cases; see text for details), the
  weak-binding scenario (dashed line) and in the hadronic phase for temperatures
  below $T_c$=180\,MeV.}
\label{fg:fig_jpsi-eq}
\end{figure}

Knowing $n_{\rm op}$, $n_{\rm hid}$ and $V_{\rm FB}$ at each 
temperature, one can solve Eq.~(\ref{eq:Ncc}) for the charm-quark 
fugacity, $\gamma_c(T)$, and apply it to compute the statistical
equilibrium limit of each charmonium state as
\begin{equation} 
N_{\Psi}^{\rm stat}=\gamma_c^2 \ V_{\rm FB} \ n_{\Psi}\ ,
\label{Npsi-stat}
\end{equation} 
in terms of its equilibrium density, $n_{\Psi}$.  In
Fig.~\ref{fg:fig_jpsi-eq} we collect the numerical results of the
charm-quark fugacity $\gamma_c(T)$ and the statistical equilibrium limit
for $J/\psi$ abundances (excluding feeddown) for central 200\,AGeV
Au+Au collisions at RHIC.
The discontinuity at 1.3\,$T_c$ for the strong-binding scenario
(dot-dashed line) is due to the inclusion of the $D$ resonances in 
the QGP medium. We smoothly interpolate around the melting temperature
for the $D$-mesons with a hyperbolic tangent function (solid line) 
to represent a more gradual (dis)appearance of the $D$ resonances
(we have checked that this procedure has negligible impact on 
the calculation of observables in Section~\ref{ch:nume_results}.\ref{sec:num_results_sps_rhic}). 

To achieve a more realistic implementation of the statistical
equilibrium limit, we apply two corrections to $N_{\Psi}^{\rm stat}$
to schematically implement off-equilibrium effects of charm quarks in
momentum and coordinate space. The former is aimed at simulating
incomplete thermalization of the charm-quark $p_t$ spectra throughout
the course of the thermally evolving bulk medium. We have shown in
Section~\ref{ch:psi_hot}.\ref{sec:open_charm_med} that the coalescence rate from non-
or partially thermalized $c$- and $\bar c$-quark spectra is smaller
than for fully thermalized ones~\cite{Grandchamp:2002wp,Greco:2003vf},
since the former are harder than the latter and thus provide less
phase-space overlap for charmonium bound-state formation. We implement
this correction by multiplying the charmonium abundances from the
statistical model with a schematic relaxation
factor~\cite{Grandchamp:2002wp},
\begin{equation}
N_{\Psi}^{\rm eq}={\mathcal R}(\tau) \  N_{\Psi}^{\rm stat} \ , \ 
{\mathcal R}(\tau)=1-\exp (-\tau/\tau^{\rm eq}_c)\ ,
\label{eq:tau_c}
\end{equation}
where $\tau^{\rm eq}_c$ is a parameter which qualitatively represents
the thermal relaxation time of charm quarks (it is one of our 2 main
adjustable parameters in our phenomenological applications in
Section~\ref{ch:nume_results}.\ref{sec:num_results_sps_rhic}). A rough estimate of this time scale may be
obtained from microscopic calculations of this quantity within the
same $T$-matrix approach as used here for charmonia, where the thermal
charm-quark relaxation time turns out to be $\tau_{\rm
  eq}^c\simeq$3-10\,fm/$c$~\cite{vanHees:2007me,Riek:2010fk}.  Such
values allow for a fair description of open heavy-flavor suppression
and elliptic flow at RHIC~\cite{vanHees:2007me,vanHees:2005wb}.  The
second correction is applied in coordinate space, based on the
realization that, after their pointlike production in hard N+N
collisions, the $c$ and $\bar c$ quarks only have a limited time to
diffuse throughout the fireball volume. At RHIC and especially at SPS
only few $c\bar c$ pairs are produced (\eg, $\dd N_{c\bar
  c}/dy\simeq1.2$ in semicentral ($b$=7\,fm) Au+Au collisions at
RHIC), and the hadronization time (QGP lifetime) is smaller than the fireball
radius. Thus, $c$ and $\bar c$ will not be able to explore the full
fireball volume but rather be restricted to a ``correlation volume'',
$V_{\rm co}$~\cite{Grandchamp:2003uw} (the analogous
concept has been successfully applied to strangeness production in
p+A and A+A collisions in the SPS energy
regime~\cite{Hamieh:2000tk}), see Section~\ref{ch:trans}.\ref{sec:cc_corr}.\ref{ssec:corr_vol} for a
more detailed discussion. In the present rate equation approach we implement this
correction by replacing the fireball volume $V_{\rm FB}$ in the
argument of the Bessel functions in Eq.~(\ref{eq:Ncc}) by the correlation
volume $V_{\rm co}$~\cite{Grandchamp:2003uw,Young:2008he}. The
latter is identified with the volume spanned by a receding $c\bar c$
pair,
\begin{equation}
V_{\rm co}(\tau)=\frac{4\pi}{3} (r_0 + \langle v_c\rangle \tau)^3 \ ,
\label{eq:vco_tau}
\end{equation}
where $r_0\simeq1.2$\,fm represents an initial radius characterizing
the range of strong interactions, and $\langle v_c\rangle$ is an
average speed with which the produced $c$ and $\bar c$ quark recede
from the production point; we estimate it from the average $p_t$ in
$D$-meson spectra in p+A
collisions~\cite{Klinksiek:2005kq,Grandchamp:2003uw,Rapp:2009my} as
$\langle v_c\rangle\simeq0.55(0.6)c$ at SPS (RHIC).  The correlation
volume leads to a significant increase of $\gamma_c$ (since $I_0/I_1$
is reduced) and thus of the modified $\Psi$ ``equilibrium limit" due
to locally increased $c\bar c$ densities. 

The $p_t$ spectra of regenerated charmonia are approximated by local thermal distributions boosted 
by the transverse flow of the medium, amounting to a standard 
blastwave description~\cite{Schnedermann:1993ws}, Eq.~(\ref{eq:blastwave_2}),
\begin{equation}
\label{pt-reg}
\frac{\dd N_{\Psi}^{\rm reg}}{p_t\dd p_t}\propto 
m_t\int^R_0 rdr K_1\left(\frac{m_t\cosh \rho}{T}\right)
I_0\left(\frac{p_t\sinh \rho}{T}\right) \ 
\end{equation}
($m_t=\sqrt{m_\Psi^2+p^2_t}$). The medium is characterized by the
transverse-flow rapidity $\rho=\tanh^{-1}v^{cell}_\perp(r)$ using a
linear flow profile $v_t(r)=v_{s}\frac{r}{R}$ with a surface velocity
$v_s=a_\perp\tau_{\rm mix}$ and transverse fireball radius
$R=R(\tau_{\rm mix})$ as given by the fireball expansion formula,
Eq.~(\ref{eq:Vfb}), at the end of the mixed phase, $\tau_{\rm mix}$.
We evaluate the blastwave expression at the hadronization transition
($T_c$) and neglect rescattering of $\Psi$'s in the hadronic phase.
Two additional effects are neglected in this treatment, which, to a
certain extent, tend to compensate each other: on the one hand, due to
incomplete charm-quark thermalization, one expects the regenerated
charmonium spectra to be harder than in equilibrium, but, on the other
hand, a good part of the regeneration occurs before the mixed
phase~\cite{Rapp:2005rr,Yan:2006ve} (as will be shown in
Section~\ref{ch:nume_results}.\ref{sec:num_results_sps_rhic}.\ref{ssec:inclu_yield}) so that the evaluation of
the blastwave expression at the end of the mixed phase presumably
overestimates the blue shift due to the flow field. We will explicitly
check these effects in
Section~\ref{ch:nume_results}.\ref{sec:full_boltz}.  Ultimately, an
explicit evaluation of the gain term with realistic (time-dependent)
charm-quark spectra within a Boltzmann equation~\cite{Zhao:2010-2}
will be able to lift these approximations.

\section{Inclusive $J/\psi$ Yield and $p_t$ Spectra at SPS and RHIC}
\label{sec:num_results_sps_rhic}

In this section we present and discuss the numerical applications
of the above framework to $J/\psi$ data in URHICs at SPS and RHIC.
For each observable, we confront the results of the strong- and 
weak-binding scenario in an attempt to discriminate qualitative
features. The feeddown to $J/\psi$ from $\chi_c$ and $\psi'$ states 
is taken into account, assuming fractions of 32\% and 8\%, 
respectively, for primordial production in $pp$ collisions.  
We have divided the discussion into the centrality dependence
of inclusive $J/\psi$ yields in Section~\ref{ch:nume_results}.\ref{sec:num_results_sps_rhic}.\ref{ssec:inclu_yield} and
the $p_t$ dependence of $J/\psi$ in Section~\ref{ch:nume_results}.\ref{sec:num_results_sps_rhic}.\ref{ssec:psi_pt}.

\subsection{Inclusive $J/\psi$ Yield}
\label{ssec:inclu_yield}

The $J/\psi$ yield in A+A collisions is usually quantified
in terms of the nuclear modification factor as a function
of centrality,
\begin{equation}
\label{eq:raa}
R_{AA}(b)=\frac {N_{J/\psi}^{AA}(b)}{N_{J/\psi}^{pp}N_{\rm coll}(b)} \ ,
\end{equation}
where $N_{\rm coll}(b)$ is the number of binary collisions of the
incoming nucleons at impact parameter $b$, recall
Eq.~(\ref{eq:ncoll}).  Before we turn to the results, we recall the
two main parameters in our approach, which are the strong coupling
constant, $\alpha_s$, and the thermal charm-quark relaxation time,
$\tau_c^{\rm eq}$.  The former controls the inelastic charmonium
reaction rate, (Eqs.~(\ref{eq:m_qfree}) and (\ref{eq:m_qfree_reg})),
and the latter the magnitude of the $\Psi$ equilibrium limits,
Eq.~(\ref{eq:tau_c}). We adjust them to reproduce the inclusive
$J/\psi$ yield for central A+A collisions at SPS and RHIC, within
reasonable bounds.  For $\alpha_s$ we find that a common value of
0.32, which is at the upper end of the value in the Coulomb term in
the $Q\bar Q$ free energy, can be used, in combination with
$\tau_c^{\rm eq}$=3.8\,fm/$c$ for the strong-binding scenario and
$\tau_c^{\rm eq}$=1.6\,fm/$c$ for the weak-binding scenario. For
simplicity, we refrain from introducing an additional temperature
dependence into these parameters. The composition of the total yield,
its centrality dependence and $p_t$ spectra can then be considered as
a prediction within each of the 2 scenarios.

\begin{figure}[!t]
\centering
\includegraphics[width=0.49\textwidth]
{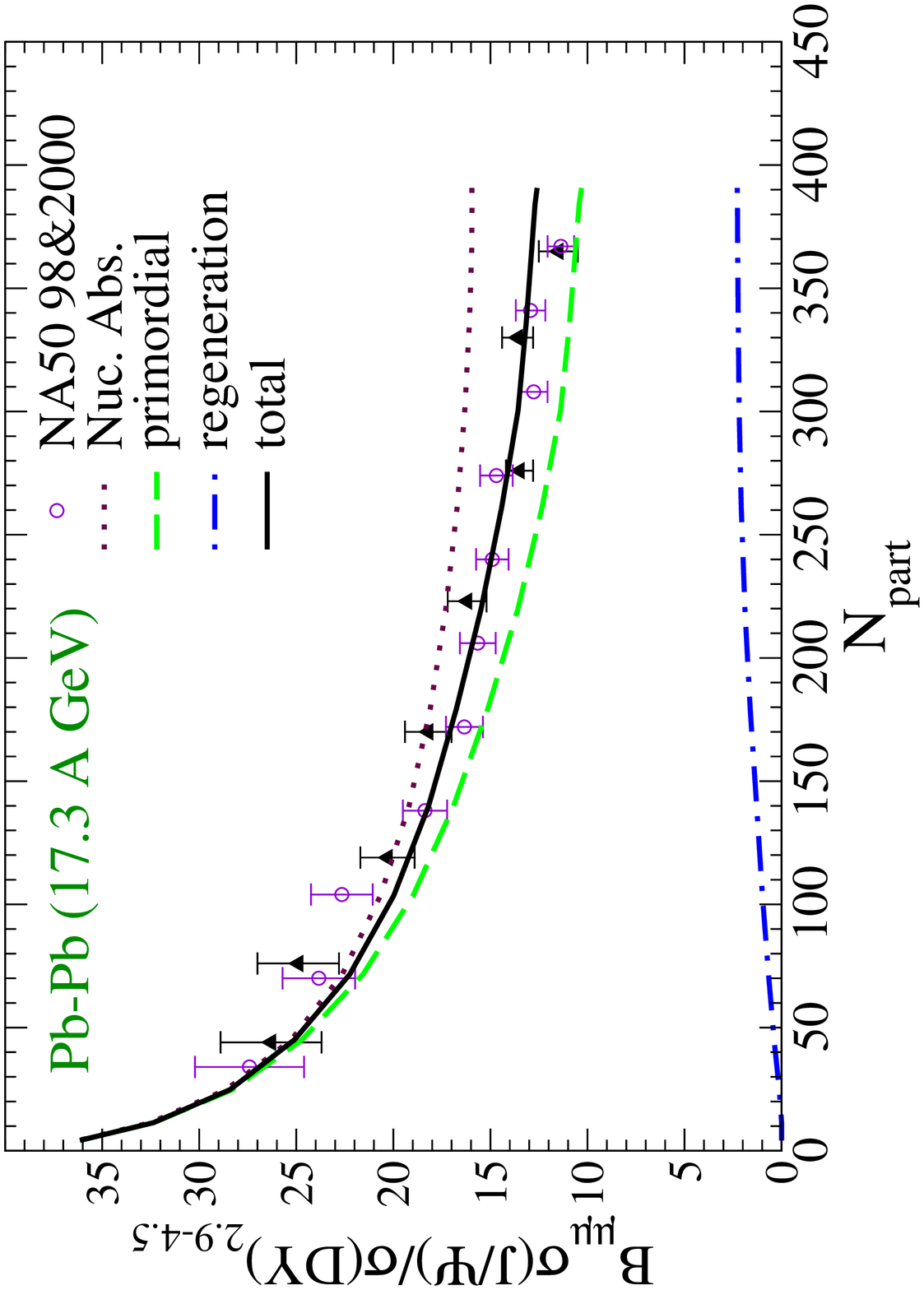}
\includegraphics[width=0.49\textwidth]
{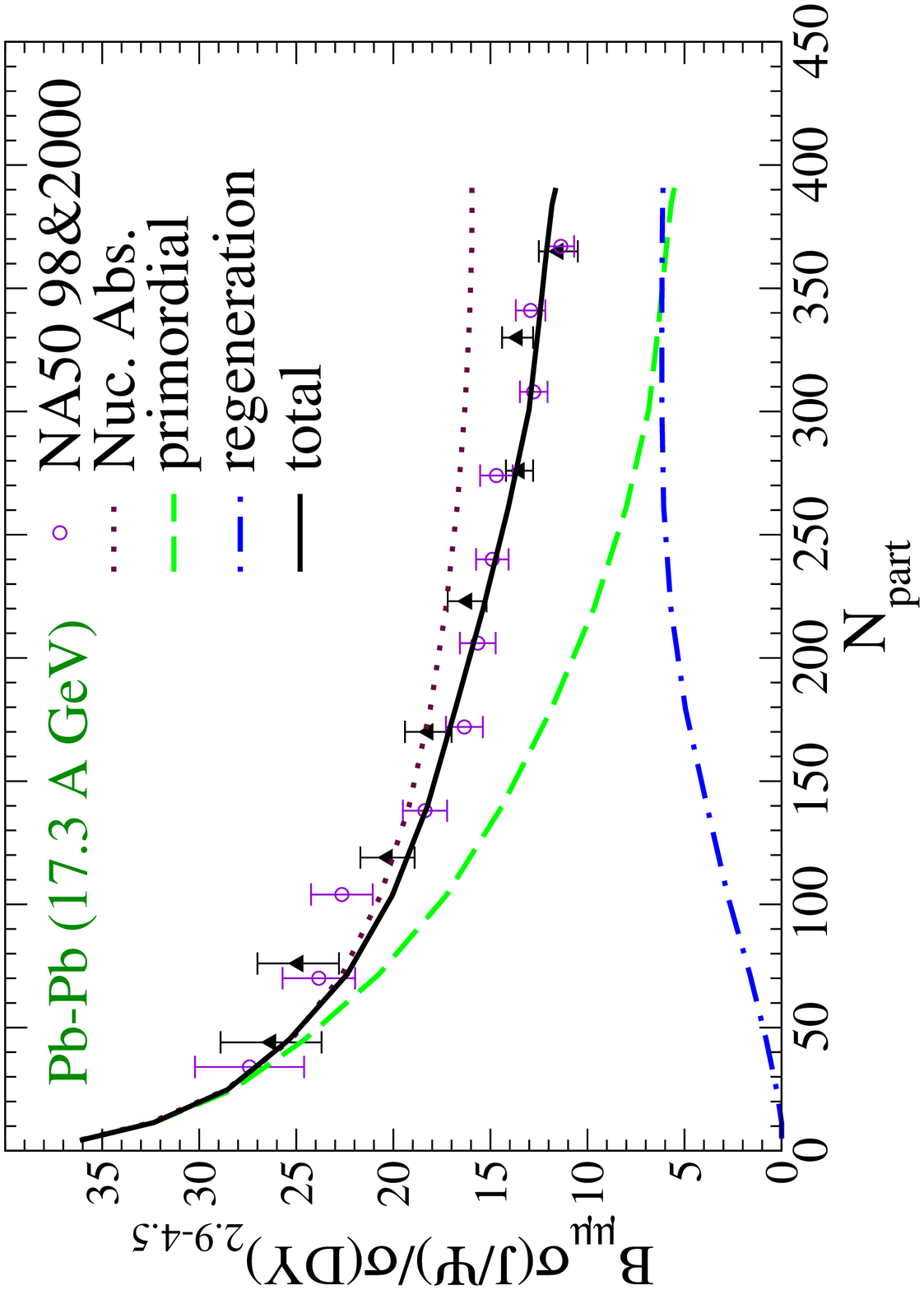}
\caption[$J/\psi$ production versus centrality at SPS evaluated with
the thermal rate-equation approach, compared to NA50 data]{$J/\psi$
  production (normalized to Drell-Yan pairs) versus centrality at SPS
  evaluated with the thermal rate-equation approach, compared to NA50
  data~\cite{Ramello:2003ig,Alessandro:2004ap}. Solid lines: total
  $J/\psi$ yield; dashed lines: suppressed primordial production;
  dot-dashed lines: regeneration component; dotted lines: primordial
  production with CNM effects only. Left panel: strong-binding
  scenario; right panel: weak-binding scenario.}
\label{fg:fig_raa-sps}
\end{figure}
\begin{figure}[!t]
\centering
\includegraphics[width=0.49\textwidth]
{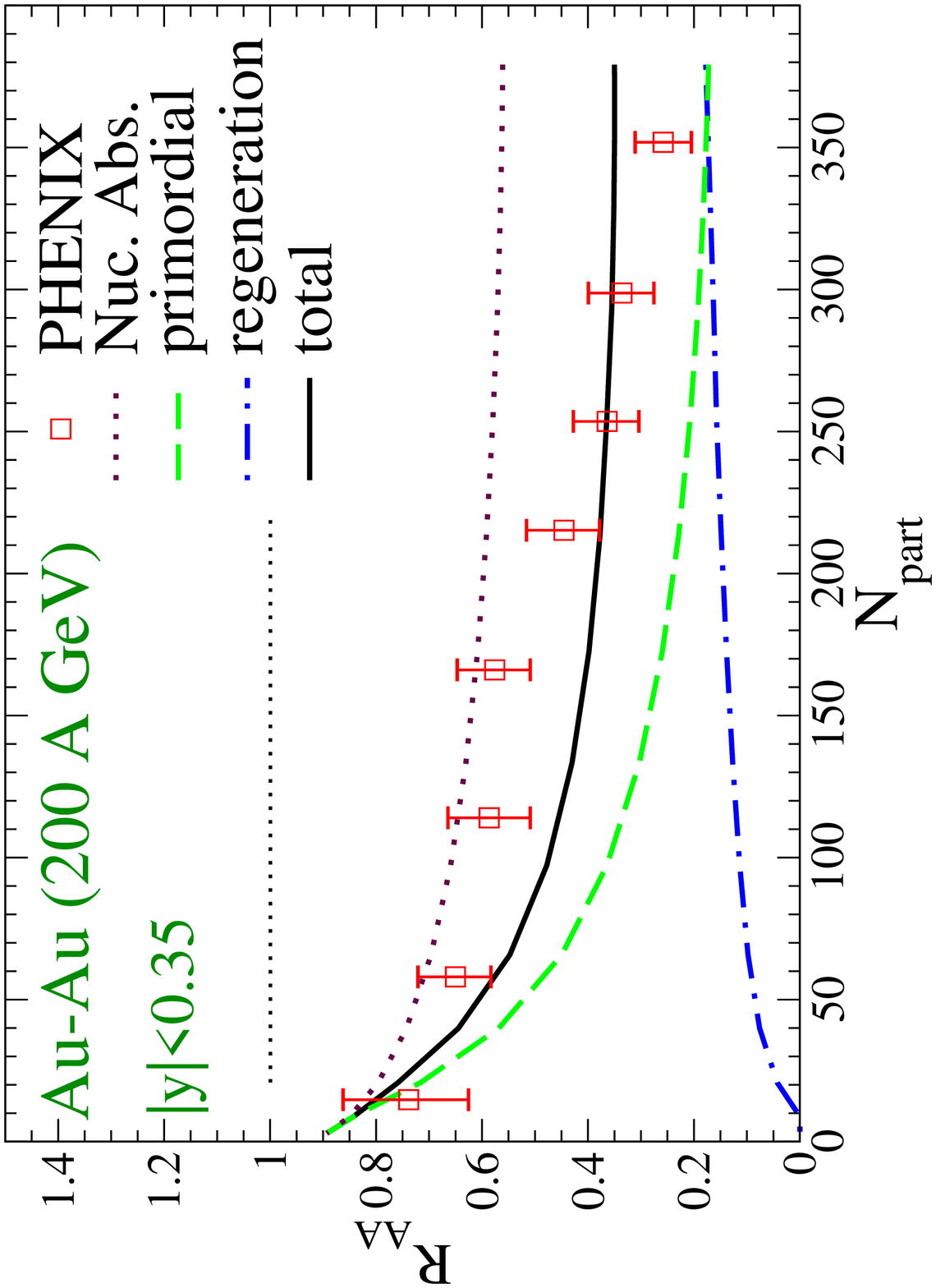}                                                                         
\includegraphics[width=0.49\textwidth]
{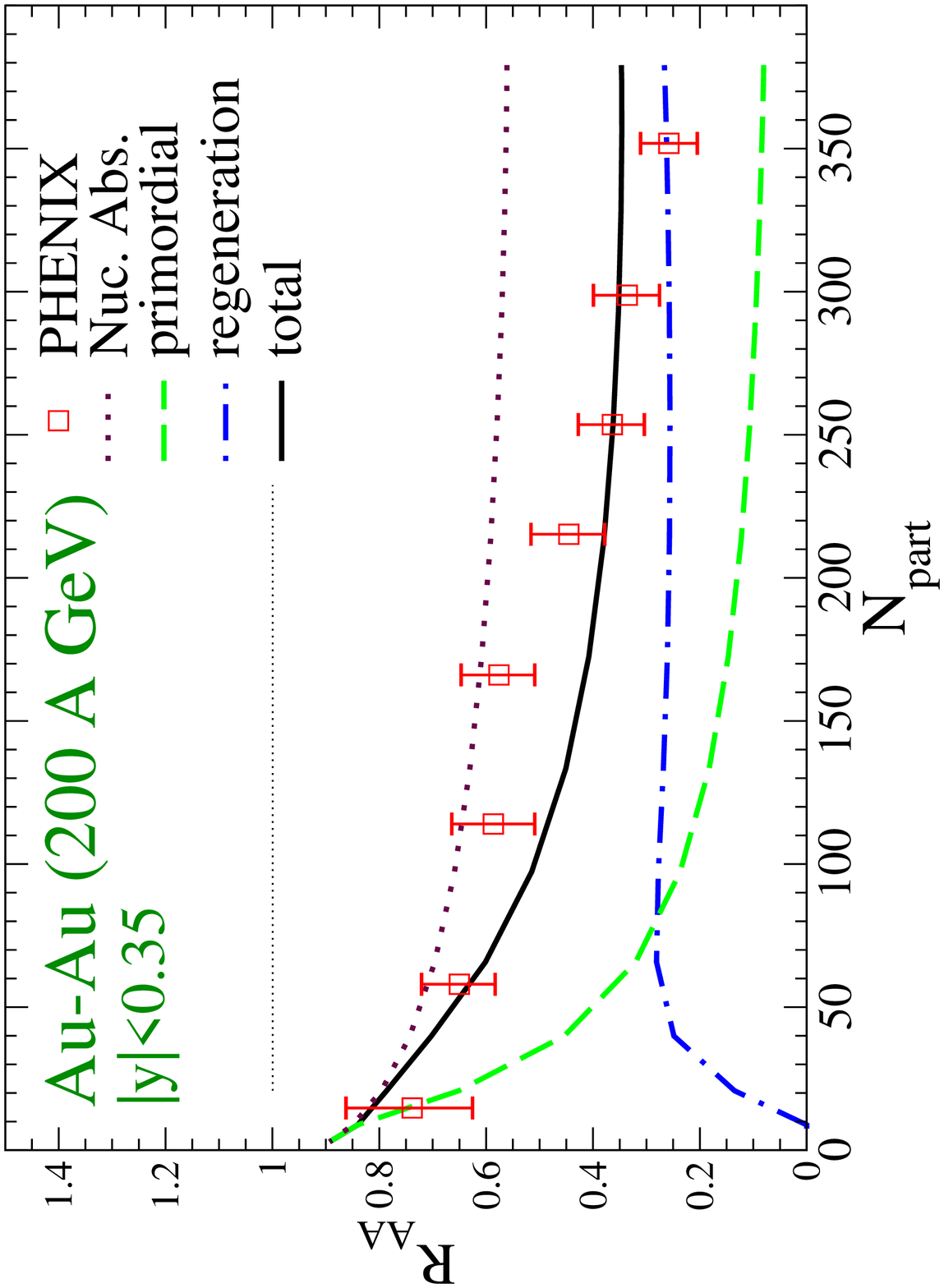}
\caption[$J/\psi$ $R_{AA}$ versus centrality at mid-rapidity at RHIC
evaluated with the thermal rate-equation approach, compared to PHENIX
data]{$J/\psi$ $R_{AA}$ versus centrality at mid-rapidity at RHIC
  evaluated with the thermal rate-equation approach, compared to
  PHENIX data~\cite{Adare:2006ns}.  Solid lines: total $J/\psi$ yield;
  dashed lines: suppressed primordial production; dot-dashed lines:
  regeneration component; dotted lines: primordial production with CNM
  effects only. Left panel: strong-binding scenario; right panel:
  weak-binding scenario.}
\label{fg:raa_rhic_mid}
\end{figure}

We begin with $J/\psi$ production in $\sqrt s$=17.3\,AGeV Pb+Pb
collisions at SPS, for which we compare our results in the strong-
and weak-binding scenario with NA50 data in Fig.~\ref{fg:fig_raa-sps}.
For these data, the denominator in Eq.~(\ref{eq:raa}) is replaced by
the number of Drell-Yan dileptons at high mass, while
the numerator includes the branching ratio into dimuons.
The pertinent proportionality factor, equivalent to the $pp$ limit
of this ratio (47.0$\pm$1.4~\cite{Roberta-priv}), and the CNM-induced
suppression (dotted line in Fig.~\ref{fg:fig_raa-sps}) are
inferred from the latest NA60 p+A measurements~\cite{Arnaldi:2010ky},
which we reproduce using the Glauber model formula, Eq.~(\ref{eq:glauber}),
with $\sigma_{\rm abs}$=7.3\,mb. The suppression of the primordial
component (dashed line) relative to nuclear absorption
(dotted line) represents the ``anomalous'' suppression by the hot
medium, which increases with centrality due to higher initial 
temperatures and longer fireball lifetimes. The regeneration 
component increases with centrality as well, mostly due to
the increase of the $\cal{R}$-factor and the larger lifetime
which facilitates the approach to the equilibrium limit 
according to Eq.~(\ref{eq:rate-eq_reg}). 
Because of detailed balance between dissociation and regeneration, 
an increase in the former also implies an increase in the latter. 
The sum (solid line) of primordial and regeneration contributions
describes the centrality dependence of the inclusive $J/\psi$ yield 
at SPS reasonably well in both scenarios. In the strong-binding 
scenario the primordial component is dominant and the majority of 
the anomalous suppression originates from the dissociation of 
$\chi_c$ and $\psi'$, since at the temperatures realized at SPS 
($T_0\simeq200$\,MeV) the quasifree dissociation rates for $\chi_c$ 
and $\psi'$ are much larger than those for $J/\psi$, recall 
Fig.~\ref{fg:rate_temp}. In the weak-binding scenario, however, the 
regeneration yield becomes comparable to the primordial one for 
semi-/central collisions due to larger dissociation rates and the 
smaller charm-quark equilibration time scale.

\begin{figure}[!t]
\includegraphics[width=0.49\textwidth]
{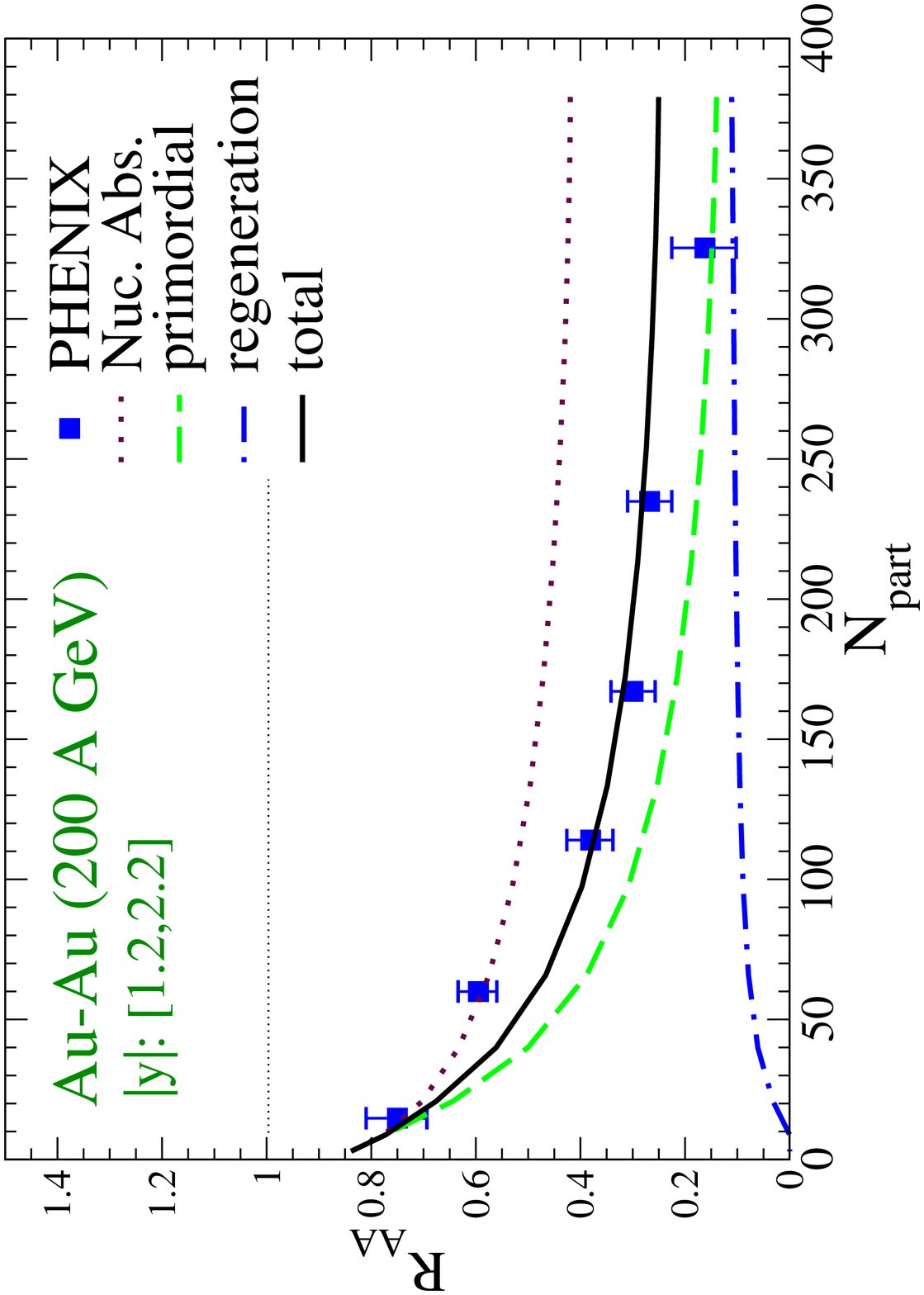}                                                                                                                                             
\includegraphics[width=0.49\textwidth]
{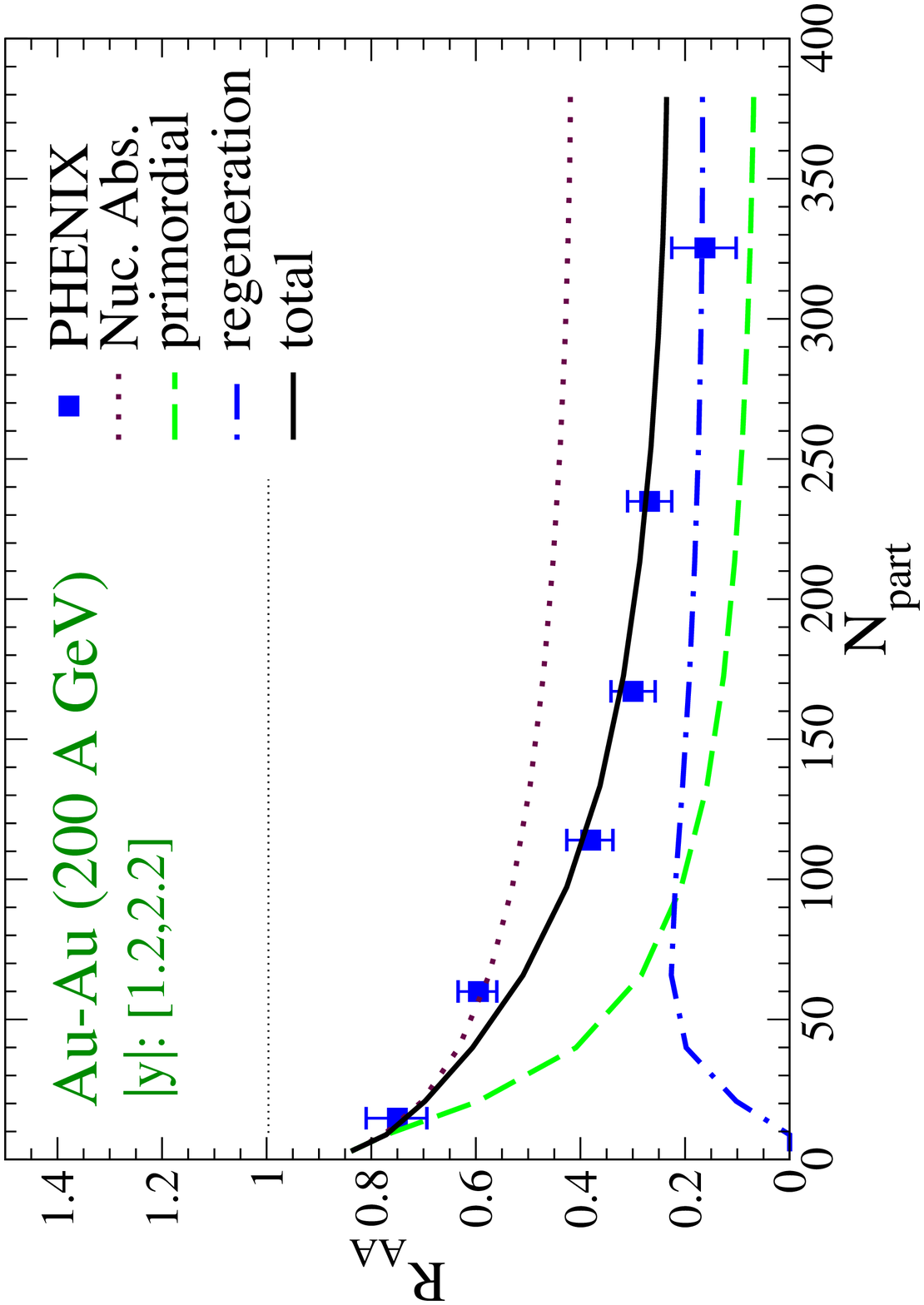}
\caption[$J/\psi$ $R_{AA}$ versus centrality at forward rapidity at
RHIC compared to PHENIX data]{$J/\psi$ $R_{AA}$ vs. centrality at
  forward rapidity compared to PHENIX data~\cite{Adare:2006ns}. Solid
  lines: total $J/\psi$ yield; dashed lines: suppressed primordial
  production; dot-dashed lines: regeneration component; dotted lines:
  primordial production with CNM effects only. Left panel:
  strong-binding scenario; right panel: weak-binding scenario.}
\label{fg:raa_rhic_forw}
\end{figure}
\begin{figure}[!t]
\centering
\includegraphics[width=0.59\textwidth]
{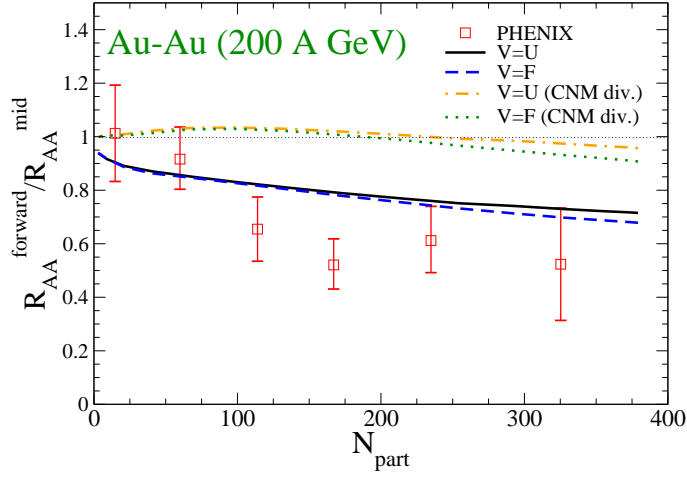}
\caption[Ratio of $J/\psi$ $R_{AA}$ between forward and mid-rapidity
versus centrality in the strong and weak binding scenarios compared to
PHENIX data]{Ratio of $R_{AA}$ for $J/\psi$ at forward and mid
  rapidity versus centrality in the strong (solid line) and weak (dashed
  line) binding scenarios compared to PHENIX
  data~\cite{Adare:2006ns}. In the upper two curves, CNM effects have
  been divided out in both numerator and denominator of the ratio.}
\label{fg:raa_centra_forw_mid}
\end{figure}

Next we examine the centrality dependence of $J/\psi$ production
in 200\,AGeV Au+Au collisions at RHIC, first focusing on mid
rapidity ($|y|<0.35$), as shown in Fig.~\ref{fg:raa_rhic_mid}.  The
suppression due to CNM effects (dotted line in Fig.~\ref{fg:raa_rhic_mid})
is inferred from latest PHENIX d+Au measurements, which we reproduce
using the Glauber model formula Eq.~(\ref{eq:glauber}) with
$\sigma_{\rm abs}$=3.5\,mb. For $N_{\rm part}\simeq0-100$, the
composition of primordial and regeneration contributions is
quite comparable to the SPS for $N_{\rm part}\simeq0-400$ within
both scenarios. Beyond $N_{\rm part}\simeq100$, suppression and
regeneration continue to increase, leveling off at an approximately
50-50\%  (20-80\%) partition for primordial and regeneration in the
strong-binding (weak-binding) scenario in central collisions.

Let us now turn to $J/\psi$ production at forward rapidity
($|y|\in[1.2,2.2]$) at RHIC, shown in Fig.~\ref{fg:raa_rhic_forw}.
Again, both strong- and weak-binding scenarios reproduce the
experimental data fairly well, with similar relative partitions 
for primordial and regeneration contributions as at mid rapidity.
However, one of the ``puzzles" about $J/\psi$ production at RHIC
is the fact that the total $J/\psi$ yield is more strongly 
suppressed at forward rapidity than at mid rapidity. In our approach 
this follows from the stronger shadowing at forward rapidity leading to
less primordial production for both $J/\psi$ and $c\bar c$ pairs. The
former (latter) leads to a reduction of the primordial (regeneration)
component. Since the thermodynamic properties of the fireball are
quite similar at mid and forward rapidity (recall Fig.~\ref{fg:fig_fb}), 
charmonium suppression and regeneration in the hot medium are 
very similar between the mid and forward rapidity as discussed in 
Ref.~\cite{Zhao:2008pp}.  
To quantify the difference at forward rapidity and mid rapidity 
we display the ratio between the corresponding $R_{AA}$'s in 
Fig.~\ref{fg:raa_centra_forw_mid} for both scenarios, which clearly
illustrates the importance of CNM effects to properly reproduce 
the data.

In Section~\ref{ch:psi_hot}.\ref{sec:qgp} we have argued that the strong- and weak-binding
scenarios discussed here may be considered as limiting cases 
for $J/\psi$ binding in the QGP, as bracketed by the identification
of the heavy-quark internal and free energies with a $Q$-$\bar Q$
potential. From the results above we believe that these scenarios
also provide a reasonably model-independent bracket on the role
of suppression and regeneration effects, in the following sense:   
At SPS, the strong-binding scenario defines a ``minimal" amount
of dissociation required to provide the anomalous suppression beyond
CNM effects (a small regeneration component is inevitable due to
detailed balance). The application to RHIC energy then implies 
an approximately equal partition of primordial and regenerated
charmonia in central Au+Au, not unlike Ref.~\cite{Yan:2006ve} where
the vacuum charmonium binding energies (``strong binding") have
been used in the QGP (together with the gluo-dissociation). 
On the other hand, in the weak-binding scenario, a large part of the 
$J/\psi$ yield in central A+A is due to regeneration even at SPS, 
limited by the constraint that for sufficiently peripheral collisions 
(and at sufficiently large $p_t$) a transition to primordial production 
compatible with p+A data should be restored. Clearly, for 
central A+A at RHIC (and certainly at LHC) the final yield is 
then dominated by regeneration. 
Since both scenarios describe the inclusive yields reasonably well,
it is mandatory to investigate more differential observables to find
discriminating evidence. This will be pursued in the following
section.  

\begin{figure}[!t]
\centering
\includegraphics[width=0.49\textwidth]
{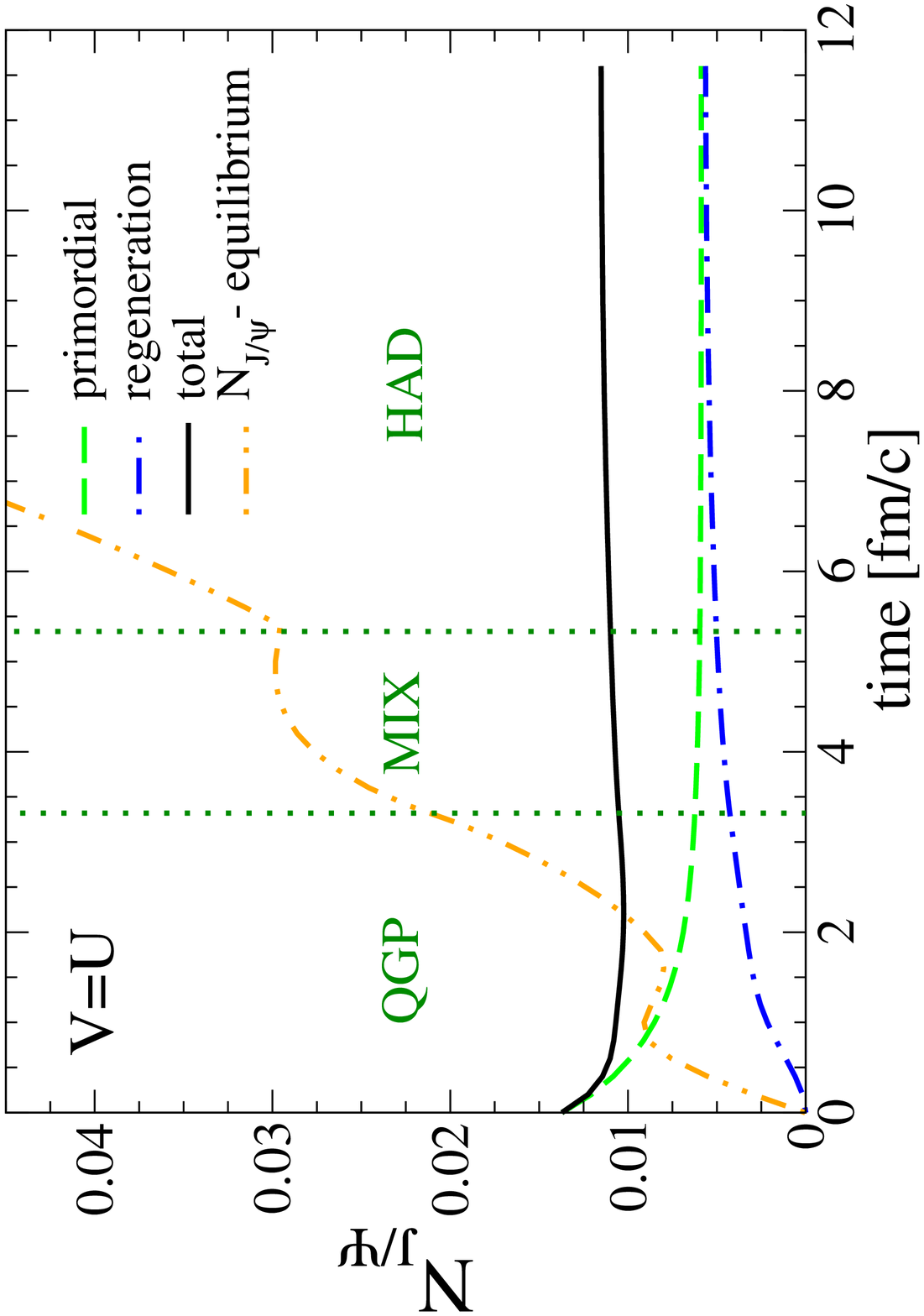}                                                                       
\includegraphics[width=0.49\textwidth]
{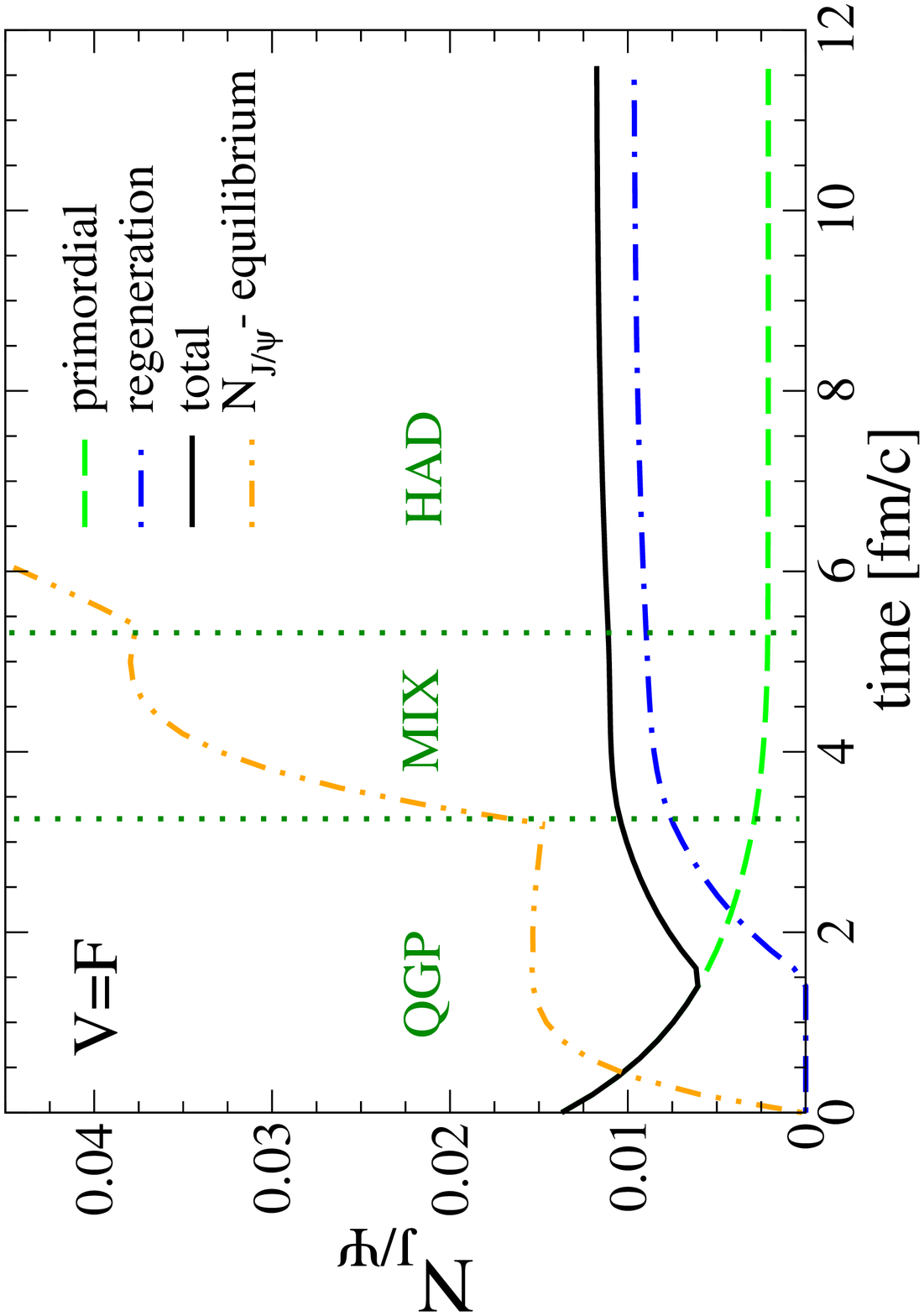}
\caption[$J/\psi$ abundance as a function of time in central collisions at
RHIC]{$J/\psi$ abundance as a function of time in central ($N_{\rm
    part}$=380) Au+Au collisions at RHIC. Solid lines: total; dashed
  line: primordial component; dot-dashed line: regeneration component;
  double dot-dashed line: equilibrium limit,
  $N_{J/\psi}^{\rm eq}$. Left panel: strong-binding scenario; right
  panel: weak-binding scenario.}
\label{fg:jpsi-time-evo}
\end{figure}
It is instructive to examine the time evolution of $J/\psi$ production
in the two scenarios, displayed in Fig.~\ref{fg:jpsi-time-evo} for 
central collisions at mid rapidity at RHIC (excluding feeddown 
from $\chi_c$ and $\psi'$).
In both scenarios most of the dissociation and regeneration indeed occur in
the QGP and mixed phase, since the hadronic reaction rates are small. 
In the weak-binding scenario the time-dependent $J/\psi$ yield 
exhibits a ``dip'' structure around $\tau\simeq1.5$\,fm/$c$ because
the large dissociation rates suppress primordial $J/\psi$ very
rapidly and regeneration only starts after the medium temperature falls
below the $J/\psi$ dissociation temperature ($T_{J/\psi}^{\rm diss}
\simeq1.25\,T_c$). This scenario is closest in spirit to the 
statistical hadronization 
model~\cite{Andronic:2006ky} where all initial charmonia are suppressed 
(or never form to begin with, except for corona effects) and are then 
produced at the hadronization transition.


\subsection{$J/\psi$ Transverse Momentum Spectra}
\label{ssec:psi_pt}

\subsubsection{Average Transverse Momentum}
\label{sssec:mean_pt_psi}
The results of the previous section suggest that, within the current
theoretical (\eg, charm-quark relaxation time, $\tau_c^{\rm eq}$) and
experimental uncertainties both of the ``limiting" scenarios can
reproduce the centrality dependence of the inclusive
$R_{AA}^{J/\psi}(N_{\rm part})$ reasonably well at both SPS and RHIC
energies. However, the composition between suppression and regeneration
yields is rather different which ought to provide a key to
distinguish the two scenarios. The obvious ``lever arm"
are charmonium $p_t$ spectra~\cite{Grandchamp:2001pf}. One expects
that the primordial component is characterized by harder $p_t$ spectra
(following a power law at high $p_t$) while the regeneration component
produces softer $p_t$ spectra characterized by phase-space
overlap of (partially) thermalized charm-quark spectra.
However, in practice, the transition from the ``soft" recombination
regime to the ``hard" primordial regime is quite uncertain; \eg, 
collective flow and incomplete thermalization
of $c$-quarks can lead to a significant hardening of the regenerated
$J/\psi$ spectra, while a dissociation rate which increases with
3-momentum~\cite{Zhao:2007hh} can induce a softening of the spectra
of surviving primordial charmonia.

\begin{figure}[tp]
\centering
\includegraphics[width=0.48\textwidth]
{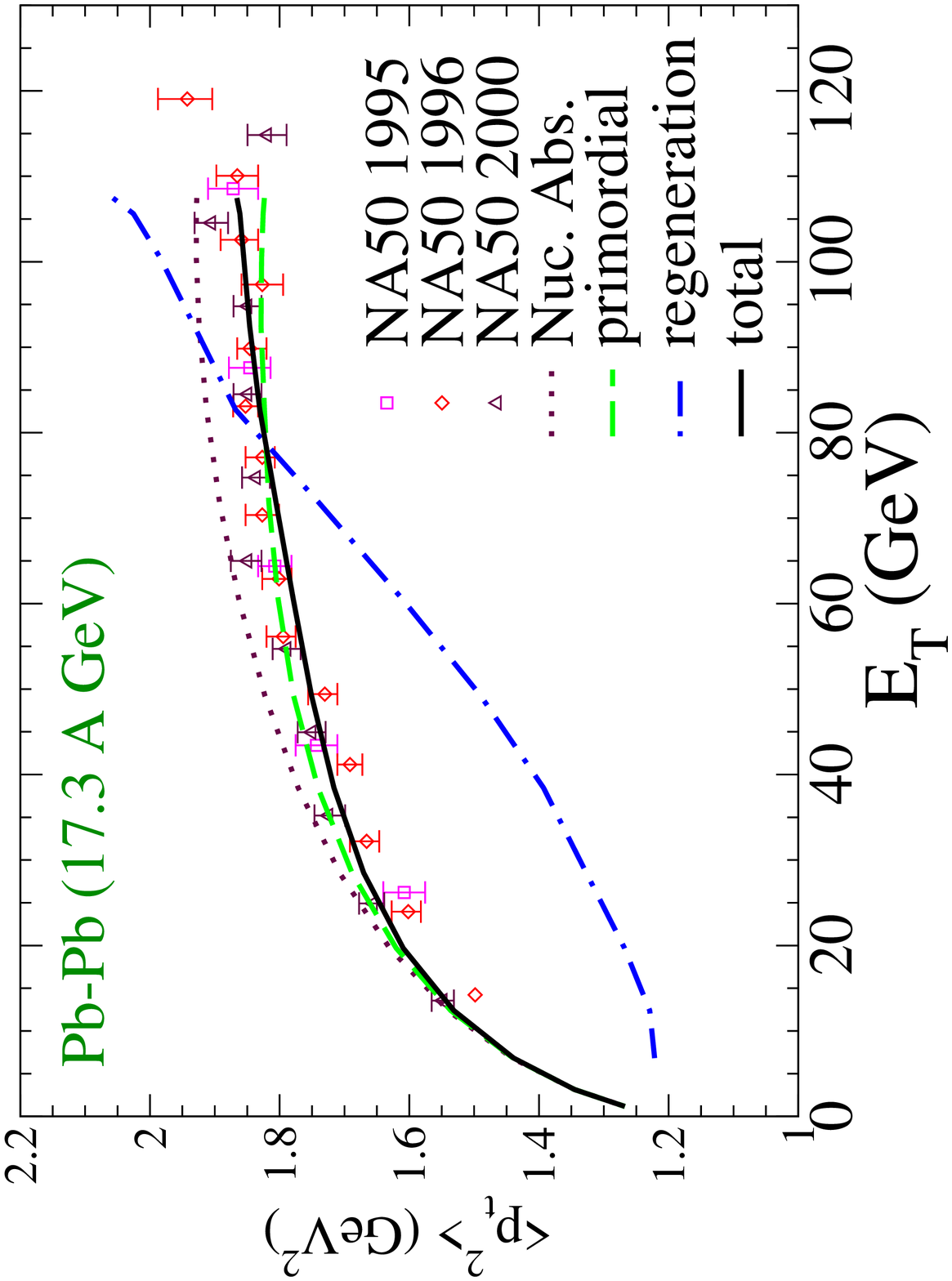}
\hspace{2mm}
\includegraphics[width=0.48\textwidth]
{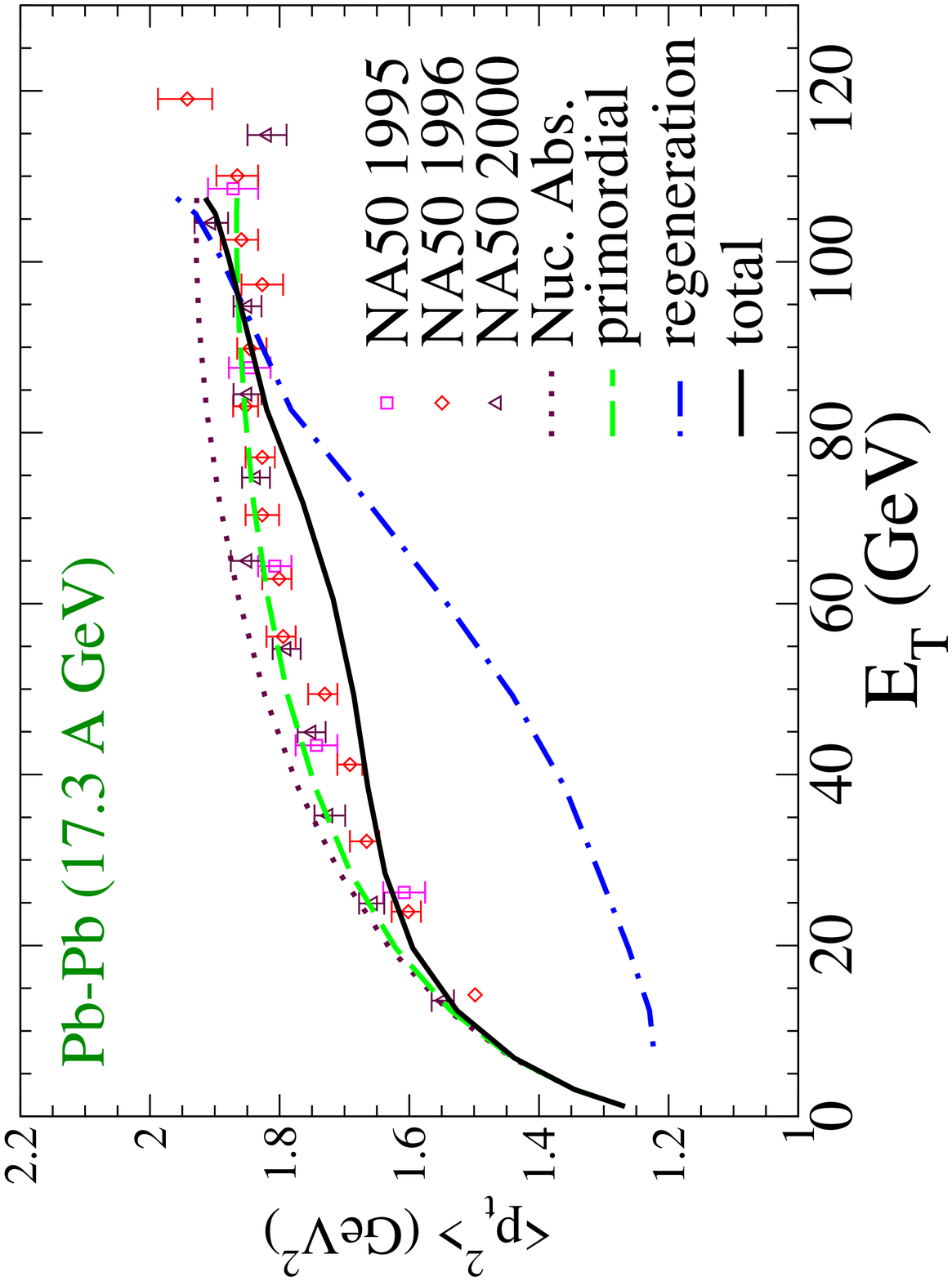}
\caption[$\langle p^2_t\rangle$ of $J/\psi$ vs. centrality at
SPS, compared to NA50 data]{$\langle p^2_t\rangle$ of $J/\psi$ vs. centrality at SPS,
  compared to NA50 data~\cite{Abreu:2000xe,Topilskaya:2003iy}.  In
  each panel, $\langle p^2_t\rangle$ is plotted for total $J/\psi$
  yield (primordial + regeneration component; solid lines), the
  suppressed primordial component (dashed line), the regeneration
  component (dash-dotted line) and primordial production with CNM
  effects only (dotted lines).  The left (right) panels correspond to
  the strong-binding (weak-binding) scenario. The transverse energy,
  $E_T$, is a measure of centrality. Its value in GeV is about
  0.275$\times N_{part}$.}
\label{fg:fig_pt2_sps}
\end{figure}

\begin{figure}[tp]
\centering
\includegraphics[width=0.49\textwidth]
{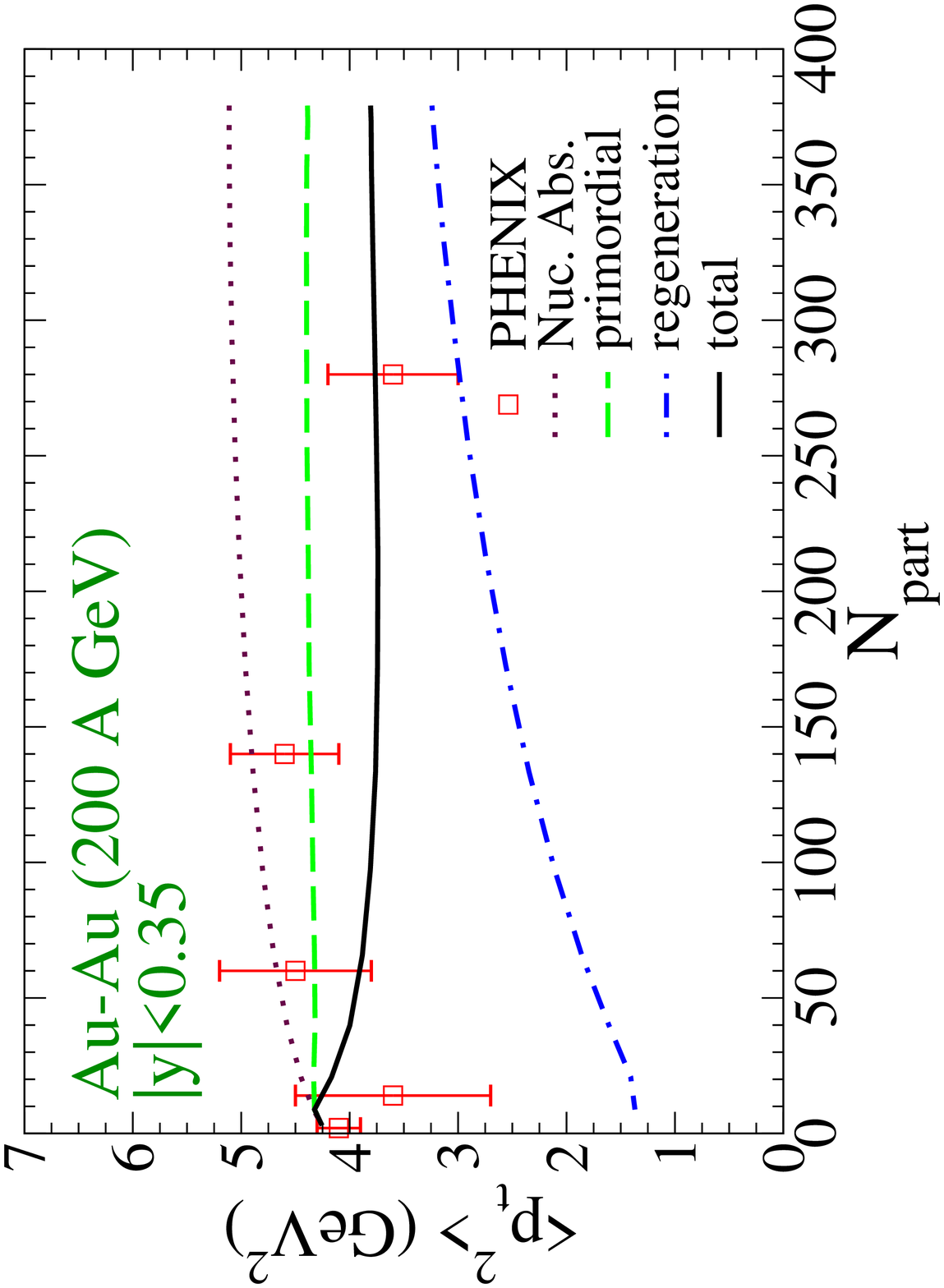}
\includegraphics[width=0.49\textwidth]
{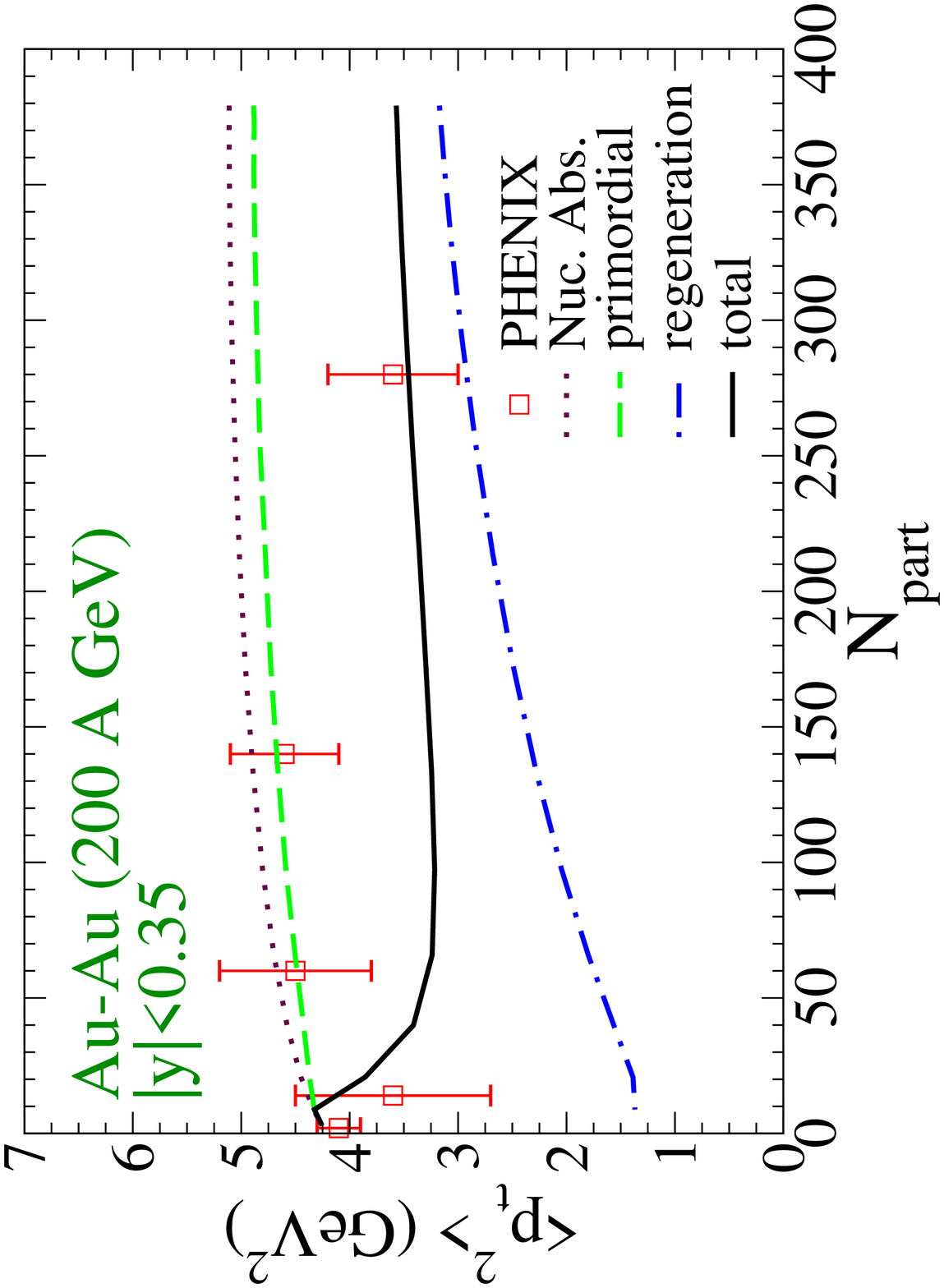}
\includegraphics[width=0.49\textwidth]
{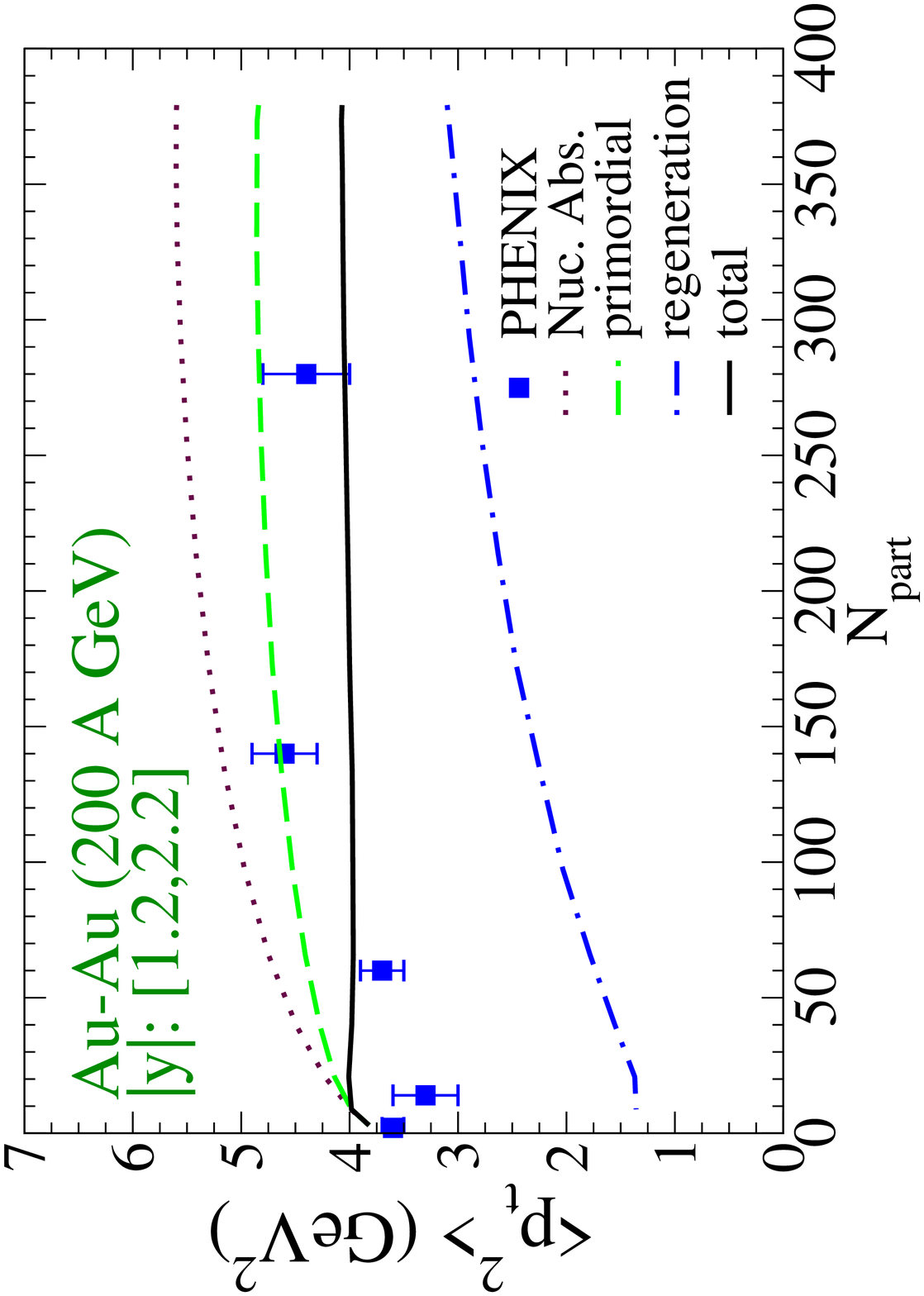}
\includegraphics[width=0.49\textwidth]
{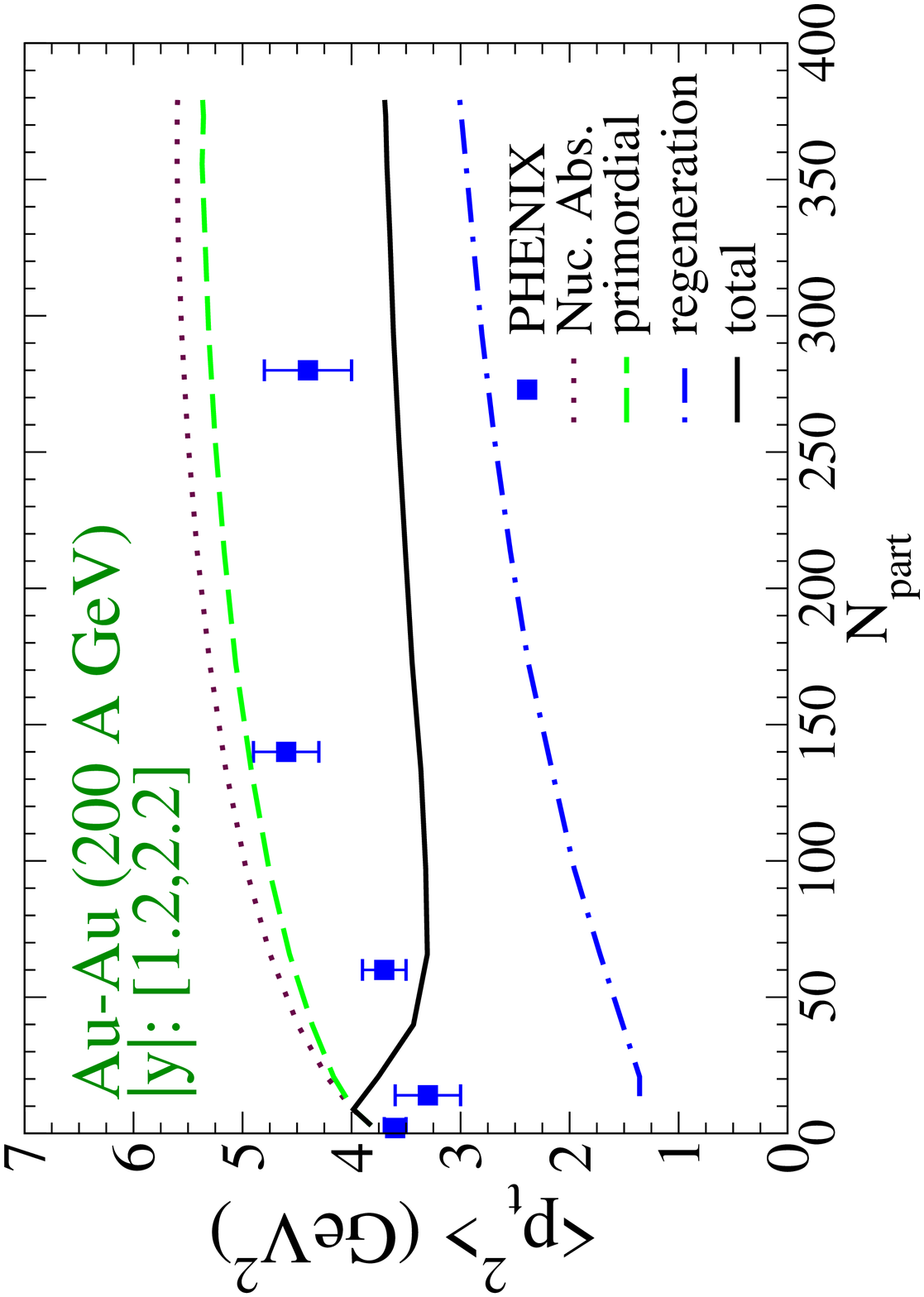}
\caption[$\langle p^2_t\rangle$ of $J/\psi$ vs. centrality at RHIC at
mid-rapidity and forward rapidity, compared to PHENIX data]{$\langle
  p^2_t\rangle$ of $J/\psi$ vs. centrality at RHIC at mid-rapidity
  (upper panels) and forward rapidity (lower panels), compared to
  PHENIX data~\cite{Adare:2006ns}.  In each panel, $\langle
  p^2_t\rangle$ is plotted for total $J/\psi$ yield (primordial +
  regeneration component; solid lines), the suppressed primordial
  component (dashed line), the regeneration component (dash-dotted
  line) and primordial production with CNM effects only (dotted
  lines).  The left (right) panels correspond to the strong-binding
  (weak-binding) scenario.}
\label{fg:fig_pt2_rhic}
\end{figure}

For a more concise discussion of the $p_t$ dependence of $J/\psi$ as a
function of centrality at SPS and RHIC we here focus on the average
$p_t^2$, as compiled in Figs.~\ref{fg:fig_pt2_sps} and \ref{fg:fig_pt2_rhic}.  At the SPS (Fig.~\ref{fg:fig_pt2_sps}), the centrality
dependence of $\langle p^2_t\rangle$ is largely dictated by the the
Cronin effect in the primordial component, especially in the
strong-binding scenario where this contribution dominates the yield at
all centralities. The momentum dependence of the dissociation rate
induces a slight suppression of $\langle p^2_t\rangle$ at large
centrality compared to the case where only CNM effects are included
(dashed vs. dotted line)~\cite{Zhao:2007hh}. In the weak-binding
scenario, larger contributions from regeneration induce a slight
``dip" structure at intermediate centralities due to a rather small
collective flow at the end of the (relatively short) mixed phase in
these collisions.
At RHIC energy (Fig.~\ref{fg:fig_pt2_rhic}), the individual primordial and
regeneration components show qualitatively similar behavior for
$\langle p^2_t\rangle(N_{\rm part})$ as at SPS, \ie, an increase due
to Cronin effect and collective flow, respectively. At mid rapidity,
the general trend is that with increasing centrality the growing
regeneration contribution pulls down the average $\langle
p^2_t\rangle$, in qualitative agreement with the data. The curvature
of the $\langle p^2_t\rangle(N_{\rm part})$ dependence, which appears
to be negative in the data, is not well reproduced, neither by the
strong- nor by the weak-binding scenario, even though the deviations
are smaller in the former.  A more microscopic calculation of the gain
term, together with more accurate estimates of the Cronin effect, are
warranted to enable more definite conclusions. For both rapidity
regions, the $\langle p^2_t\rangle$ of the suppressed primordial
component is slightly larger in the weak- than in the strong-binding
scenario. This is caused by the stronger 3-momentum increase of the
dissociation rate in the strong-binding scenario, recall the right
panel of Fig.~\ref{fg:rate_temp}.

The overall comparison to SPS and RHIC data for the $p_t$ dependence 
of $J/\psi$'s seems to indicate a slight preference for the 
strong-binding scenario. This is mostly derived from the observation
that for  peripheral collisions the experimentally observed 
$\langle p_t^2 \rangle$ essentially follows the extrapolation of the 
Cronin effect, suggesting $J/\psi$ production of predominantly
primordial origin (the collective flow imparted on the regeneration
component appears to be too small at these centralities, rendering the weak-binding scenario problematic). 


\subsubsection{High $p_t$ $J/\psi$ Production and Elliptic Flow}
\label{sssec:high_pt_psi}
The average $p_t^2$ essentially characterizes the momentum dependence
of charmonium production at low and moderate $p_t$ where most of the
yield is concentrated. Recent RHIC
data~\cite{Adare:2008sh,Abelev:2009qaa} have triggered considerable
interest in $J/\psi$ production at high $p_t\simeq5-10$\,GeV which is
expected to provide complementary information. It was found that the
suppression in $R_{AA}^{J/\psi}(p_t\gsim5\,{\rm GeV})$ in Cu-Cu
collisions is reduced compared to the low-$p_t$ region, with
$R_{AA}$-values of $\sim$0.7-1 or even
larger~\cite{Abelev:2009qaa}. This is quite surprising in light of the
light-hadron spectra measured thus far at RHIC which all exhibit
stronger suppression of $R_{AA}\simeq0.25$ for $p_t\gsim6$\,GeV (even
electron spectra from open heavy flavor, \ie, charm and bottom
decays). It also appears to be at variance with the thermal $J/\psi$
dissociation rates which, if anything, increase with momentum (recall
Fig.~\ref{fg:rate_temp}) and thus imply a stronger suppression at
higher $p_t$. Furthermore, the leakage effect mentioned in
Section~\ref{ch:nume_results}.\ref{sec:trans_rate_eq} is not strong enough to produce the
experimentally observed increase in $R_{AA}^{J/\psi}(p_T\gsim5\,{\rm
  GeV})$~\cite{Zhao:2007hh}.

\begin{figure} [!t]
\centering
\includegraphics[width=0.59\textwidth]{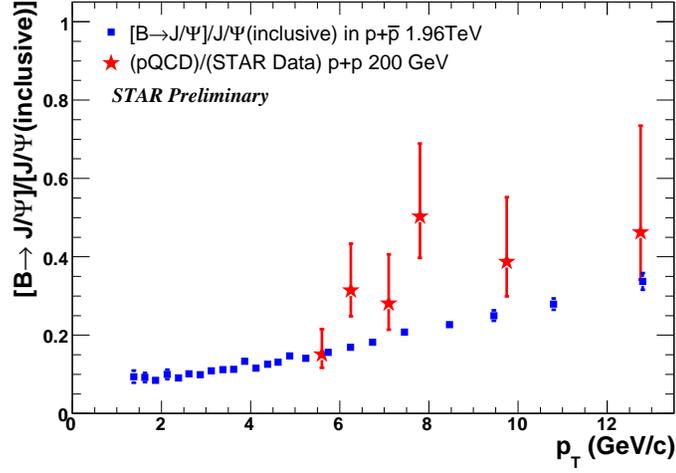}
\caption[Ratio of $J/\psi$ from $B$-meson feeddown to inclusive
$J/\psi$ indicated by Tevatron and STAR data]{Ratio of $J/\psi$ from
  $B$-meson feeddown to inclusive $J/\psi$ indicated by
  Tevatron~\cite{Acosta:2004yw} and STAR
  data~\cite{Tang:2008private}.}
\label{fg:fig_form}
\end{figure}
Therefore we consider the following two effects primarily relevant at
high $p_t$~\cite{Zhao:2008vu}: (1) Finite formation times for the
charmonium states~\cite{Blaizot:1988ec,Karsch:1988,Gavin:1990gm} and
(2) Bottom feeddown.  Concerning (1), one expects reduced geometrical
sizes for a ``pre-resonance" $c\bar{c}$ pair relative to a fully
formed charmonium due to a finite formation time, $\tau_{f}$, required
to build up the hadronic wave function. If the ``pre-resonance"
$c\bar{c}$ pair is in a color-singlet state, as suggested by the color
singlet model (CSM) (recall Section~\ref{ch:pre-eq}.\ref{sec:psi_pp}),
its smaller geometrical size will translate into a smaller
dissociation cross section, since it appears color-neutral for
incoming partons whose momenta are not large enough to resolve its
inner structure. For a schematic estimate we parameterize the
evolution of the pre-hadronic dissociation rate as
\begin{equation}
\Gamma_{pre-\Psi}(\tau)=\Gamma_{\Psi} \tau/\tau_{f}^{lab} 
 \quad , \quad \tau \le \tau_{f}^{lab} = \tau_{f} m_t/m_\Psi  
\label{eq:form_time}
\end{equation}
with $\Gamma_{\Psi}$: (nuclear, partonic or hadronic) dissociation
rates for a formed charmonium, $\tau$: fireball proper time,
$\tau_{f}$=0.89(2.01,1.50)~fm/$c$: formation time of
$J/\psi$($\chi_c$, $\psi^{\prime}$) in its rest frame, and
$m_t=(m_\Psi^2+p_t^2)^{1/2}$. Essentially, (pre-) charmonium
dissociation rates acquire an additional momentum dependence through
Lorentz time dilation, being reduced at high $p_t$.  Also note that
the longer formation times of $\chi_c$ and $\psi^{\prime}$ imply less
suppression relative to $J/\psi$, quite contrary to standard
dissociation and regeneration mechanisms: as higher excited $c\bar c$
states than $J/\psi$ they have smaller binding energy and therefore are
more easily destroyed, and they are heavier so that
their equilibrium abundances are suppressed compared to $J/\psi$ by the
thermal Boltzmann factor leading to a reduced regeneration. Therefore high-$p_t$
$\chi_c$ and $\psi^{\prime}$ could provide a rather unique signature of
the formation time effect.  Concerning (2), Fig.~\ref{fg:fig_form}
shows recent data on the $B\to J/\psi$ feeddown fraction in elementary
p+p($\rm{\bar{p}}$) collisions, which is quite significant.  As an
estimate of this contribution, we use the Tevatron
data~\cite{Acosta:2004yw} and replace the corresponding fraction of
primordial component.

\begin{figure}[t]
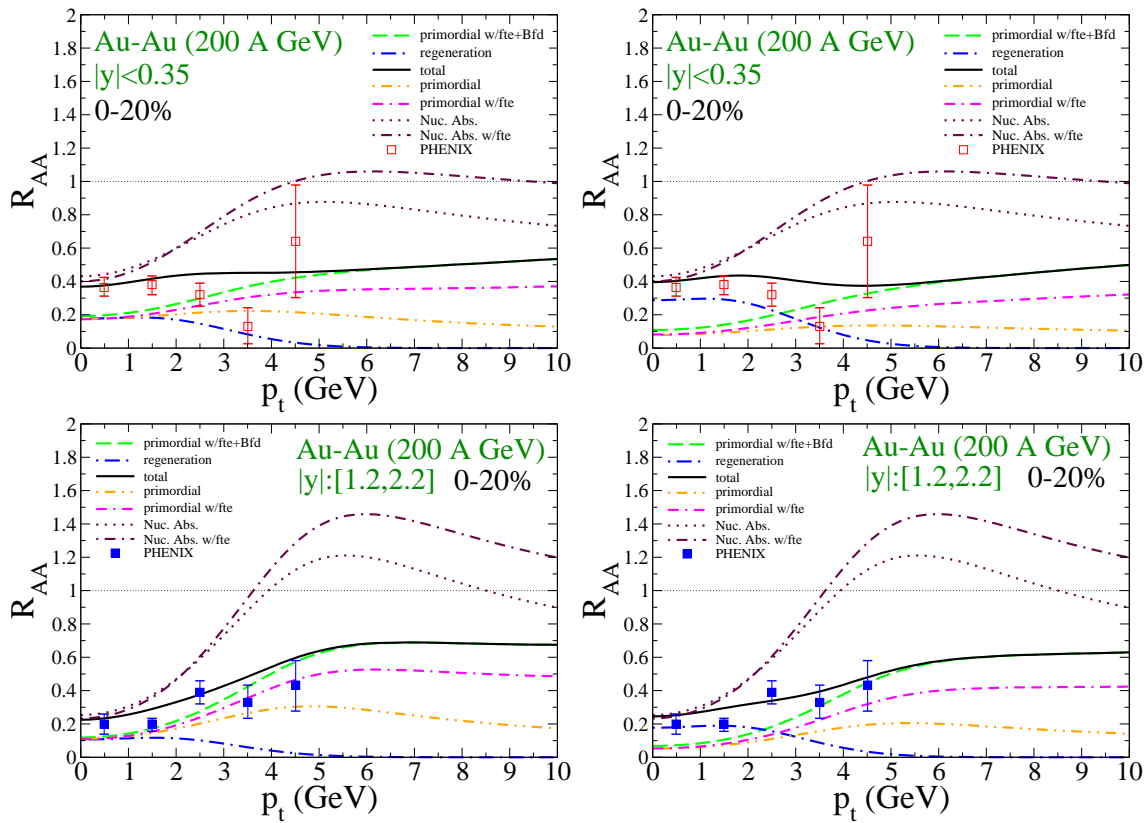

\centering
\includegraphics[width=0.49\textwidth]{raa_pt_rhic_u_0826.eps}
\includegraphics[width=0.49\textwidth]{raa_pt_rhic_f_0826.eps}
\includegraphics[width=0.49\textwidth]{raa_pt_rhicf_u_0910.eps}
\includegraphics[width=0.49\textwidth]{raa_pt_rhicf_f_0910.eps}
\caption[$J/\psi$ $R_{AA}$ versus $p_t$ in central 200\,AGeV Au+Au
collisions including formation-time effects and $B$-meson feeddown
contributions]{$J/\psi$ $R_{AA}$ versus $p_t$ in central 200\,AGeV
  Au+Au collisions including formation-time effects (fte) and
  $B$-meson feeddown (Bfd) contributions. PHENIX
  data~\cite{Adare:2006ns} are compared to our rate-equation
  calculations in the strong- and weak-binding scenarios (left and
  right panels, respectively). Upper panels: mid-rapidity; lower
  panels: forward rapidity.}
\label{fg:fig_raa-pt}
\end{figure}
\begin{figure} [!t]
\centering
\includegraphics[width=0.59\textwidth]{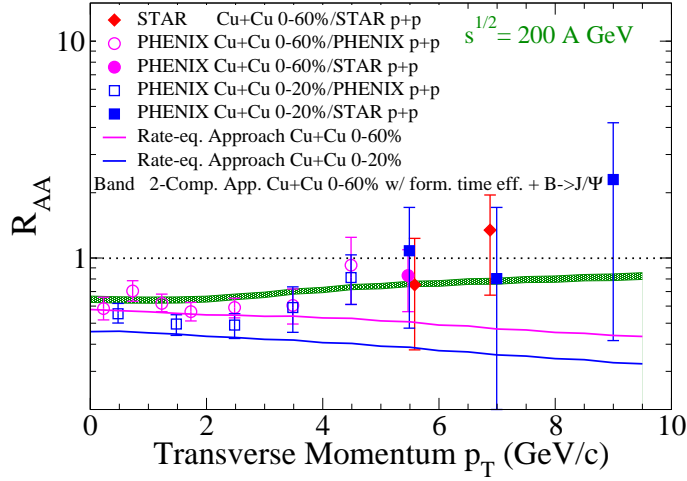}
\caption[$J/\psi$ $R_{AA}$ vs. $p_t$ in 0-60\% Cu+Cu at RHIC, compared
RHIC data]{$J/\psi$ $R_{AA}$ vs. $p_t$ in 0-60\% Cu+Cu collisions,
  compared RHIC data~\cite{Adare:2008sh,Tang:2008uy}. The formation
  time effects and $B$-meson feeddown (green band) are included. The
  partition between the primordial and regeneration components is
  based on the thermal rate-equation approach in
  Ref.~\cite{Zhao:2007hh}.}
\label{fg:high_pt_raa_cucu}
\end{figure}
Combining formation time effects and $B$-meson feeddown we obtain 
results for $R_{AA}^{J/\psi}(p_t)$, as displayed in
Fig.~\ref{fg:fig_raa-pt} up to $p_t=10$\,GeV. We find that the
suppression is reduced to about 0.5 at the highest $p_t$, compared to
about 0.4 at low $p_t$. This is similar to the moderate enhancement we
found in the Cu-Cu case~\cite{Zhao:2008vu}, as shown in
Fig.~\ref{fg:high_pt_raa_cucu}. Surprisingly, the high-$p_t$ suppression turns out to be very similar
in both strong- and weak-binding scenarios, despite the fact that the
high-$p_t$ yield is exclusively due to the primordial component whose
strength is very different in the 2 scenarios at low $p_t$. The reason
is the 3-momentum dependence of the dissociation rates, which become
quite similar in the 2 scenarios at large 3-momentum: at $p\simeq
10$~GeV, the difference in the energy-threshold due to binding
energies of several 100\,MeV becomes less relevant so that a collision
with almost any thermal parton is energetic enough for dissociating
the bound state. For $J/\psi$ at forward rapidity (shown in lower
panels of Fig.~\ref{fg:fig_raa-pt}), we additionally include the
effect that the nuclear shadowing (responsible for the extra
CNM-induced suppression relative to mid-rapidity) decreases with the
$p_t$ of the primordially produced $J/\psi$. We assume that it ceases
to exist at around $p_t$=10\,GeV, recall the discussion in
Section~\ref{ch:pre-eq}.\ref{sec:psi_aa}.\ref{ssec:cnm}.\ref{sssec:shadow}.

\begin{figure}[!t]
\centering
\includegraphics[width=0.59\textwidth]{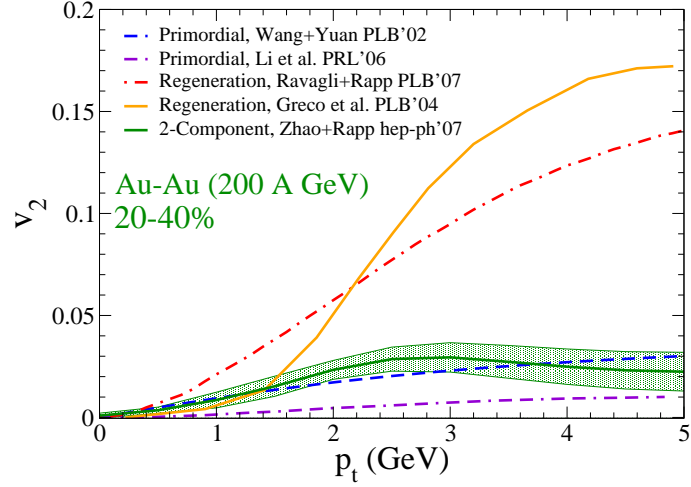}
\caption[$J/\psi$ $v_2(p_t)$ for 20-40\% central Au+Au collisions
at RHIC]{$J/\psi$ $v_2(p_t)$ for 20-40\% central Au+Au collisions at
  RHIC. Figure taken from Ref.~\cite{Zhao:2008vu}. The band represents
  the sum of uncertainties based on differences between two
  independent ``input'' $v_2(p_t)$ calculations for each of the two
  components.}
\label{fg:v2-pt}
\end{figure}

Let us also estimate the elliptic flow, $v_2(p_t)$, of the $J/\psi$,
(recall Eq. (\ref{eq:v2})), which is hoped to be another good
discriminator of primordial and regenerated production. For the
former, a nonzero $v_2$ is basically due to the path-length difference
when traversing the azimuthally asymmetric fireball, typically not
exceeding 2-3\%~\cite{Yan:2006ve,Wang:2002ck}. For the latter, much
larger values can be obtained if the coalescing charm quarks are close
to thermalized~\cite{Ravagli:2007xx,Greco:2003vf}. However, as pointed
out in Ref.~\cite{Zhao:2008vu}, the well-known mass effect suppresses
the $v_2(p_t)$ for heavy particles at $p_t\lsim m$; it is precisely in
this momentum regime where the regeneration component is prominent. In
Ref.~\cite{Zhao:2008vu} we estimate the total $J/\psi$ $v_2(p_t)$ by
combining the $v_2$ from the primordial production computed in
Refs.~\cite{Yan:2006ve,Wang:2002ck} and that from the regeneration
computed in Refs.~\cite{Ravagli:2007xx,Greco:2003vf} with relative
fractions determined from our thermal rate-equation approach. In
Fig.~\ref{fg:v2-pt} we show the resulting total $J/\psi$ $v_2(p_t)$
for 20-40\% central Au+Au collisions, where 
neither the formation time effect nor
$B$-meson feeddown is included. These two effects would further increase the
fraction of the primordial components in the intermediate to high
$p_t$ region leading to an even smaller total $v_2$ in this region. 
Thus, we predict that in both strong- and
weak-binding scenarios the total $J/\psi$ $v_2(p_t)$ does not exceed
$\sim$5\% at any $p_t$, similar to what is found in
Refs.~\cite{Yan:2006ve,Song:2010er}. The only alternative option we
can envision are strong elastic interactions of the $J/\psi$ which are
only conceivable in the strong binding scenario to avoid break-up in
scattering off thermal partons~\cite{Rapp:2009my}.


\section{Explicit Calculations of the Regeneration Component}
\label{sec:full_boltz}

In this section we explicitly calculate $J/\psi$ regeneration from input charm-quark spectra using the Boltzmann equation (\ref{eq:boltz-eq_reg}) instead of the rate equation (\ref{eq:rate-eq_reg}) to check the sensitivity of $\Psi$ regeneration to off-equilibrium effects in charm-quark spectra.

\subsection{Sensitivity of $J/\psi$ Regeneration to Charm-Quark $p_t$ Spectra}
\label{ssec:raa_centra_cquark}
\begin{figure}[tp]
\centering                                                                      
\includegraphics[width=0.59\textwidth]
{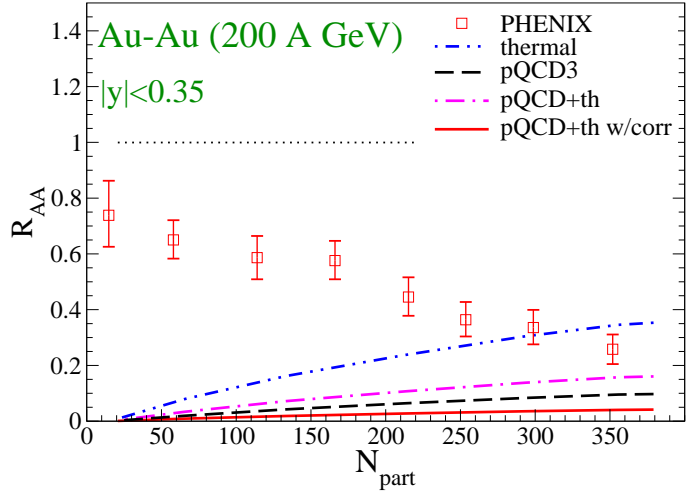}
\caption[$J/\psi$ regeneration from different charm-quark spectra]{$J/\psi$ regeneration from different charm-quark spectra. Double-dot-dashed line: thermal charm-quark spectra. Dashed line: 3-dimensionally isotropic pQCD charm-quark spectra. Dot-dashed line: pQCD charm-quark spectra in transverse plane + thermal spectra in longitudinal direction. Solid line: pQCD charm-quark spectra in transverse plane with angular correlation between $c$ and $\bar{c}$ + thermal spectra in longitudinal direction. }
\label{fg:raa_c_spectra}
\end{figure}

We first compare the inclusive yield of $J/\psi$'s regenerated from the
three types of charm-quark momentum spectra introduced in
Section~\ref{ch:psi_hot}.\ref{sec:open_charm_med}: 1) thermal charm-quark
spectra; 2) 3-dimensionally isotropic pQCD spectra; 3) 
transversely pQCD + longitudinally thermal spectra.  Our calculation
is performed at RHIC energy where the regeneration component takes up
a significant fraction. For comparison purpose we only consider
$J/\psi$ regeneration in QGP phase with the quasifree process $i + c +
\bar{c}\rightarrow i+J/\psi$ $(i={g,q,\bar{q}})$ within the strong binding scenario. Neither the
canonical ensemble effect nor the correlation volume effect (see
Section~\ref{ch:trans}.\ref{sec:cc_corr}) are applied; these two effects will be
discussed in the next section. The resulting centrality dependence of
inclusive yields of regenerated $J/\psi$ is compared in
Fig.~\ref{fg:raa_c_spectra}.
Clearly, the thermal charm-quark spectra are most efficient in
regenerating $J/\psi$ due to the large phase space overlap between
$c$, $\bar{c}$ quarks and light partons in the medium. The
regeneration from longitudinally thermal and transversely pQCD spectra
is reduced to about half relative to the thermal spectra. The
regeneration from 3-dimensionally isotropic pQCD charm-quark spectra
is even reduced by a factor of 4. This result is similar to what is
found in a quark coalescence model~\cite{Greco:2003vf} where the
number of $J/\psi$ coalesced from pQCD charm-quark spectra is smaller
than that from thermal spectra by a factor of 3. However in
Ref.~\cite{Yan:2006ve} a similar Boltzmann transport approach was
employed for $J/\Psi$ regeneration with the (inverse)
gluo-dissociation process as the regeneration mechanism and it is
found that the inclusive yield of regenerated $J/\psi$ from pQCD charm
spectra is quite comparable with that from thermal spectra (within a
$\sim$30\% difference). Further investigations are needed to clarify
the discrepancy. We also find that if the angular correlation between
the transverse momentum of $c$ and $\bar c$ is included, to reflect the
back-to-back charm production in initial hard collisions, the
regeneration is further reduced to about 1/10 of the regeneration from
fully thermalized charm-quark spectra, as a result of further
reduction of the $c\bar{c}$ overlap in momentum space.

\begin{figure}[tp]
\centering                                                                      
\includegraphics[width=0.75\textwidth]
{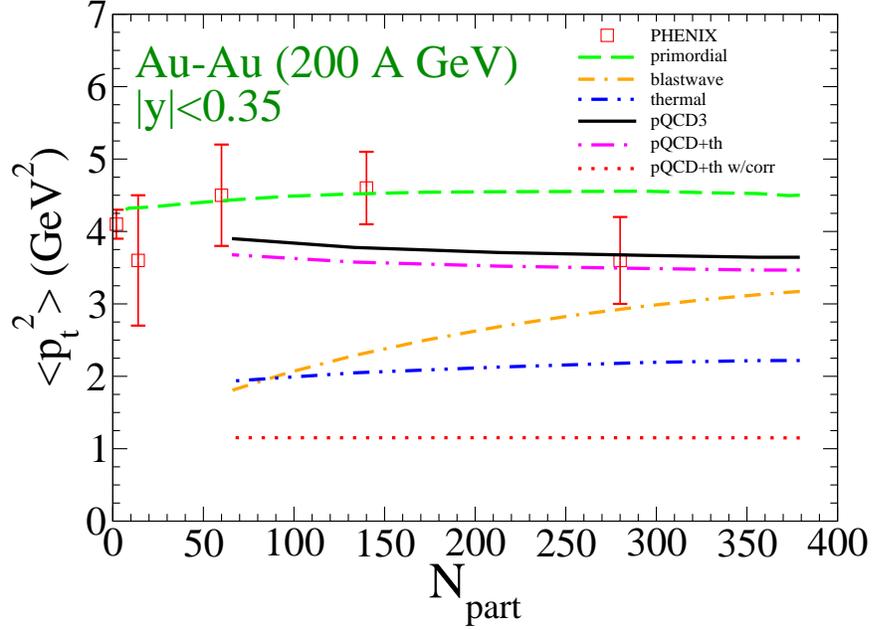}
\caption[$\langle p^2_t\rangle$ of $J/\psi$'s regenerated from different
charm-quark spectra]{$\langle p^2_t\rangle$ of $J/\psi$'s regenerated
  from different charm-quark spectra. Double-dot-dashed line: thermal
  charm-quark spectra; solid line: 3-dimensional isotropic pQCD charm-quark spectra; dot-dashed line: longitudinal thermal + transverse
  pQCD charm-quark spectra; dotted line: longitudinal thermal +
  transverse pQCD charm-quark spectra with angular correlation between
  $c$ and $\bar{c}$; double dash-dotted line: the $\langle
  p^2_t\rangle$ from the rate equation + blastwave treatment; dashed line: the $\langle
  p^2_t\rangle$ of the primordial component.}
\label{fg:pt2_c_spectra}
\end{figure}

Next we proceed to the $\langle p^2_t\rangle$ of regenerated $J/\psi$, displayed in Fig.~\ref{fg:pt2_c_spectra}.
As expected, the $\langle p^2_t\rangle$ of $J/\psi$ regenerated from
pQCD charm-quark spectra is larger than that from thermal spectra. We
note that the $\langle p^2_t\rangle$ from thermal charm-quark spectra
is lower than that estimated from the blastwave formula. This is due
to the fact that in the strong-binding scenario most of the
regeneration processes occur at an early stage of the medium evolution
(see Fig.~\ref{fg:jpsi-time-evo}), when the collective flow has not
yet fully built up. Here the (transverse) medium flow effect is
included only for the thermal charm-quark spectra, for the other two
types of pQCD spectra we assume the medium is at rest in the lab
frame. We also note that the introduction of the angular correlation
between $c$ and $\bar{c}$ momenta significantly lowers the $\langle
p^2_t\rangle$ of regenerated $J/\psi$. This is because high $p_t$
$J/\psi$'s are more likely to be regenerated from two charm quarks with
a small angle between them, the probability of which is significantly
suppressed by the back-to-back correlation of charm pairs. Therefore
the regeneration of $J/\psi$'s at high $p_t$ is more reduced than at
low $p_t$ leading to a reduction of $\langle p^2_t\rangle$.
\subsection{Sensitivity of $J/\psi$ Regeneration to Charm-Quark Correlation Volume}
\label{ssec:raa_centra_vcorr}
In this section we explicitly evaluate $J/\psi$ regeneration from
individual events with integer numbers of $c\bar c$ pairs produced,
recall Fig.~\ref{fig:ncc_distrib}. The correlation volume effect (see
Section~\ref{ch:trans}.\ref{sec:cc_corr}.\ref{ssec:corr_vol}) is
applied separately to events with integer ($k$) charm quark pairs.  At
the end we average the regenerated $J/\psi$ over these events.
\begin{figure}[htp]
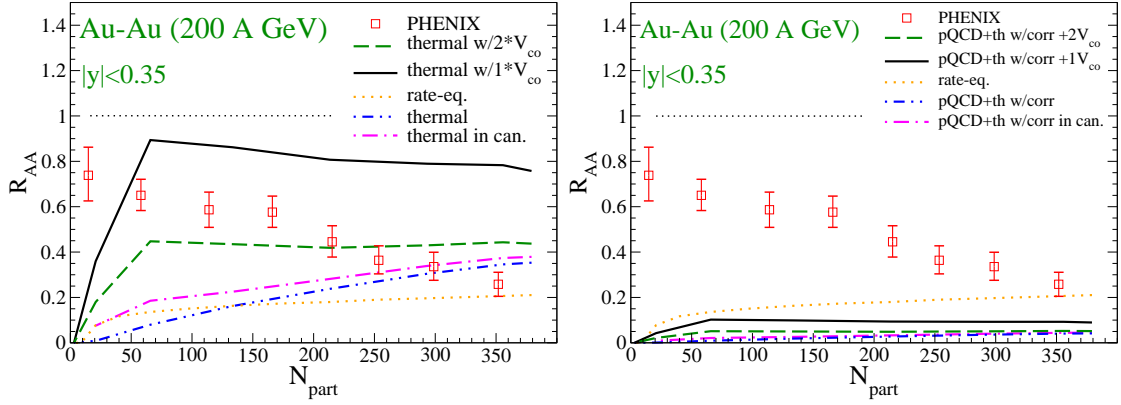

\centering                                                                      
\includegraphics[width=0.48\textwidth]
{raa_centra_rhic_u_th_1021.eps}
\includegraphics[width=0.48\textwidth]
{raa_centra_rhic_u_pQCD_1021.eps}
\caption[$J/\psi$ regeneration with different charm-quark correlation volumes]{$J/\psi$ regeneration with different charm-quark correlation volumes. Left panel: thermal charm-quark spectra; right panel: longitudinal thermal + transverse pQCD charm-quark spectra with angular correlation between $c$ and $\bar{c}$. Double-dot-dashed line: regeneration in grand-canonical ensemble; dot-dashed line: regeneration in canonical ensemble, but without correlation volume effect; solid line: regeneration with 1$\times V_{co}$; dashed line: regeneration with 2$\times V_{co}$; dotted line: regeneration from the thermal rate equation approach. }
\label{fg:raa_centra_vco}
\end{figure}
The time dependence of the correlation volume $V_{co}(\tau)$ is
modelled using Eq.~(\ref{eq:vco_tau}). A rather simplified
prescription is adopted to treat the merging of correlation volumes:
If at any given moment $\tau$ the total correlation volume (sum over
all ``bubbles'') $kV_{co}(\tau)$ is larger than $V_{FB}(\tau)$,
$kV_{co}(\tau)$ is set to $V_{FB}(\tau)$ for the subsequent
evolution. In order to check the sensitivity of $J/\psi$ regeneration
to different sizes of the correlation volume, we multiply
$V_{co}(\tau)$ from Eq.~(\ref{eq:vco_tau}) with different overall
scaling factors and compare the resulting inclusive yield. The results
are presented in Fig.~\ref{fg:raa_centra_vco}. First we see that the
charmonium regeneration is stronger in the canonical ensemble than in
the grand-canonical ensemble. Including the correlation volume effect
further enhances the regeneration due to the effectively larger
probability for one charm quark to find its partner (by a factor of
$V_{FB}/kV_{co}$). Because $V_{co}$ grows faster than $V_{FB}$ this
effect is more pronounced for peripheral collisions where the medium
lifetime is relatively shorter and the early stage regeneration has
greater impact on the final yield. We also note that with $V_{co}$
doubled to $2V_{co}$ the regeneration yield almost drops by a factor
of two (dashed line). However, further increasing $V_{co}$ the
regeneration yield only decreases by a limited amount due to the fact
that for large correlation volume $kV_{co}\sim V_{FB}$ the correlation
volume ``bubbles'' begin merging and their maximal size is restricted
to the entire fireball volume, $V_{FB}$.  This can be seen from the
fact that for central collisions the regeneration with $2V_{co}$ is
already quite close to the limit (Dot-dashed line) where the
$c\bar{c}$ pairs are correlated inside the entire fireball
(essentially no correlation volume effect). With the correlation
volume effect included the inclusive yields of $J/\psi$ regeneration
from thermal and from longitudinal thermal + transverse pQCD
charm-quark spectra (with $c\bar{c}$ angular correlation) span a
rather large range. The rate equation result (with 1$\times V_{co}$)
is inside these two limits.

Our studies in this section demonstrate the fact that $J/\psi$
regeneration is very sensitive to the charm quark off-equilibrium
effects in both momentum (partial thermalization) and 
coordinate space (correlation volume effect). It is therefore of
crucial importance to implement these effects in a more systematic and
more realistic way. Ideally one would obtain the time-dependent
joint $c$ and $\bar{c}$ phase space distribution from, \eg,
Langevin simulations, as input to address these issues. Work in
this direction is planned.

\section{$\psi'$ and $\chi_c$ Production}
\label{sec:psip}

In addition to providing a feeddown contribution to $J/\psi$
production, excited charmonia can give valuable complementary
information on the medium created in heavy-ion collisions. Being more
loosely bound states they usually have larger dissociation rates than
$J/\psi$, and due to their heavier masses, their thermal equilibrium
abundances are smaller than $J/\psi$. Therefore we expect stronger
suppression for excited charmonia compared to $J/\psi$. However, if
formation time effects~\cite{Blaizot:1988ec,Gavin:1990gm,Karsch:1988}
are important, one may observe less suppression for $\chi_c$ than for
$J/\psi$.

\begin{figure}[tp]
  \centering
  \includegraphics[width=0.59\textwidth]{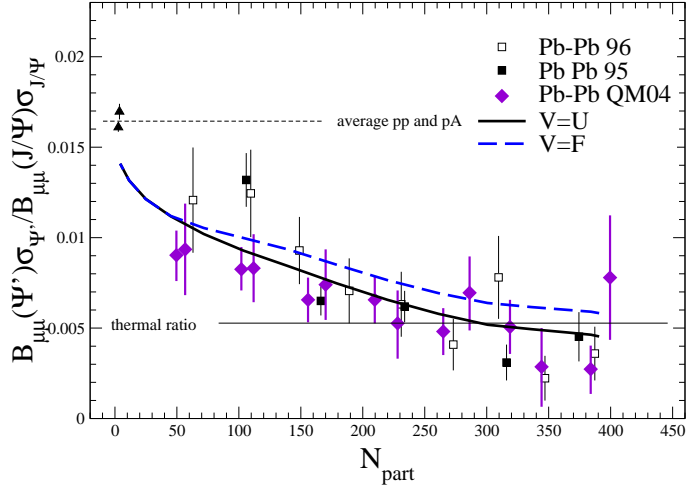}
  \caption[Inclusive $\psi'$ to $J/\psi$ ratio at SPS compared to NA50 data]{Inclusive $\psi'$ to $J/\psi$ ratio at SPS compared to NA50 data~\cite{Abreu:1998vw,Sitta:2004hj}. Solid line: strong-binding scenario; dashed line weak-binding scenario.}
\label{fg:psip_psi}
\end{figure}

With the thermal rate equation approach we calculate the $\psi'$ to
$J/\psi$ ratio in $\sqrt{s}$=17.3A Pb+Pb collisions and compare to
NA50 measurements in Fig.~\ref{fg:psip_psi}. In p+p
collisions at SPS energies, the ratio of produced $\psi'$ to
$J/\psi$ mesons amounts to a value of about 0.017 (the branching
ratios from $\psi'$ and $J/\psi$ into dimuons are included). The nuclear
absorption cross section for $\psi'$, $\sigma^{\psi'}_{\rm abs}$, extracted
by NA50 is 7.9\,mb compared to $\sigma^{J/\psi}_{\rm abs}$=4.4mb for
$J/\psi$. Both these two numbers are measured in 400\,GeV p+A
collisions. 
With the updated $\sigma^{J/\psi}_{\rm abs}$=7.3\,mb at 158\,GeV and
assuming that the ratio between $\sigma^{\psi'}_{\rm abs}$ and
$\sigma^{J/\psi}_{\rm abs}$ is the same for p+A collisions at both
158 \,GeV and 400\, GeV, we obtain $\sigma^{\psi'}_{\rm abs}$=13mb for
$\sqrt{s}$=17.3\,AGeV Pb+Pb
collisions. 
The data from NA50 suggest additional suppression in $\psi'$ relative
to $J/\psi$ in Pb+Pb collisions on top of the larger nuclear
absorption cross section for $\psi'$. In the strong-binding scenario
this can be explained by the larger quasifree dissociation rate in QGP
due to the smaller binding for $\psi'$. In the weak binding scenario a
significant fraction of charmonia is from regeneration, less $\psi'$
than $J/\psi$ are regenerated due to the lower dissociation
temperature of $\psi'$, $T^{\psi'}_{diss}\simeq T_c$
vs. $T^{J/\psi}_{diss}\sim 1.25T_c$, implying a later onset of
regeneration for $\psi'$.

\begin{figure}[tp]
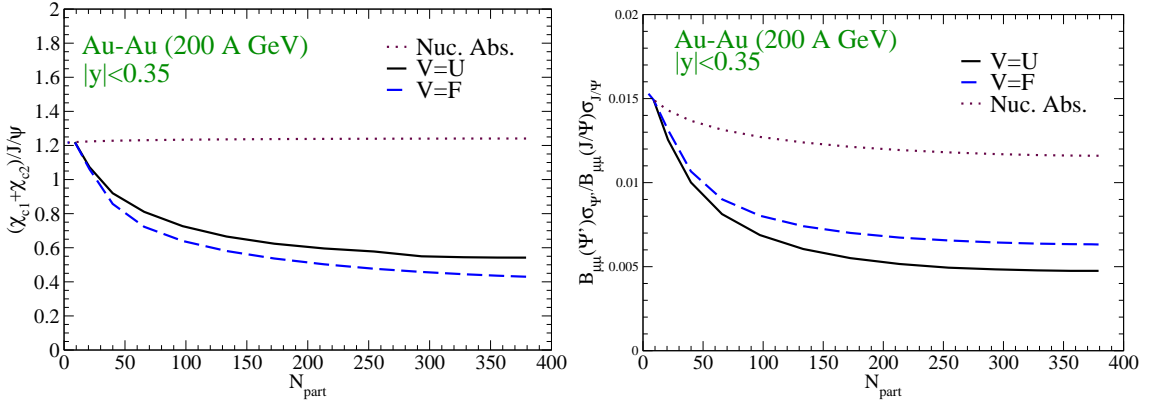

\centering
\includegraphics[width=0.49\textwidth]{chi_psi_ratio_rhic_0922.eps}
\includegraphics[width=0.49\textwidth]{psip_psi_ratio_rhic_0922.eps}
\caption[Inclusive $\psi'/(J/\psi)$ and
$(\chi_{c1}+\chi_{c2})/(J/\psi)$ ratio at RHIC evaluated with the
thermal rate-equation approach]{Inclusive
  $(\chi_{c1}+\chi_{c2})/(J/\psi)$ ratio (left panel) and
  $\psi'/(J/\psi)$ ratio (right panel) vs. centrality at mid-rapidity
  at RHIC evaluated with the thermal rate-equation approach. Solid
  line: strong binding scenario; dashed line: weak binding
  scenario. Dotted line: nuclear absorption only. The $J/\psi$'s in
  the denominator include feeddown from $\chi_c$ and $\psi'$.}
\label{fg:chi_c_psi}
\end{figure}

For $\Psi$ production in $\sqrt{s}$=200\,AGeV Au+Au collisions at
mid-rapidity at RHIC we use $\sigma^{J/\psi}_{\rm
  abs}$=$\sigma^{\chi_c}_{\rm abs}$=3.5\,mb and $\sigma_{\rm
  abs}^{\psi'}\simeq$6.5\,mb. The $\psi'$ to $J/\psi$ ratio and the
$\chi$ to $J/\psi$ ratio are displayed in Fig.~\ref{fg:chi_c_psi}. For
$\chi_c$ states, we constrain ourselves to $\chi_{c1}$ and $\chi_{c2}$
with a combined average branching ratio of 27\% into
$J/\psi$'s. 
Both drop with centrality below the ratios obtained from CNM-induced
suppression.


\section{Charmonium Production at FAIR}
\label{sec:psi_fair}

Finally we briefly discuss charmonium production at the forthcoming
FAIR accelerator. The typical collision energy for heavy nuclei is up
to $\sqrt{s}\sim$10\,AGeV. The medium created at FAIR is expected to
have lower initial temperature and larger baryon density than at SPS,
recall Figs.~\ref{fg:mub_vs_s} and \ref{fg:evo_traj}. In central
collisions QGP is still expected to form.  

We calculate charmonium production in $\sqrt s$=8.8\,AGeV Pb+Pb
collisions using the thermal rate equation approach. For the
charmonium production cross in p+p collisions we take $\dd
\sigma_{pp}^{\Psi}/dy$=2.4\,nb~\cite{Abt:2005qr}. We use $\sigma_{\rm
  abs}^{J/\psi}$=7.3\,mb~\cite{Arnaldi:2010ky} and $\sigma_{\rm
  abs}^{\psi'}$=13.0\,mb by assuming the nuclear absorption at FAIR is
the same as at SPS.
The input charm-quark cross section in p+p collisions is taken as $\dd
\sigma_{\bar cc}/\dd y$ ($y$=0)=0.2\,$\mu$b according to the
extrapolation in Ref.~\cite{Andronic:2007zu}.

\begin{figure}[tp]
  \centering
  \includegraphics[width=0.59\textwidth]{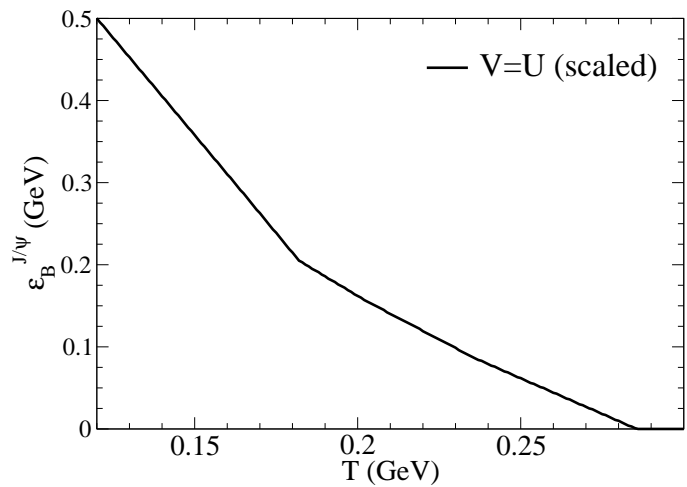}
  \caption[Temperature dependence of $J/\psi$ binding energy at FAIR
  in the strong binding scenario]{Temperature dependence of $J/\psi$
    binding energy at FAIR in the strong binding scenario. It is
    obtained by rescaling the $J/\psi$ binding energy at $\mu_B$=0
    according to Eq.~(\ref{eq:binding_fair}).}
\label{fg:binding_fair}
\end{figure}
\begin{figure}[hbp]
\centering
\includegraphics[width=0.59\textwidth]{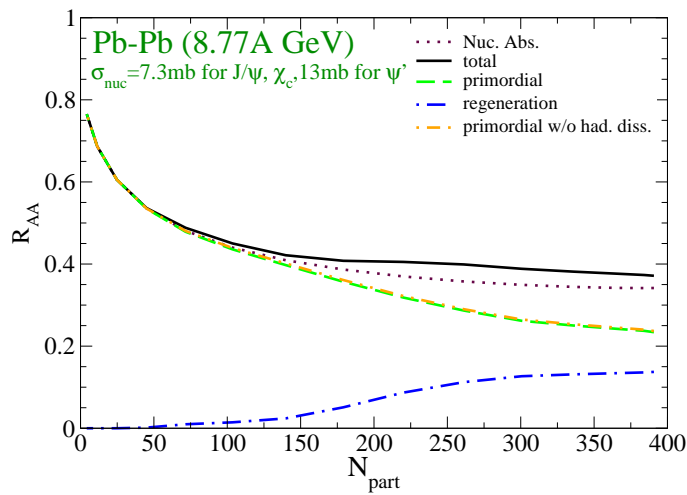}
\caption[Results of the thermal rate-equation approach for $R_{AA}^{J/\psi}$
  vs. centrality at FAIR]{Results of the thermal rate-equation approach for $R_{AA}^{J/\psi}$
  vs. centrality at FAIR. Solid line: total $J/\psi$ yield; dashed
  line: suppressed primordial production; dot-dashed line: regeneration component; dotted line:
  primordial production with nuclear absorption
  only; double-dash-dotted line: suppressed primordial production with
  quasifree suppression in the QGP only (without hadronic
  suppression).}
\label{fg:raa_centra}
\end{figure}
Since currently lQCD can only yield reliable results in the low baryon
density regime, we lack information on in-medium $\Psi$ properties from first
principle calculations for FAIR conditions. For exploratory
studies we obtain the in-medium $\Psi$ binding energies at FAIR by
rescaling those at RHIC energy according to the ratio between the
critical temperatures, $T^{\rm FAIR}_c$ and $T^{\rm RHIC}_c$, namely,
\begin{align}
\left .\epsilon_B(T)\right|_{\rm FAIR}=\left .\epsilon_B\left(\frac{T^{\rm
    RHIC}_c}{T^{\rm FAIR}_c}\,T\right)\right|_{\rm RHIC},
\label{eq:binding_fair}
\end{align} 
where $T^{\rm RHIC}_c$ is around 180\,MeV and $T^{\rm RHIC}_c$ is
around 135\,MeV, recall Fig.~\ref{fg:evo_traj}. In this dissertation
we restrict ourselves to the strong binding scenario. The resulting in-medium
binding energy for $J/\psi$ is illustrated in
Fig.~\ref{fg:binding_fair}.

We present the resulting centrality dependence of $J/\psi$ $R_{AA}$ in
Fig.~\ref{fg:raa_centra}.  For central collisions the regeneration
component takes up a significant fraction and compensates for the
anomalous suppression of the primordial component, rendering the total
yield even slightly above that resulting solely from CNM-induced
suppression.  Most of the suppression of the primordial component is
due to nuclear absorption. The majority of the anomalous suppression
is from partonic dissociation in the QGP. The suppression in hadronic
matter is negligible. Note however that the $SU$(4) effective theory
employed here for hadronic dissociation takes into account only
mesons-induced suppression, while the medium at FAIR energy is rather
baryon-dense. Therefore the hadronic suppression shown in
Fig.~\ref{fg:raa_centra} must be considered as a lower limit,
susceptible to significant corrections from inelastic $\Psi$
collisions with baryons as studied, \eg, in Ref.~\cite{Liu:2001ce}.


\begin{figure}[tp]
\centering
\includegraphics[width=0.59\textwidth]{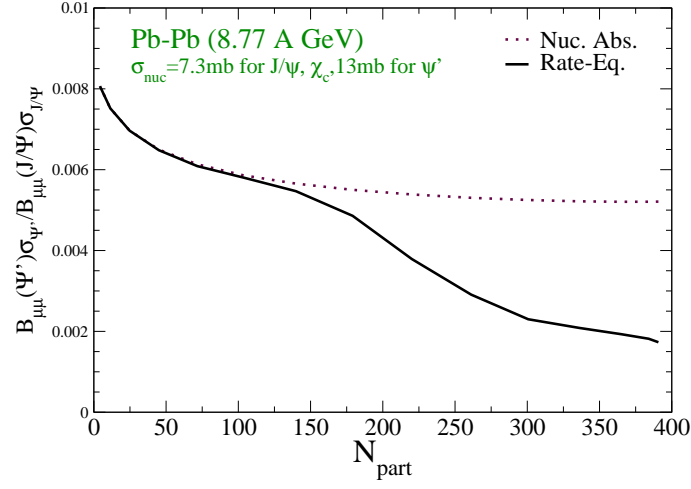}
\caption[Results of the thermal
  rate-equation approach for the $\psi'/(J/\psi)$ ratio vs. centrality
  at FAIR]{Results of the thermal
  rate-equation approach for the $\psi'/(J/\psi)$ ratio vs. centrality
  at FAIR. Solid line: full result; dotted line: primordial component
  with CNM effects only.}
\label{fg:psip_psi_ratio}
\end{figure}
\begin{figure}[hbp]
\centering
\includegraphics[width=0.59\textwidth]{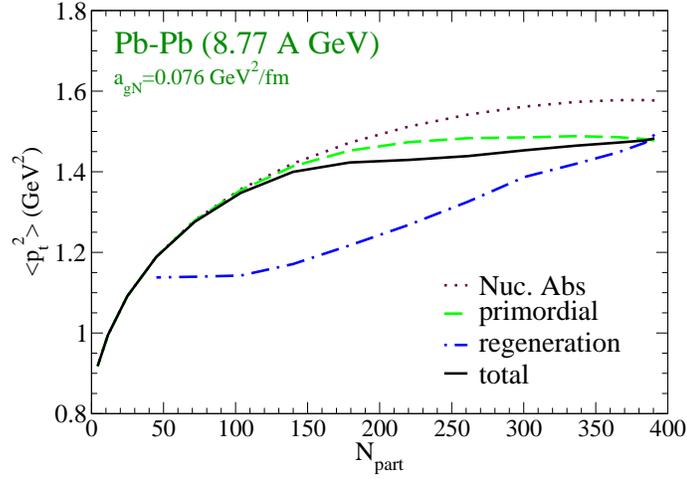}
\caption[Results of the thermal rate-equation approach for $\langle
  p_t^2 \rangle$ vs. centrality at FAIR]{Results of the thermal rate-equation approach for $\langle
  p_t^2 \rangle$ vs. centrality at FAIR. Solid line: total $J/\psi$
  yield. dashed line: suppressed primordial production; dot-dashed
  line: thermal regeneration; dotted line: primordial production with
  CNM effects only. }
\label{fg:pt2_centra}
\end{figure}
Next we evaluate the centrality dependence of the $\psi'$ to $J/\psi$ ratio.
In p+p collisions in the FAIR energy regime this ratio is
extrapolated to be around 0.01~\cite{Linnyk:2006ti} (The branching
ratios from $\psi'$ and $J/\psi$ into dimuons are included). The
results of the rate-equation approach, presented in
Fig.~\ref{fg:psip_psi_ratio}, show that the $\psi'$/($J/\psi$) drops
with centrality well below the ratios obtained from CNM-induced suppression
as a result of anomalous suppression.



\begin{figure}[tp]
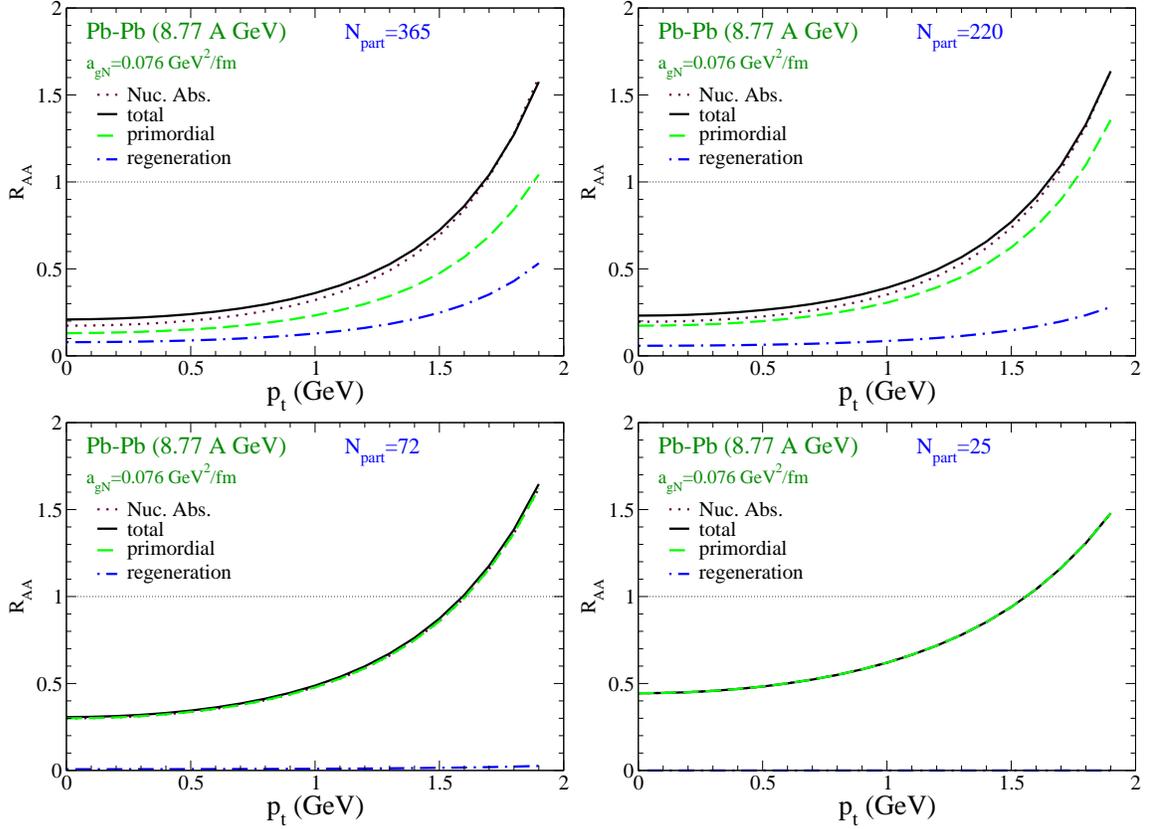

\centering
\includegraphics[width=0.49\textwidth]{raa_pt_CBM_b2_u_1017.eps}
\includegraphics[width=0.49\textwidth]{raa_pt_CBM_b6_u_1017.eps}
\includegraphics[width=0.49\textwidth]{raa_pt_CBM_b10_u_1017.eps}
\includegraphics[width=0.49\textwidth]{raa_pt_CBM_b12_u_1017.eps}
\caption[$R^{J/\psi}_{AA}$ vs. transverse momentum for different
centrality selections of Pb+Pb collisions at FAIR]{$R^{J/\psi}_{AA}$
  vs. transverse momentum for different centrality selections of Pb+Pb
  collisions at FAIR.  Solid line: total $J/\psi$ yield; dashed line:
  suppressed primordial production; dot-dashed line: thermal
  regeneration; dotted line: primordial production with CNM effects
  only.}
\label{fg:raa_pt}
\end{figure}

We finally turn to $J/\psi$ transverse momentum spectra.  The $\langle
p^2_t\rangle_{pp}$ in p+p collisions at FAIR energy regime is
extrapolated to be around 0.8\,GeV$^2$~\cite{Chaudhuri:2006zg}. We
assume a Cronin effect at FAIR similar to that of SPS, with
$a_{gN}$=0.076GeV$^2$/fm. The resulting centrality dependence of
$\langle p_t^2 \rangle$ is displayed in
Fig.~\ref{fg:pt2_centra}. Similar to SPS energies the $\langle p_t^2
\rangle$ for the primordial and regeneration component exhibits a
different centrality dependence, where the former is largely
determined by the Cronin effect. 
The anomalous suppression in QGP induces a small suppression of
$\langle p_t^2 \rangle$ due to the larger dissociation rate for $\Psi$
with higher momentum, recall the right panel of
Fig.~\ref{fg:rate_temp}. The $\langle p_t^2 \rangle$ for the
regeneration component increases with centrality due to the growing
collective flow.
The $p_t$ dependence of $R^{J/\psi}_{AA}$ for selected centralities is
summarized in Fig.~\ref{fg:raa_pt}. Again, the Cronin effect prevails
via the primordial component and the collective flow entails the
$R_{AA}$ of the regeneration component to increase with 
$p_t$.

\chapter{Conclusions and Outlook}
\label{ch:end}

Quantum Chromodynamics (QCD) has been established as the underlying
theory for the strong interaction decades ago, but several open
problems persist in the non-perturbative regime of the strong
interaction to date. One of the prominent problems is to understand
the phase structure of the QCD matter which ultimately consists of
quarks and gluons. At low temperatures quarks and gluons are confined
into hadrons, which become the effective degrees of freedom of the QCD
matter. First principle numerical calculations of discretized
(lattice) QCD predict that at a temperature of $\sim$170\,MeV
(10$^{12}K$), ordinary hadronic matter will undergo a transition into
a deconfined phase where quarks and gluons are the relevant degrees of
freedom, forming the so-called Quark-Gluon Plasma (QGP). One expects
to gain insights into the strong interaction and many-body dynamics of
QCD through investigating the properties of QGP. However, since QGP
exists only at very high temperature, its natural occurrence is
rare. Fortunately QGP can be created in present-day's laboratories
through ultra-relativistic heavy-ion collisions (URHICs). Since in
URHICs QGP exists for a very short time (several fm/$c$), its
properties can only be inferred from finally observed particles. Among
these particles the charmonia as the bound states of charm ($c$) and
anti-charm ($\bar{c}$) quarks turn out to be an excellent probe for
QGP in URHICs: The strong force binding $c$ and $\bar{c}$ in vacuum is
expected to be screened in QGP by surrounding colored quarks and
gluons. As a result, charmonium bound states are easier to dissociate
by collisions with particles in the medium, leading to a reduction of
its experimentally observed yield.


In order to utilize charmonium as a quantitative probe for the
properties of the medium created in URHICs, one needs a framework
associating experimental observables with theoretical calculations of
in-medium charmonium properties. In this work we have constructed
such a framework in which a Boltzmann transport equation is employed
to describe the time evolution of the charmonium phase space
distribution in URHICs. The in-medium charmonium properties figuring
into the transport equation are constrained by thermal lattice QCD
(lQCD): We have estimated the charmonium in-medium binding energies
from a potential model with input 2-body potentials inferred from the
free energy of a static pair of heavy quark and antiquark computed in
lQCD. Based on the obtained charmonium in-medium binding energies a
``quasifree'' approximation has been employed to evaluate the
charmonium dissociation and regeneration rates, which are the main
inputs to the transport equation. Finally, the consistency between the
charmonium binding energy and dissociation rates has been verified
with the quarkonium current-current correlation function, another
quantity calculated by lQCD independently from the heavy quark free
energy. In this way we have established a link between the
equilibrium properties of charmonia calculated from first principles,
but in euclidean spacetime, and the off-equilibrium evolution of
charmonium phase space distributions in heavy-ion collisions.

In the Boltzmann equation the dissociation and regeneration rates
play the role of the interface between the microscopic dynamics and
the macroscopic observables. In this dissertation we have worked out the momentum
dependence of the charmonium dissociation rate, which turns out to be
sensitive to the charmonium binding energies. 
We have also explicitly worked out the charmonium regeneration rates
based on the detailed balance between the dissociation and
regeneration processes, not restricted to equilibrium conditions. We
have found that the dominant regeneration process in QGP is a 3-to-2
process with the initial states consisting of $c$, $\bar c$ and a
light parton. In the ``quasifree'' approximation this 3-to-2 process
is found to be factorizable into a 2-to-2 scattering process and a
2-to-1 coalescence process. We have calculated the regeneration rates
with different input charm quark momentum spectra. We have found that,
the thermal charm-quark spectra turn out to be much more efficient in
regenerating charmonia than the charm-quark spectra from initial hard
collisions, due to a larger phase space overlap between $c$ and
$\bar{c}$ quarks. Both the inclusive yield of regenerated charmonia
and their $p_t$ spectra appear to be very sensitive to the level of
thermalization of the input charm quark spectra.

Within the current uncertainties from various inputs our results from
the transport equation agree reasonably well with the $J/\psi$
production data measured at the Super-Proton-Synchrotron (SPS) and the
Relativistic Heavy-Ion Collider (RHIC), thus corroborating the picture
of the deconfining phase transition as predicted by QCD. Moreover we
have demonstrated that the $J/\psi$ transverse momentum data exhibit
promise for discriminating power for different scenarios for the
heavy-quark potential as extracted from lQCD data: The scenario with
the internal energy (rather than the free energy) of the $c\bar{c}$
system identified as the potential is slightly favored, though further
theoretical studies and more precise experimental data are required to
draw definitive conclusions.

To reduce the current uncertainties of the theoretical approach,
further developments are in order. First, our calculations show that
in URHICs the partition between the primordial and regenerated
charmonia is sensitive to their in-medium binding energy, therefore a
reliable determination of the in-medium $c$-$\bar{c}$ 2-body potential
is needed for an accurate evaluation of charmonium in-medium binding
energy, and thus of the dissociation and regeneration rates.

Second, an explicit calculation of charmonium regeneration from
(time-dependent) charm-quark phase-space distributions in $c$-$\bar c$
recombination reactions should be performed. One should use realistic
charm-quark spectra as following, \eg, from Langevin
simulations with constraints from the $T$-matrix formalism and from
open-charm observables, to reduce the currently large uncertainty
in this part of the input.

Third, hydrodynamic simulation of the medium evolution could be 
employed for a more detailed and realistic description of 
the temperature and the flow field of the underlying medium, 
especially in coordinate space.

Fourth, a microscopic model for primordial $c\bar c$ and charmonium
production is warranted to better disentangle nuclear shadowing and
absorption in the pre-equilibrium stage, including formation-time
effects. This would improve the initial conditions for the transport
approach in the hot medium.

These developments will ultimately lead to a comprehensive approach
which can serve as a quantitative bridge between charmonium
phenomenology in heavy-ion collisions and theoretical studies of
charmonia in the QGP (and hadronic matter). It will enable us to
deduce from experimental observables effective degrees of freedom for
the hot and dense medium created at various collision energies,
improving our knowledge of the phase structure of hot and dense QCD
matter. Furthermore it will provide insights into basic properties of the
strong force in terms of color screening of Coulomb and confining
interactions. All these aspects of information will eventually
contribute to the establishment of a coherent picture of the strong
interaction in the non-perturbative regime.








\vita
{\vspace{2cm}
\begin{tabular}{@{}p{2in}p{3in}}
Name:             & Xingbo Zhao \\            
Address:          & Department of Physics and Astronomy, \\
                  & 12 Physics Hall, \\
                  & Ames, Iowa, 50011-3160 \\        
Email Address:    & {xbzhao@tamu.edu} \\       
Education:        & B.S., Physics, University of Science and Technology of China, 2005 \\         & Ph.D., Physics, Texas A\&M University, 2010 \\
\end{tabular}
}

\end{document}